%% file: main.tex
\documentclass[a4paper]{article}
\usepackage{amsmath}
\usepackage{amssymb}
\usepackage{paralist}
\usepackage{tikz}\usetikzlibrary{positioning,3d,shapes.geometric, arrows, fit, arrows.meta, backgrounds}
\usepackage[misc]{ifsym}
\usepackage{hyperref}
\usepackage{placeins}
\usepackage{subcaption}
\usepackage{authblk}
\usepackage[margin=3cm]{geometry}
\allowdisplaybreaks

\newcommand{\R}{\mathbb{R}}

\newcommand{\Bset}{\mathcal{B}}
\newcommand{\Cset}{\mathcal{C}}

\def \beq {\begin {eqnarray}}
\def \eeq {\end {eqnarray}}
\def \ba {\begin {eqnarray*}}
\def \ea {\end  {eqnarray*}}

\title{Learned enclosure method for experimental EIT data}
\author[1]{Sara~Sippola}
\author[1]{Siiri~Rautio}
\author[2,3]{Andreas~Hauptmann}
\author[4]{Takanori~Ide}
\author[1]{Samuli~Siltanen}

\affil[1]{\footnotesize Department of Mathematics and Statistics, University of Helsinki, Finland}
\affil[2]{Research Unit of Mathematical Sciences, University of Oulu, Finland}
\affil[3]{Department of Computer Science, University College London, UK}
\affil[4]{Department of Mathematics and Information Science, Faculty of Science, Josai University, Japan}

\date{}

\begin{document}
\maketitle

\begin{abstract}
Electrical impedance tomography (EIT) is a non-invasive imaging method with diverse applications, including medical imaging and non-destructive testing. The inverse problem of reconstructing internal electrical conductivity from boundary measurements is nonlinear and highly ill-posed, making it difficult to solve accurately. In recent years, there has been growing interest in combining analytical methods with machine learning to solve inverse problems. In this paper, we propose a method for estimating the convex hull of inclusions from boundary measurements by combining the enclosure method proposed by Ikehata with neural networks. We demonstrate its performance using experimental data. Compared to the classical enclosure method with least squares fitting, the learned hull achieves superior performance on both simulated and experimental data.
\end{abstract}

\section{Introduction}
In electrical impedance tomography (EIT), one feeds electric currents to a conductive body $\Omega$ through electrodes placed on the boundary $\partial \Omega$. Voltages at the electrodes are then measured, and the goal is to recover information about the conductivity distribution inside $\Omega$. EIT has applications in medical imaging \cite{isaacson2004reconstructions,zablah2025feasibility}, nondestructive testing \cite{smyl2018detection,tallman2020structural}, and process monitoring \cite{seppanen2001state}.  
The mathematical model of EIT was formulated in 1980 by Calder\'on \cite{Calderon1980}. Reconstruction methods have been thoroughly studied from both analytical and numerical perspectives, which has led to many variations of approaches with mathematically different characteristics \cite{mueller2012linear}. In particular, EIT reconstruction is a severely ill-posed inverse problem \cite{alessandrini1988stable}, making it extremely sensitive to modeling error and measurement noise. 

In this work, we study the enclosure method: an analytical EIT reconstruction method proposed by Ikehata \cite{Ikehata1999a,Ikehata2000c}. The enclosure method aims to reconstruct the \emph{convex hull} of inclusions in the target, based on asymptotic behavior encoded in an indicator function. Practical numerical implementations of the enclosure method \cite{ikehata2000numerical, ikehata2004electrical} are based on least squares fitting of lines to noise-robust values. We note that a related inclusion detection approach is the factorization method \cite{bruhl2001explicit,bruhl2000numerical}.

Lately, there has been an increasing interest in combining analytical methods and machine learning for ill-posed inverse problems such as EIT. For example, Tanyu et al. \cite{tanyu2023} investigated deep learning and analytical approaches, such as deep D--bar \cite{hamilton2018deep}, deep direct sampling \cite{guo2021construct}, and the dominant current learning approach \cite{wei2019dominant}. For the enclosure method, previous work \cite{siltanen2020electrical} proposed a reconstruction method combining the enclosure method with machine learning, where the least squares fitting is replaced with a shallow neural network. The method is demonstrated with simulated data in the continuum measurement model, and the network is able to clearly outperform the least squares approach.

This paper investigates the reconstruction method of combining the enclosure method and neural networks in EIT for experimental measurement data. We focus on reconstructing the convex hull of inclusions embedded in a constant background conductivity from boundary measurements. The previous work in \cite{siltanen2020electrical} considered the continuum model in EIT and simulated data. To show the validity of the proposed method, we demonstrate it with two-dimensional simulated data and experimental measurements from the open dataset published in \cite{hauptmann2017open}. The proposed neural network is able to recover the convex hull from noisy indicator function values significantly more accurately than the least squares approach used for comparison. Potential real-world applications for the proposed method could include skin cancer detection \cite{pathiraja2020clinical}, and detecting cracks in the interior of a conductive object or in conductive paint on the surface of a structure \cite{hauptmann2018revealing, Hallaji_2014}.

This paper is organized as follows. In Section 2, we describe the enclosure method for the continuum measurement model in EIT. Then, we describe connecting the continuum model to the complete electrode model in order to deal with two-dimensional experimental measurement data. In Section 3, we describe the generation of simulated training data and the architecture of our neural network. In Section 4, we present numerical results of the proposed method for both simulated and experimental data. Section 5 provides a discussion and conclusion.

\section{Enclosure method for EIT}

Electrical impedance tomography is a powerful non-destructive imaging method. Full reconstruction of a conductivity function ${\sigma:\Omega\rightarrow \R}$ from boundary measurements is a severely ill-posed problem \cite{mueller2012linear}. Therefore, it is interesting to study the extraction of partial information about the conductivity distribution $\sigma$. Consider conductivities of the form 
\begin{equation}\label{inclusion_cond}
      \sigma(x)=1+\chi_D(x) h(x),
\end{equation}
where the set $D\subset\Omega$ is called an {\it inclusion} and $\chi_D$ is the characteristic function of $D$. Here, the background conductivity is known to be constant 1, and the set $D$ may consist of several components.  The goal of inclusion detection is to recover the shape or size of $D$ approximately from EIT measurements. In this work, we let $\Omega\subset\R^2$ to be the unit disc. 

\begin{figure}[htbp]
\centerline{
\begin{picture}(280,215)(80,-8)
\put(10,-6){\includegraphics[height=7.1cm]{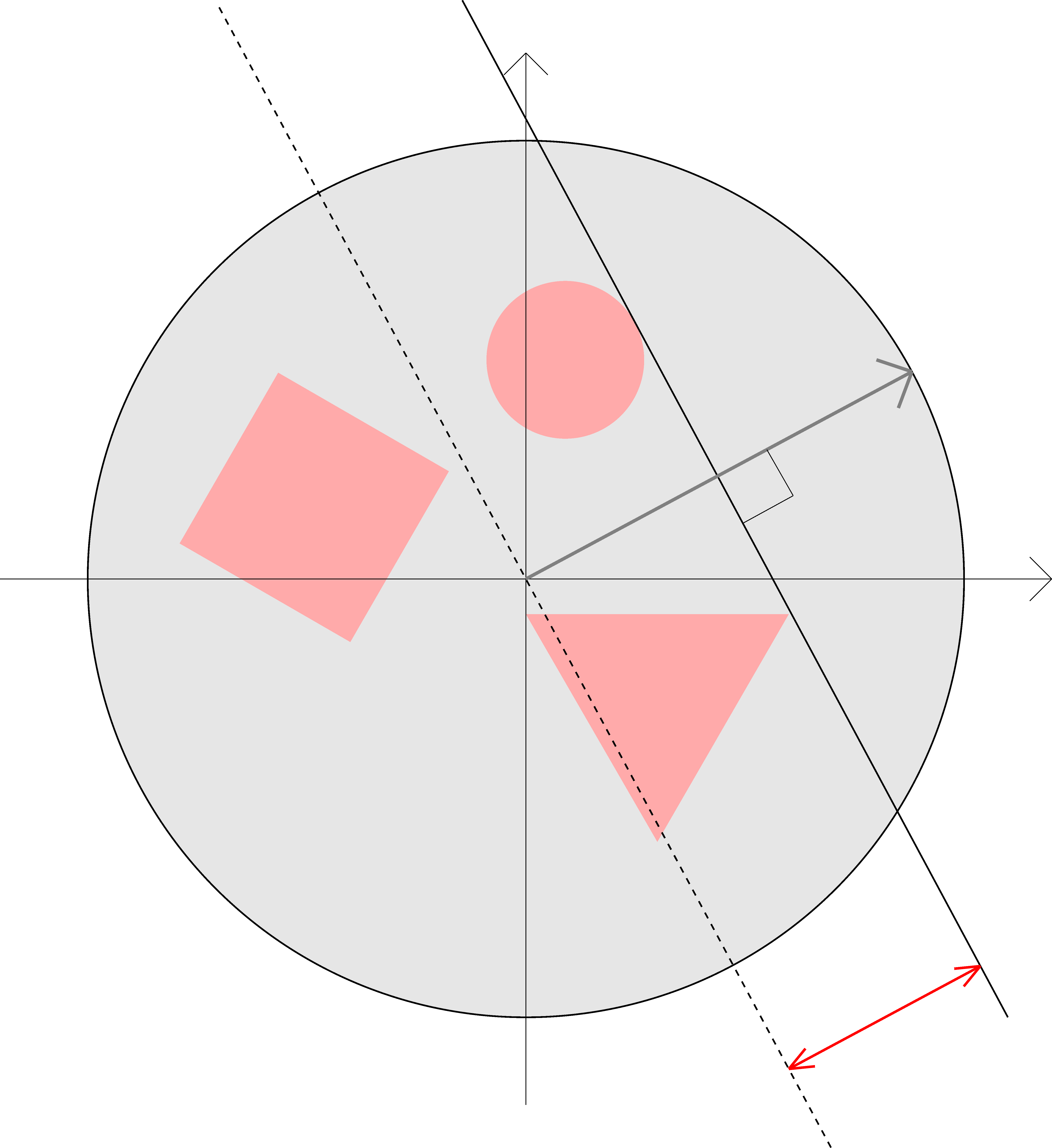}}
\put(164,10){\large \bf \rotatebox{-58}{{$h_D(\omega)$}} }
\put(65,50){\large \bf $\Omega$} 
\put(60,100){\large \bf $D$} 
\put(105,129){\large \bf $D$} 
\put(120,69){\large \bf $D$} 
\put(200,92){$x_1$} 
\put(99,191){$x_2$} 
\put(150,125){\Large \bf $\omega$} 
\put(260,10){\includegraphics[height=6cm]{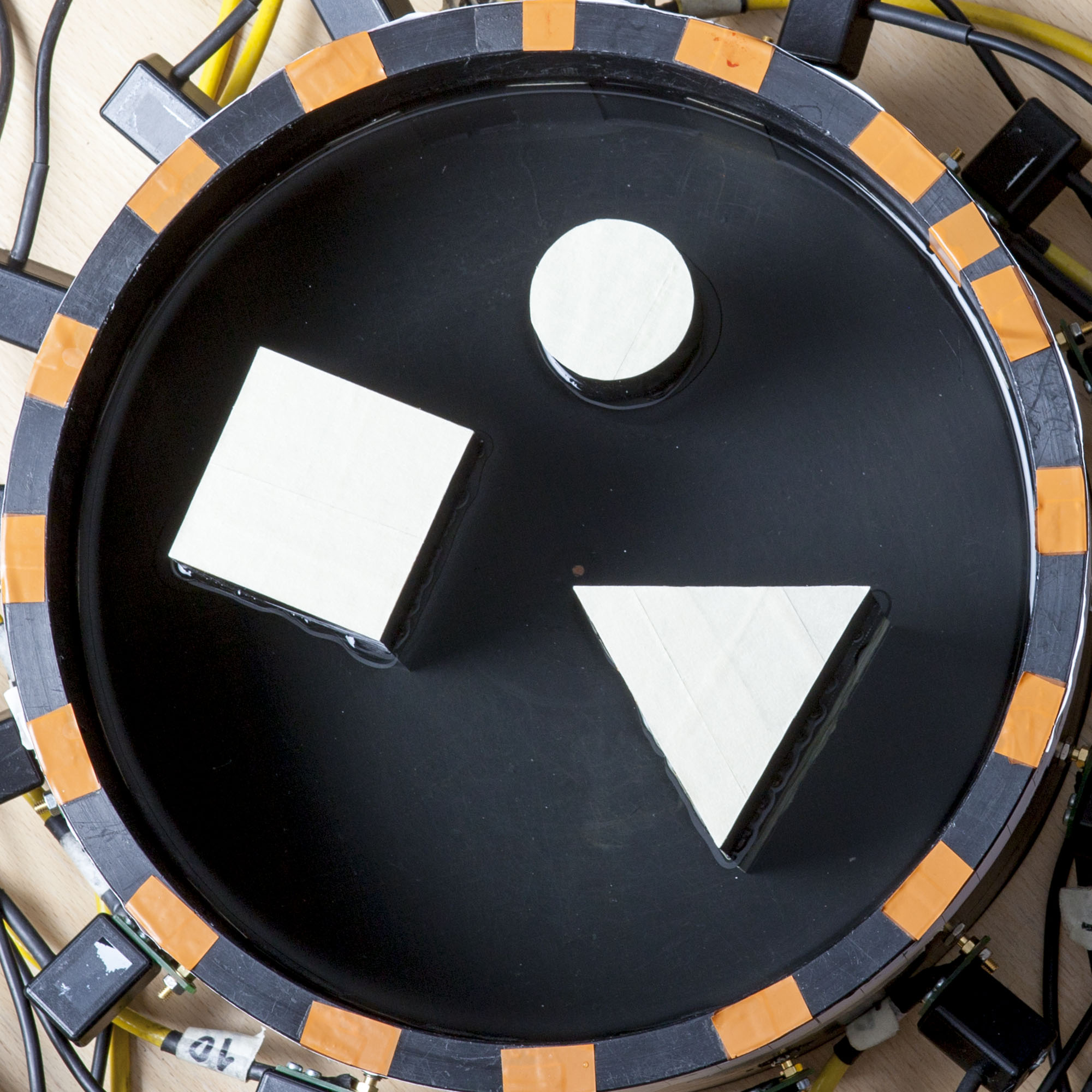}}
\end{picture}
}
\caption{The image on the right shows an experimental phantom with plastic shapes placed in saline (salt water). On the left is a simulated phantom resembling the experimental setting. The image shows the line determined by the support function ${{h_D(\omega)}:=\sup_{x\in D}x\cdot\omega}$ in the direction $\omega\in S^1$ for the inclusions $D\subset\Omega$, where $\Omega$ is the unit disk.}
\label{fig:suppfun}
\end{figure}

The {\it enclosure method } was introduced by Masaru Ikehata in \cite{Ikehata2000c}. Define the support function of the inclusion $D\subset \R^2$, with direction specified by a unit vector $\omega\in S^1$, by
\begin{equation}\label{suppfun}
  h_D(\omega) :=\sup_{x\in D} x\cdot \omega.
\end{equation}
See Figure \ref{fig:suppfun} for an illustration of the support function. 

The enclosure method specifies how to use EIT measurements to calculate a so-called {\it indicator function}  $I_\omega(\tau)$, defined for positive numbers $\tau>0$. By Theorem 2.1 of \cite{ikehata2000numerical}, we have
\begin{equation}\label{asymptotic}
h_D(\omega)=\lim_{\tau\longrightarrow\infty}\frac{\log |I_{\omega}(\tau)|}{2\tau}.
\end{equation}

We assume that $D$ is a union of finitely many components with piecewise smooth boundaries. In \cite{Ikehata2000c}, it is assumed that $h:\Omega\rightarrow \R$  satisfies the so-called {\it jump condition} along with other assumptions, ensuring that the conductivity $\sigma$ has a jump on the boundary $\partial D$. In practice, the jump condition means that the components of $D$ should be either all conductive or all resistive compared to the background conductivity. The theory does not cover cases where $D$ consists of both conductive and resistive inclusions, and using the enclosure method in such cases can lead to errors in the reconstructed convex hull \cite{siltanen2020electrical}.

\subsection{Least squares fit}

Fitting a line to the points
\begin{equation}\label{LSfit}
    \left(\tau,\frac{1}{2}\log|I_{\omega}(\tau)|\right) \in \R^2,
\end{equation}
in the least squares sense yields the so-called least squares (LS) fit. For small $\tau$ values, the slope of the least squares fit approximates the support function $h_D(\omega)$ in the direction $\omega$. 

As an example, consider $I_{\omega_i}(\tau_j)$ for
\begin{equation}\label{omegatau}
    \begin{array}{lc}
         \omega_i = [\cos{\theta_i},\sin{\theta_i}]^T,&  \\
         \theta_i = 2\pi(i-1)/M, & i=1,\dots,M, \\
         \tau_j = \frac{j}{2}, & j=1,\dots,N.
    \end{array}
\end{equation}

\begin{figure}[!b]
\centerline{
\begin{picture}(260,150)(80,-8)
\put(60,-6){\includegraphics[height=5cm]{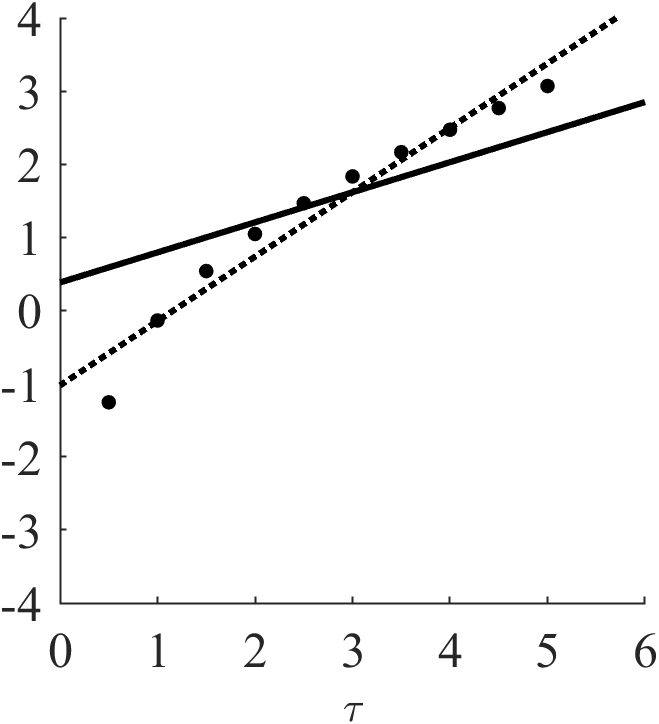}}

\put(85,55){$\frac{1}{2} \text{log} | I_{\omega_4} (\tau_j)|$} 

\put(140,90){\rotatebox{15}{$h_D(\omega_4)\tau + b$}}
\put(135,115){\rotatebox{35}{LS fit}}
\put(240,10){\includegraphics[height=4cm]{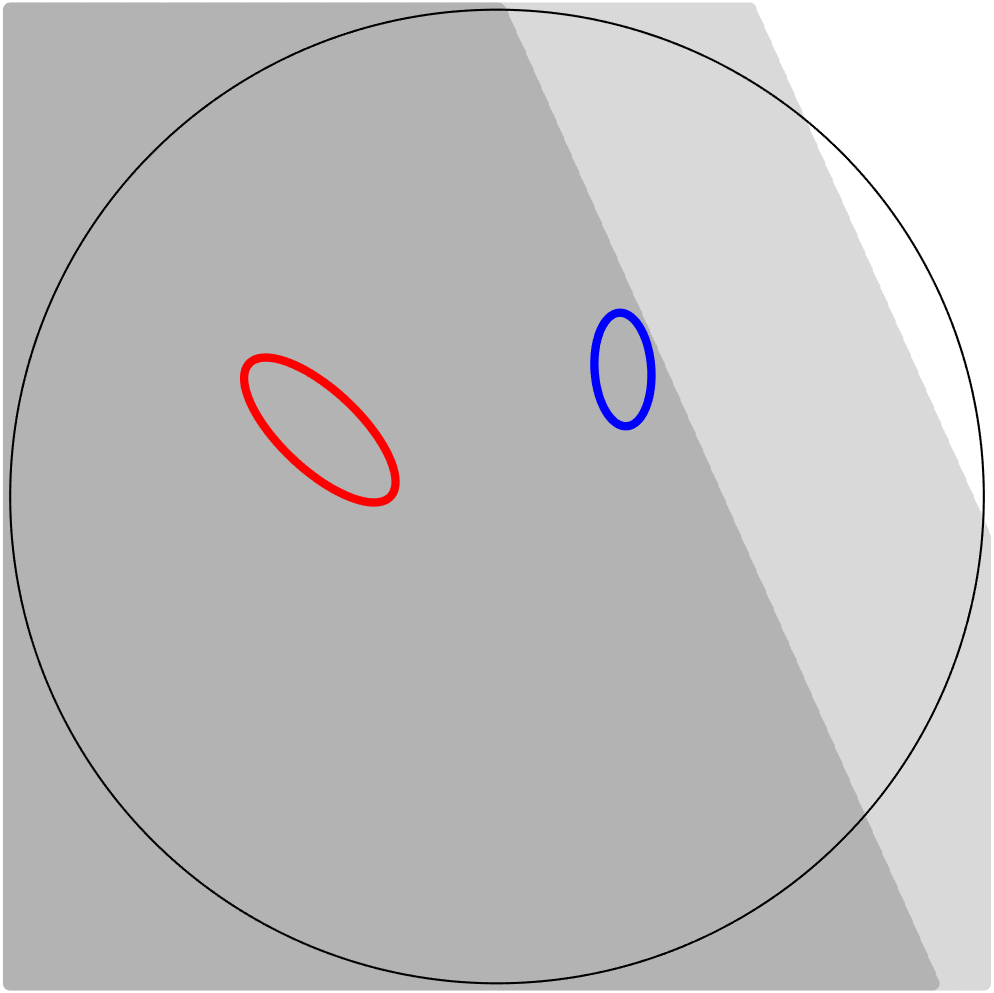}}
 \put(297,66){\vector(2,1){50}}
 \put(330,75){$\omega_4$} 
\end{picture}
}
\caption{An example of the slope of the least squares fit approximating the true support function of the phantom on the right in the direction $\omega_4$. The values of $\frac{1}{2} \text{log} | I_{\omega_4} (\tau_j)|$, $j=1,\dots,10$, are shown on the left as dots. The LS fit is plotted as a dotted and $h_D(\omega_4)\tau+b$ as a solid line. The true support function $h_D(\omega_4)$ corresponds to the distance of the probing half-plane from the origin, shown on the right in dark gray. The slope of the least squares fit approximates this distance but overestimates it, as can be seen from the light gray half-plane. To get the full convex hull of inclusions, this process is repeated for all directions $\omega_i$.}
\label{fig:LS_example}
\end{figure}

Set $M=45$ and $N=10$, meaning that the indicator function is computed over 45 directions uniformly distributed in the interval $[0,2\pi]$ and over ten $\tau$ values ranging from $0.5$ to $5$ with step size $0.5$. As an example, fixing $i=4$ allows us to compute $\frac{1}{2}\log|I_{\omega_4}(\tau_j)|$ for ten $\tau_j$ values and fit a line through the points. The slope of this line approximates $h_D(\omega_4)$. See Figure \ref{fig:LS_example} for an illustration.

\subsection{Continuum measurement model}

We follow the  mathematical formulation of Calder\'on \cite{Calderon1980} for building our EIT model. 
Applying voltage $f$ at the boundary leads to the following elliptic partial differential equation (PDE):
\begin{equation}\label{condeq}
\left\{
\begin{array}{rcl}
\nabla\cdot\sigma\nabla u &=& 0 \mbox{ in }\Omega,\\
u|_{\partial\Omega} &=& f.
\end{array}
\right.
\end{equation}
Here, $u$ is the electric voltage potential and $-\sigma\nabla u$ is the electric current vector field that has no sinks or sources inside $\Omega$. The quantity $\sigma\partial u/\partial \vec{n}|_{\partial\Omega}=\sigma\vec{n}\cdot \nabla u|_{\partial\Omega}$ is the electric current density along the boundary, where $\vec{n}$ is the outward normal vector field on $\partial\Omega$.
The {\it  Dirichlet-to-Neumann (DN) map}
\begin{equation}\label{DNmap}
  \Lambda_\sigma: f\mapsto \sigma\frac{\partial u}{\partial \vec{n}}|_{\partial\Omega}
\end{equation}
is an ideal infinite-precision model for the set of all possible static voltage-to-current boundary measurements. 

Let $\omega$ be a unit vector, and denote by $\omega^\perp$ the vector $\omega$ rotated counterclockwise by angle $\pi/2$.
Then, $  \omega\cdot\omega^\perp=0.$
For each $\tau>0$ and $x\in\R^2$, define an exponential function
\begin{equation}\label{calderon}
  f_\omega(x;\tau):=e^{\tau x\cdot\omega+i\tau x\cdot\omega^\perp}.
\end{equation}
The indicator function $I_\omega(\tau)$ is defined by
\begin{equation}\label{indicator}
I_\omega(\tau)= \int_{\partial\Omega} f_\omega(\,\cdot\,;\tau) \Big( ( \Lambda_\sigma-\Lambda_1)\overline{f_\omega(\,\cdot\,;\tau)}\Big) dS,
\end{equation}
where $dS$ is the Lebesgue measure on the boundary.

For numerical computations, we need to represent the operators $\Lambda_\sigma$ and $\Lambda_1$ as matrices. To that end, we define the trigonometric basis functions 
\begin{equation}\label{Fbasis_cont}
\phi_n(\theta)= 
\left \{ \begin{array}{lc}
\displaystyle \pi^{-1/2}\cos\left((n+1)\theta/2\right), \quad \mbox{ for odd }n, \\
\displaystyle \pi^{-1/2}\sin\left(n\theta/2\right), \quad \mbox{ for even } n.
\end{array}\right.     
\end{equation}
Also, we need to use the constant basis function 
\begin{equation}\label{Fbasis_cont_constant}
\phi_0=(2\pi)^{-1/2}
\end{equation}
Note that the functions (\ref{Fbasis_cont}) and (\ref{Fbasis_cont_constant}) form an orthonormal basis for $L^2(\partial\Omega)$.
We can now approximate functions using a truncated Fourier series:
\begin{equation}\label{Fseries_truncated}
    f(\theta) \approx \sum_{n=0}^N \langle f,\phi_n\rangle \phi_n(\theta),
\end{equation}
using the $L^2(\partial\Omega)$ inner product
$
  \langle v,w \rangle = \int_0^{2\pi} v(\theta)\overline{w(\theta)}\,d\theta.
$

Also, $\Lambda_\sigma$ can be represented approximately as the $(N+1)\times (N+1)$ matrix $L_\sigma$ defined as follows:
\begin{equation}\label{Lsigma_matrix}
    [L_\sigma]_{k\ell} = \langle  \phi_k ,\Lambda_\sigma \phi_\ell\rangle.
\end{equation}
Here, $k$ is the row index and $\ell$ is the column index. 

\subsection{Complete electrode model}

We describe the complete electrode model (CEM) \cite{Cheney1999,somersalo1992existence}.
Inside $\Omega$, we have $\nabla\cdot\sigma\nabla u=0$. Applied current is zero off the electrodes, and on the $\ell$th electrode $e_\ell$ the current is given by
\begin{equation}
    I_\ell = \int_{e_\ell}\sigma\frac{\partial u}{\partial \nu} ds, \quad \ell=1,2,\ldots, L.
\end{equation}
Conservation of charge dictates that the current pattern has $\sum_{\ell=1}^L I_\ell = 0$. Measured voltages $V_\ell$ are constant on electrodes:
\begin{equation}
    V_\ell = u(z) \quad\mbox{for} ~ z  ~\mbox{ on}~ e_\ell,  \quad \ell=1,2,\ldots,L.
\end{equation}
The CEM problem has a unique solution with Robin boundary condition containing contact impedances $\zeta_\ell$:
\begin{equation}
    u+\zeta_\ell\,\sigma\frac{\partial u}{\partial\nu} = V_\ell \quad \mbox{on} \quad e_\ell, \quad \ell=1,2,\ldots, L,
\end{equation}
and requirement
\begin{equation}
    \sum_{\ell=1}^L V_\ell = 0.
\end{equation}

Given a trigonometric current pattern $I^n\in\R^L$, we solve CEM for voltage vector $V^n\in\R^L$. Let's denote this process with a linear operator $\mathcal{R}_\sigma^{\mbox{\tiny CEM}}$:
\begin{equation}
    V^n := \mathcal{R}_\sigma^{\mbox{\tiny CEM}} I^n.
\end{equation}
Collect $L-2$ measurements together as columns in a matrix $V$:
\begin{equation}
    V = 
    \left[\!\!
    \begin{array}{llcl}
    V^1_1 & V^2_1 & \dots & V^{(L-2)}_1\\
    V^1_2 & V^2_2 & \dots & V^{(L-2)}_2\\
    \vdots & \vdots & \ddots & \vdots\\
    V^1_L & V^2_L & \dots & V^{(L-2)}_L\\
    \end{array}
    \!\!\right]
    =
    \mathcal{R}_\sigma^{\mbox{\tiny CEM}}
    \left[\!\!
    \begin{array}{llcl}
    I^1_1 & I^2_1 & \dots & I^{(L-2)}_1\\
    I^1_2 & I^2_2 & \dots & I^{(L-2)}_2\\
    \vdots & \vdots & \ddots & \vdots\\
    I^1_L & I^2_L & \dots & I^{(L-2)}_L\\
    \end{array}
    \!\!\right]
\end{equation}

Note: in anticipation of matching CEM measurements with continuum model measurements, we omit the $(L-1)$th current pattern  $I^{(L-1)}_\ell =\cos\left(L\theta_\ell/2\right)$.

Assume we have $L$ electrodes on the unit circle with centers at angles $\theta_\ell:=(\ell-1)2\pi/L$ for $\ell=1,\dots,L$. Analogously to the trigonometric basis of the continuum model, we work with trigonometric current patterns $I^n\in\R^L$, where $n=1,2,\dots,L-1$:
\begin{equation}
    I^n_\ell= 
    \left \{ \begin{array}{lc}
    \displaystyle \cos\left((n+1)\theta_\ell/2\right), \quad \mbox{ for odd }n, \\
    \displaystyle \sin\left(n\theta_\ell/2\right), \quad \mbox{ for even } n.
    \end{array}\right.
\end{equation}
Note that $n=L$ is excluded as $\sin\left(L\theta_\ell/2\right)=0$ for all $1\leq\ell\leq L$, and therefore $I^L$ would be the zero vector. Indeed, there can be only $L-1$ linearly independent nontrivial zero-mean vectors in $\R^L$.

\subsection{Obtaining the DN map from adjacent measurements}

In the following, we shortly discuss how to transform real-world data measurements into a discrete DN map $\boldsymbol{\Lambda}_\sigma$ used for the enclosure method. 
The experimental system only allows for pair-wise injection patterns, and in particular, we will use the adjacent current pattern as reported in \cite{hauptmann2017open}.
Thus, the first task is to transform the adjacent current patterns into the trigonometric basis from Equation \eqref{Fbasis_cont}.

We achieve this by solving a simple linear equation. Given the experimental measured adjacent voltages $V_{ad}$ and a coefficient transform matrix $C$ we obtain the desired trigonometric voltages $V_{tri}$ by solving 
\[
C V_{tri} = V_{ad}. 
\]
The basis transform matrix is defined element-wise as
\[
C_{i,j}= \langle c^{tri}_i , c^{ad}_j \rangle,
\]
where $c^{ad}_j$ is the $j$-th adjacent current pattern and $c^{tri}_i$ is the $i$-th basis from \eqref{Fbasis_cont}. The transformed measurements can then be used to construct the DN matrix in the trigonometric basis.

We note that a more advanced conversion has been developed in \cite{garde2021mimicking}, which could be further used to reduce the gap between CEM and continuum model, as done in \cite{alsaker2024ct}.

\section{Neural network model}
The aim is to learn the support function $h_D(\omega)$ from the  points in (\ref{LSfit}) using a convolutional neural network, and to recover the convex hull of inclusions based on this.

\subsection{Training data}
We simulated 10 000 training data phantoms consisting of the unit disk with 1 to 3 disjoint elliptical inclusions. The distance between inclusions is at least 0.05 and the distance between each inclusion and the unit circle is at least 0.1, but otherwise the placement of the inclusions is random. Each elliptical inclusion has a random orientation and the length of the semi-major axis $R$ is uniformly distributed in the interval $[0.05,0.3]$, and the length of the semi-minor axis is uniformly distributed in the interval $[0.05,R]$. The conductivity values of the inclusions were chosen to be either more conductive or resistive than the constant background conductivity value 1, with probability $1/2$. The specific conductive and resistive values are uniformly distributed in the interval $[2,4]$ for conductive inclusions and $[0.2,0.8]$ for resistive inclusions. Notably, the training data set contains both phantoms that satisfy the jump condition and phantoms that don't.

After generating the conductivity phantoms, we simulated the DN map $\boldsymbol{\Lambda}_\sigma$ for each phantom and the unit disk without inclusions by solving the forward EIT problem as explained in \cite[Section 13.2]{mueller2012linear}, using the continuum model and finite element method in 2D. We used a finite element mesh with 33025 points and 65536 triangles with the maximum edge length 0.0123, providing solutions with 6 significant digits of accuracy.  We added white Gaussian noise of magnitude $10^{-4.5}$ to the resulting {\it  Neumann-to-Dirichlet (ND)} matrices. The magnitude of the noise was chosen so that significant information in the ND matrices is not lost. We then inverted the matrices to obtain approximations of the DN maps and used equation (\ref{indicator}) to compute the indicator functions $I_{\omega_i}(\tau_j)$ for $\omega_i$ and $\tau_j$ defined as in the example (\ref{omegatau}). Although $\tau$ is taken to infinity in equation (\ref{asymptotic}) to reveal the support function, severely restricting $\tau$ is necessary in practical settings because measurement noise makes the indicator function unstable with higher $\tau$ values \cite{siltanen2020electrical}. 

In accordance with the least squares fit, the values of the indicator functions were used to compute the points
\begin{equation}
    \frac{1}{2} \log|I_{\omega_i}(\tau_j)|
\end{equation}
for each $\omega_i$ and $\tau_j$. The resulting values were gathered into a $10\times 45$ matrix with rows corresponding to the $\tau$ values and columns corresponding to the 45 directions $\omega$. To accentuate the periodic nature of the indicator function, we followed a similar padding approach as in \cite{siltanen2020electrical}, so that each indicator function matrix is reflect-padded to size $30\times50$. 

\begin{figure}[!b]
    \centering
    \begin{subfigure}{0.29\textwidth}
        \centering
        \includegraphics[height=2.6cm]{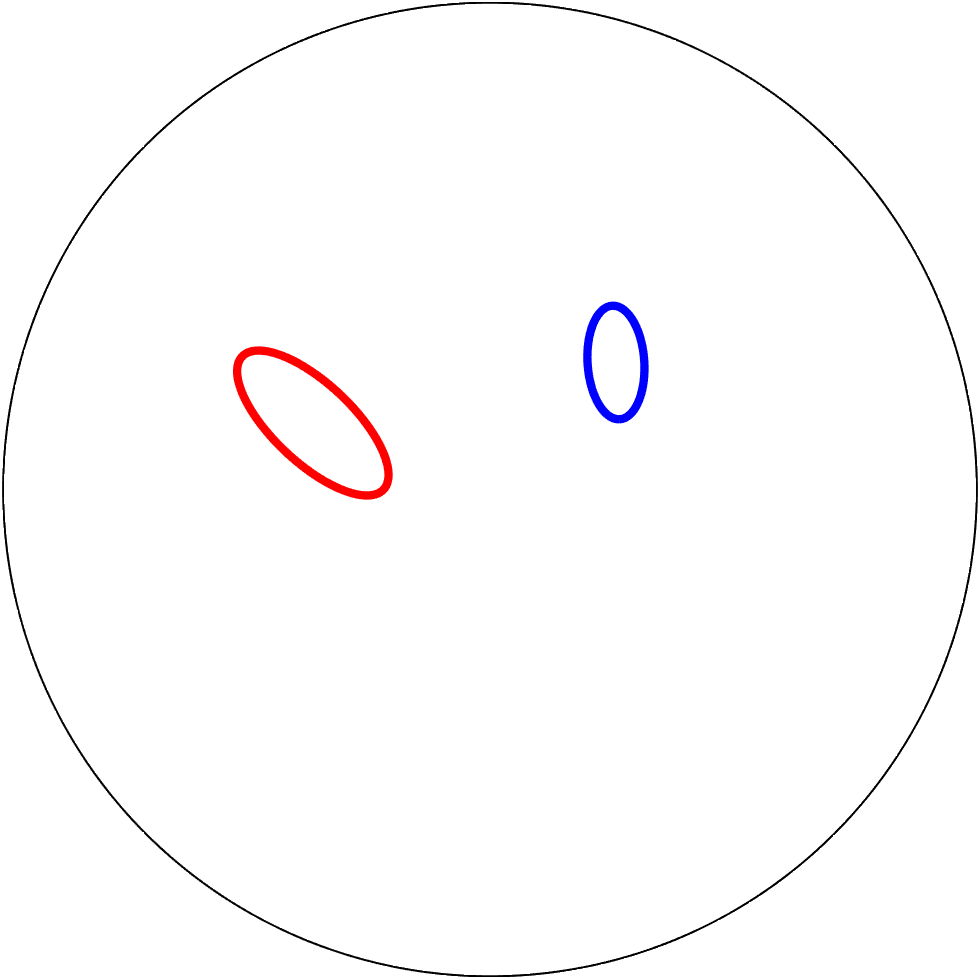}
        \caption{}
        \label{fig:training_conductivity}
    \end{subfigure}
    \hfill
    \begin{subfigure}{0.4\textwidth}
        \includegraphics[width=6cm]{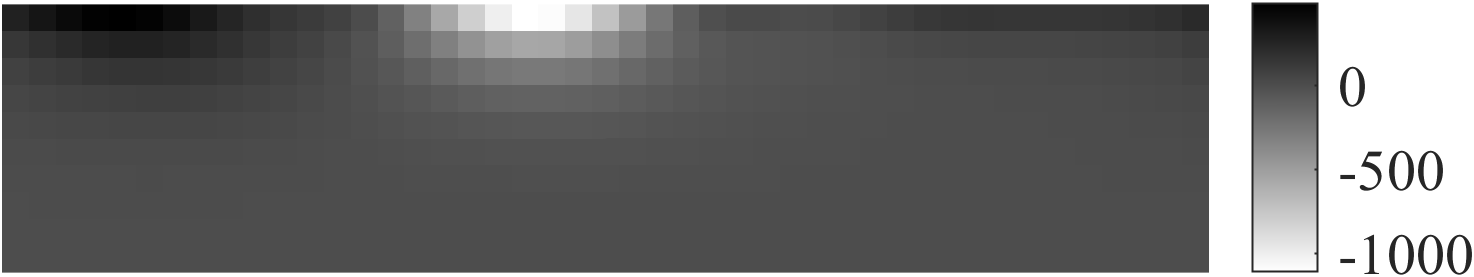} \\[0.1cm]
        \includegraphics[width=5.7cm]{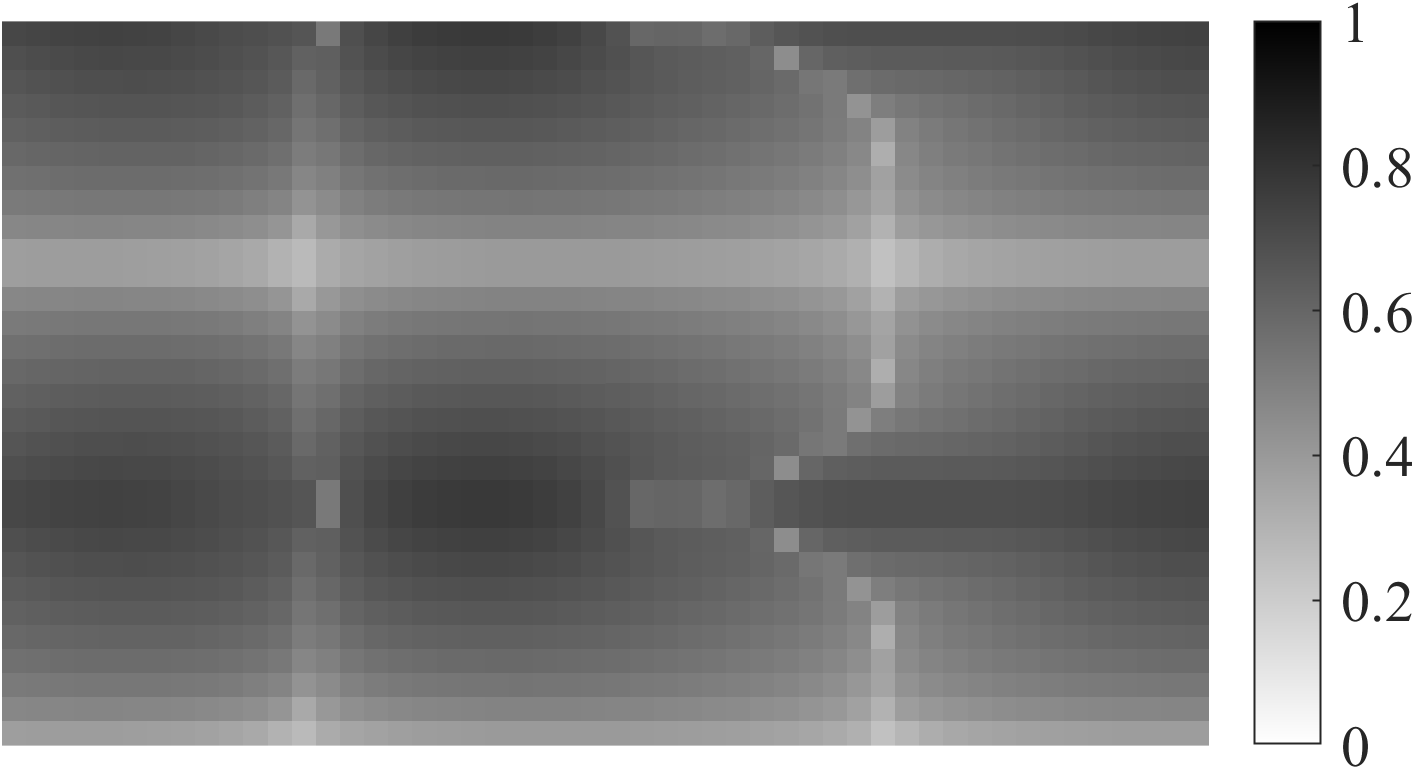}
        \caption{}
    \end{subfigure}
    \hfill
    \begin{subfigure}{0.29\textwidth}
        \centering
        \includegraphics[height=4.45cm]{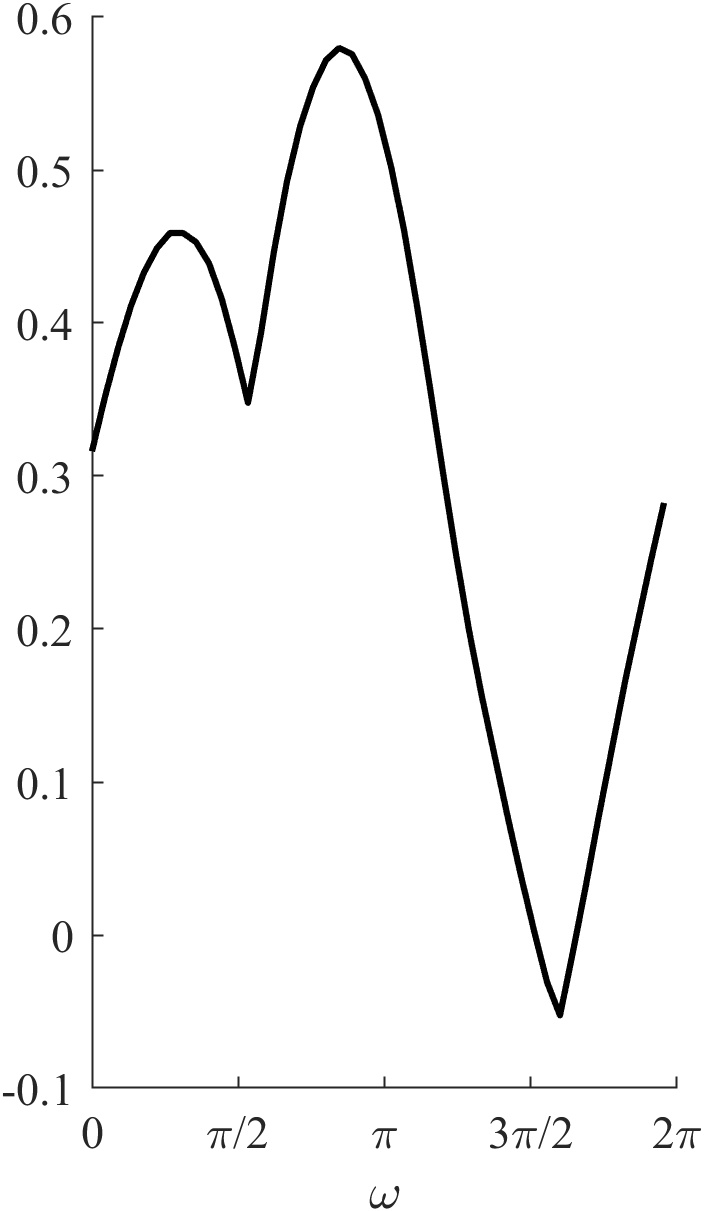}
        \caption{}
    \end{subfigure}
    \caption{An example of the indicator function matrix (B, top) and the input (B, bottom) given to the neural network for one conductivity phantom (A) from the training data set. The output (C) is a vector, plotted here as a line over $\omega$, where each element corresponds to the true support function in the direction $\omega_i$.}
    \label{fig:training_input_output}
\end{figure}

We used the same 45 directions $\omega_j$ to compute the 45 true support functions directly from each conductivity phantom according to equation (\ref{suppfun}), resulting in a vector of size $45\times 1$. We refer to this vector as the true support vector. 

The input given to the neural network is the set of 10000 matrices of size $30\times50$, and the output is the corresponding set of 10000 true support vectors. For an illustration of the indicator matrix, input and output for one conductivity phantom, see Figure \ref{fig:training_input_output}. We split the data $80/20$ into training and validation sets.

\subsection{Network architecture}
We use a shallow feed-forward convolutional network implemented with MATLAB's Deep Learning Toolbox to learn the support vectors based on the inputs. See Figure \ref{fig:nn_arch} for an illustration of the network architecture. In the convolutional layers, we use a three-pixel zero padding. The network was trained using the half-mean-squared-error loss function for 200 epochs. The optimization method was Stochastic gradient descent with momentum of $0.9$, with an initial learning rate of $0.01$, which was reduced every 5 epochs by a factor of $0.2$. 

The benefit of using a shallow network is that it is inexpensive to train. The training set can be smaller as the number of learnable parameters is low. If the method works already with a smaller network, it is unnecessary to train a larger model to solve the same problem. In this work, we also tested architectures with more convolutional layers, but this did not significantly improve the overall results on the experimental test set.

\begin{figure}[!t]
\input{NN_arch}
\label{fig:nn_arch}
\end{figure}

\section{Results}
We test the proposed method of learned hulls with both simulated and experimental data. We use the above neural network model to learn the support vectors based on the inputs, and compute the learned hull of the inclusions from each learned support vector.

For experimental data, we created phantoms by segmenting the objects in the photographs of the measurements available in \cite{hauptmann2017open}. We computed the true support functions for both simulated and experimental data directly from the simulated or segmented conductivity phantoms, and used the true support functions to compute the true convex hulls. We use the true convex hulls as a ground truths.

As demonstrated in Figure \ref{fig:LS_example}, the least squares fit (\ref{LSfit}) can be used to compute the least squares hull, a rough approximation of the true convex hull. In order to assess the performance of the neural network, we compare the errors of the least squares hulls and learned hulls relative to the ground truths.

\begin{figure} [!ht]
\begin{picture}(320,100)
 \put(50,10){\includegraphics[height=3cm]{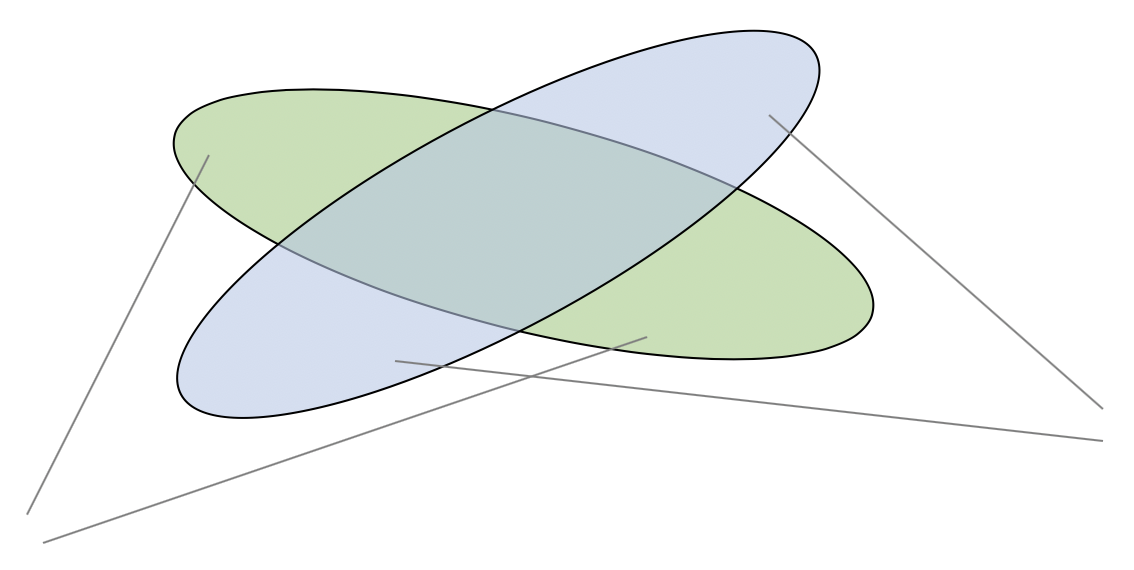}}
 \put(85,40){$\Bset$}
 \put(90,70){$\Cset$}
 \put(30,5){$ \Cset\setminus\Bset$}
 \put(57,0){{(false negatives)}}
 \put(218,27){$ \Bset\setminus\Cset$}
 \put(245,22){{(false positives)}}
\end{picture}
\caption{\label{fig:reöerr}We measure the quality of reconstructions of convex hulls using formula (\ref{relerr}). This picture illustrates how the error can be seen as a sum of false negatives and false positives.}
\end{figure}

Denote the true convex hull by $\Cset\subset\R^2$ and the reconstructed convex hull by $\Bset\subset\R^2$. To measure the quality of $\Bset$ as an approximation to $\Cset$ quantitatively, we use the relative error formula
\begin{equation}\label{relerr}
  \frac{|\Cset\setminus\Bset| +   |\Bset\setminus\Cset| }{|\Omega|} \cdot 100\%,
\end{equation}
where $|\,\cdot\,|$ denotes the area of a planar set. Here $|\Omega|=\pi$ since we work with the unit disc. Figure \ref{fig:reöerr} illustrates our choice of measure for reconstruction quality.

\subsection{Simulated data} We simulated 1000 phantoms consisting of the unit disk with 1 to 3 elliptical, triangular, or convex quadrilateral inclusions. The random properties of the inclusions are the same as in the training data set, except for shape and size. Each inclusion is independently chosen to be elliptical, triangular, or quadrilateral with probability $1/3$. The size of the elliptical inclusions is the same as in the training data set, but the triangular and quadrilateral inclusions are inscribed in a circle with a radius $r$ uniformly distributed in the interval $[0.15,0.3]$, meaning that very small triangles and quadrilaterals are omitted. To avoid triangular and quadrilateral inclusions that are too narrow, the distance between the vertices is at least $r$. The indicator function matrices, input matrices and true support vectors were formed in the same way as in the training data, and with the same amount of noise. 

Histogram comparisons of the relative errors of the least squares solution versus the proposed method are presented in Figure \ref{fig:relerr_histograms}, and Figure \ref{fig:simulated_examples} shows examples of the simulated data results. The example cases were chosen to demonstrate the performance of the method in a diverse set of conductivity phantoms.

\begin{figure} [!ht]
  \begin{subfigure}{0.99\textwidth}
    \includegraphics[width=0.95\linewidth]{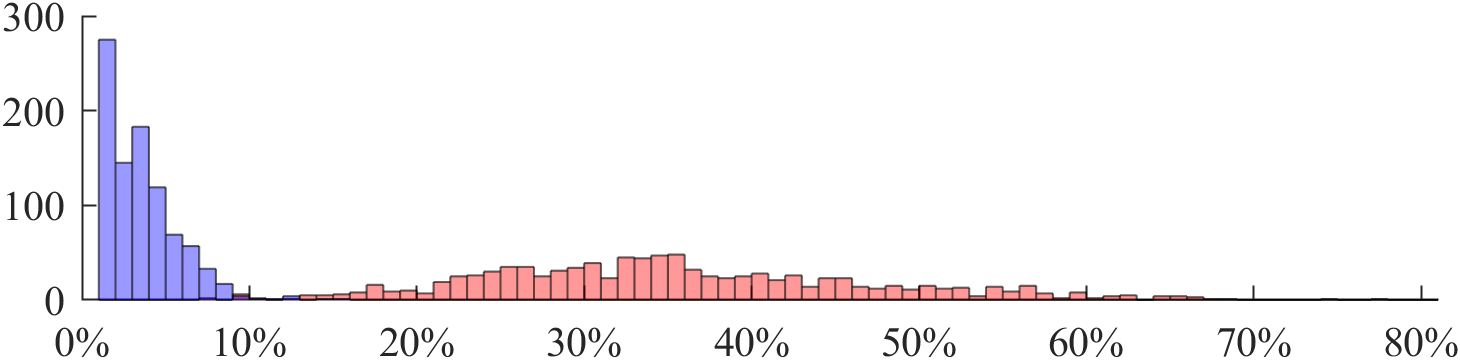}
    \centering
    \caption{Convex hull relative errors}
    \label{fig:relerr_histograms_a}
  \end{subfigure}
  \newline
  \begin{subfigure}{0.99\textwidth}
    \includegraphics[width=0.95\linewidth]{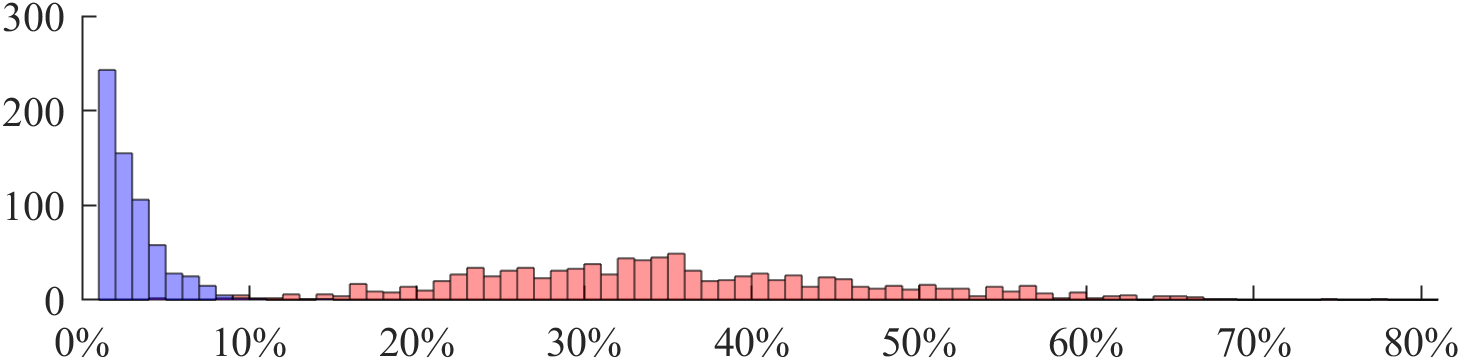}
    \centering
    \caption{False positives }
    \label{fig:relerr_histograms_b}
  \end{subfigure}
  \newline
  \begin{subfigure}{0.99\textwidth}
    \includegraphics[width=0.95\linewidth]{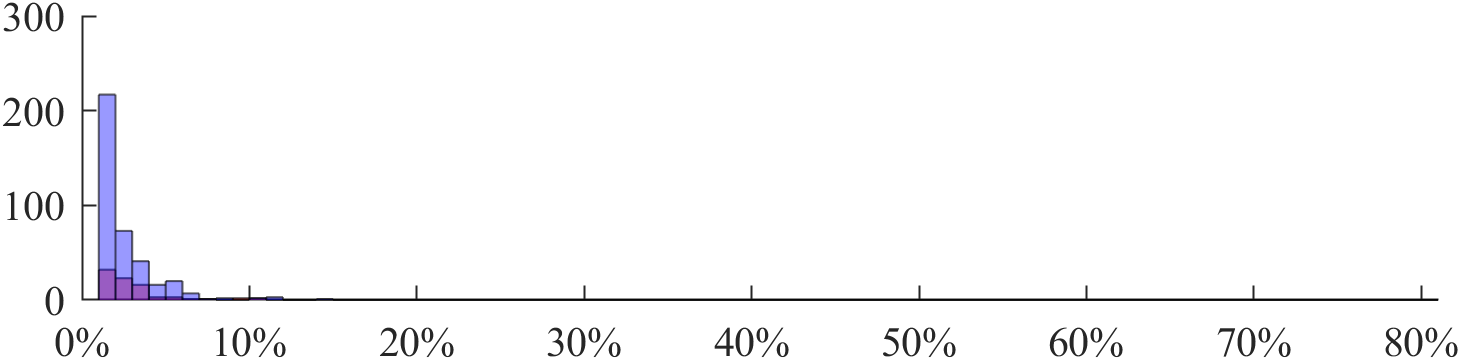}
    \centering
    \caption{False negatives}
    \label{fig:relerr_histograms_c}
  \end{subfigure}
  \caption{Histograms showing the distribution of the relative errors of the learned hulls (blue) and least squares hulls (red) in the simulated testing data set. The horizontal axis represents the relative error (\ref{relerr}) and the vertical axis represents the frequency. Most of the learned hulls have a relative error that falls under 10\%, whereas the relative errors of the least squares hulls range from 7\% to 78\%. False positives account for most of the errors, meaning that both approaches tend to overestimate the hull.}
    \label{fig:relerr_histograms} 
\end{figure}

\begin{figure}[!ht]
    \begin{tabular}[t]{lccccc}
        Case & Support vectors & Ground truth & Learned & Least squares \\[.2cm]
        I
        & \begin{minipage}{0.25\textwidth}
        \centering \includegraphics[height=3cm]{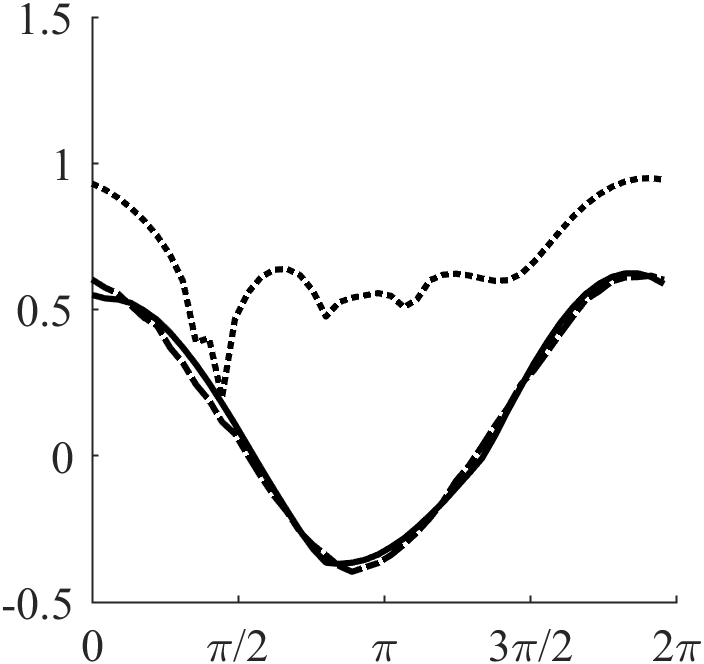} 
        \end{minipage}
        & \begin{minipage}{0.18\textwidth}
        \centering \includegraphics[height=2.3cm]{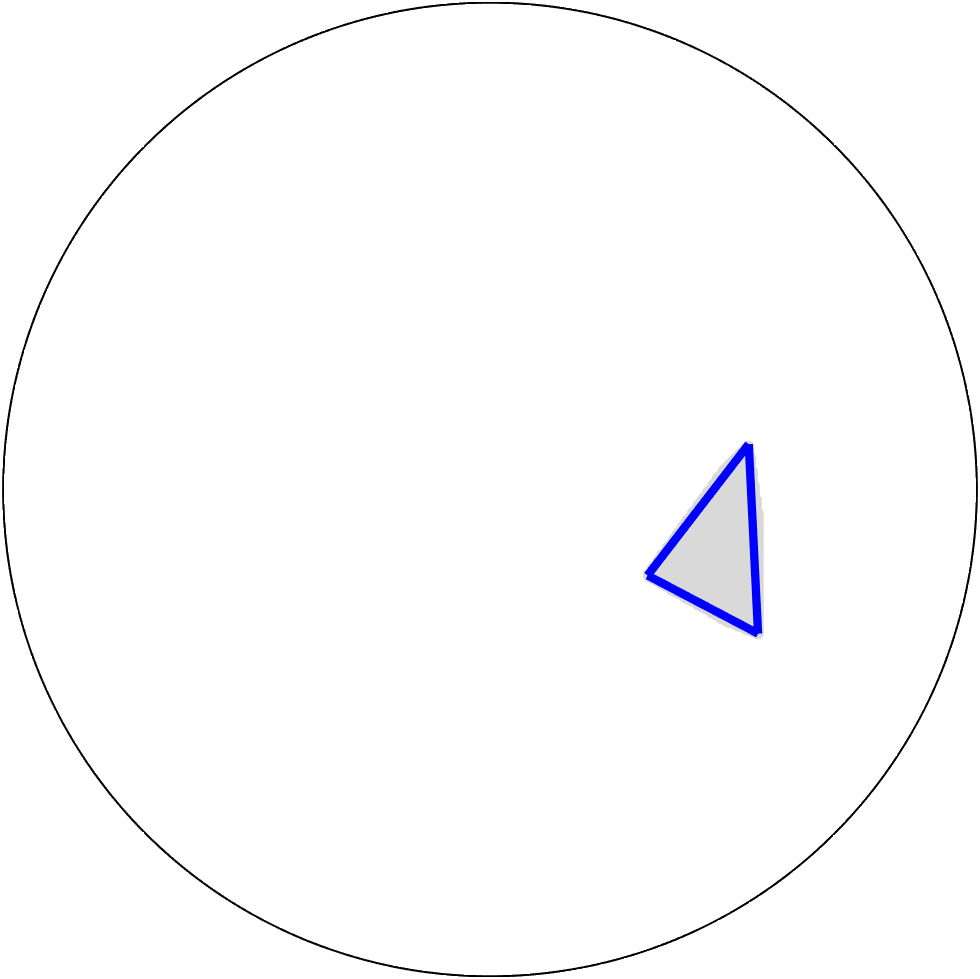} \newline
        \end{minipage}
        & \begin{minipage}{0.18\textwidth}
        \centering \includegraphics[height=2.3cm]{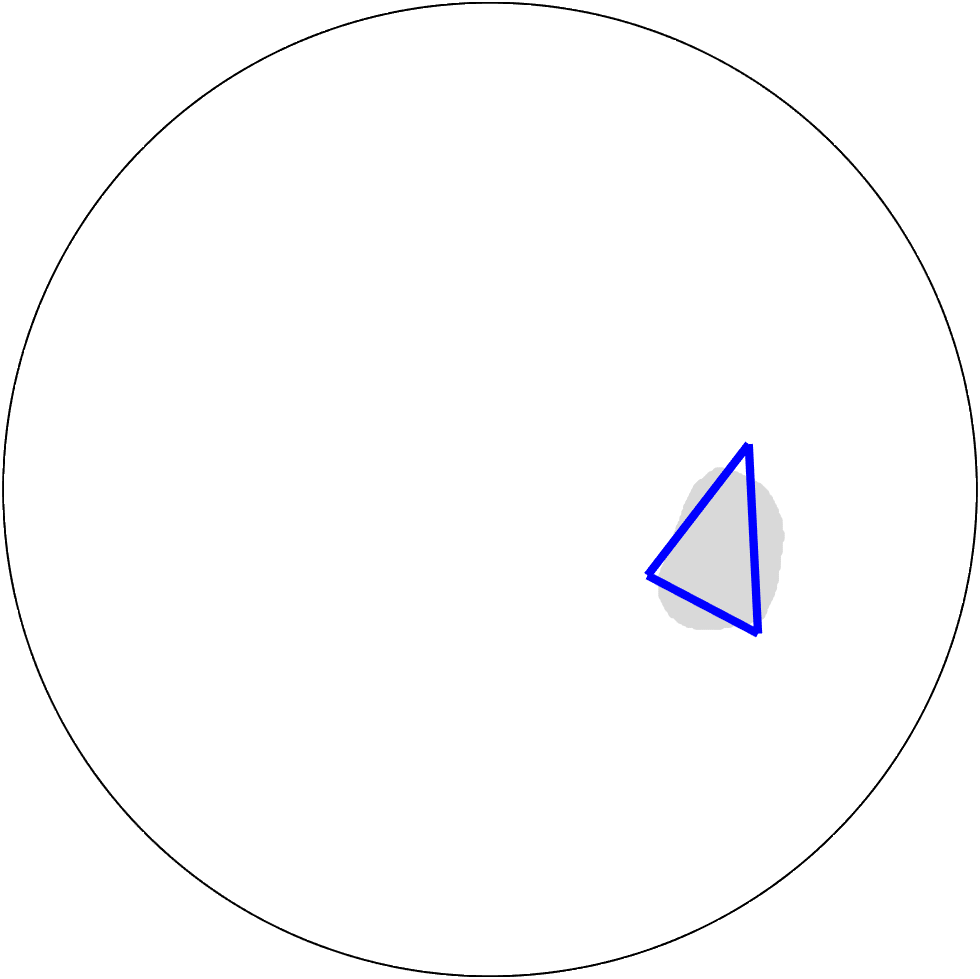} \\ 0.7\%
        \renewcommand{\thefigure}{I}
         \captionlistentry{} 
         \label{fig:simulatedresults_I}
        \end{minipage}
        & \begin{minipage}{0.18\textwidth}
        \centering \includegraphics[height=2.3cm]{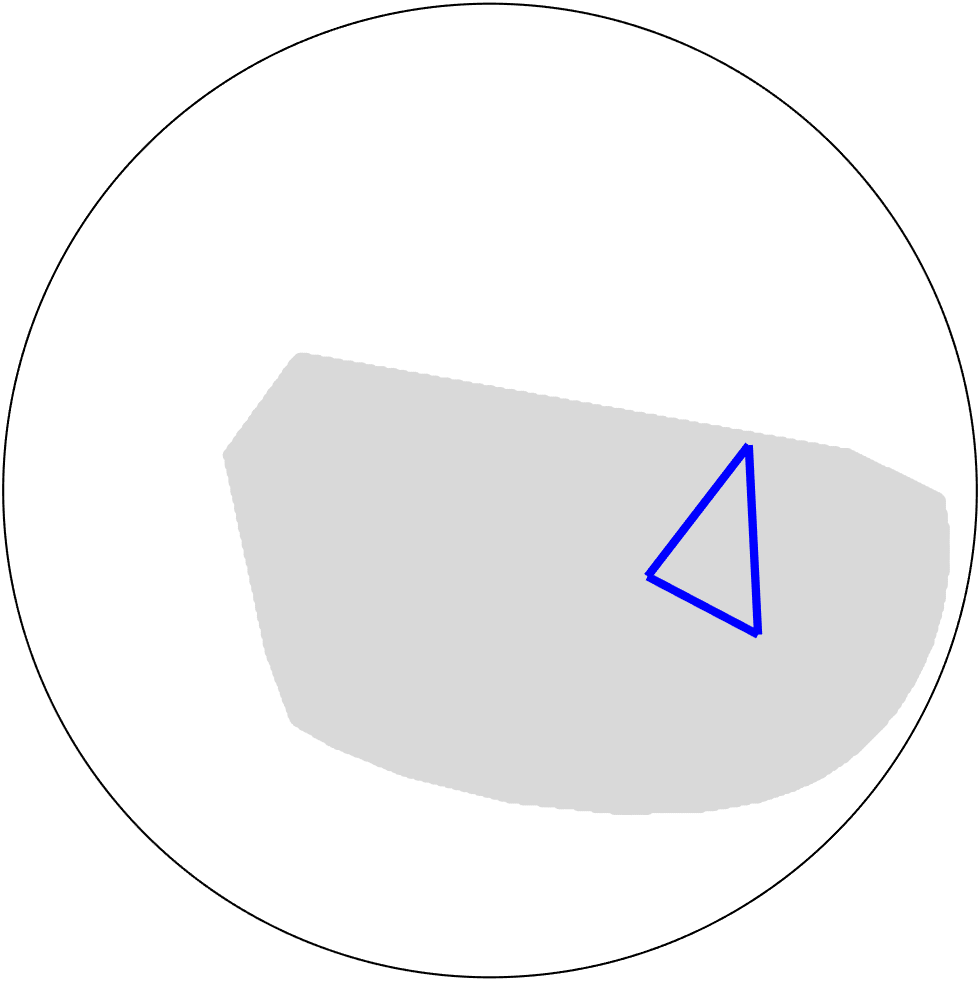}\\34.2\%
        \end{minipage} \\[1.3cm]
        II
         & \begin{minipage}{0.25\textwidth}
        \centering \includegraphics[height=3cm]{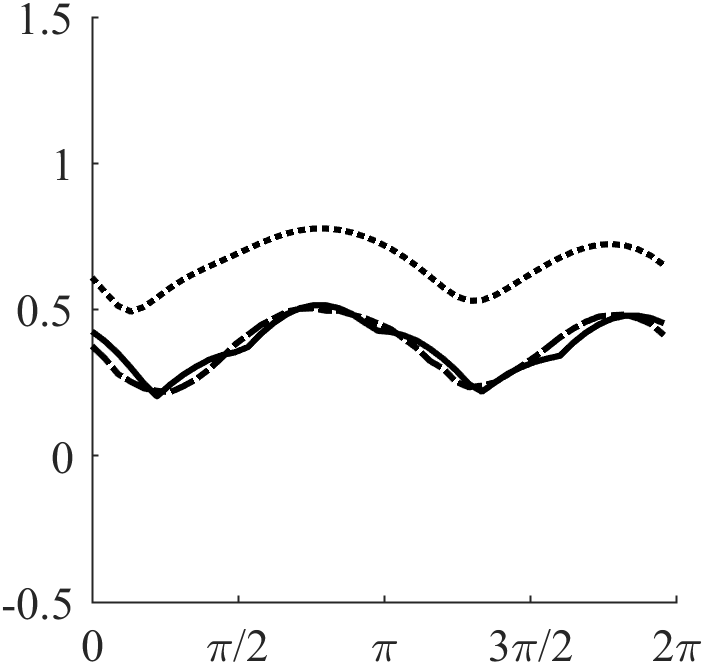}
        \end{minipage}
        & \begin{minipage}{0.18\textwidth}
        \centering \includegraphics[height=2.3cm]{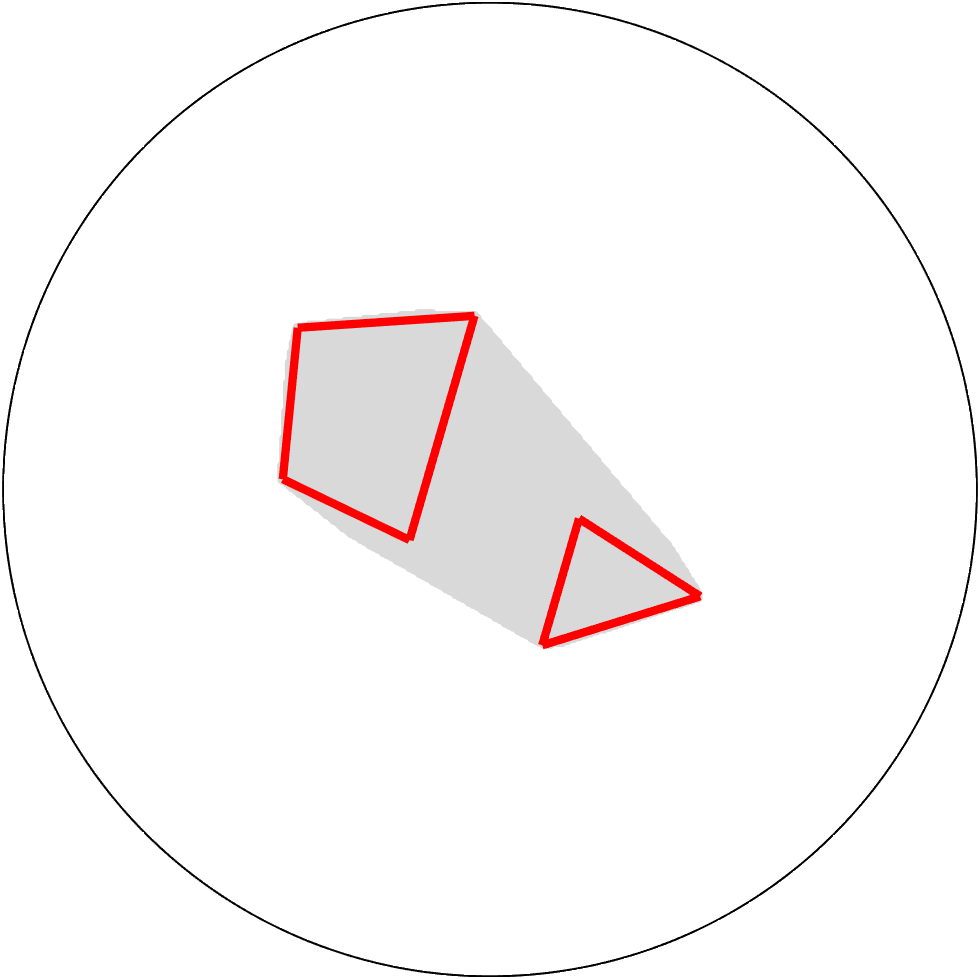} \newline
        \end{minipage}
        & \begin{minipage}{0.18\textwidth}
        \centering \includegraphics[height=2.3cm]{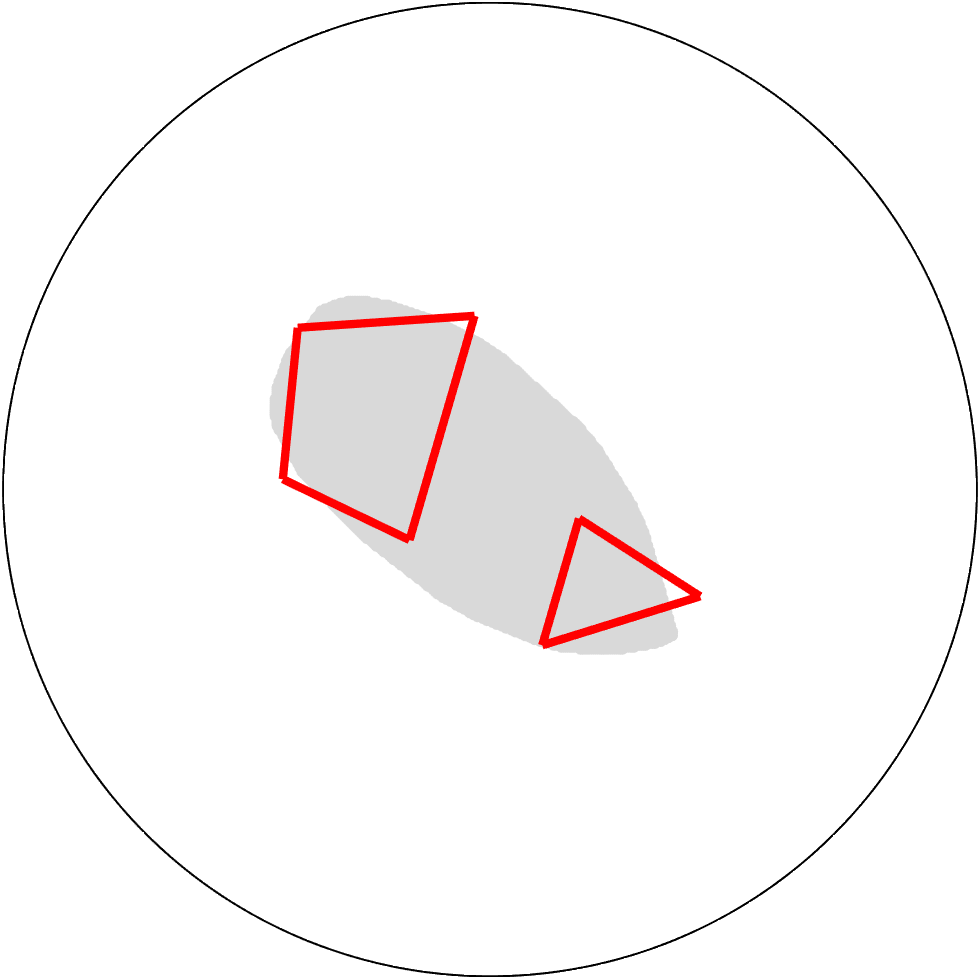} \\1.7\%
        \renewcommand{\thefigure}{II}
         \captionlistentry{} 
         \label{fig:simulatedresults_II}
        \end{minipage}
        & \begin{minipage}{0.18\textwidth}
        \centering \includegraphics[height=2.3cm]{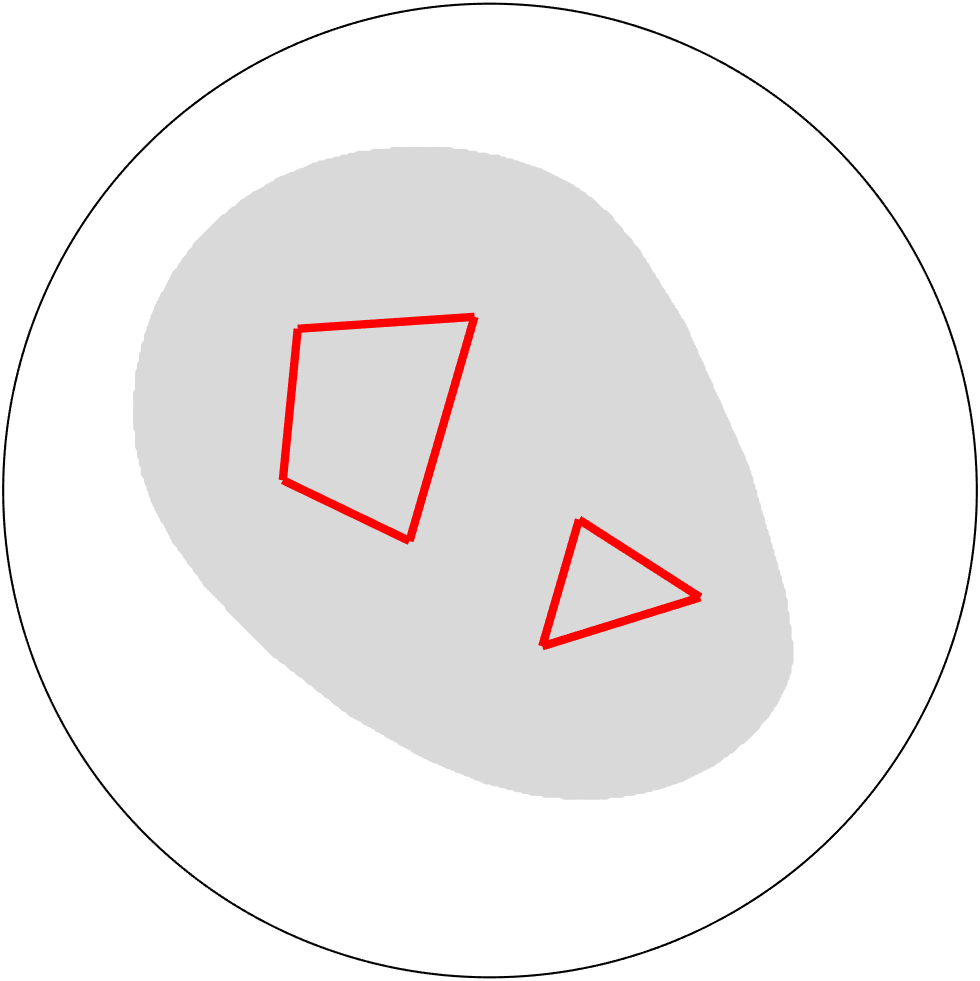}\\32.9\%
        \end{minipage} \\[1.3cm]
        III 
        & \begin{minipage}{0.25\textwidth}
        \centering \includegraphics[height=3cm]{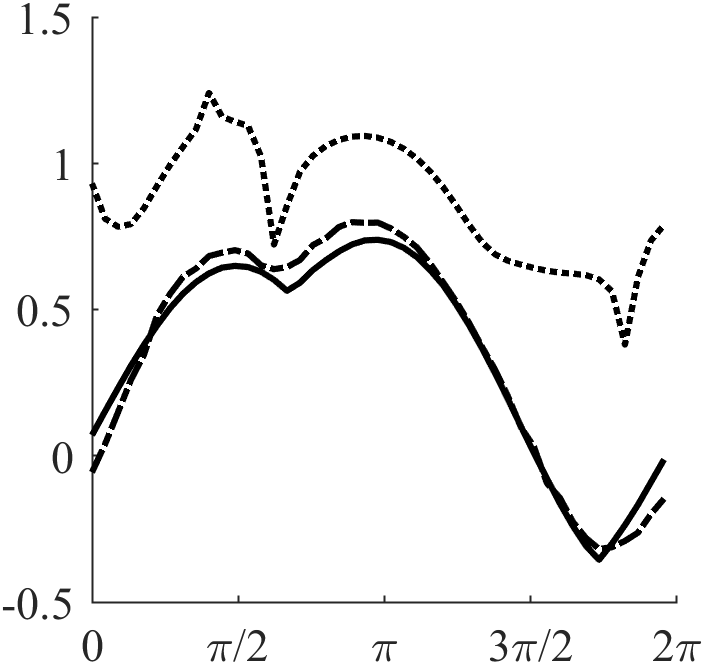}
        \end{minipage}
        & \begin{minipage}{0.18\textwidth}
        \centering \includegraphics[height=2.3cm]{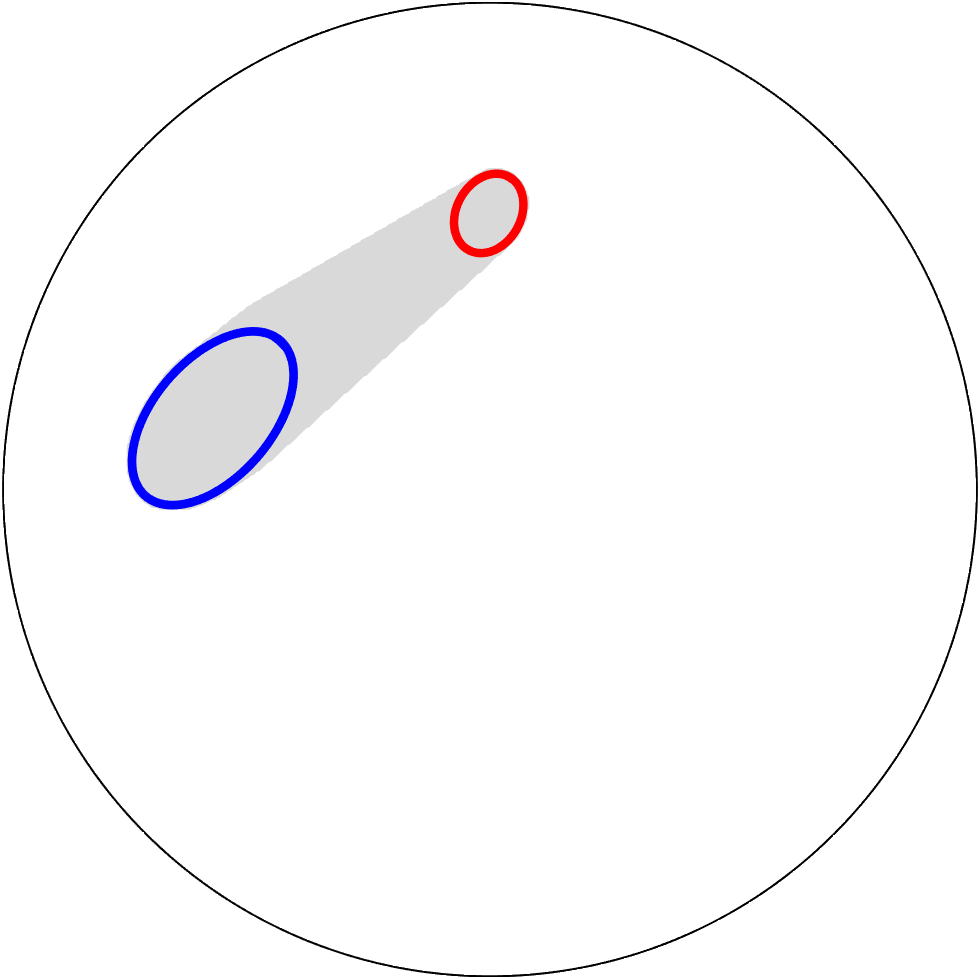} \newline
        \end{minipage}
        & \begin{minipage}{0.18\textwidth}
        \centering \includegraphics[height=2.3cm]{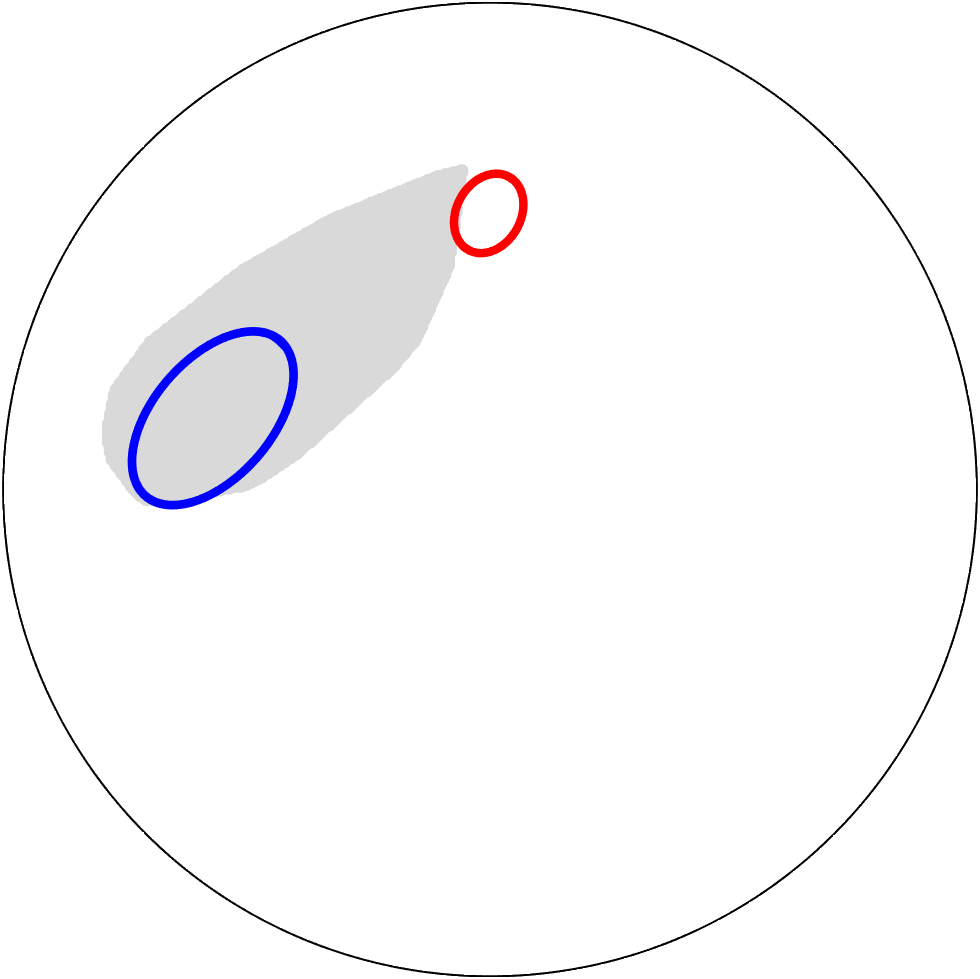} \\3.9\%
        \renewcommand{\thefigure}{III}
         \captionlistentry{} 
         \label{fig:simulatedresults_III}
        \end{minipage}
        & \begin{minipage}{0.18\textwidth}
        \centering \includegraphics[height=2.3cm]{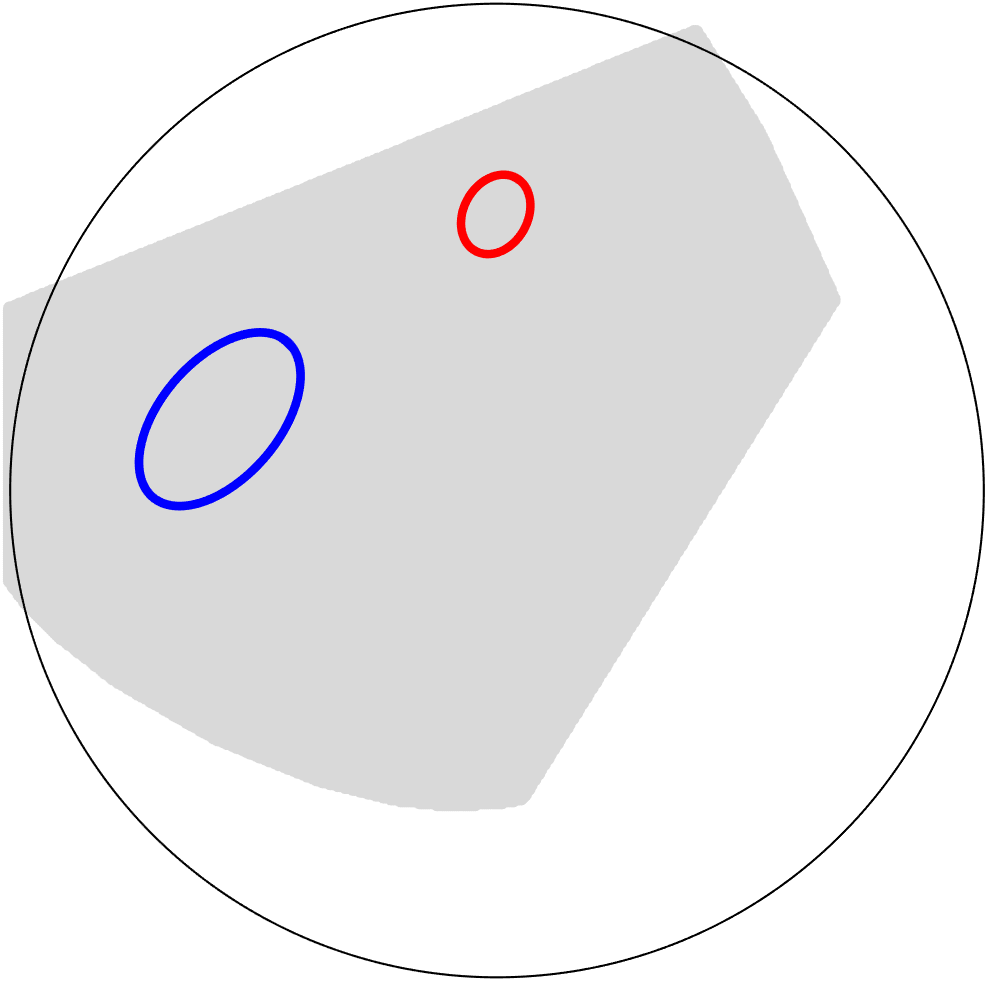}\\54.3\%
        \end{minipage} \\[1.3cm]
        IV 
        & \begin{minipage}{0.25\textwidth}
        \centering \includegraphics[height=3cm]{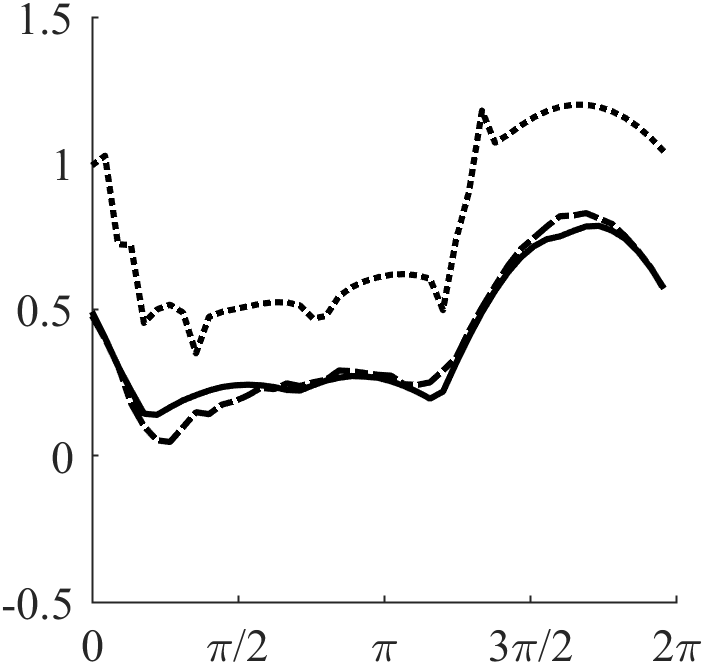}
        \end{minipage}
        & \begin{minipage}{0.18\textwidth}
        \centering \includegraphics[height=2.3cm]{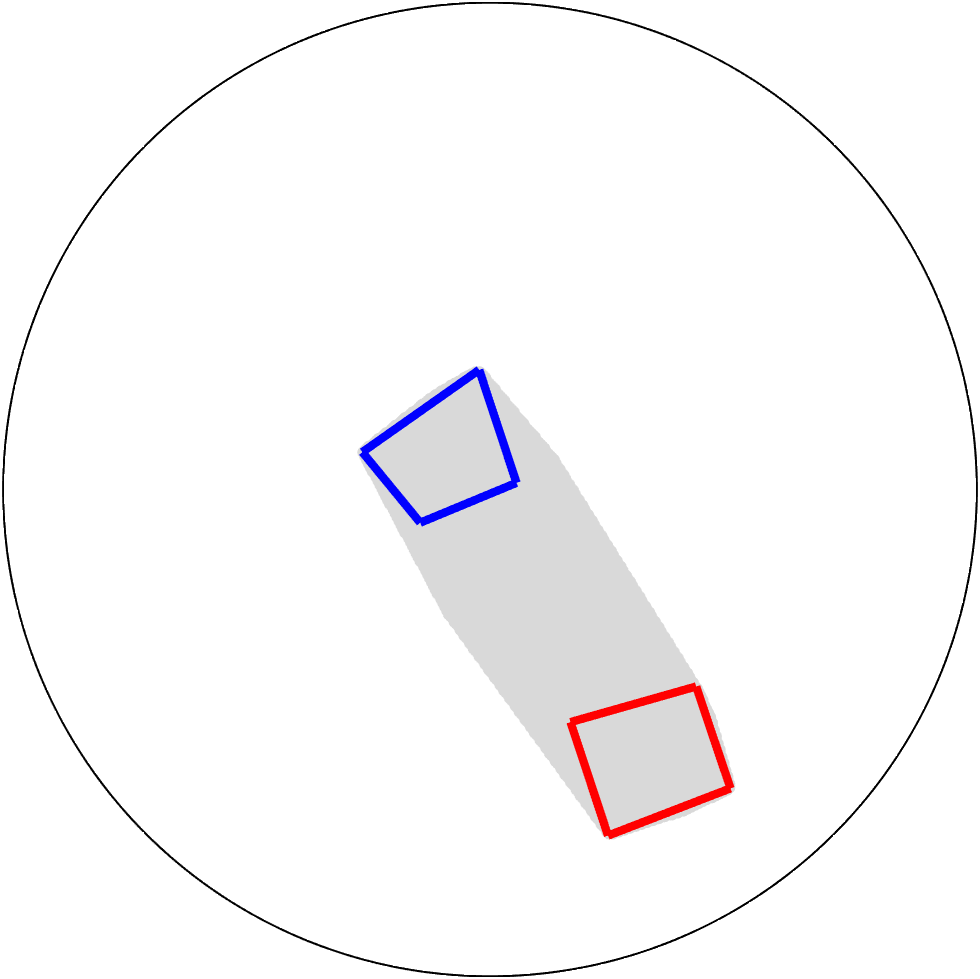} \newline
        \end{minipage}
        & \begin{minipage}{0.18\textwidth}
        \centering \includegraphics[height=2.3cm]{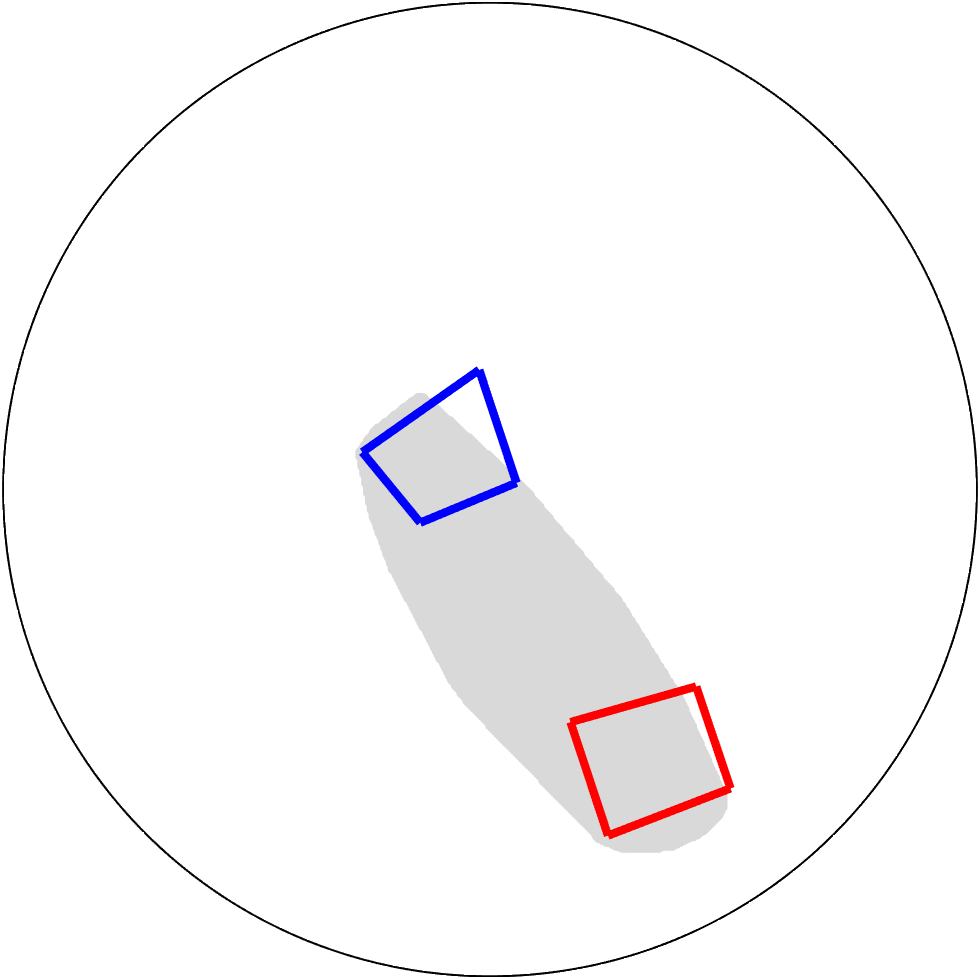} \\4.4\%
        \renewcommand{\thefigure}{IV}
         \captionlistentry{} 
         \label{fig:simulatedresults_IV}
        \end{minipage}
        & \begin{minipage}{0.18\textwidth}
        \centering \includegraphics[height=2.3cm]{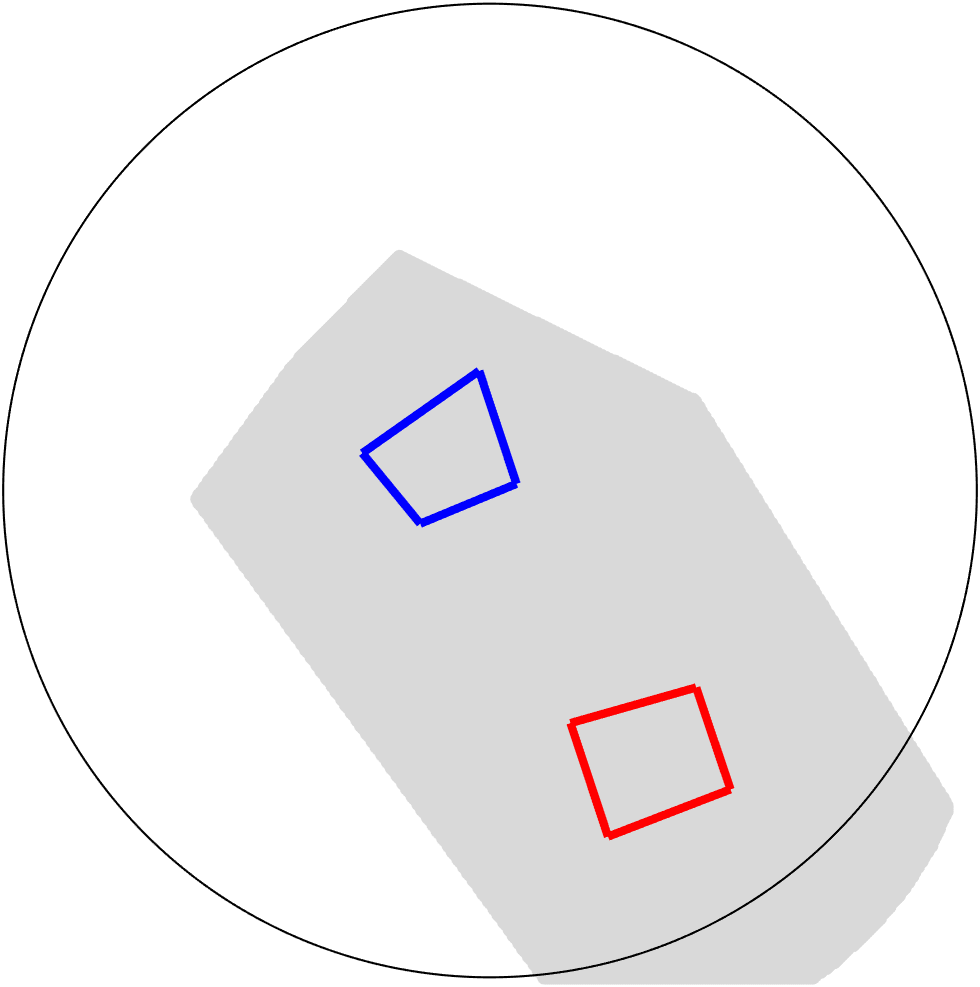}\\36.0\%
        \end{minipage} \\[1.3cm]
        V 
        & \begin{minipage}{0.25\textwidth}
        \centering \includegraphics[height=3cm]{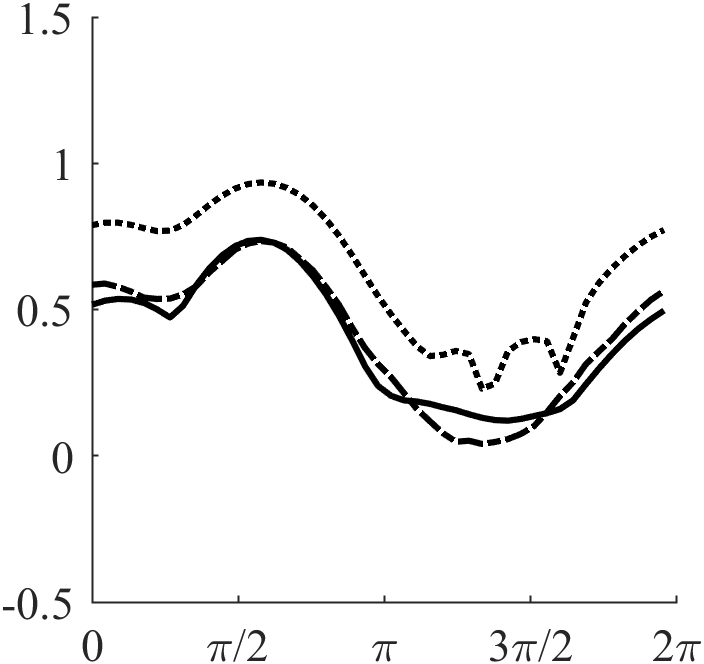}
        \end{minipage}
        & \begin{minipage}{0.18\textwidth}
        \centering \includegraphics[height=2.3cm]{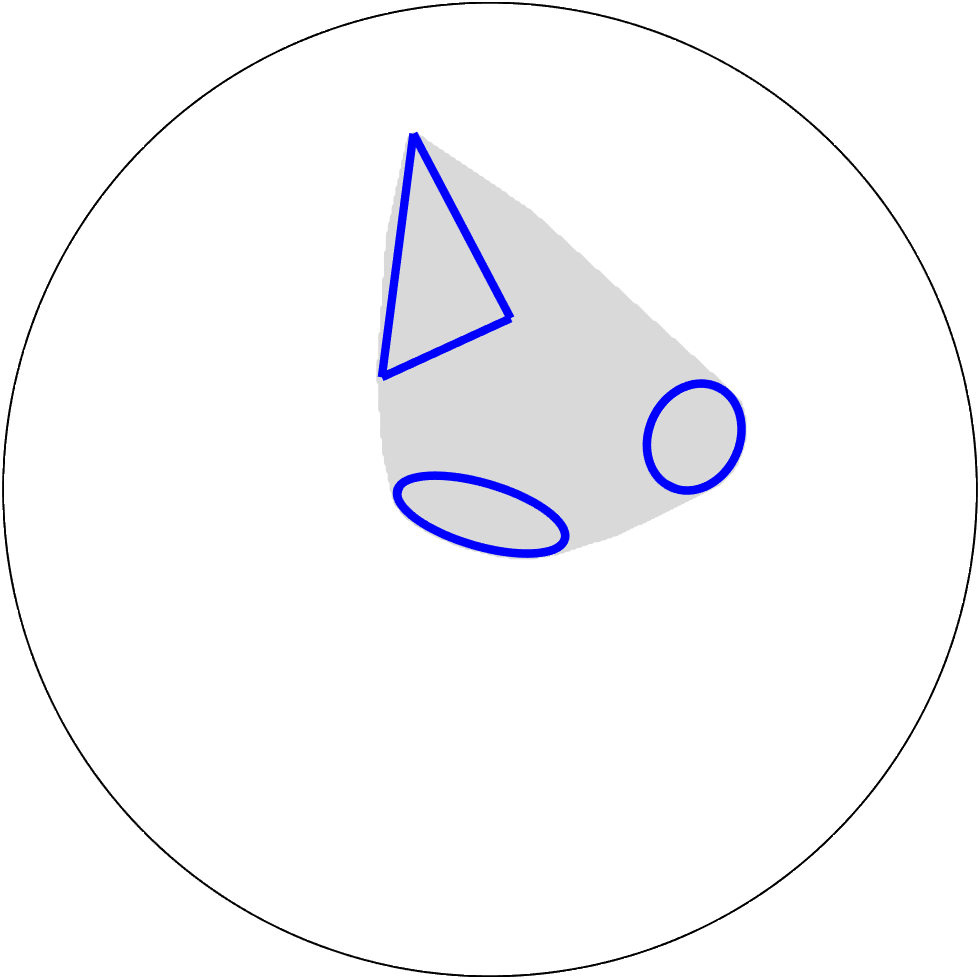} \newline
        \end{minipage}
        & \begin{minipage}{0.18\textwidth}
        \centering \includegraphics[height=2.3cm]{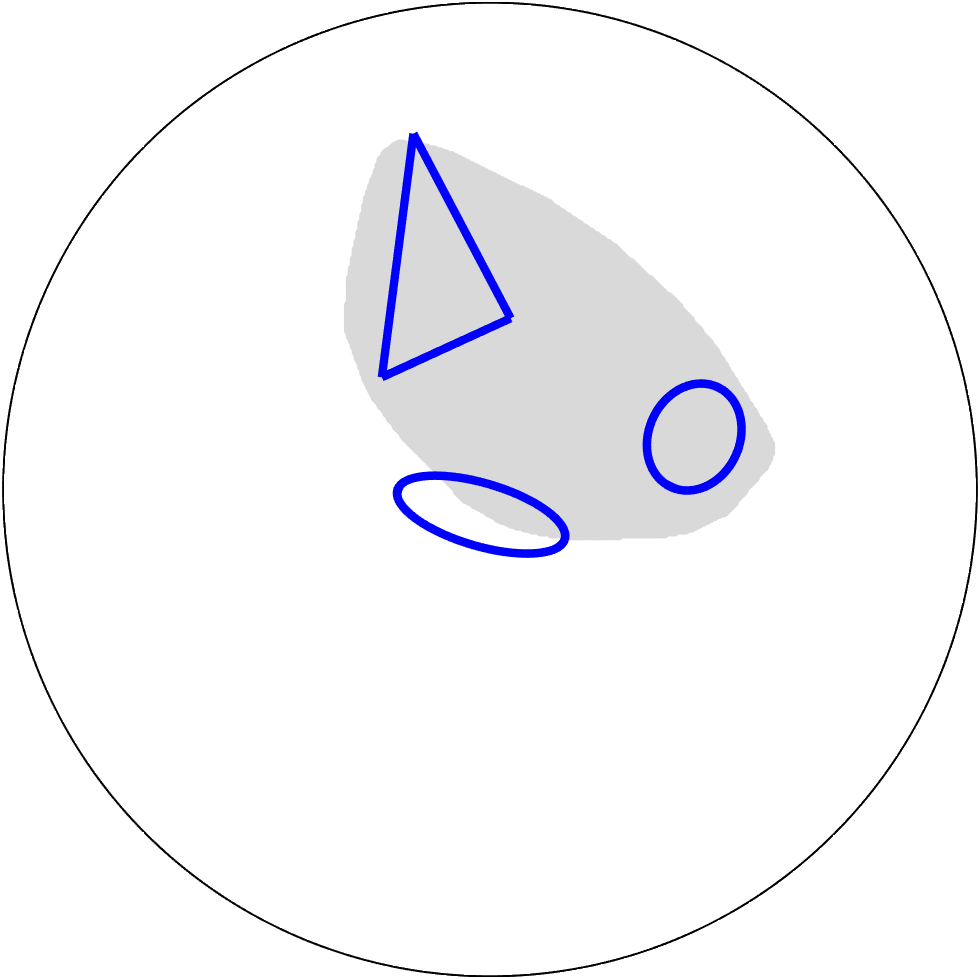} \\4.3\%
        \renewcommand{\thefigure}{V}
         \captionlistentry{} 
         \label{fig:simulatedresults_V}
        \end{minipage}
        & \begin{minipage}{0.18\textwidth}
        \centering \includegraphics[height=2.3cm]{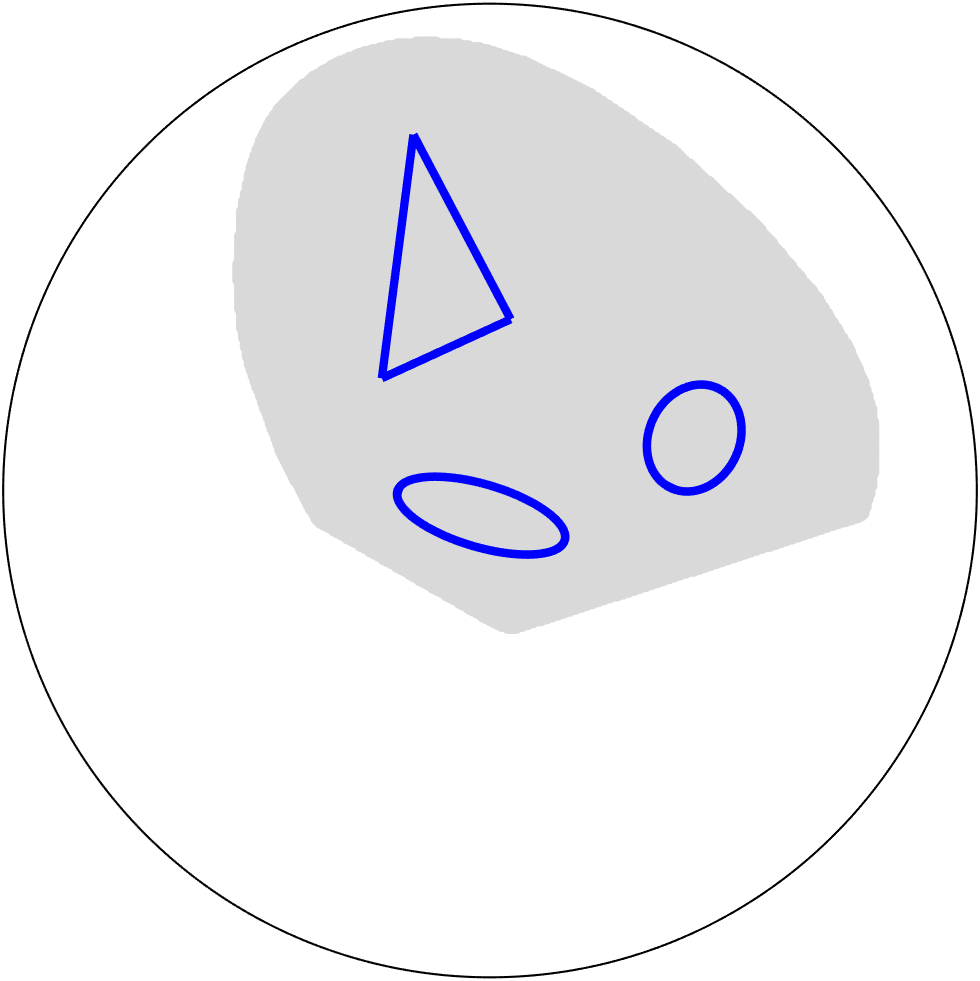}\\25.2\%
        \end{minipage} \\[1.3cm]
        VI
        & \begin{minipage}{0.25\textwidth}
        \centering \includegraphics[height=3cm]{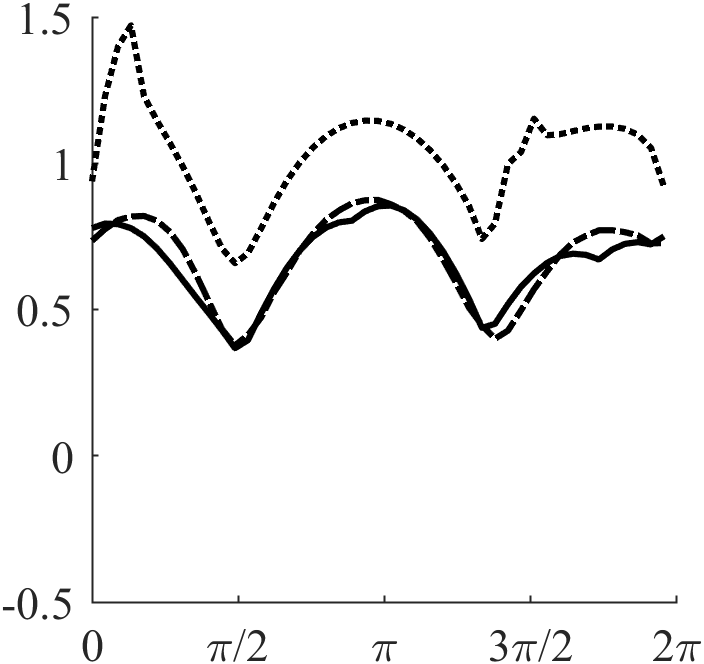}
        \end{minipage}
        & \begin{minipage}{0.18\textwidth}
        \centering \includegraphics[height=2.3cm]{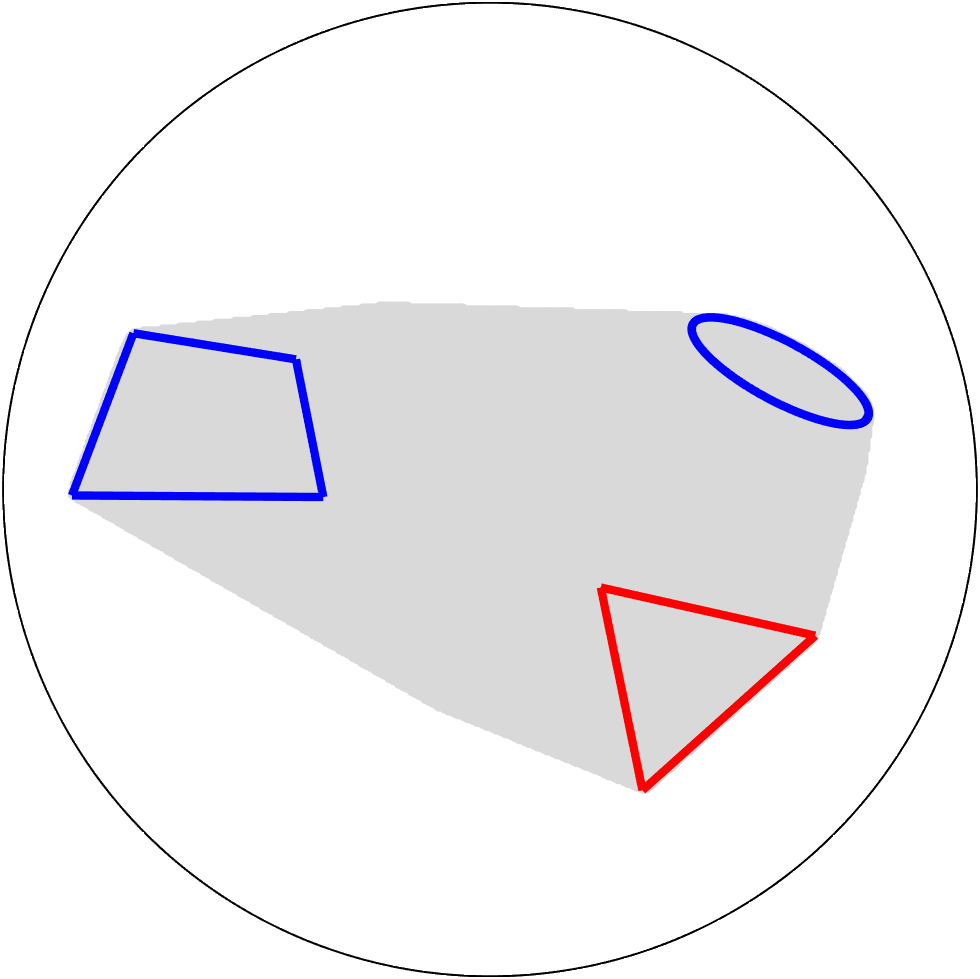} \newline
        \end{minipage}
        & \begin{minipage}{0.18\textwidth}
        \centering \includegraphics[height=2.3cm]{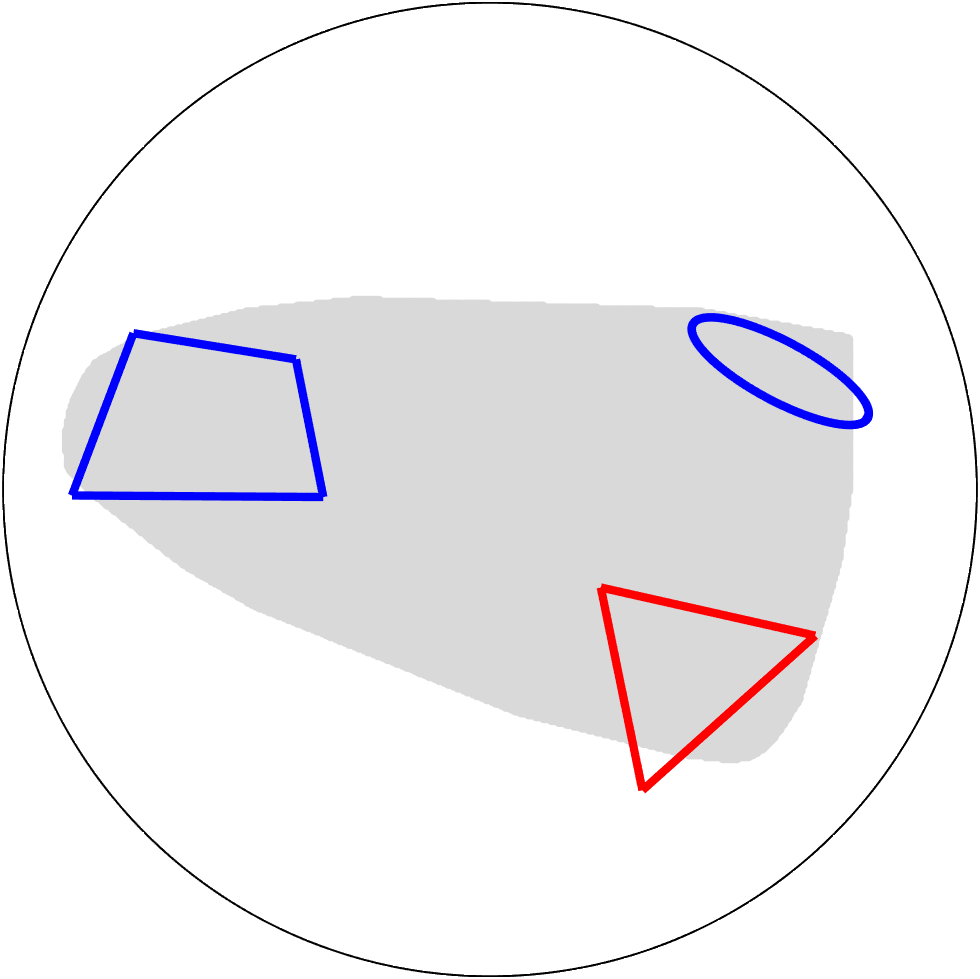} \\3.9\%
        \renewcommand{\thefigure}{VI}
         \captionlistentry{} 
         \label{fig:simulatedresults_VI}
        \end{minipage}
        & \begin{minipage}{0.18\textwidth}
        \centering \includegraphics[height=2.3cm]{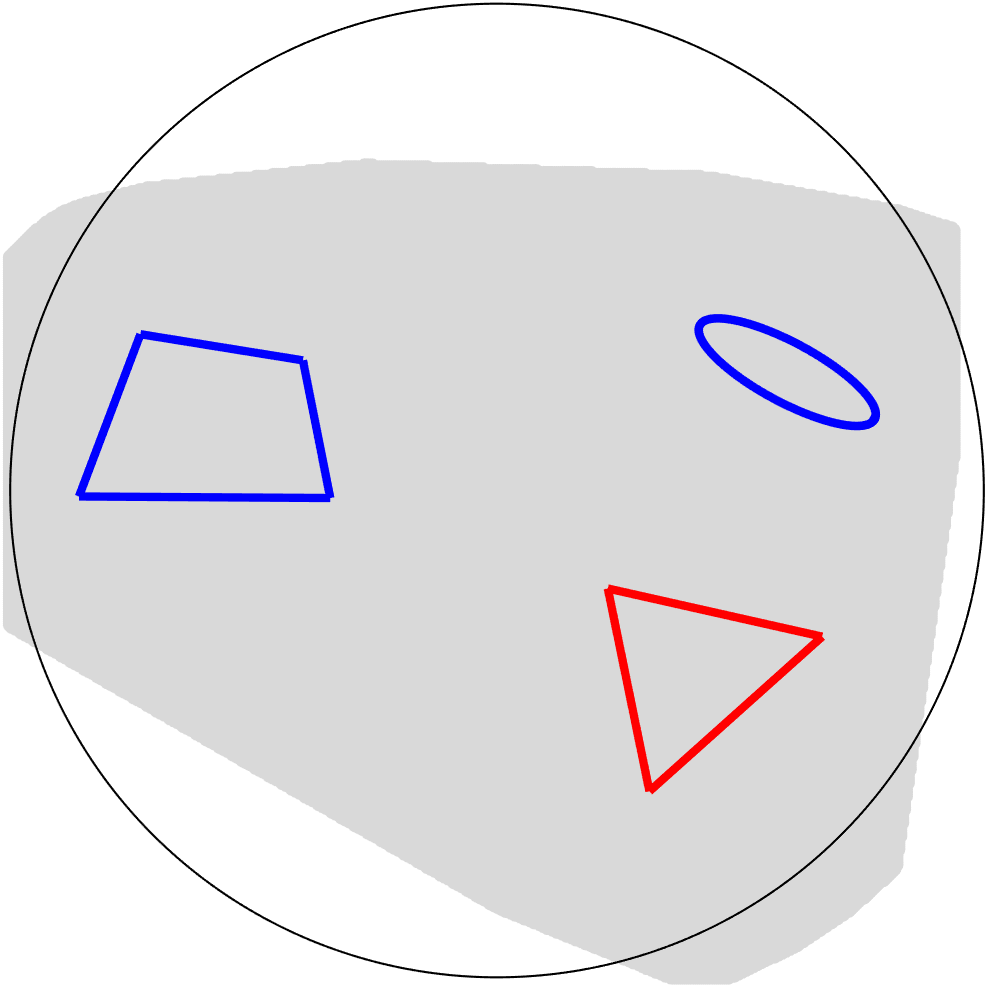}\\52.1\%
        \end{minipage} \\[1.3cm]
    \end{tabular} 
  \addtocounter{figure}{-6}
    \caption{Comparison of the support vectors and the hulls of the simulated phantoms, computed using LS (dotted line), learned approach (dashed line), and the ground truth (solid line). The error relative to the ground truth is shown below each phantom. }
    \label{fig:simulated_examples}
\end{figure}

\subsection{Experimental data}
We use a total of 22 experimental phantoms from the open dataset \cite{hauptmann2017open} to test the method. The experimental setup consists of a circular cylinder tank (diameter 28cm) filled with saline, with either plastic (resistive) or metal (conductive) objects. Each phantom has 1-3 circular, triangular, or square objects. The KIT4 measurement system \cite{kourunen2008suitability} at the University of Eastern Finland used to collect the measurements has 16 electrodes.

We use conductivity phantoms created from the photographs of the measurement setting as ground truths. To combat the distortion in the images caused by saline and the camera lens, we obtained ground truths by segmenting the objects in the images according to their highest point. Since the support function detects jumps on the boundaries of the inclusions and not the actual conductivities, it is not necessary for the inclusions to match the true conductivities of the plastic and metal objects. With this in mind, we assigned the conductive inclusions a value of 10 and the resistive inclusions a value of 0.1. The conductivity of the background was set to 1. We created a conductivity phantom based on the segmentation, and computed true support functions and true convex hulls according to (\ref{suppfun}). See Figure \ref{fig:segmentation_example} for an example of the segmentation.

See Figures \ref{fig:experimental_results_1}-\ref{fig:experimental_results_4} for the experimental data results. The results follow the numbering scheme from the original open data set and they are presented roughly in ascending order of difficulty.

\begin{figure}[!ht]
    \begin{tabular}{lccccc}
        Case & Phantom & Ground truth & Learned & Least squares \\ [.2cm]
        2.1 
        &
        \begin{minipage}{0.25\textwidth}
        \centering \includegraphics[height=2.8cm]{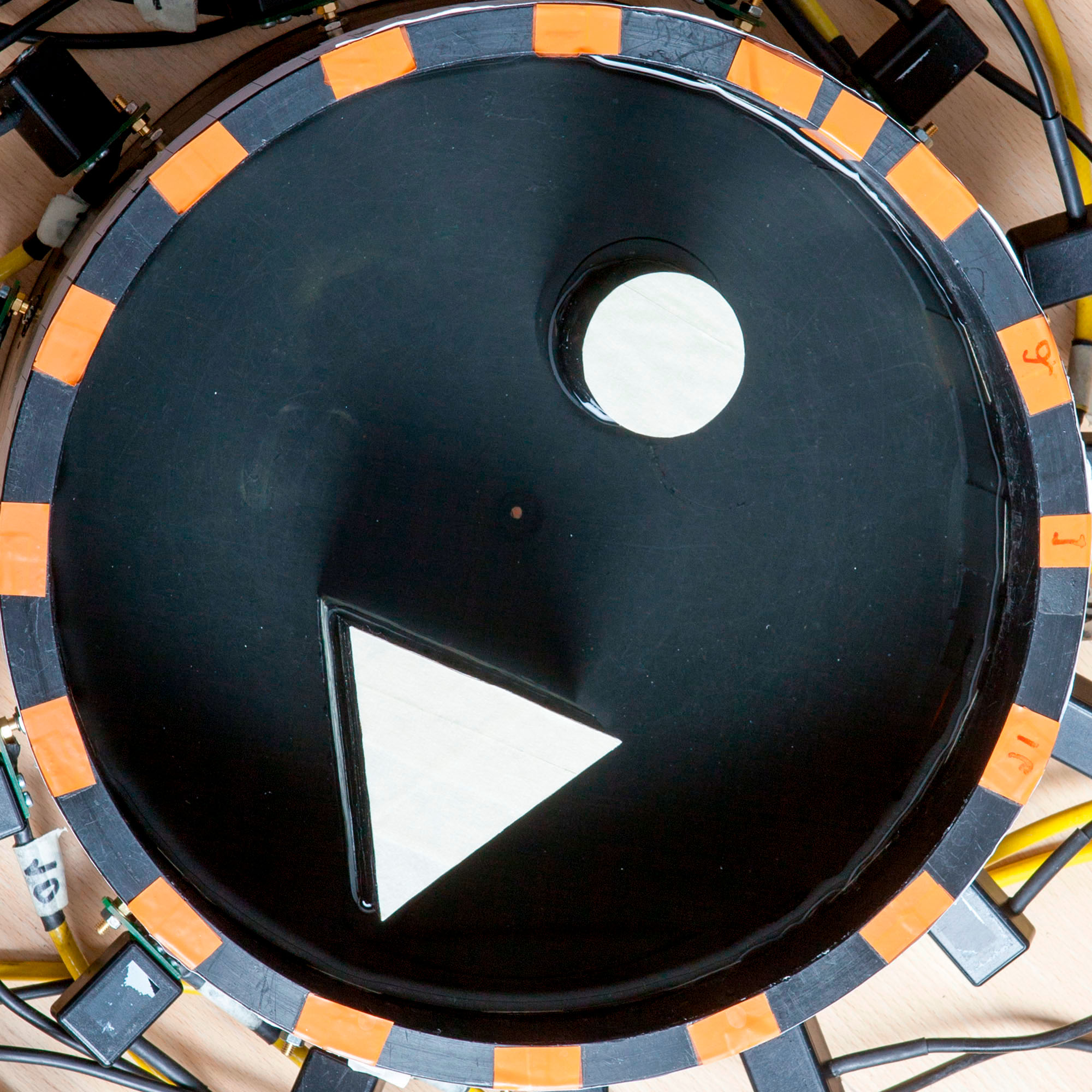} \newline
        \end{minipage}
        &\begin{minipage}{0.18\textwidth}
        \centering \includegraphics[height=2.3cm]{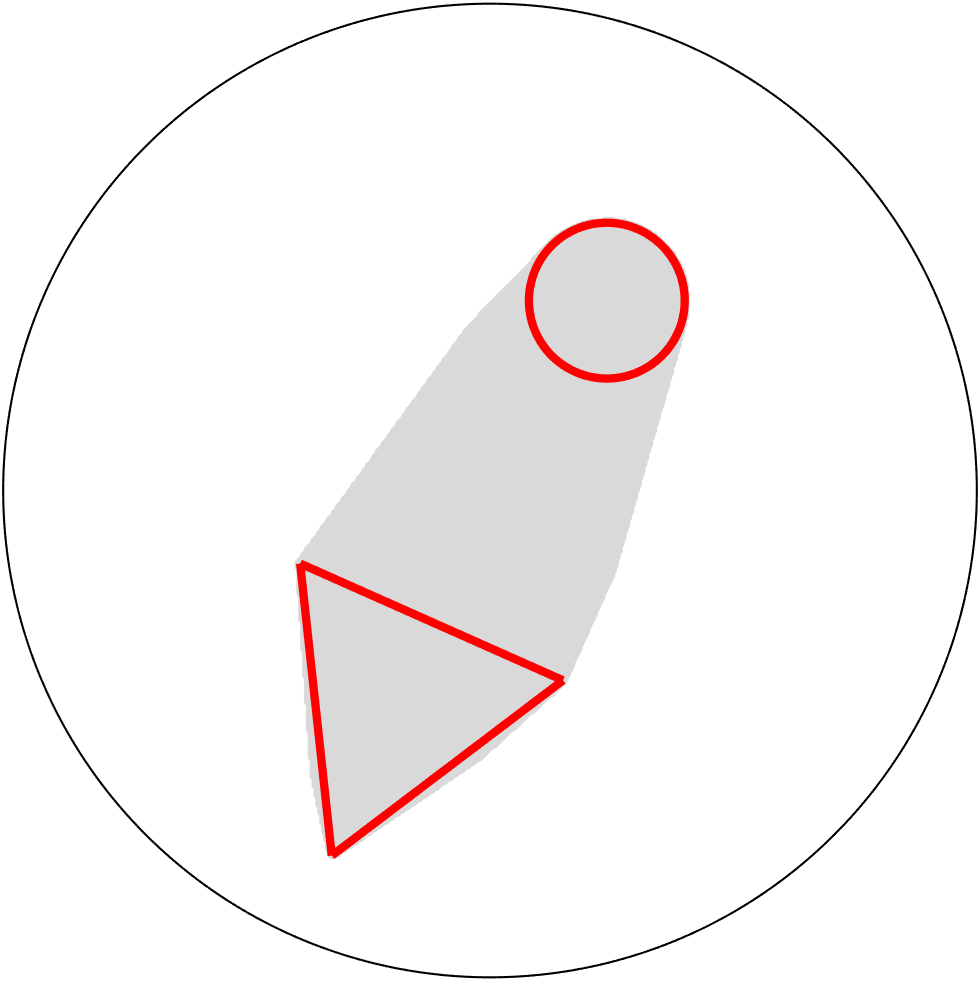} \newline
        \end{minipage}
        & \begin{minipage}{0.18\textwidth}
        \centering \includegraphics[height=2.3cm]{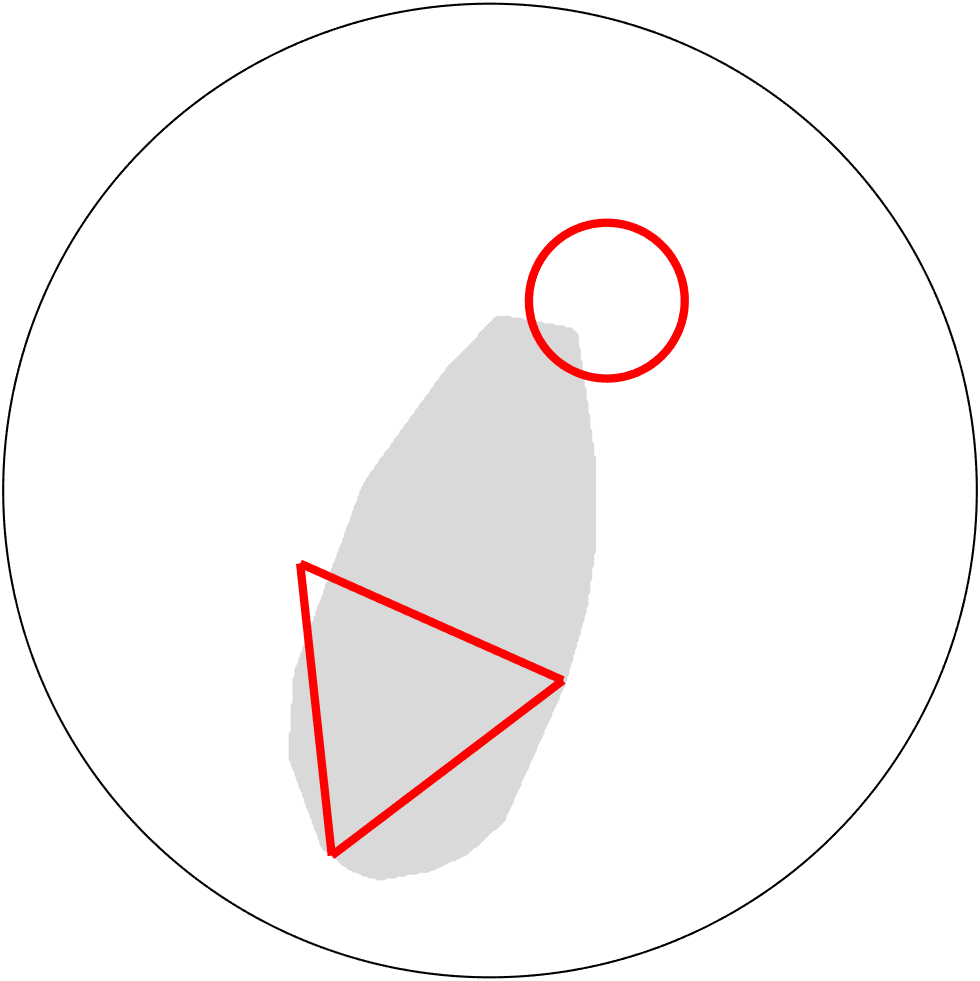} \newline 7.7\%
        \renewcommand{\thefigure}{2.1}
         \captionlistentry{} 
         \label{fig:experimental_results_2-1}
        \end{minipage}
        & \begin{minipage}{0.18\textwidth}
        \centering \includegraphics[height=2.3cm]{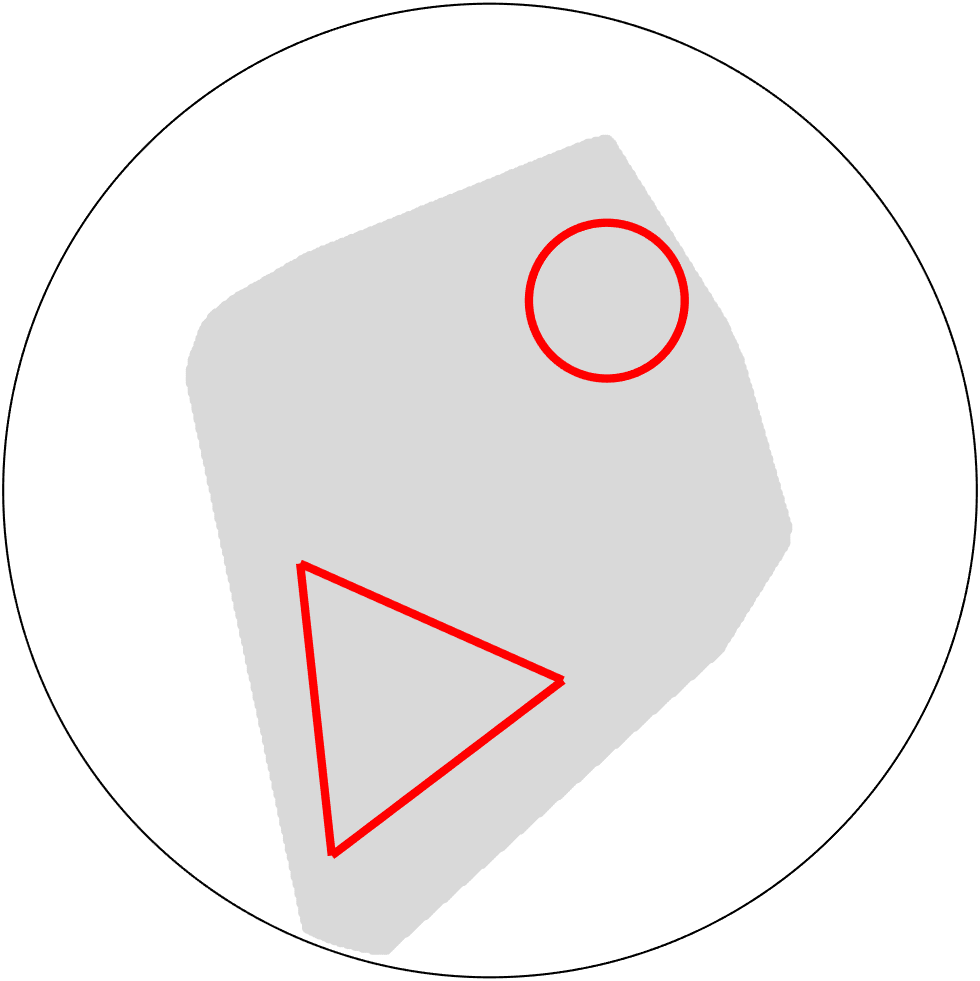} \newline 26.0\%
        \end{minipage}\\[1.3cm]
        2.2 
        &
        \begin{minipage}{0.25\textwidth}
        \centering \includegraphics[height=2.8cm]{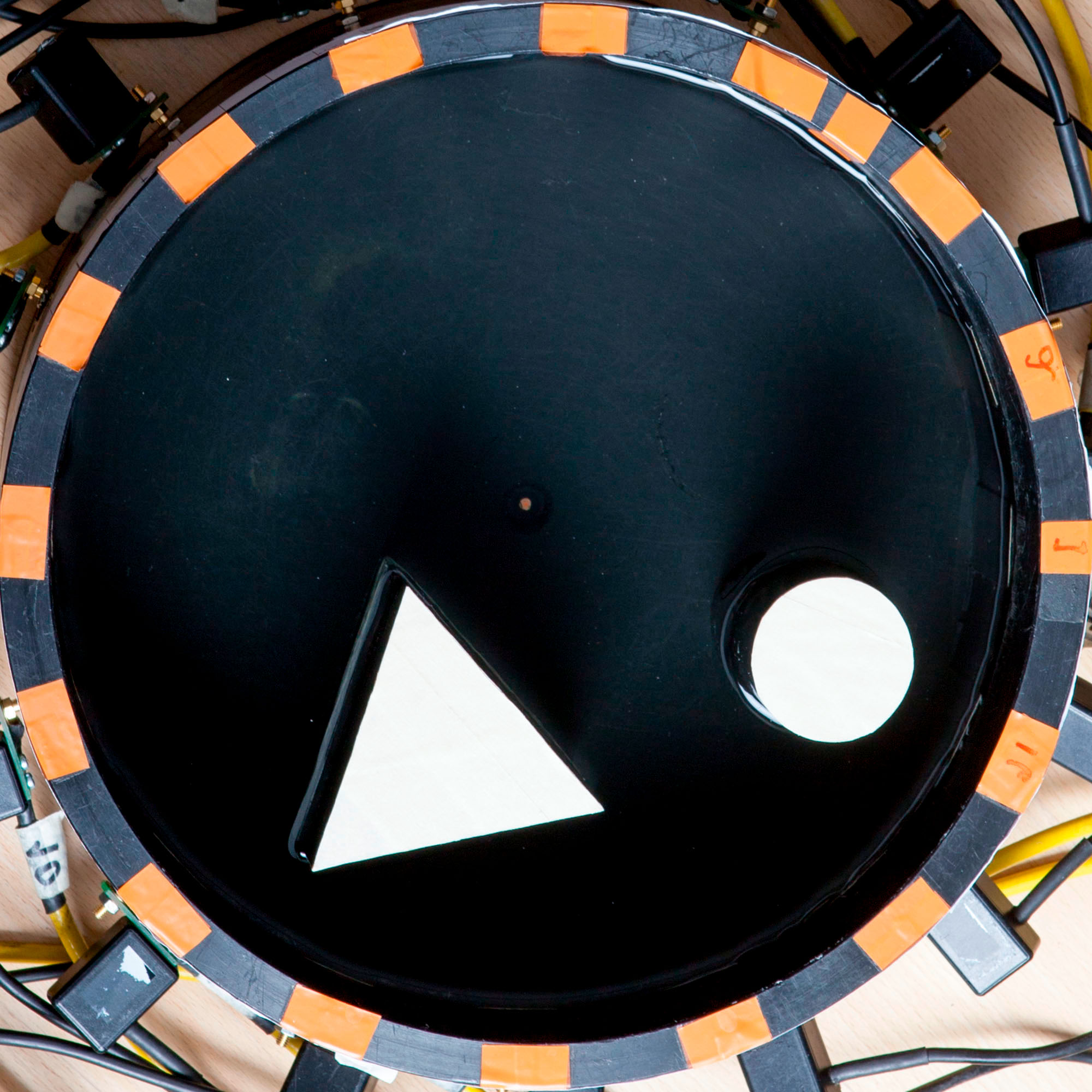} \newline
        \end{minipage}
        &\begin{minipage}{0.18\textwidth}
        \centering \includegraphics[height=2.3cm]{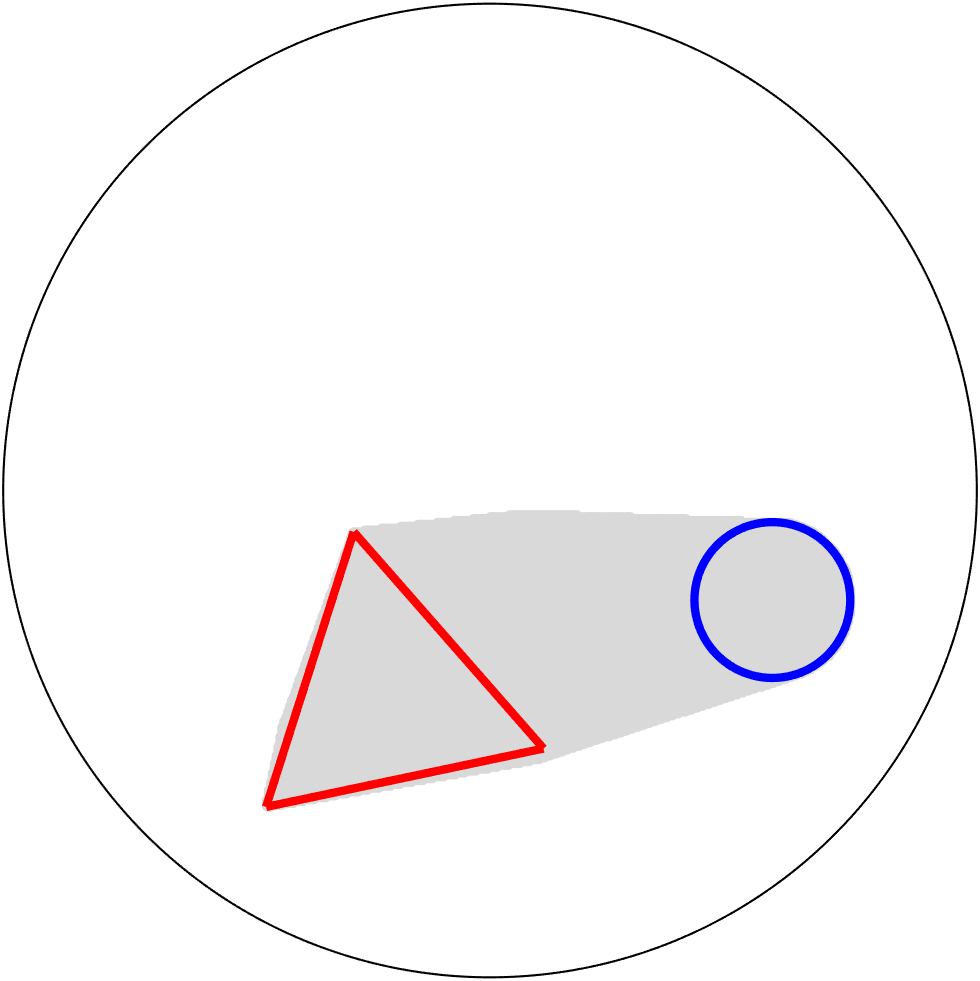} \newline
        \end{minipage}
        & \begin{minipage}{0.18\textwidth}
        \centering \includegraphics[height=2.3cm]{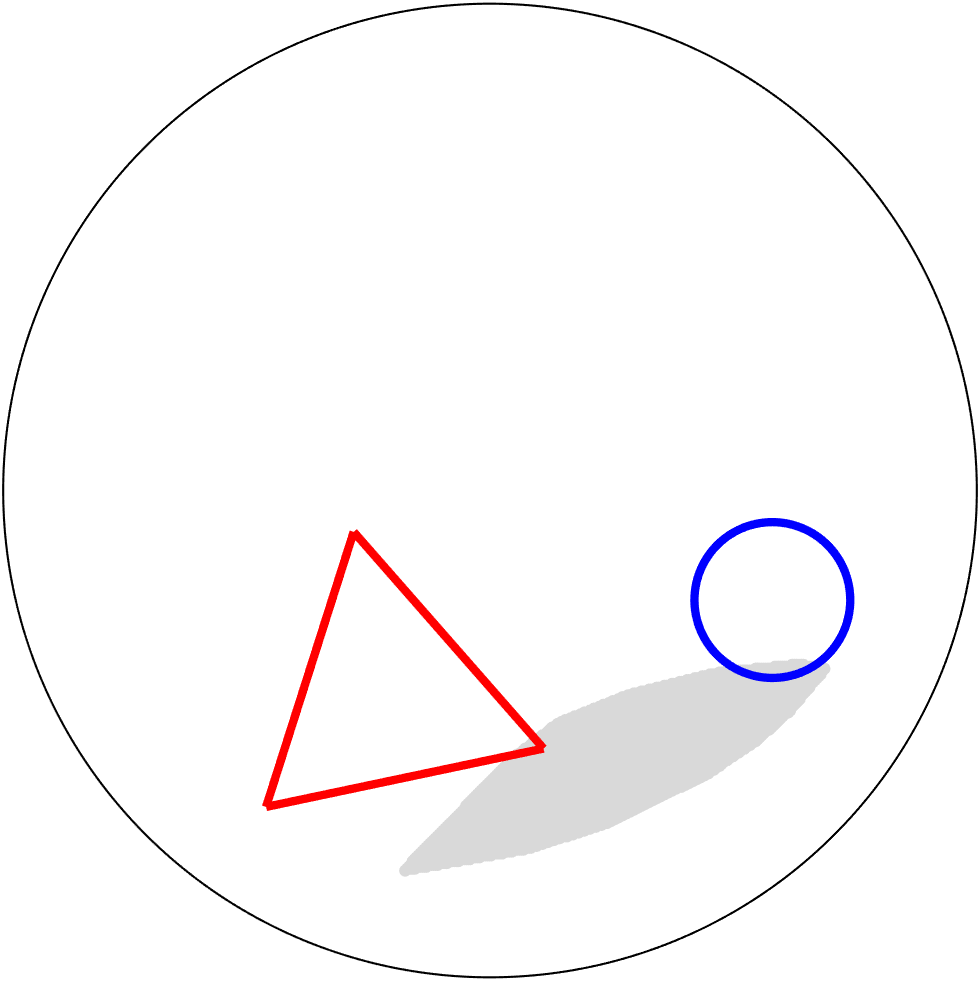} \newline 20.0\%
        \renewcommand{\thefigure}{2.2}
         \captionlistentry{} 
         \label{fig:experimental_results_2-2}
        \end{minipage}
        & \begin{minipage}{0.18\textwidth}
        \centering \includegraphics[height=2.3cm]{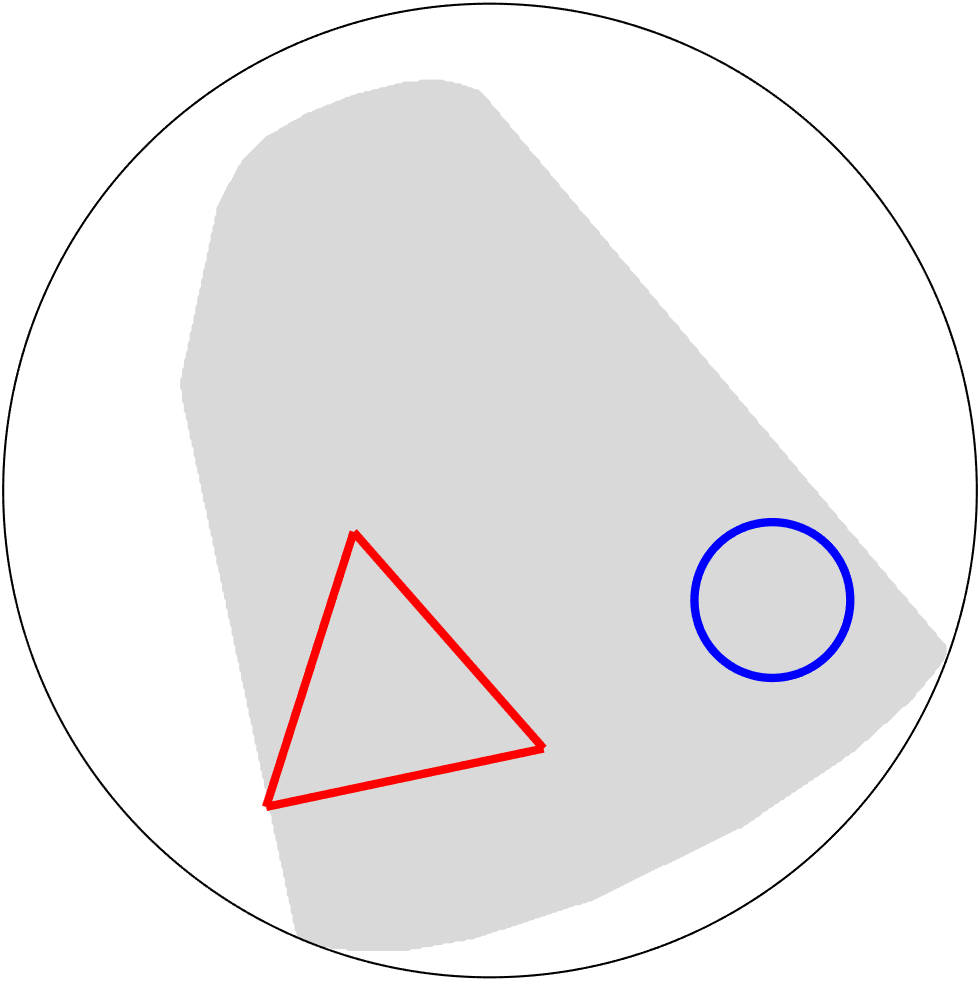} \newline 44.1\%
        \end{minipage}\\[1.3cm]
        2.3 
        &
        \begin{minipage}{0.25\textwidth}
        \centering \includegraphics[height=2.8cm]{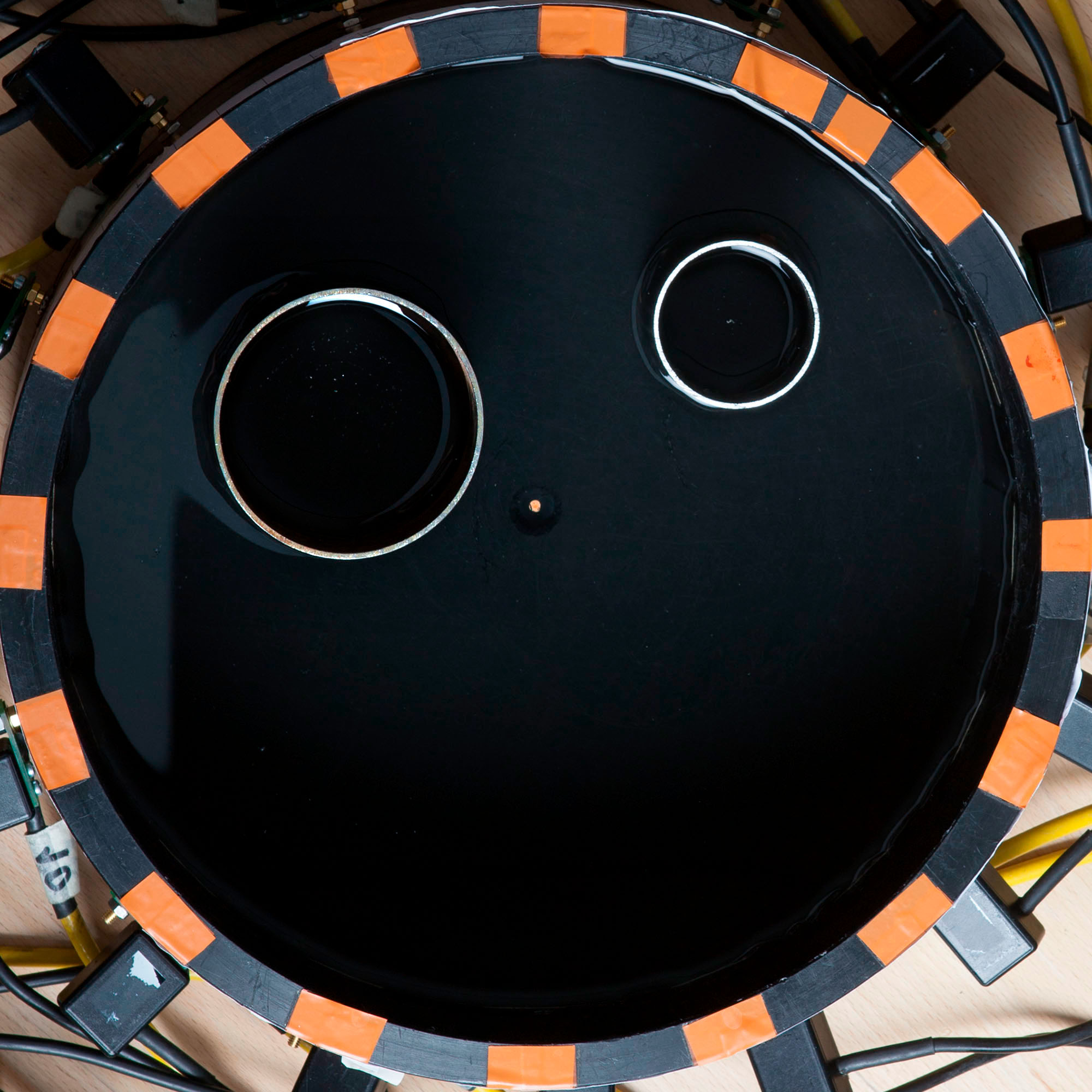} \newline
        \end{minipage}
        &\begin{minipage}{0.18\textwidth}
        \centering \includegraphics[height=2.3cm]{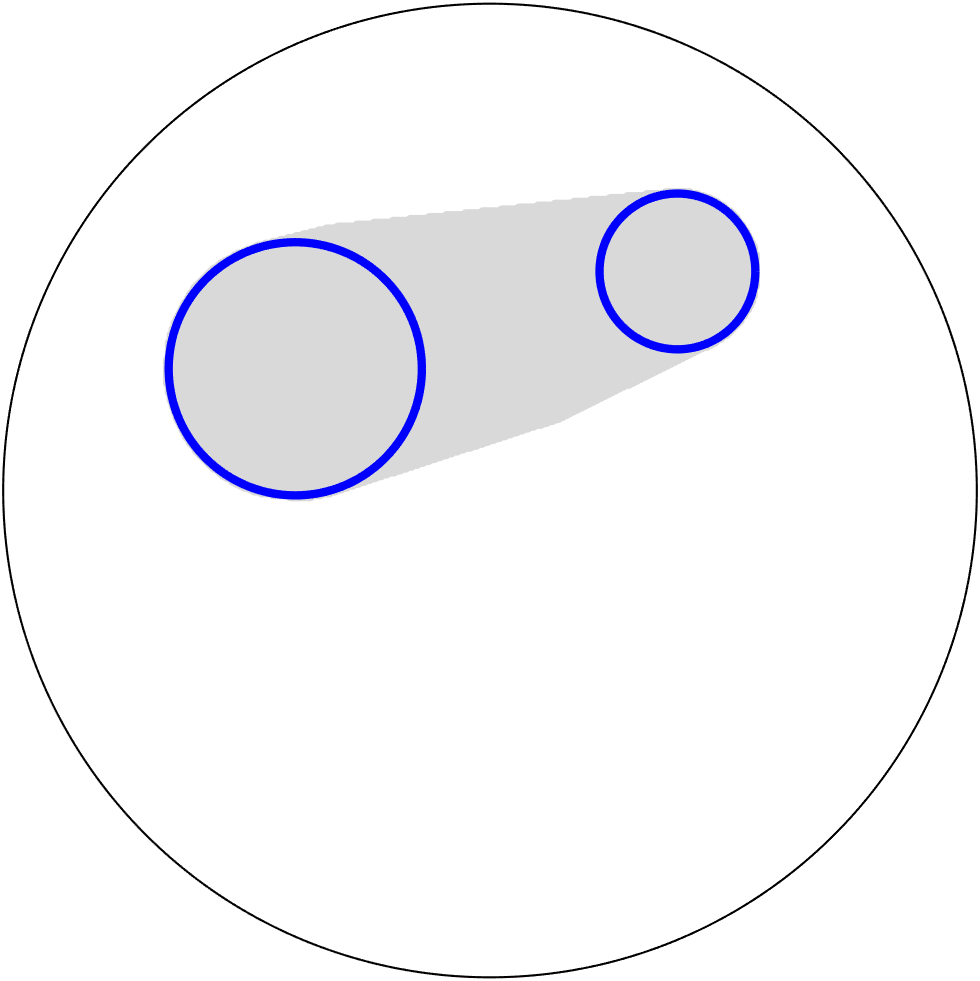} \newline
        \end{minipage}
        & \begin{minipage}{0.18\textwidth}
        \centering \includegraphics[height=2.3cm]{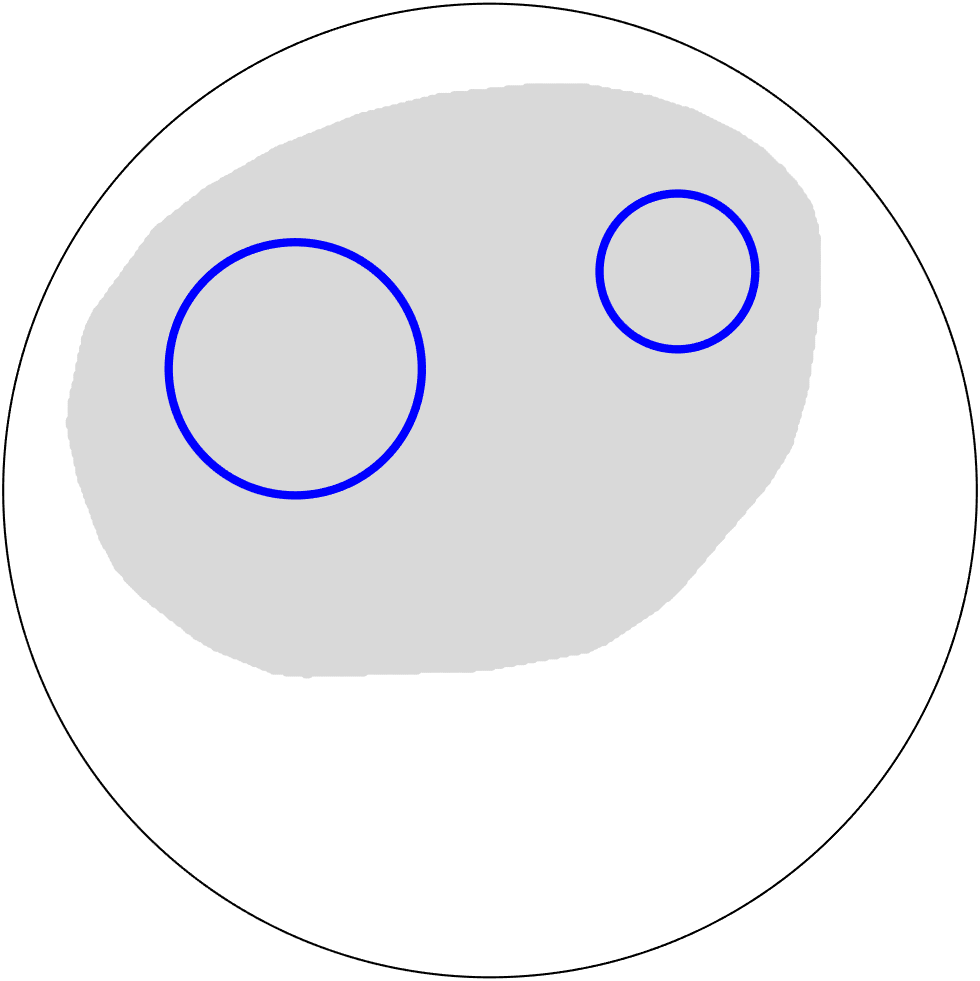} \newline 33.0\%
        \renewcommand{\thefigure}{2.3}
         \captionlistentry{} 
         \label{fig:experimental_results_2-3}
        \end{minipage}
        & \begin{minipage}{0.18\textwidth}
        \centering \includegraphics[height=2.3cm]{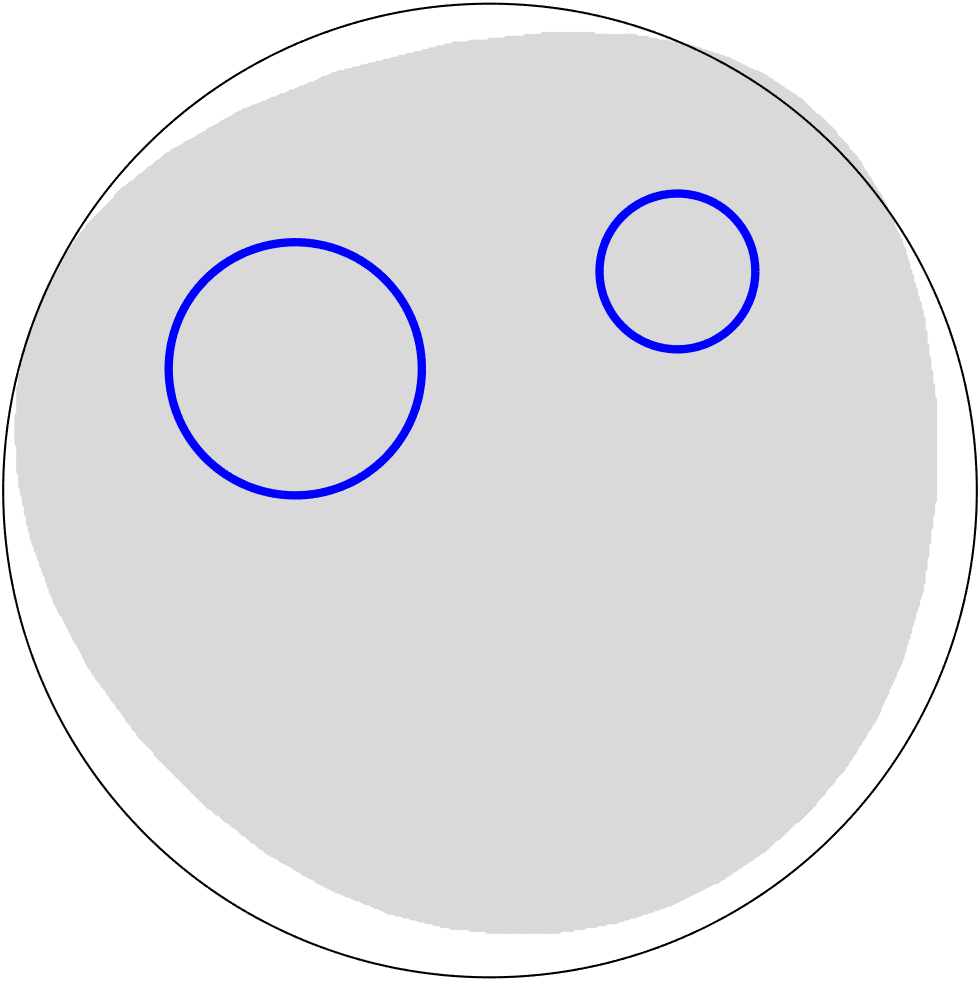} \newline 77.5\%
        \end{minipage}\\[1.3cm]
        2.4 
        &
        \begin{minipage}{0.25\textwidth}
        \centering \includegraphics[height=2.8cm]{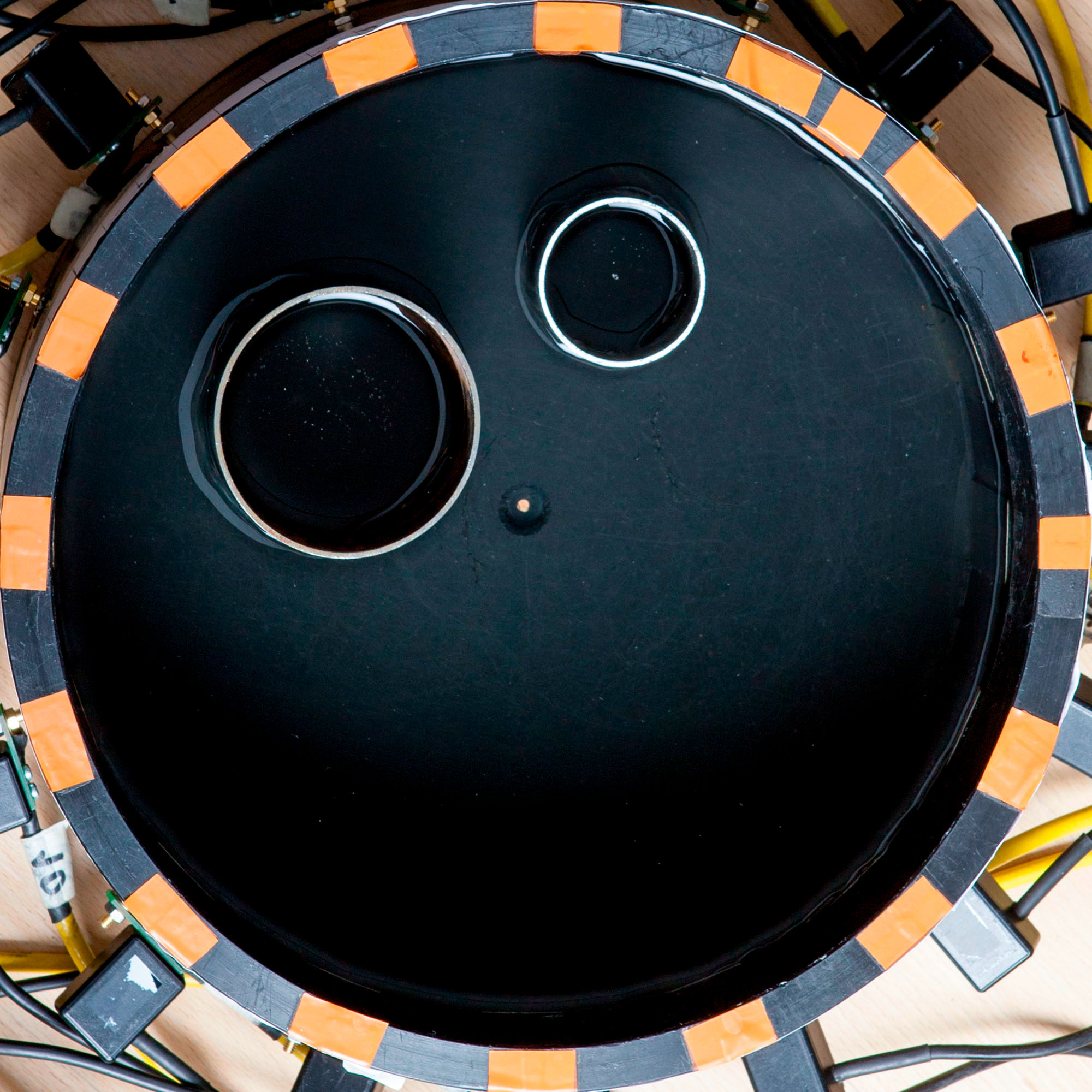} \newline
        \end{minipage}
        &\begin{minipage}{0.18\textwidth}
        \centering \includegraphics[height=2.3cm]{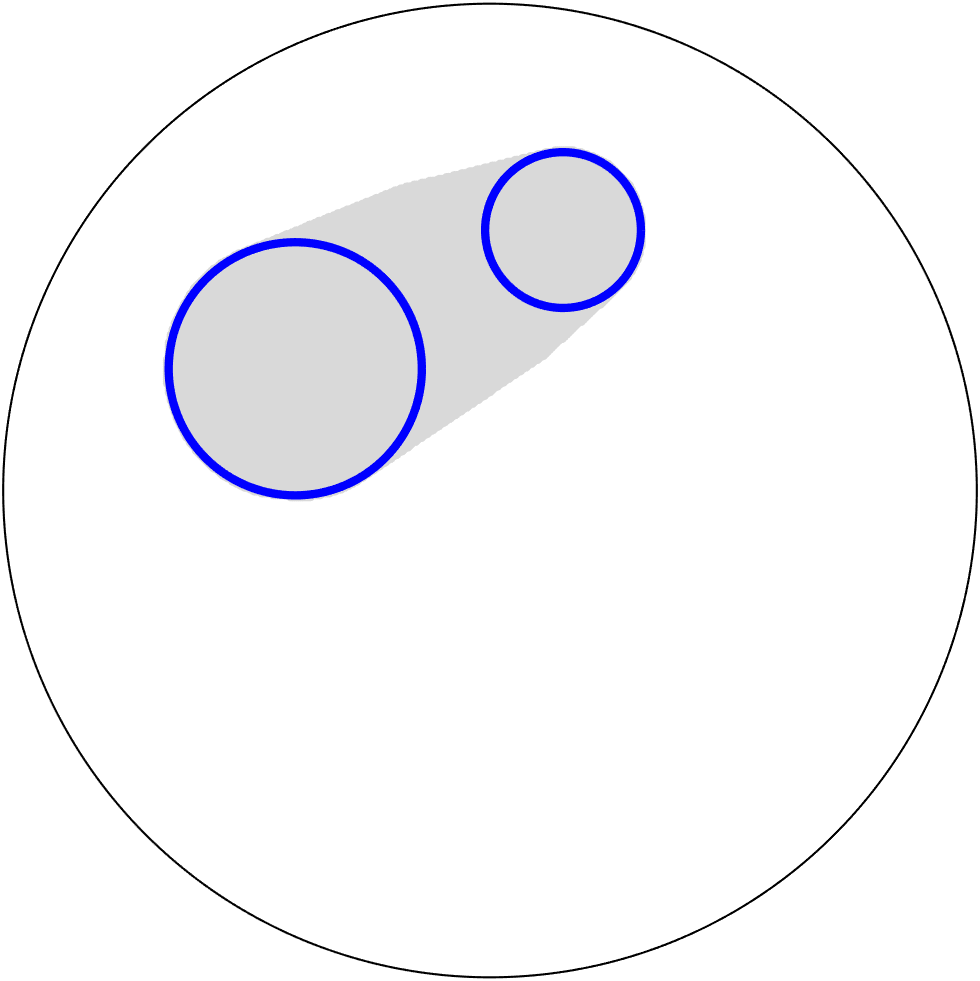} \newline
        \end{minipage}
        & \begin{minipage}{0.18\textwidth}
        \centering \includegraphics[height=2.3cm]{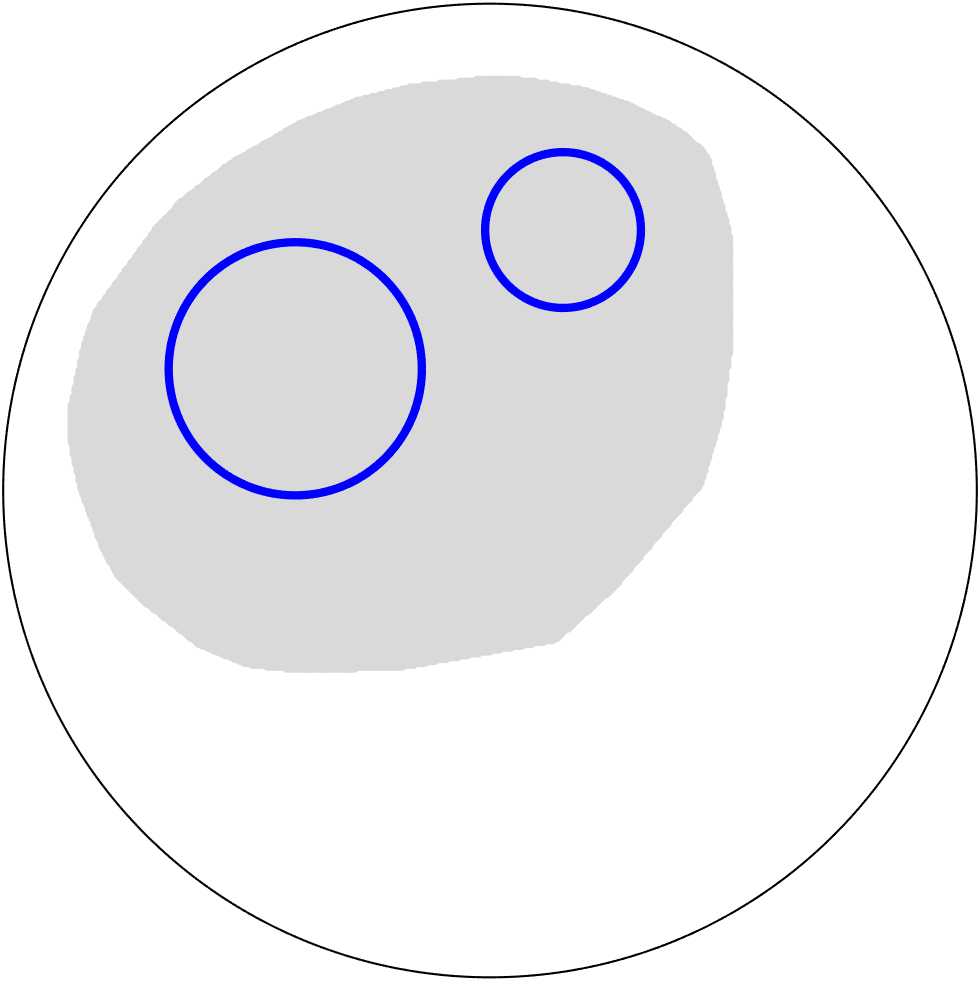} \newline 30.6\%
        \renewcommand{\thefigure}{2.4}
         \captionlistentry{} 
         \label{fig:experimental_results_2-4}
        \end{minipage}
        & \begin{minipage}{0.18\textwidth}
        \centering \includegraphics[height=2.3cm]{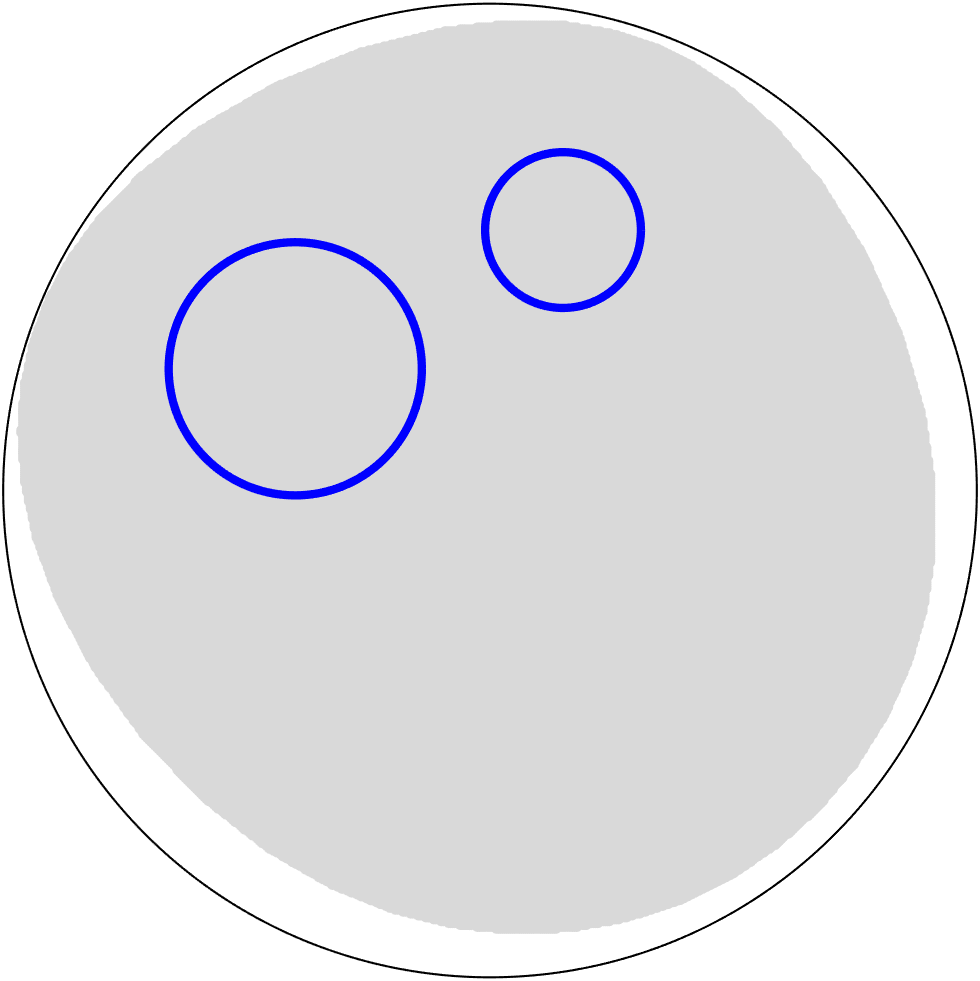} \newline 79.3\%
        \end{minipage}\\[1.3cm]
        2.5 
        &
        \begin{minipage}{0.25\textwidth}
        \centering \includegraphics[height=2.8cm]{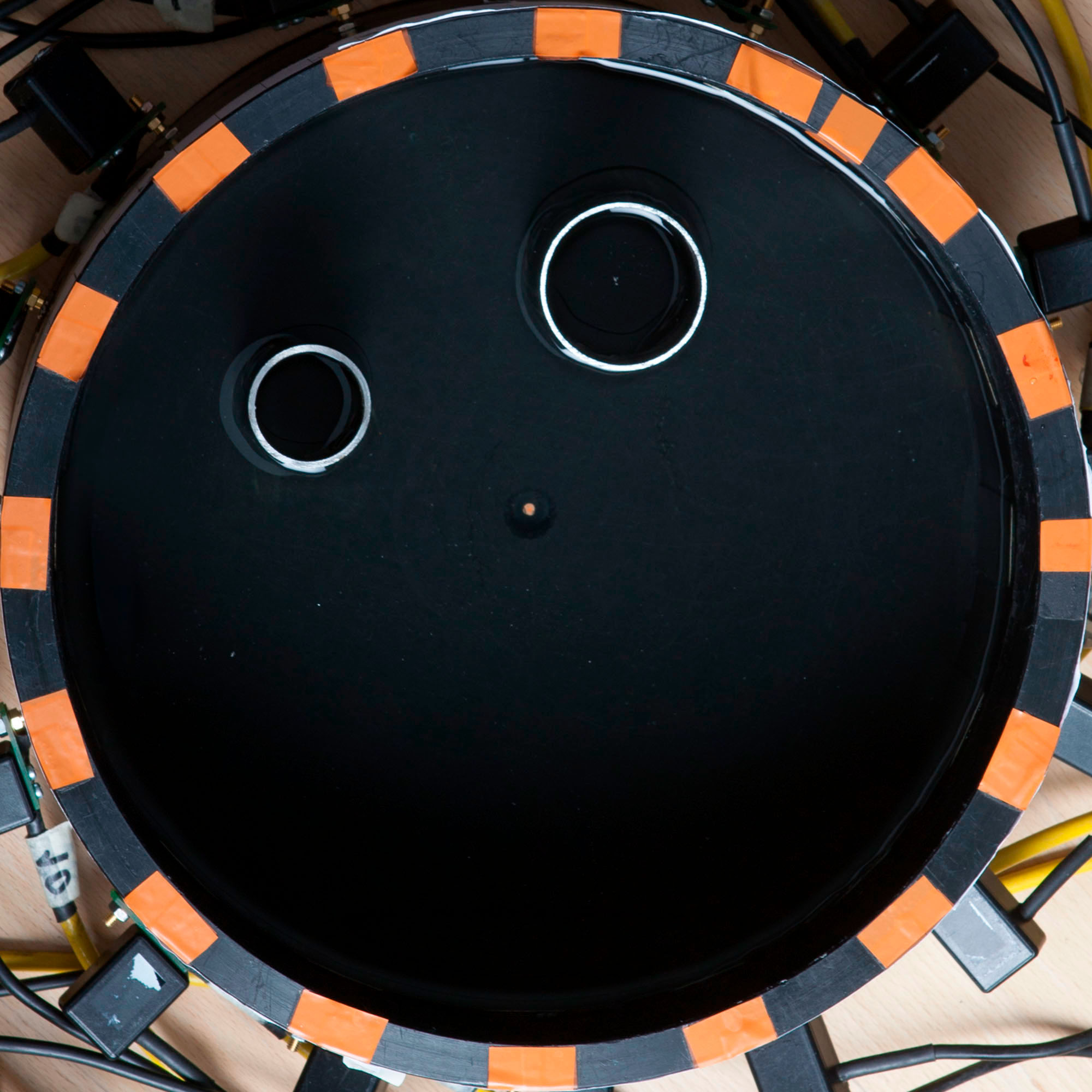} \newline
        \end{minipage}
        &\begin{minipage}{0.18\textwidth}
        \centering \includegraphics[height=2.3cm]{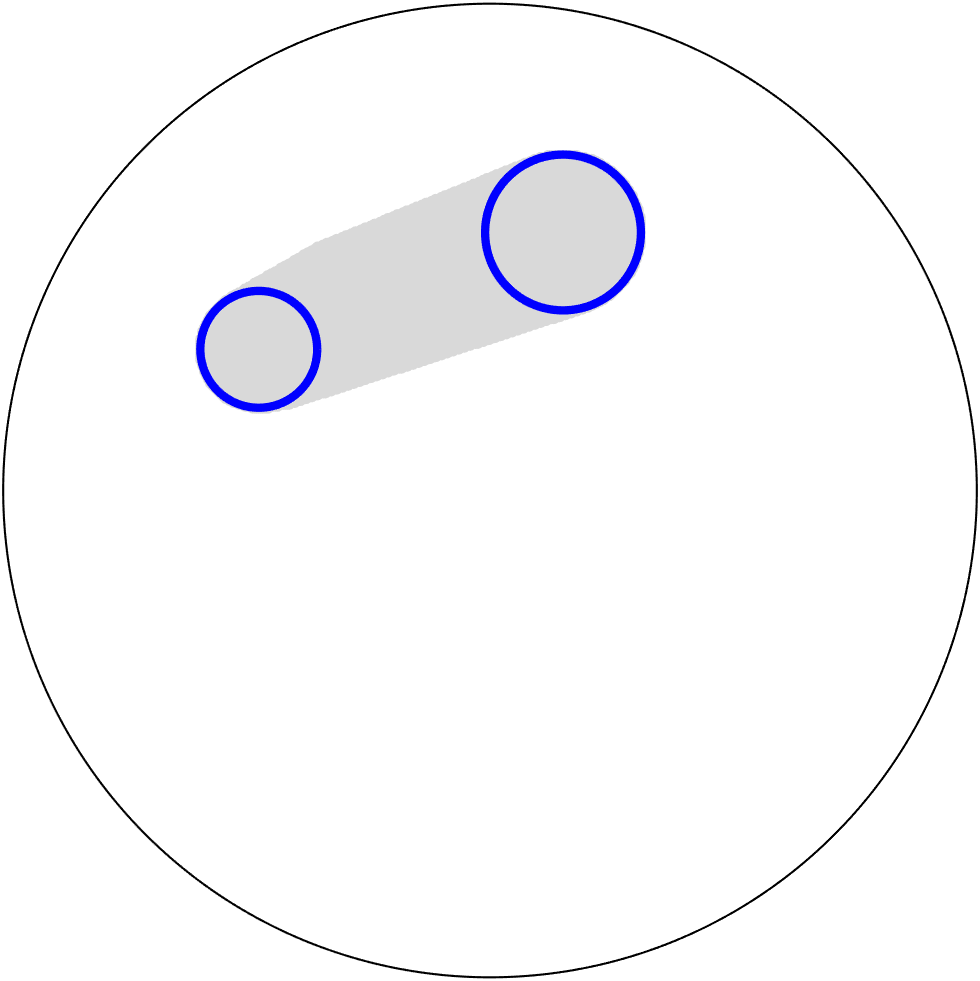} \newline
        \end{minipage}
        & \begin{minipage}{0.18\textwidth}
        \centering \includegraphics[height=2.3cm]{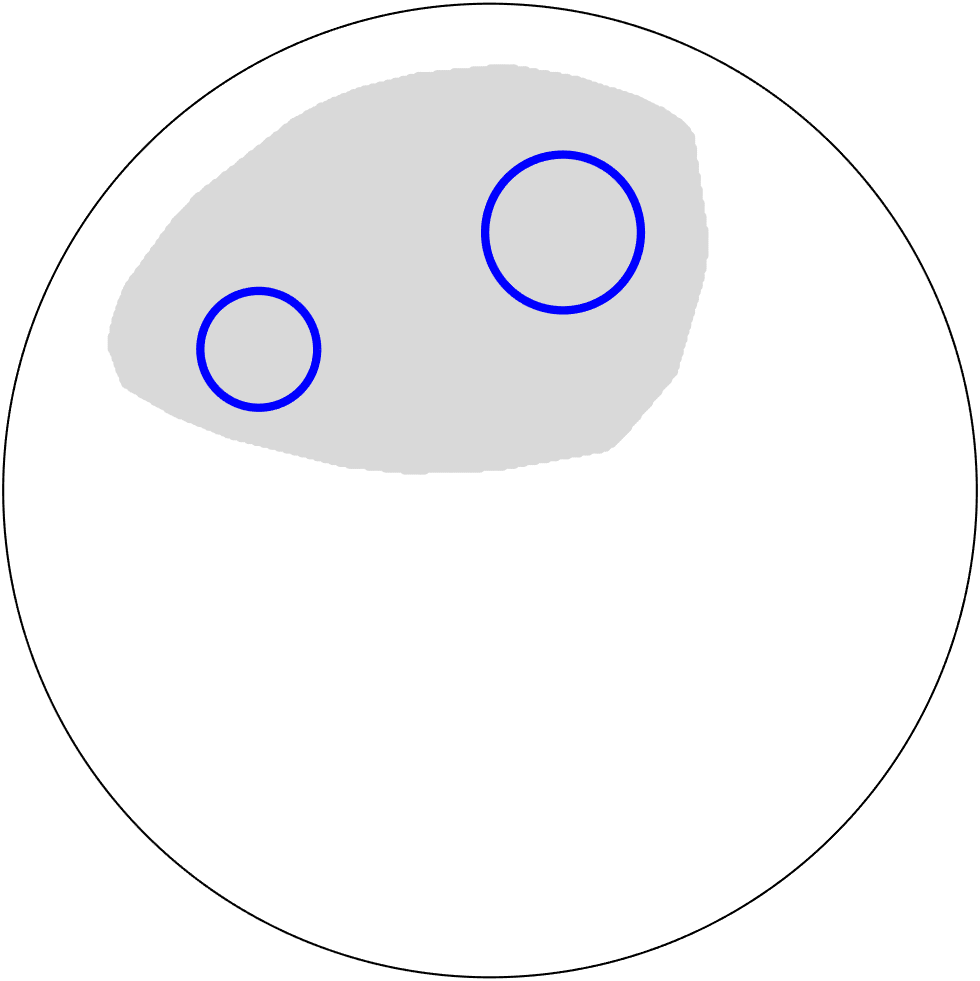} \newline 17.8\%
        \renewcommand{\thefigure}{2.5}
         \captionlistentry{} 
         \label{fig:experimental_results_2-5}
        \end{minipage}
        & \begin{minipage}{0.18\textwidth}
        \centering \includegraphics[height=2.3cm]{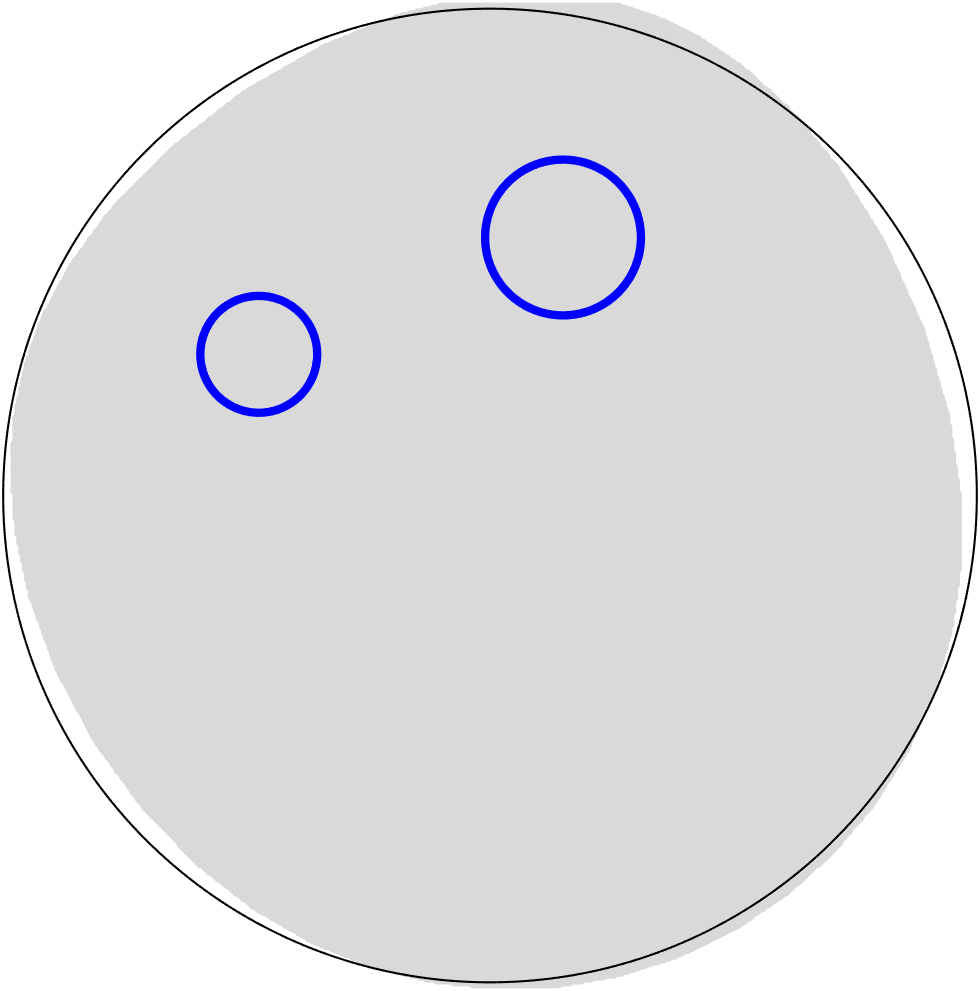} \newline 97.2\%
        \end{minipage}\\[1.3cm]
        2.6 
        &
        \begin{minipage}{0.25\textwidth}
        \centering \includegraphics[height=2.8cm]{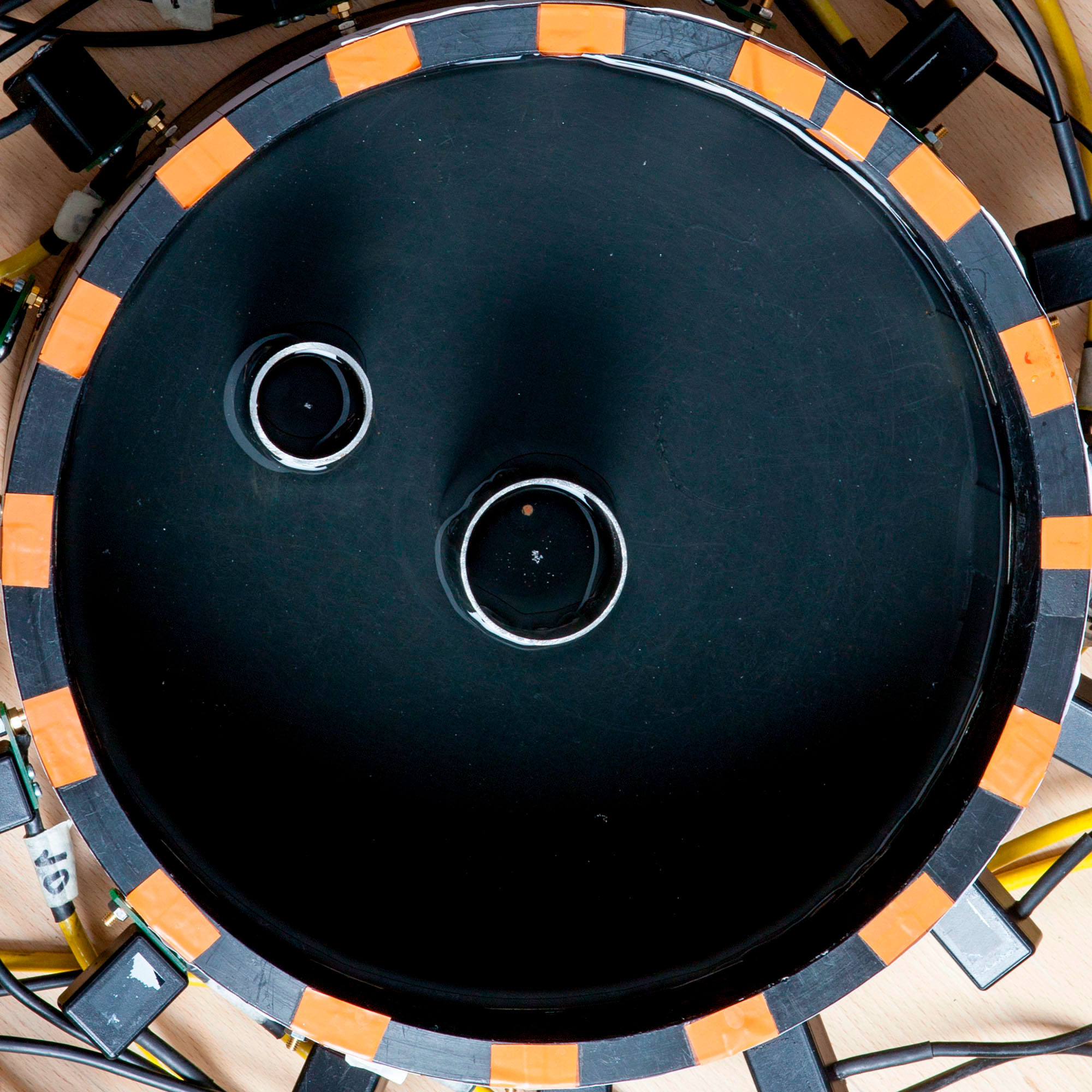} \newline
        \end{minipage}
        &\begin{minipage}{0.18\textwidth}
        \centering \includegraphics[height=2.3cm]{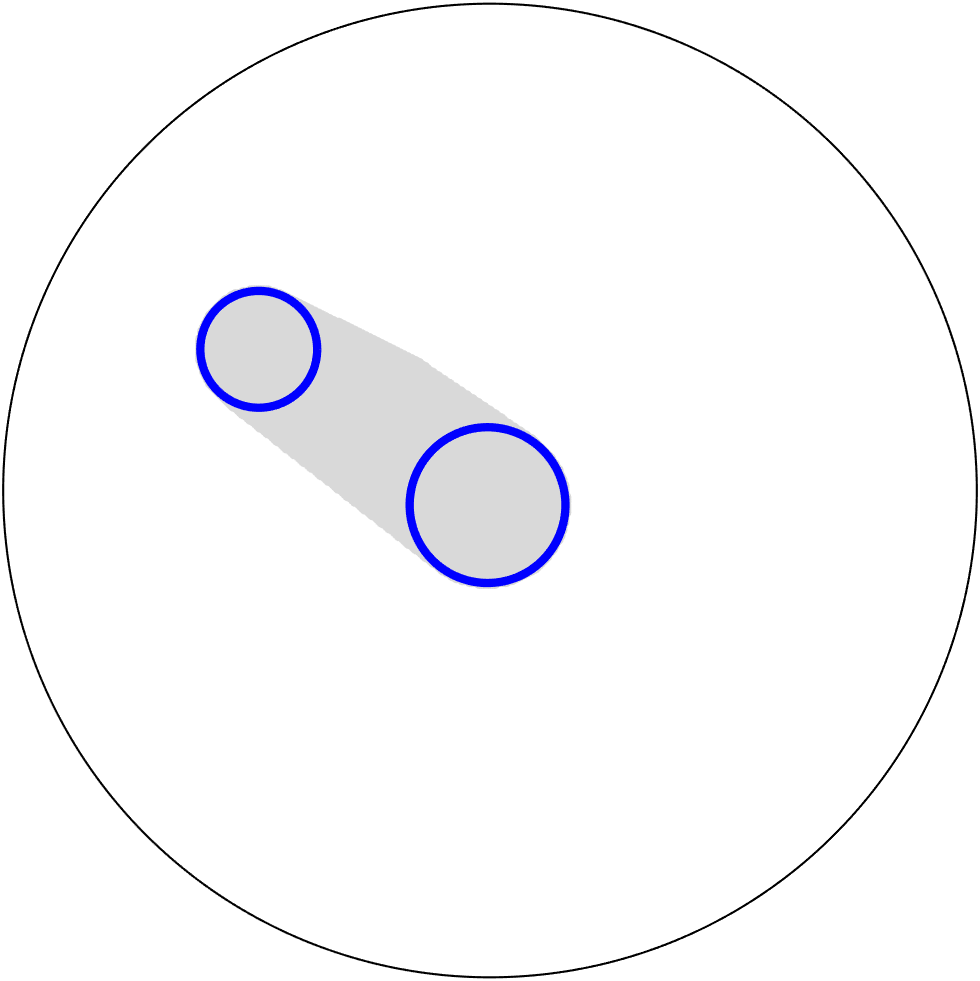} \newline
        \end{minipage}
        & \begin{minipage}{0.18\textwidth}
        \centering \includegraphics[height=2.3cm]{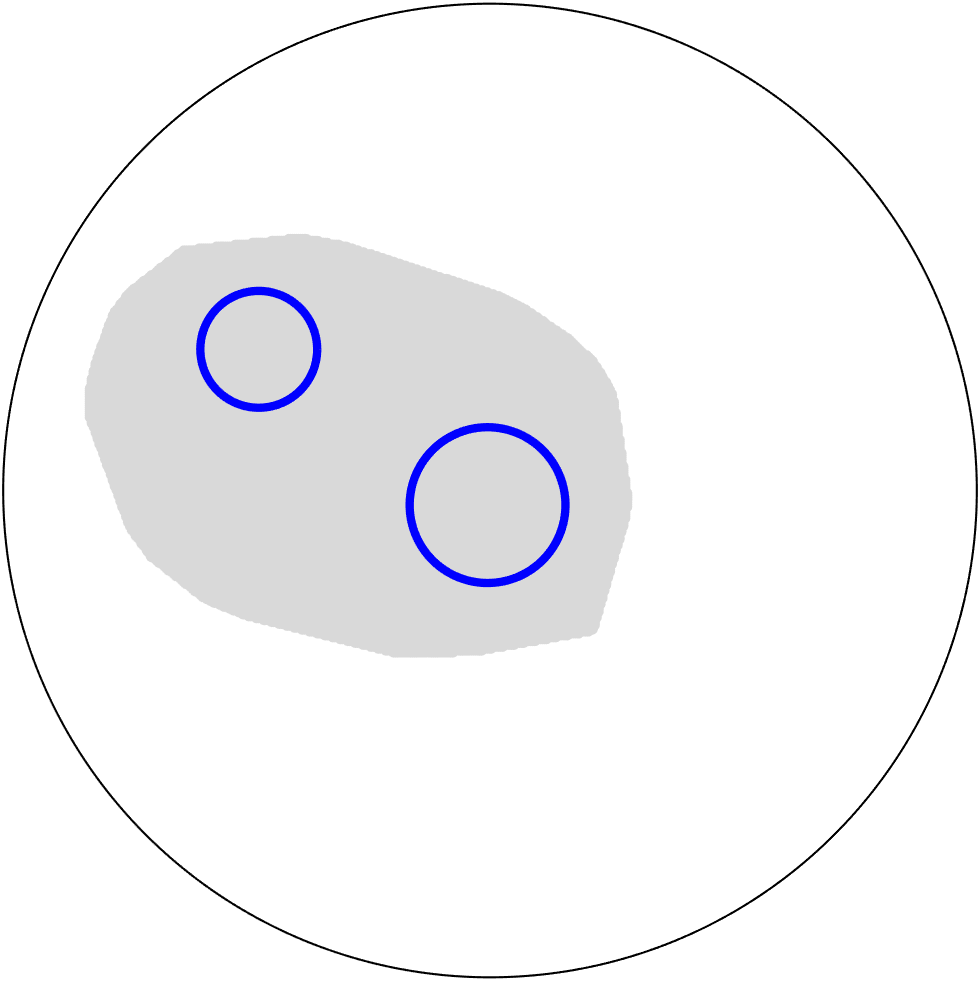} \newline 17.8\%
        \renewcommand{\thefigure}{2.6}
         \captionlistentry{} 
         \label{fig:experimental_results_2-6}
        \end{minipage}
        & \begin{minipage}{0.18\textwidth}
        \centering \includegraphics[height=2.3cm]{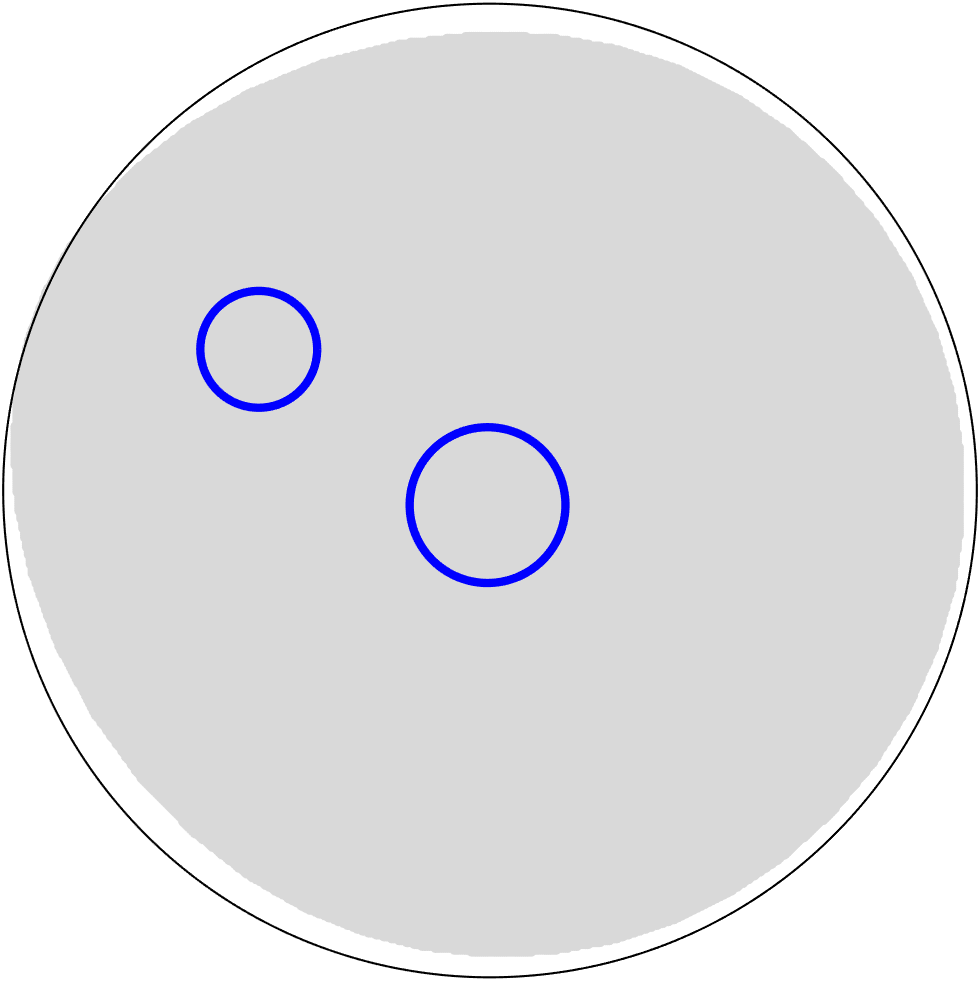} \newline 92.5\%
        \end{minipage}
    \end{tabular} 
  \addtocounter{figure}{-6}
    \caption{Comparison of the ground truth, learned and least squares hulls of the experimental phantoms. The error relative to the ground truth is shown below each phantom.}
    \label{fig:experimental_results_1}
\end{figure}

\clearpage
\begin{figure}[ht]
    \begin{tabular}{lccccc}
        Case & Phantom & Ground truth & Learned & Least squares \\ [.2cm]
        3.1 
        &
        \begin{minipage}{0.25\textwidth}
        \centering \includegraphics[height=2.8cm]{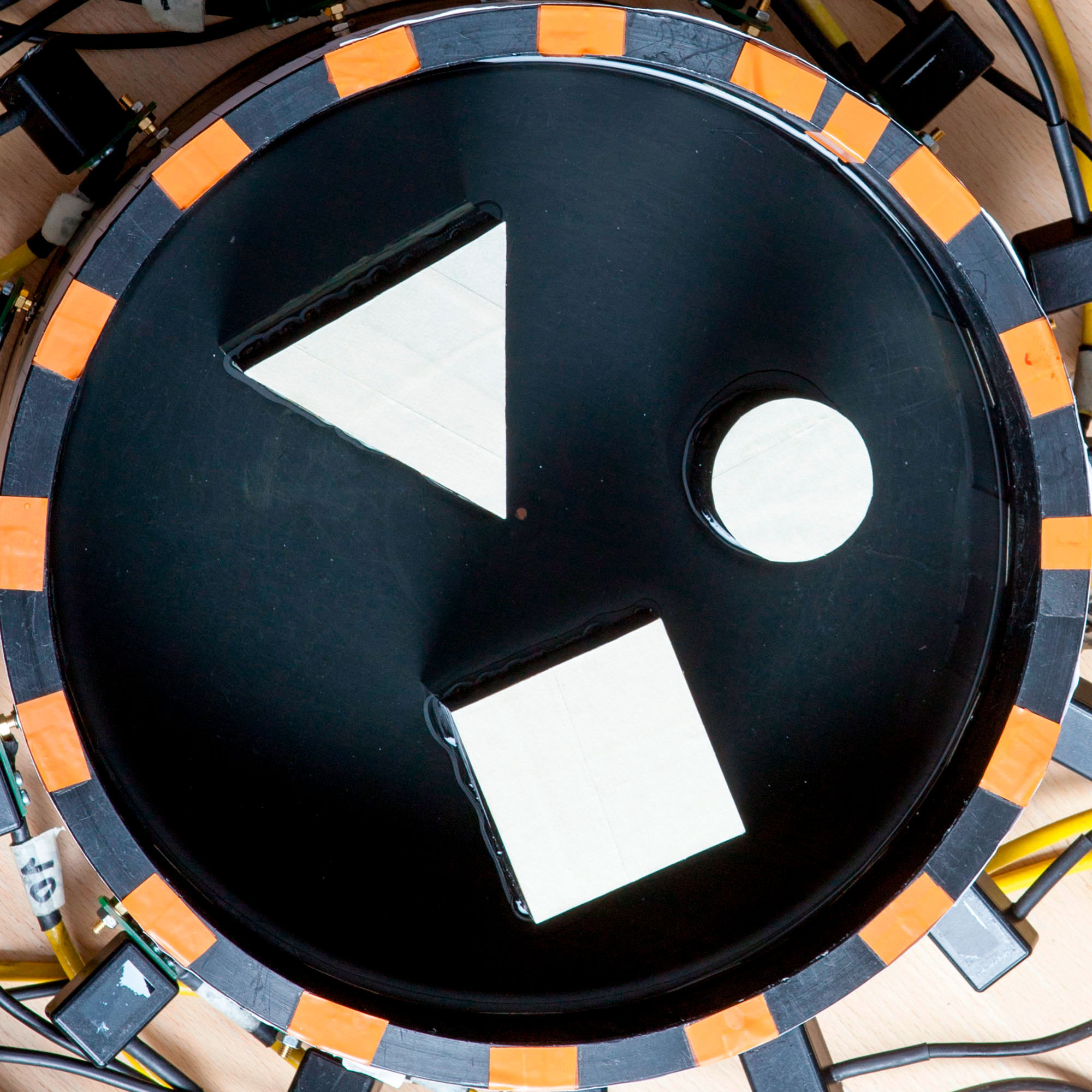} \newline
        \end{minipage}
        &\begin{minipage}{0.18\textwidth}
        \centering \includegraphics[height=2.3cm]{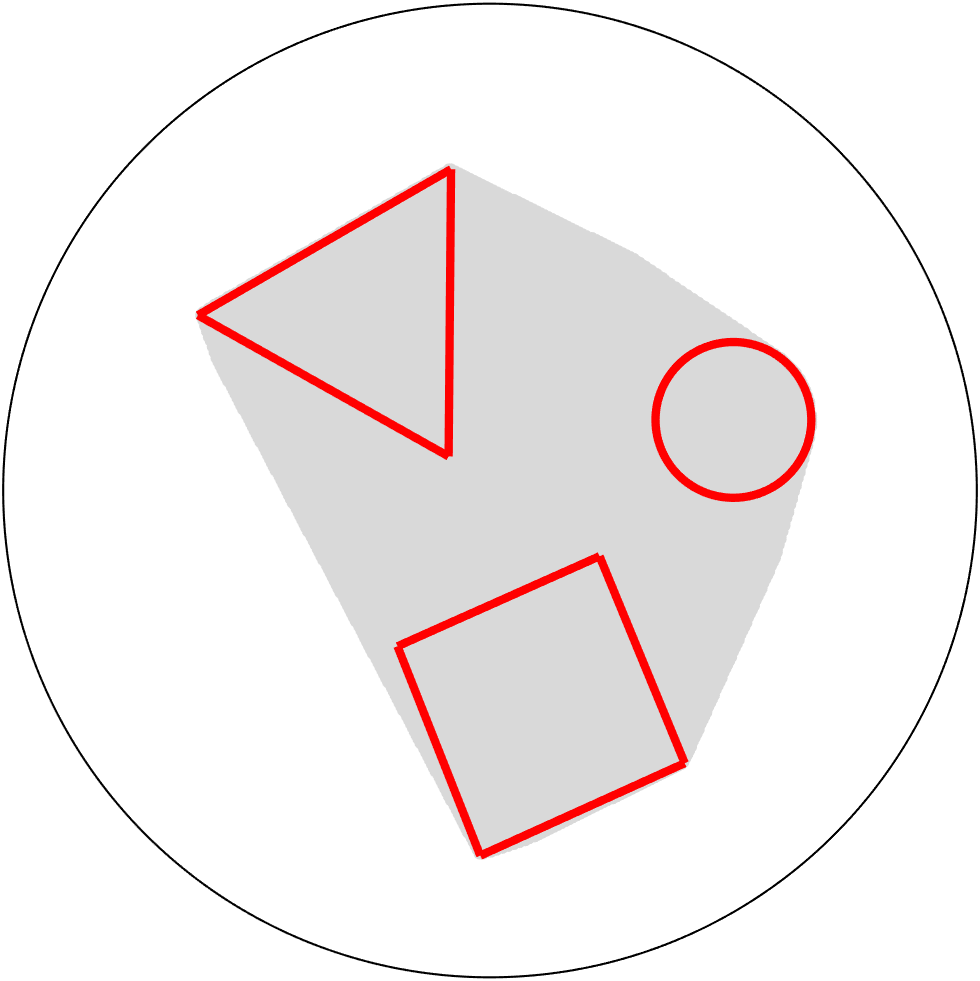} \newline
        \end{minipage}
        & \begin{minipage}{0.18\textwidth}
        \centering \includegraphics[height=2.3cm]{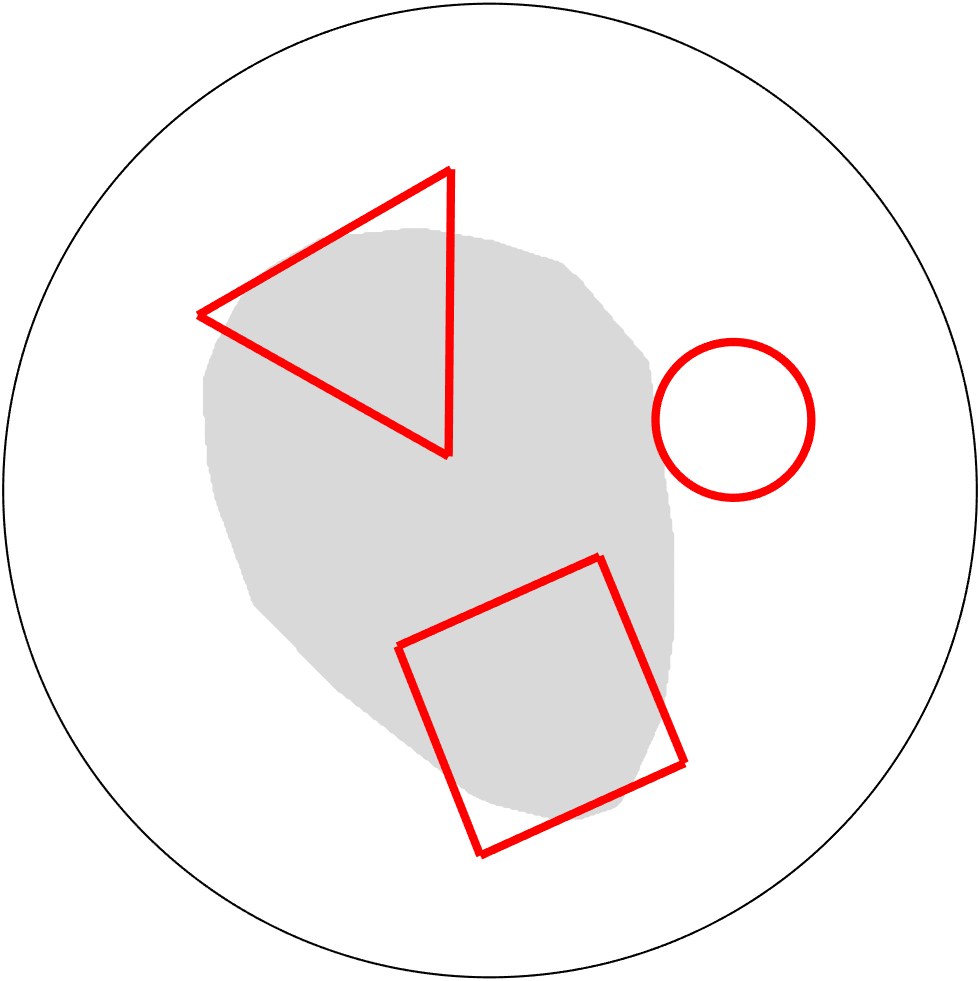} \newline 13.8\%
        \renewcommand{\thefigure}{3.1}
         \captionlistentry{} 
         \label{fig:experimental_results_3-1}
        \end{minipage}
        & \begin{minipage}{0.18\textwidth}
        \centering \includegraphics[height=2.3cm]{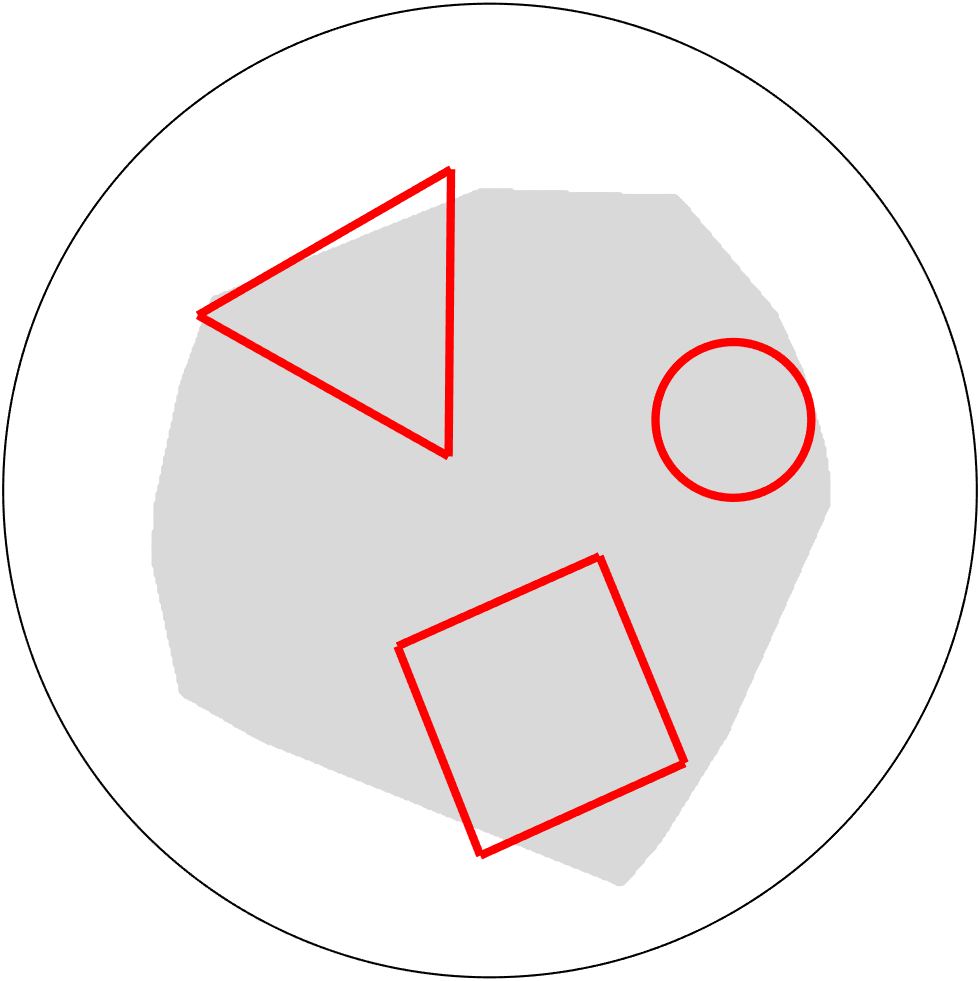} \newline 14.3\%
        \end{minipage}\\[1.3cm]
        3.2 
        &
        \begin{minipage}{0.25\textwidth}
        \centering \includegraphics[height=2.8cm]{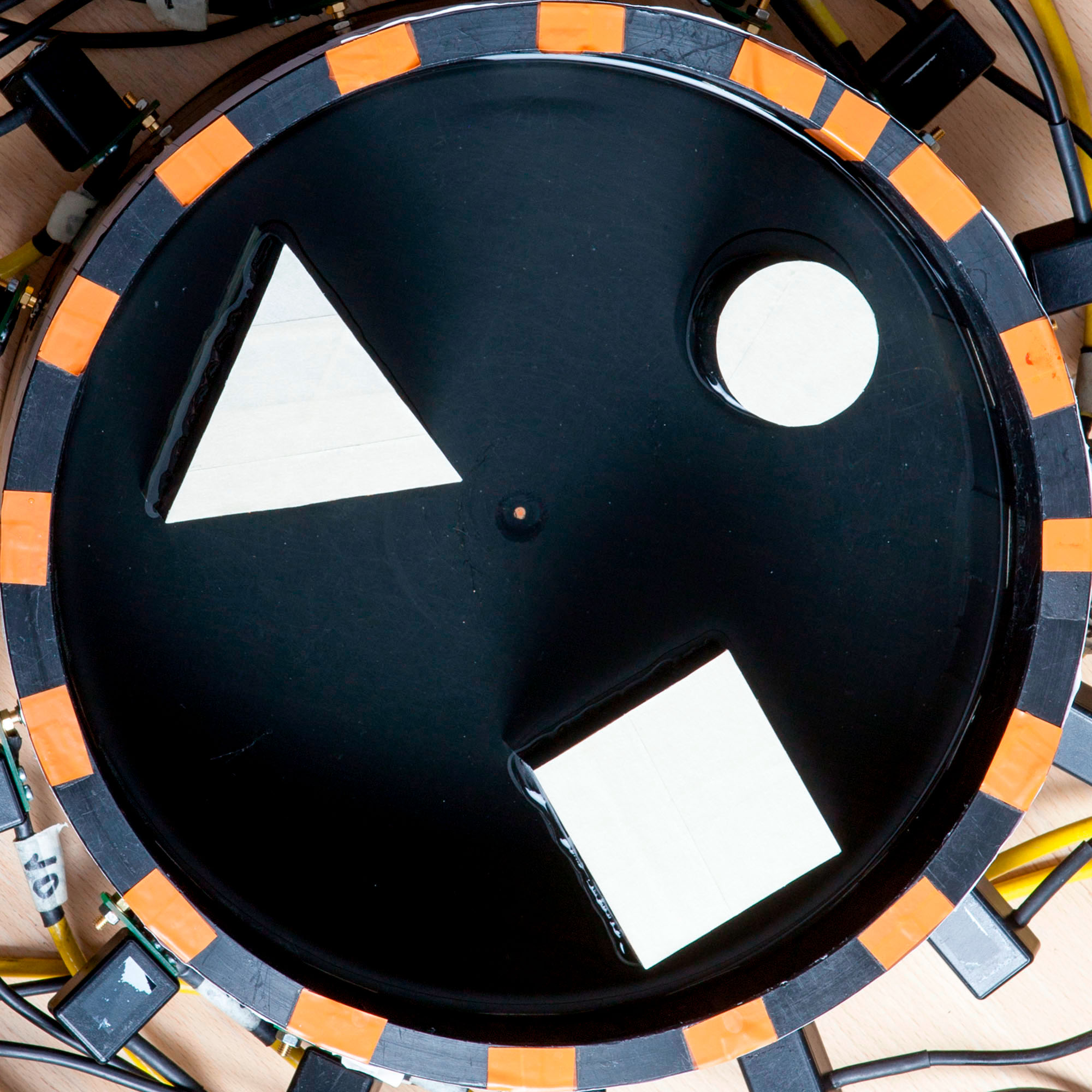} \newline
        \end{minipage}
        &\begin{minipage}{0.18\textwidth}
        \centering \includegraphics[height=2.3cm]{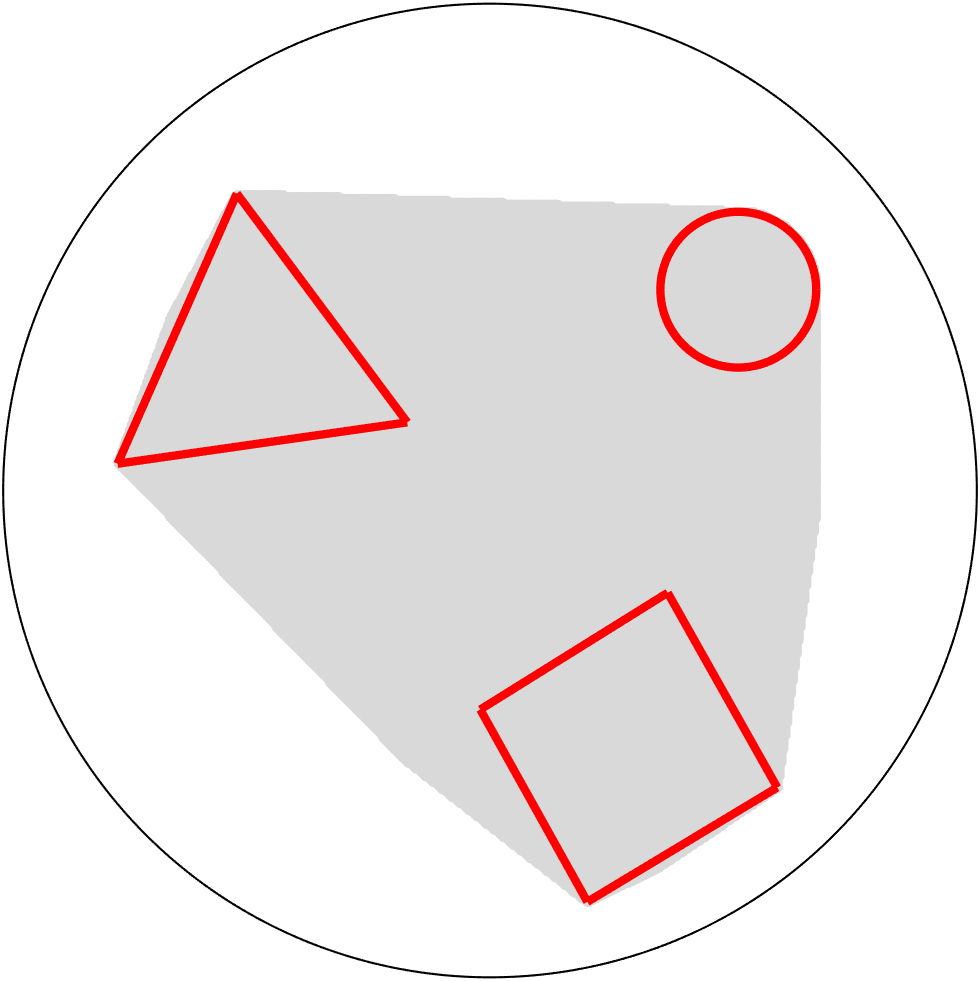} \newline
        \end{minipage}
        & \begin{minipage}{0.18\textwidth}
        \centering \includegraphics[height=2.3cm]{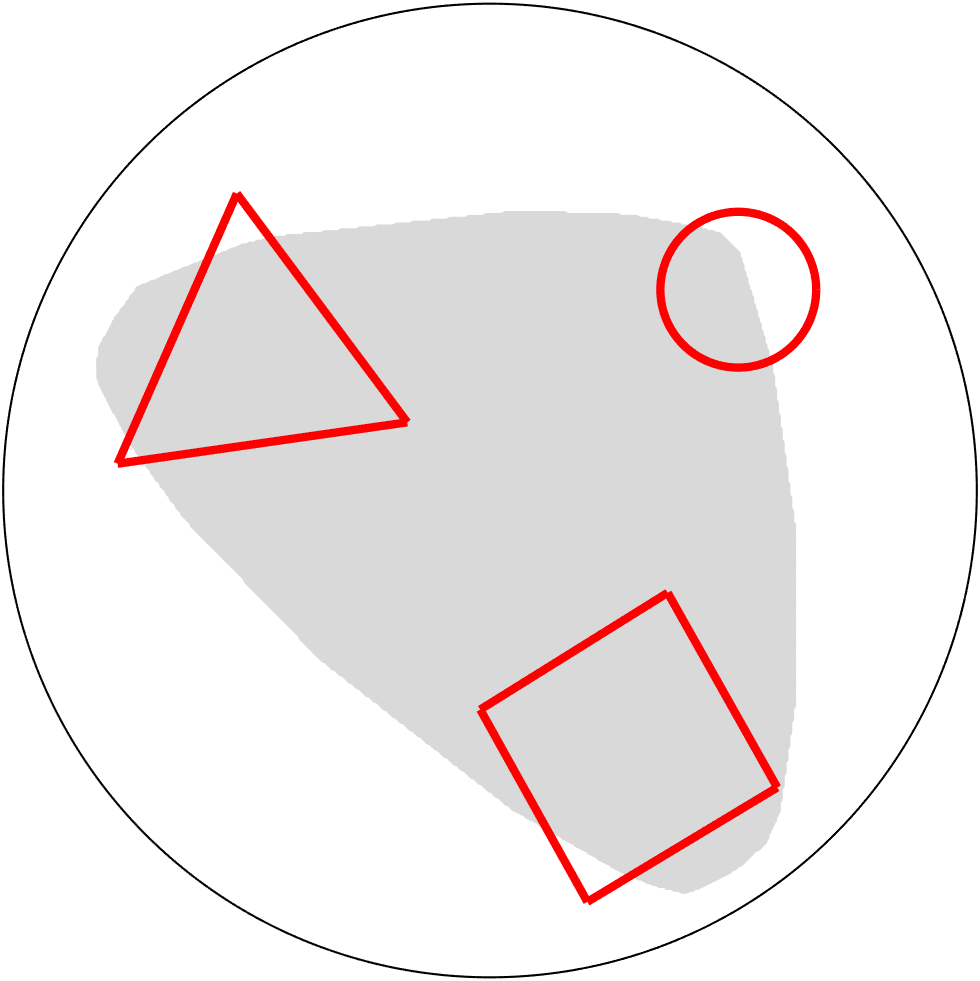} \newline 8.5\%
        \renewcommand{\thefigure}{3.2}
         \captionlistentry{} 
         \label{fig:experimental_results_3-2}
        \end{minipage}
        & \begin{minipage}{0.18\textwidth}
        \centering \includegraphics[height=2.3cm]{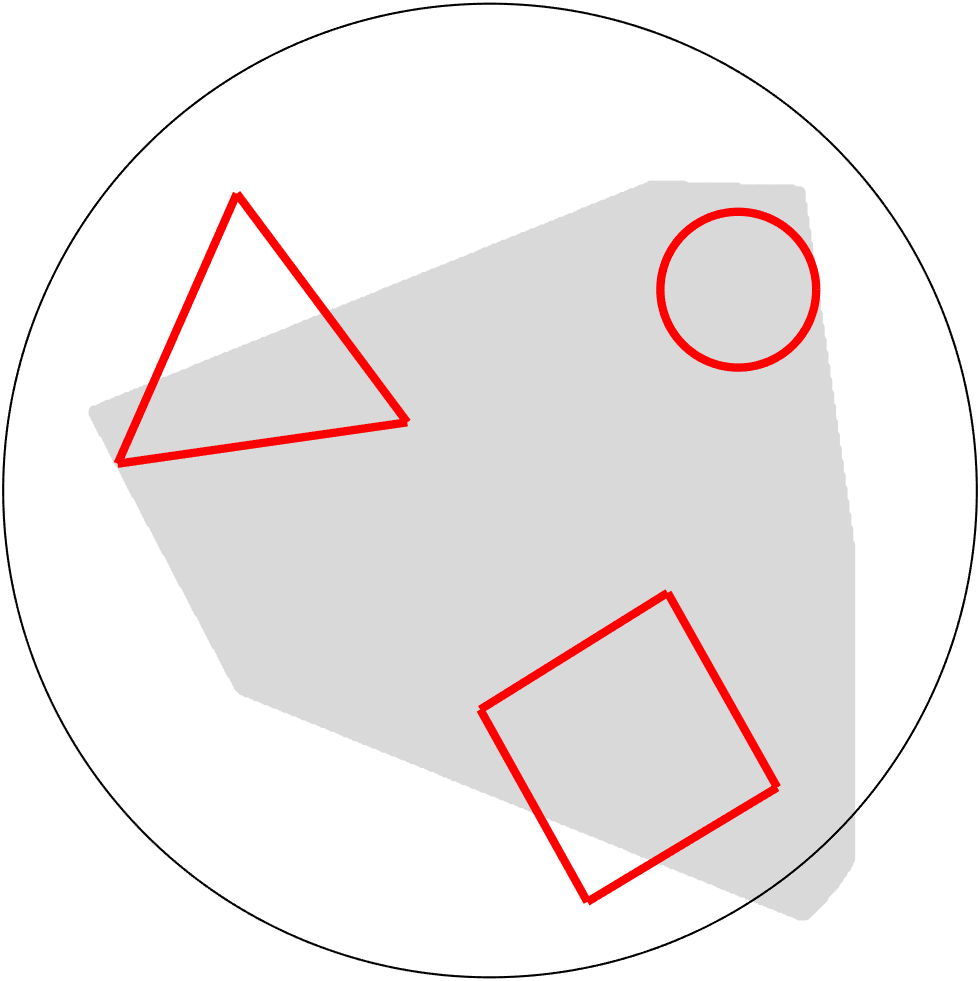} \newline 14.8\%
        \end{minipage}\\[1.3cm]
        3.3 
        &
        \begin{minipage}{0.25\textwidth}
        \centering \includegraphics[height=2.8cm]{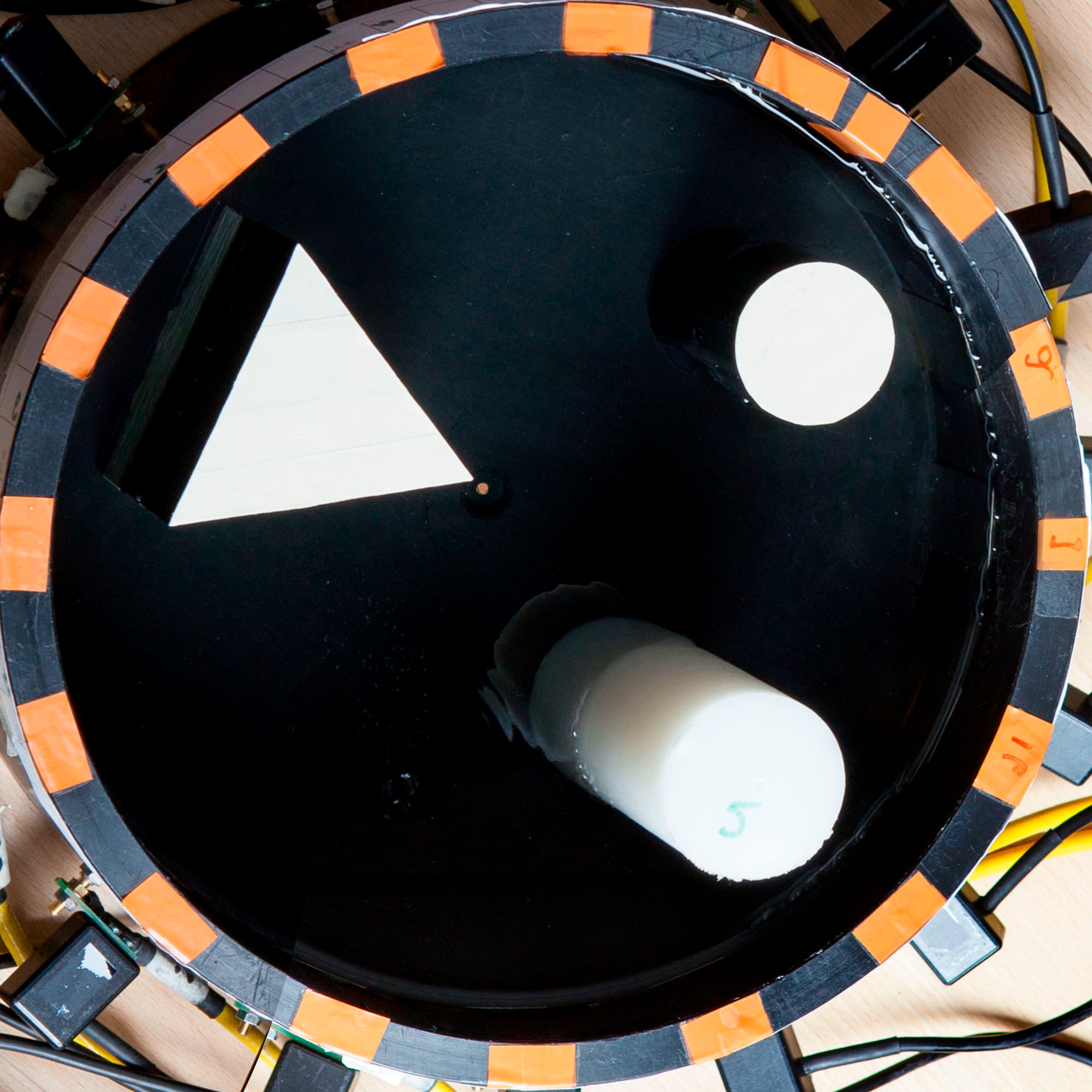} \newline
        \end{minipage}
        &\begin{minipage}{0.18\textwidth}
        \centering \includegraphics[height=2.3cm]{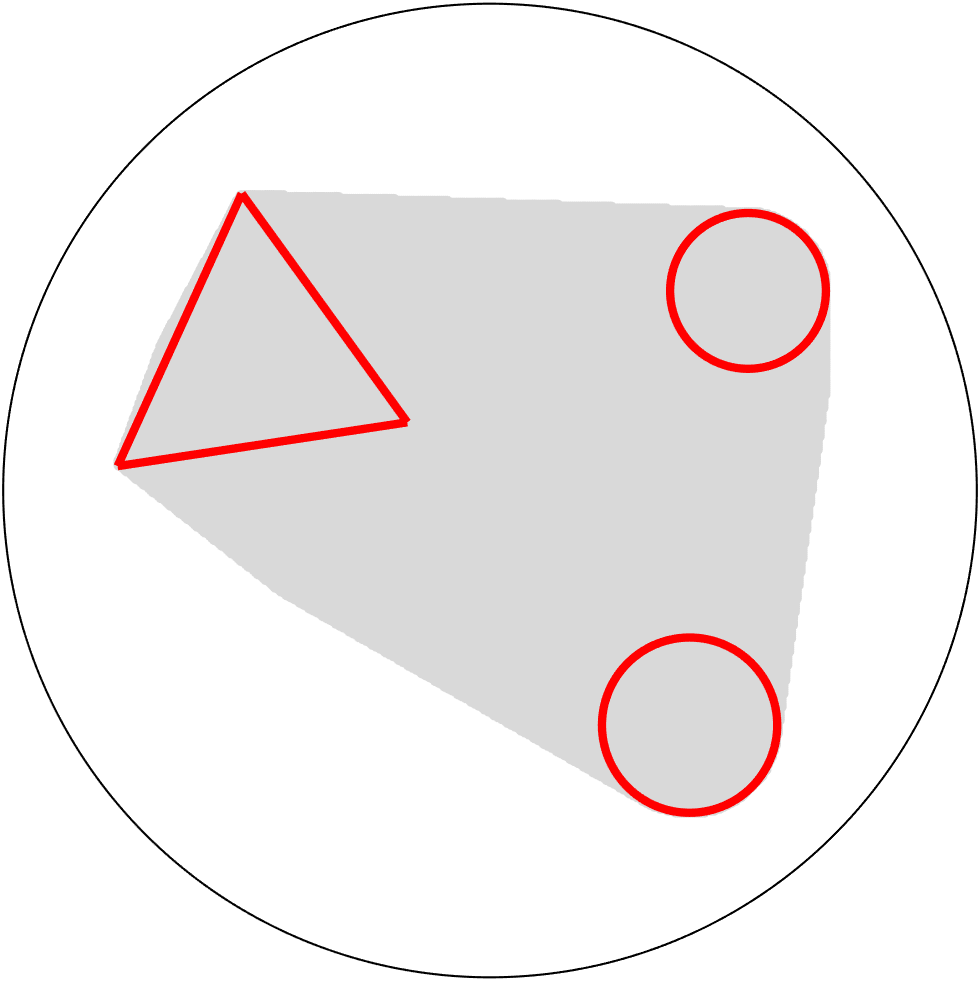} \newline
        \end{minipage}
        & \begin{minipage}{0.18\textwidth}
        \centering \includegraphics[height=2.3cm]{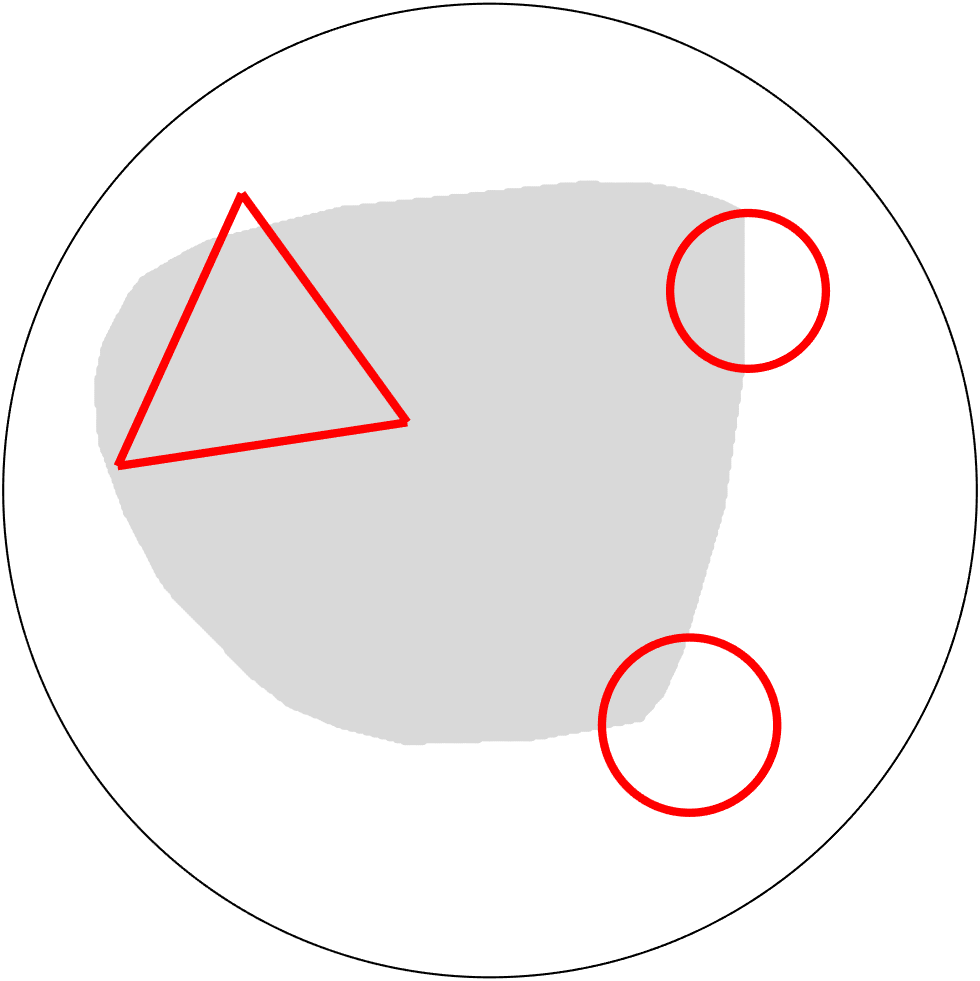} \newline 16.3\%
        \renewcommand{\thefigure}{3.3}
         \captionlistentry{} 
         \label{fig:experimental_results_3-3}
        \end{minipage}
        & \begin{minipage}{0.18\textwidth}
        \centering \includegraphics[height=2.3cm]{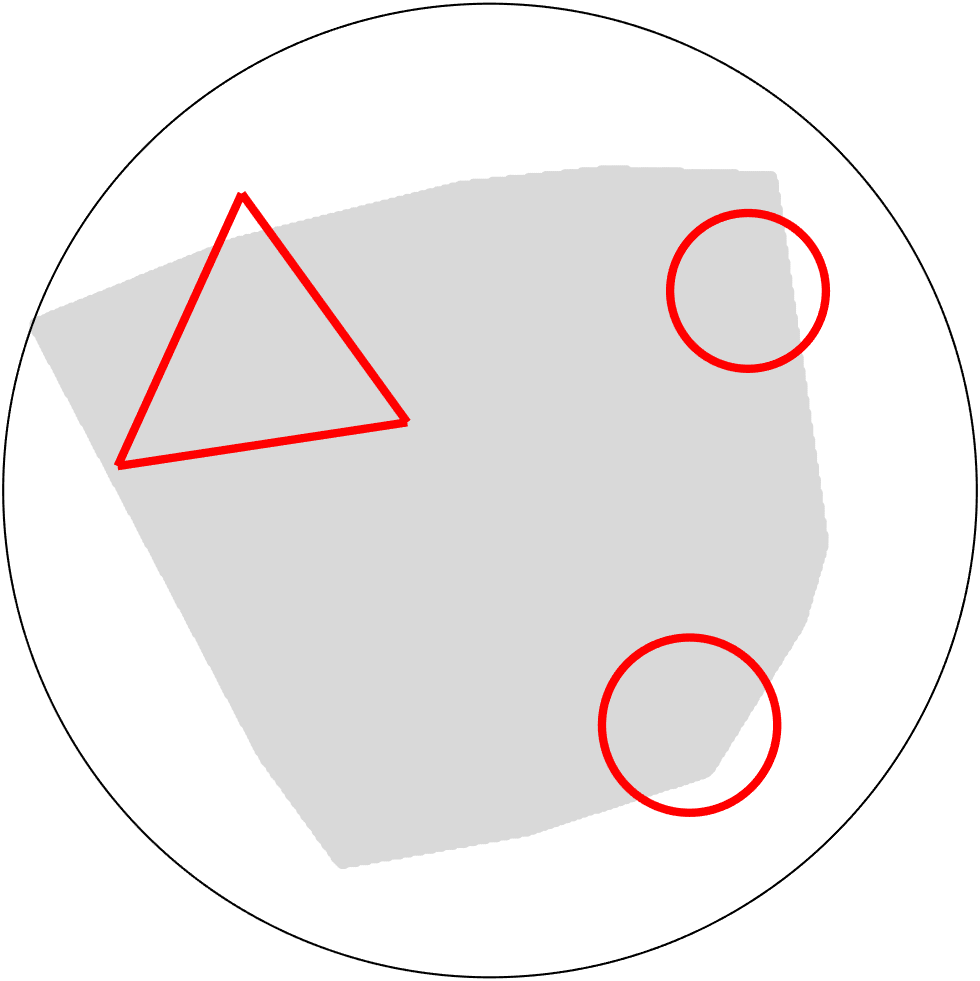} \newline 16.7\%
        \end{minipage}\\[1.3cm]
        3.4 
        &
        \begin{minipage}{0.25\textwidth}
        \centering \includegraphics[height=2.8cm]{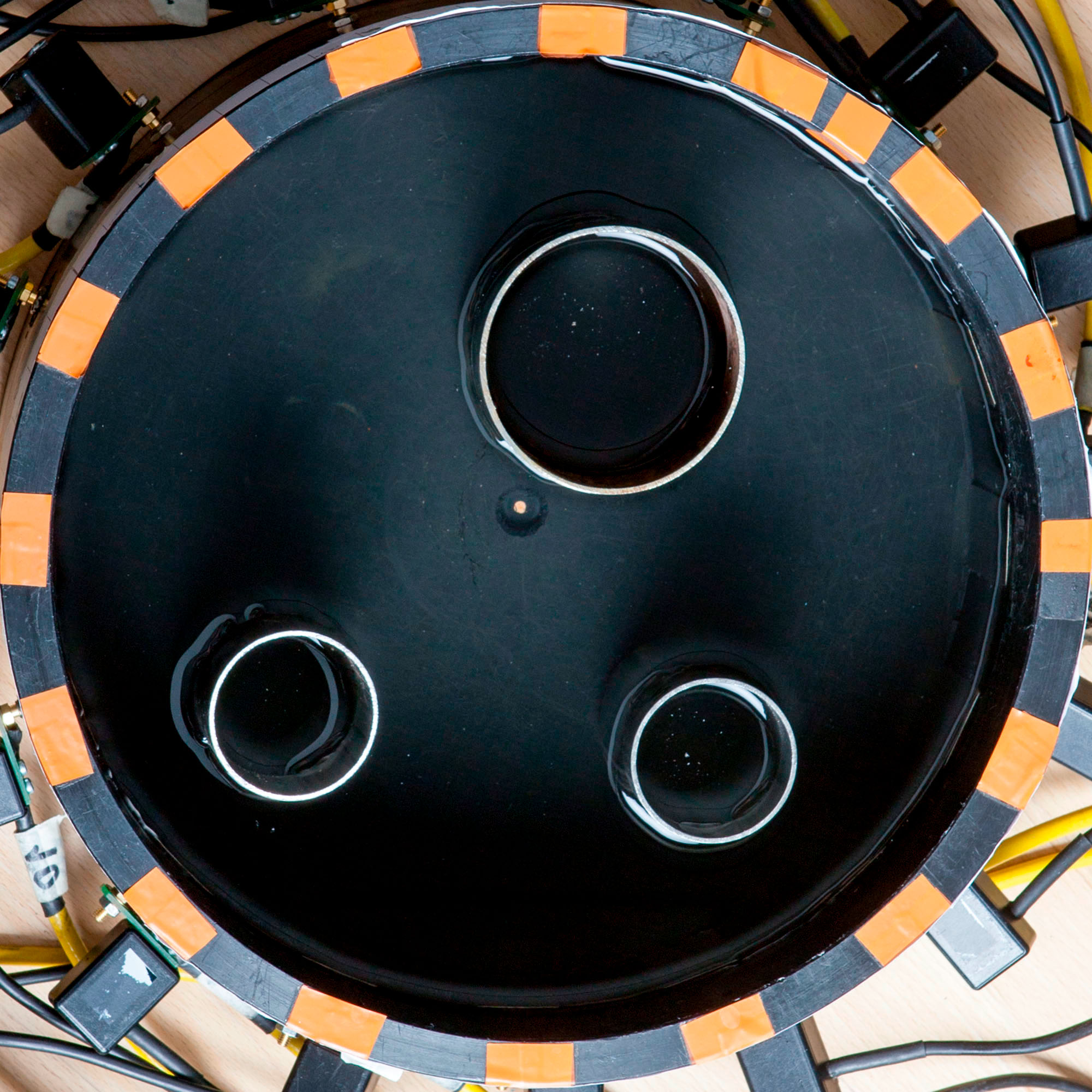} \newline
        \end{minipage}
        &\begin{minipage}{0.18\textwidth}
        \centering \includegraphics[height=2.3cm]{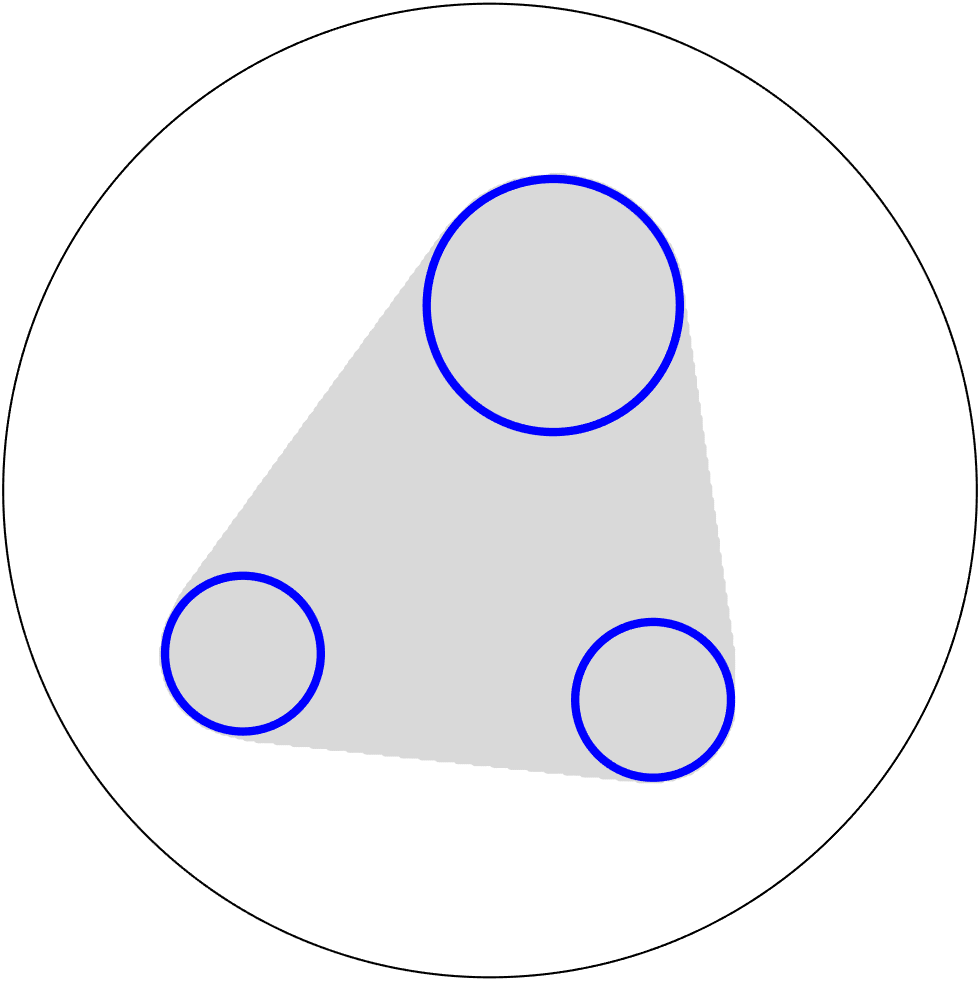} \newline
        \end{minipage}
        & \begin{minipage}{0.18\textwidth}
        \centering \includegraphics[height=2.3cm]{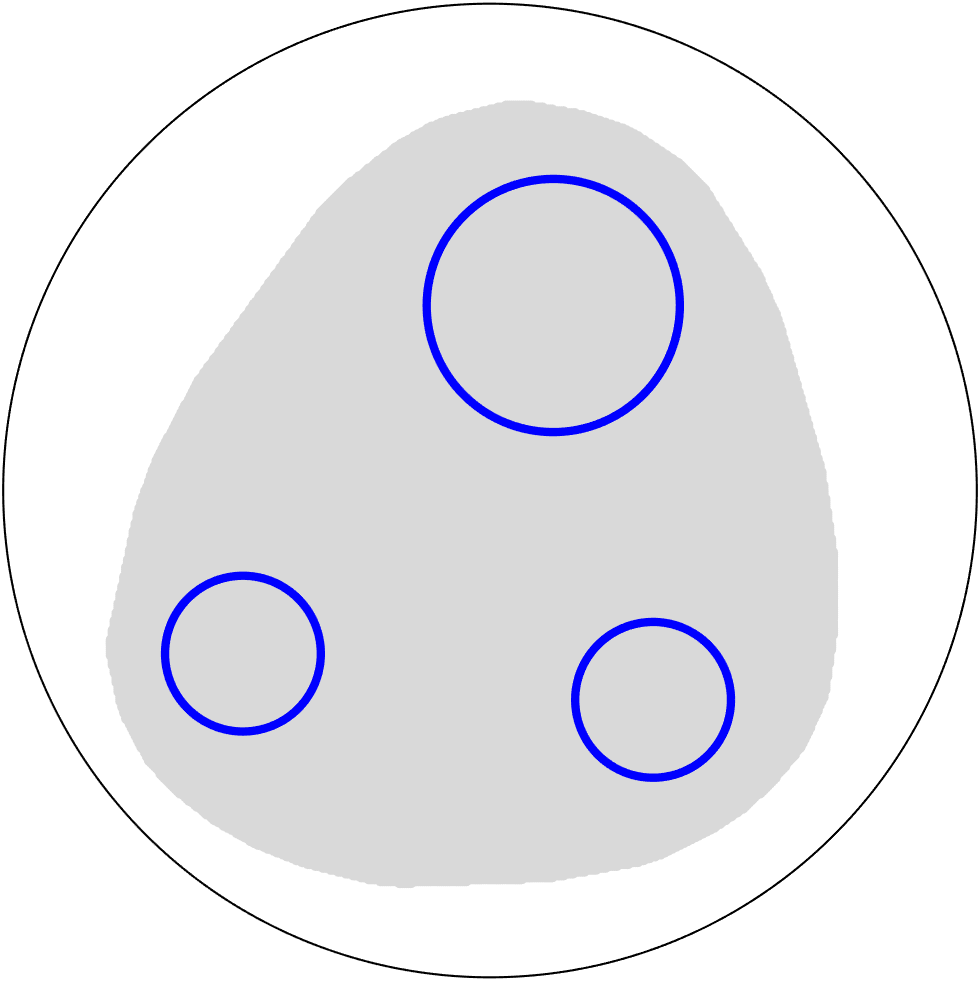} \newline 29.2\%
        \renewcommand{\thefigure}{3.4}
         \captionlistentry{} 
         \label{fig:experimental_results_3-4}
        \end{minipage}
        & \begin{minipage}{0.18\textwidth}
        \centering \includegraphics[height=2.3cm]{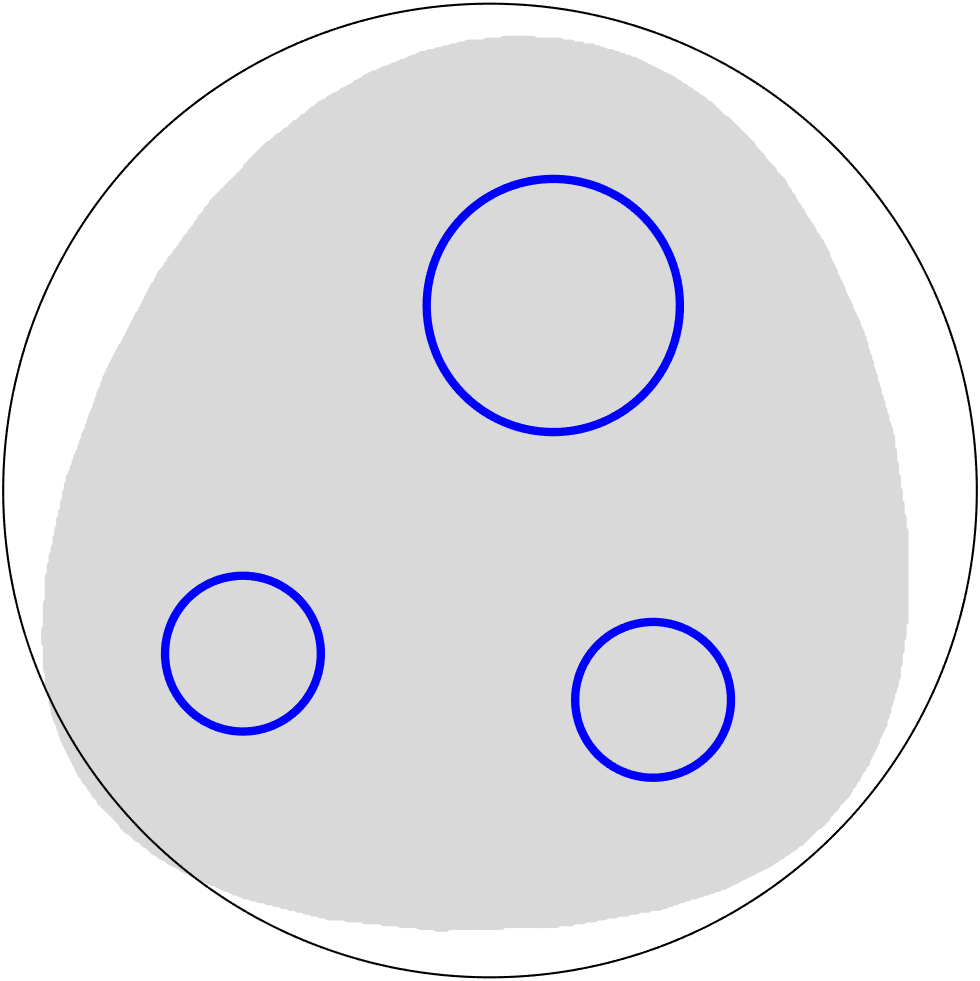} \newline 54.3\%
        \end{minipage}\\[1.3cm]
        3.5 
        &
        \begin{minipage}{0.25\textwidth}
        \centering \includegraphics[height=2.8cm]{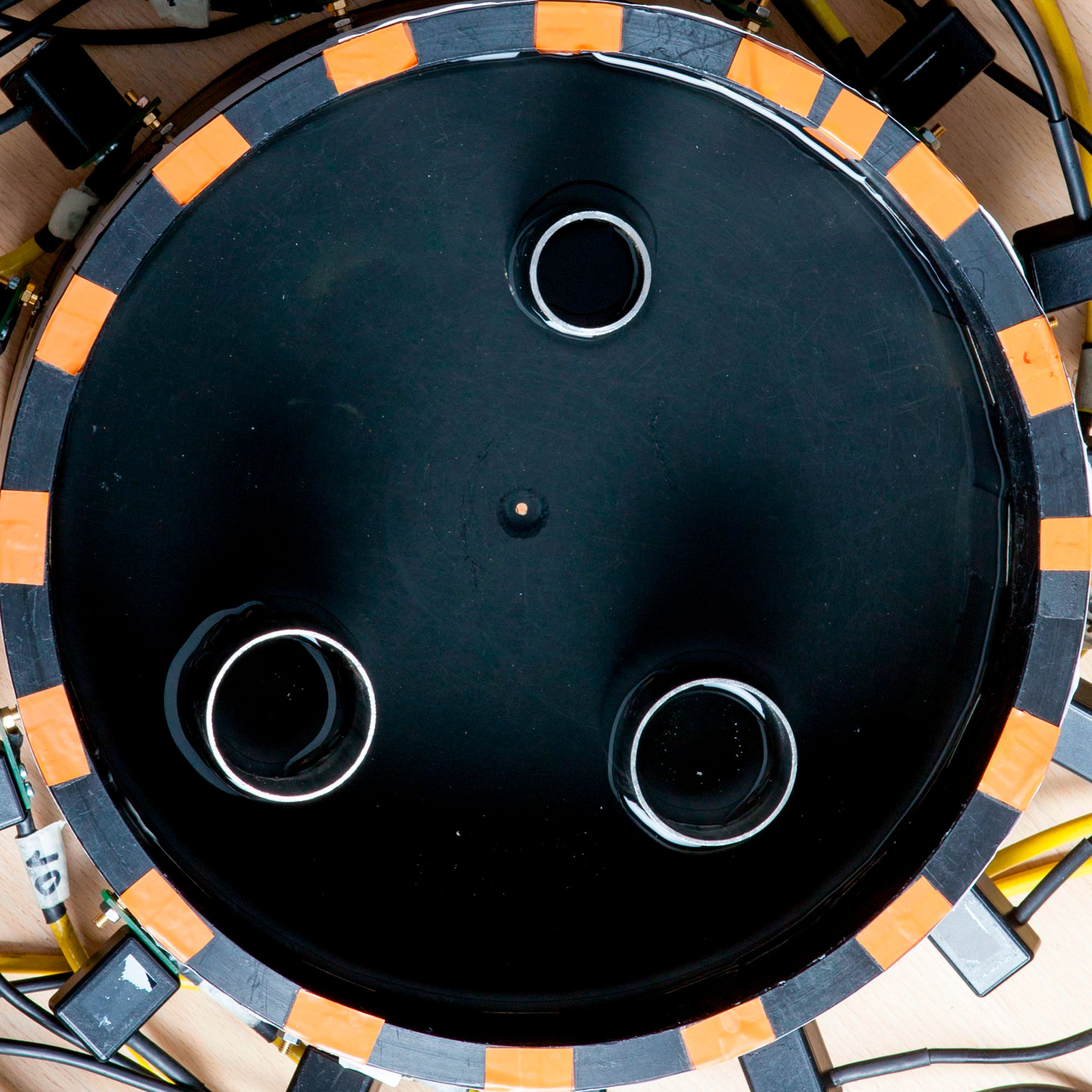} \newline
        \end{minipage}
        &\begin{minipage}{0.18\textwidth}
        \centering \includegraphics[height=2.3cm]{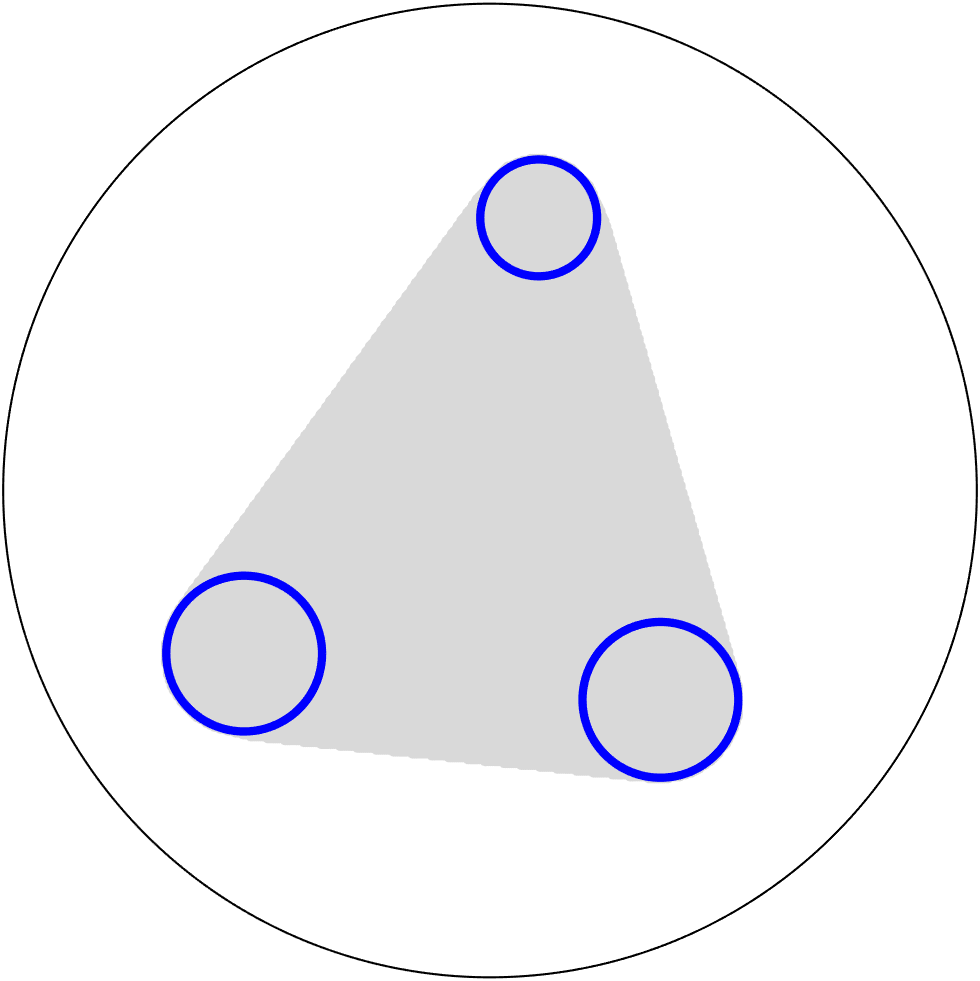} \newline
        \end{minipage}
        & \begin{minipage}{0.18\textwidth}
        \centering \includegraphics[height=2.3cm]{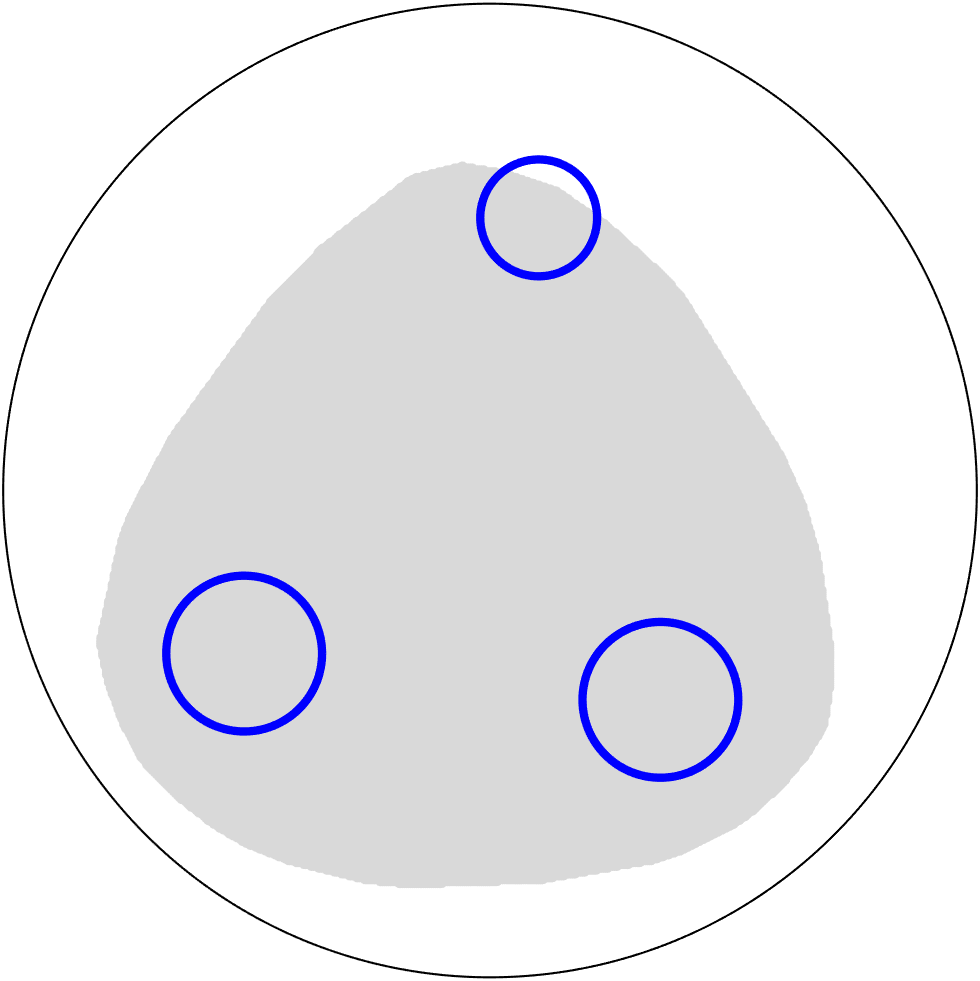} \newline 17.4\%
        \renewcommand{\thefigure}{3.5}
         \captionlistentry{} 
         \label{fig:experimental_results_3-5}
        \end{minipage}
        & \begin{minipage}{0.18\textwidth}
        \centering \includegraphics[height=2.3cm]{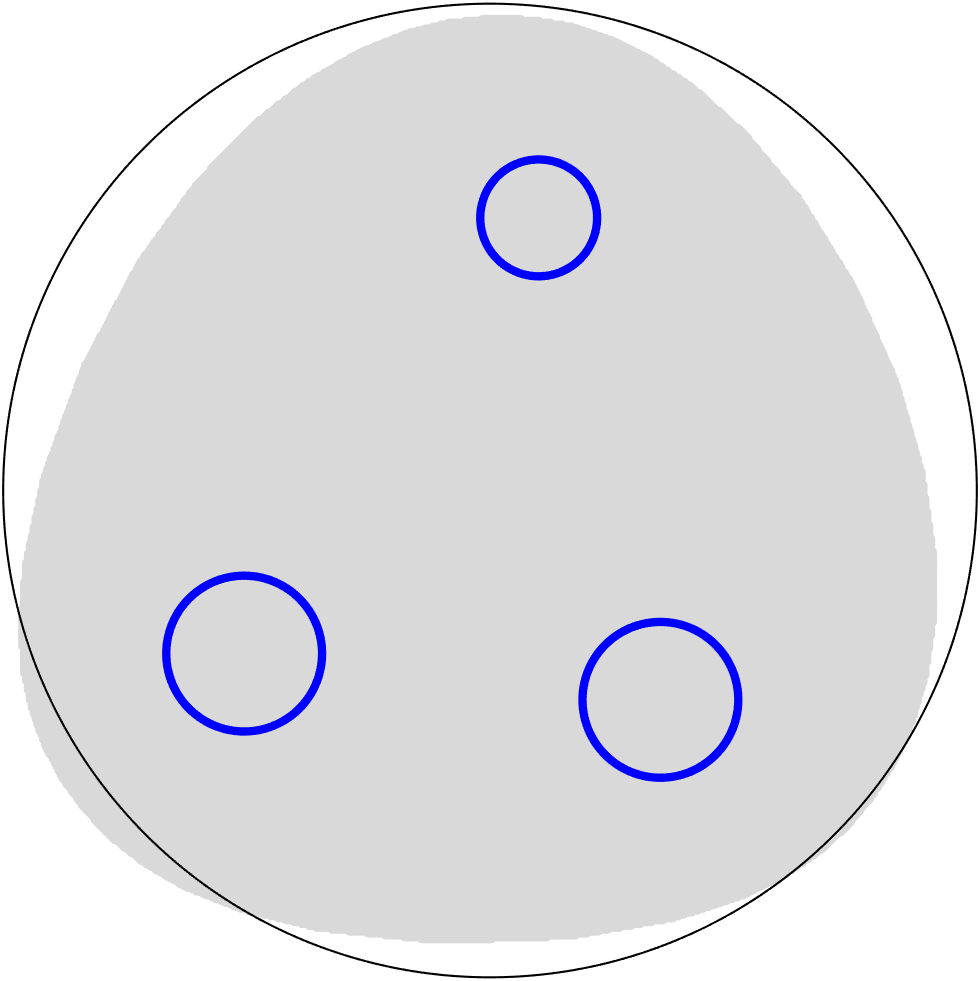} \newline 64.4\%
        \end{minipage}\\[1.3cm]
        3.6 
        &
        \begin{minipage}{0.25\textwidth}
        \centering \includegraphics[height=2.8cm]{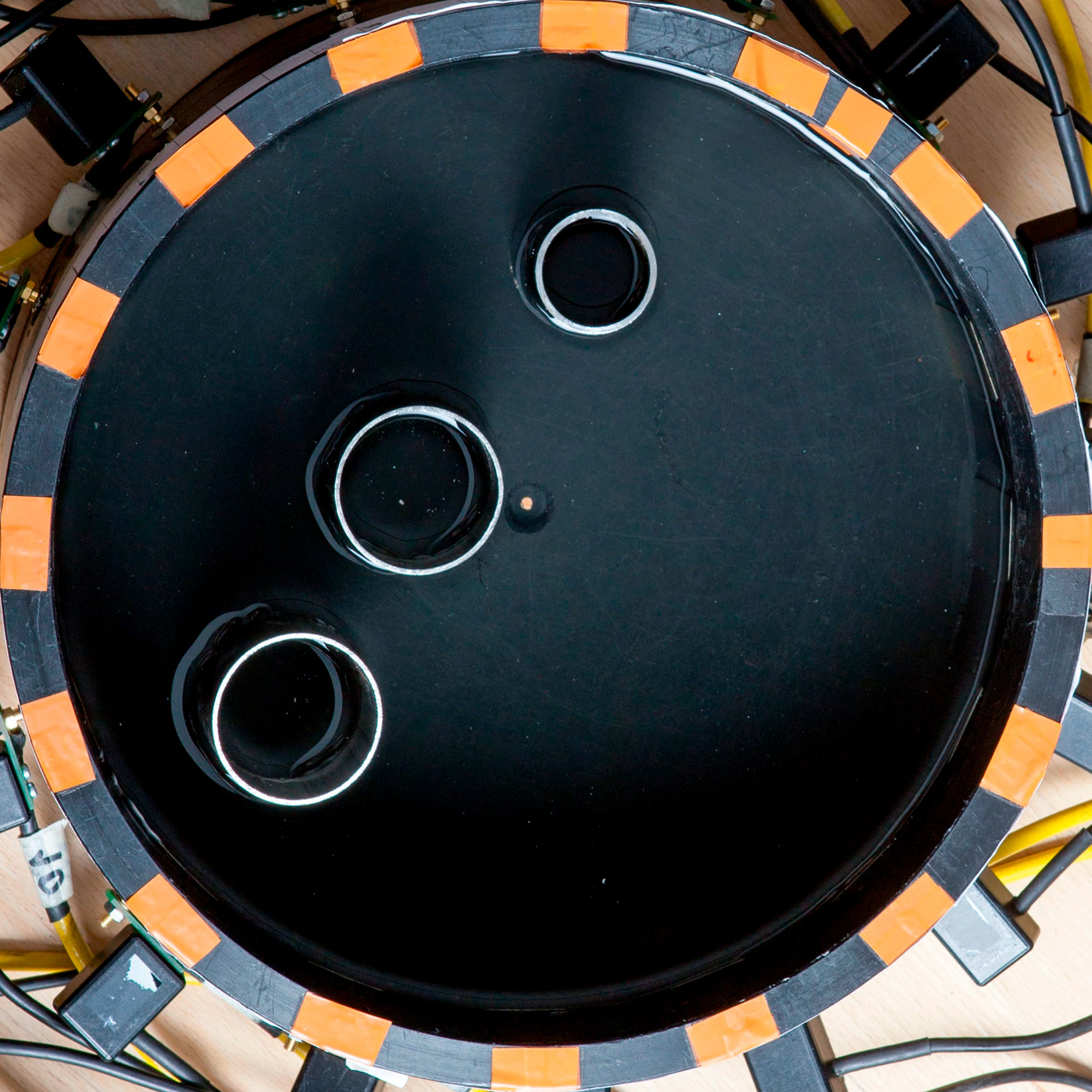} \newline
        \end{minipage}
        &\begin{minipage}{0.18\textwidth}
        \centering \includegraphics[height=2.3cm]{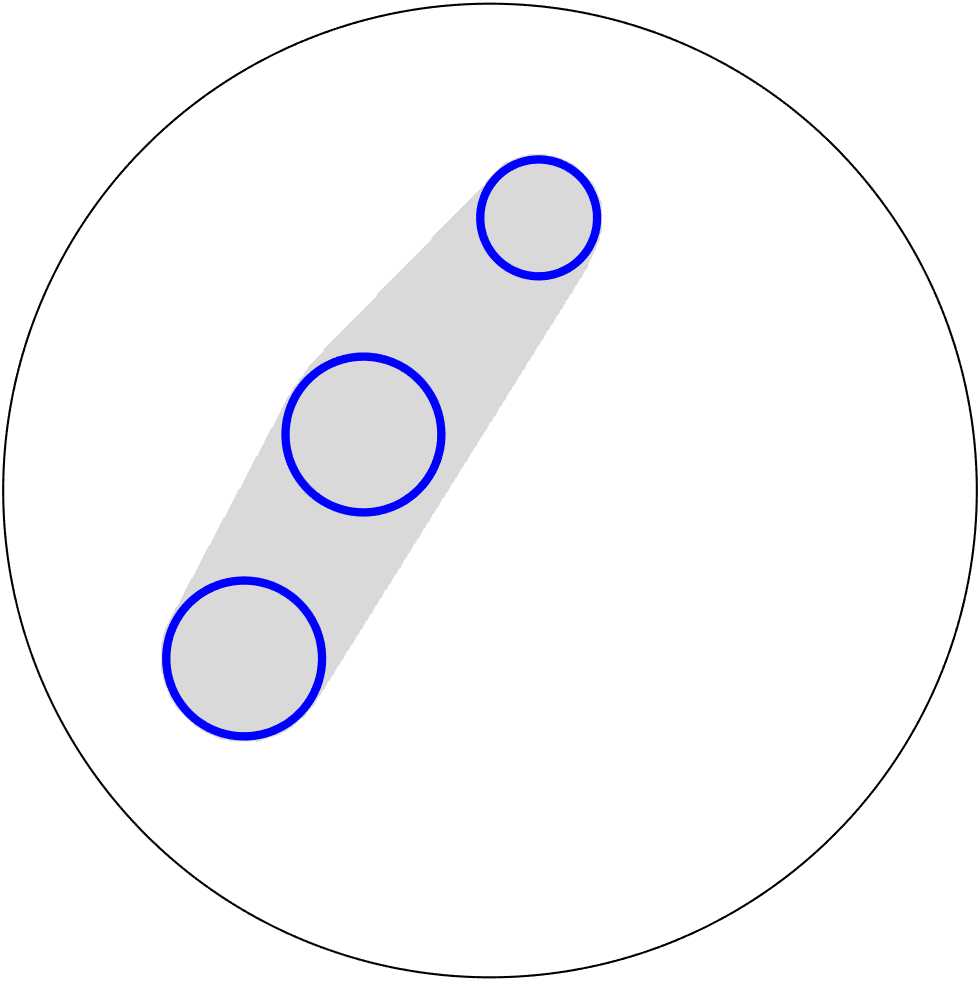} \newline
        \end{minipage}
        & \begin{minipage}{0.18\textwidth}
        \centering \includegraphics[height=2.3cm]{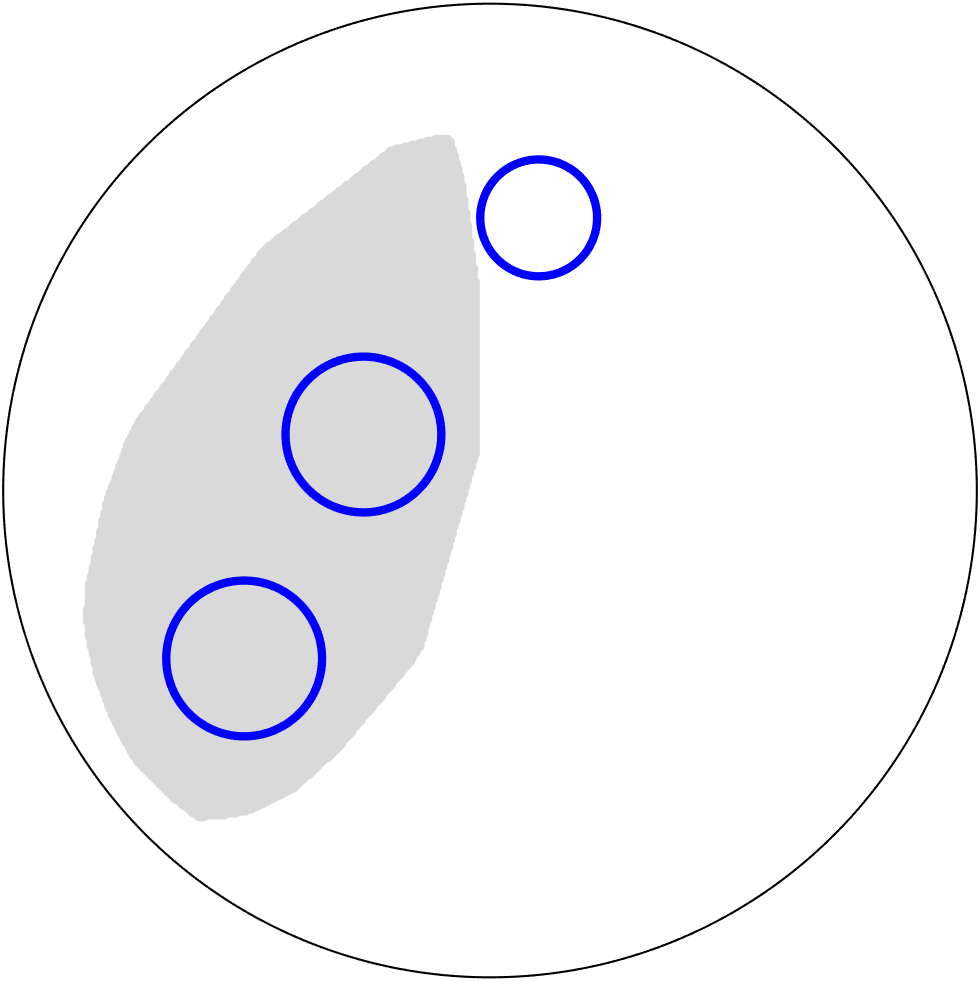} \newline 15.7\%
        \renewcommand{\thefigure}{3.6}
         \captionlistentry{} 
         \label{fig:experimental_results_3-6}
        \end{minipage}
        & \begin{minipage}{0.18\textwidth}
        \centering \includegraphics[height=2.3cm]{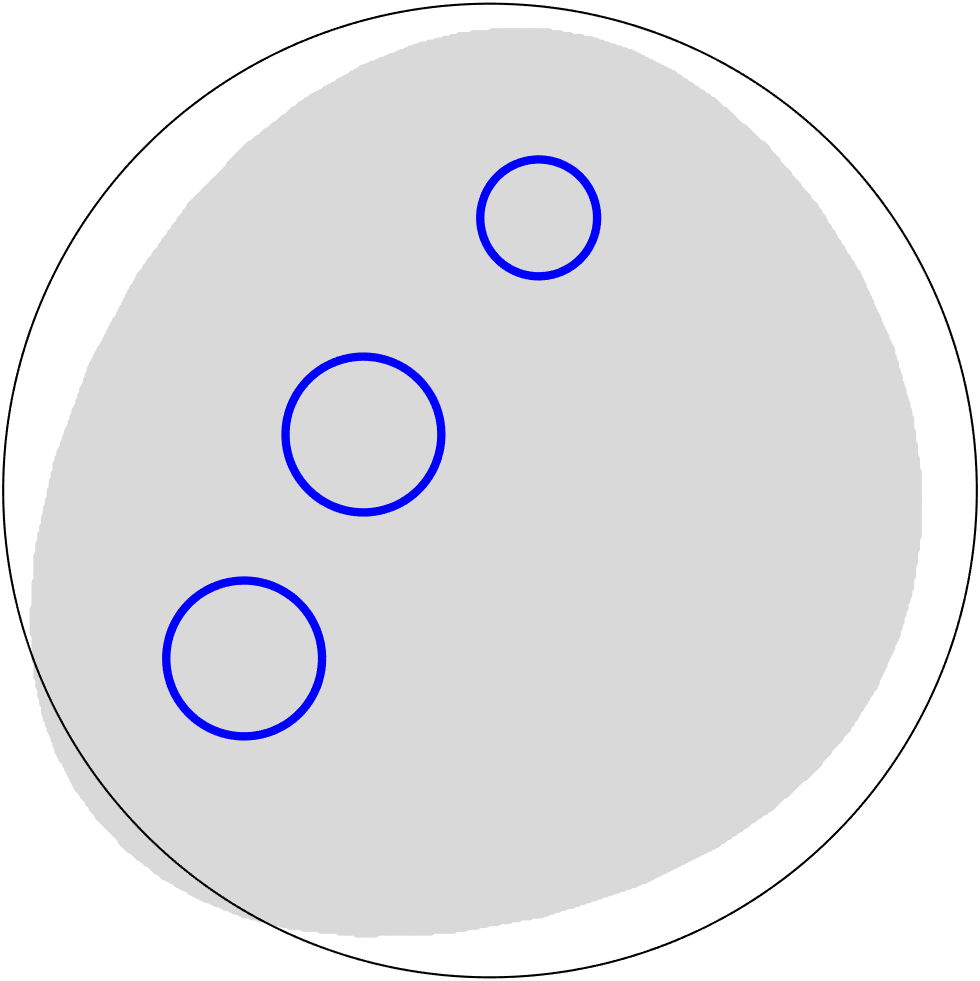} \newline 69.6\%
        \end{minipage}
    \end{tabular} 
  \addtocounter{figure}{-6}
    \caption{Comparison of the ground truth, learned and least squares hulls of the experimental phantoms. The error relative to the ground truth is shown below each phantom.}
    \label{fig:experimental_results_2}
\end{figure}

\clearpage
\begin{figure}[ht]
    \begin{tabular}{lccccc}
        Case & Phantom & Ground truth & Learned & Least squares \\ [.2cm]
        4.1 
        &
        \begin{minipage}{0.25\textwidth}
        \centering \includegraphics[height=2.8cm]{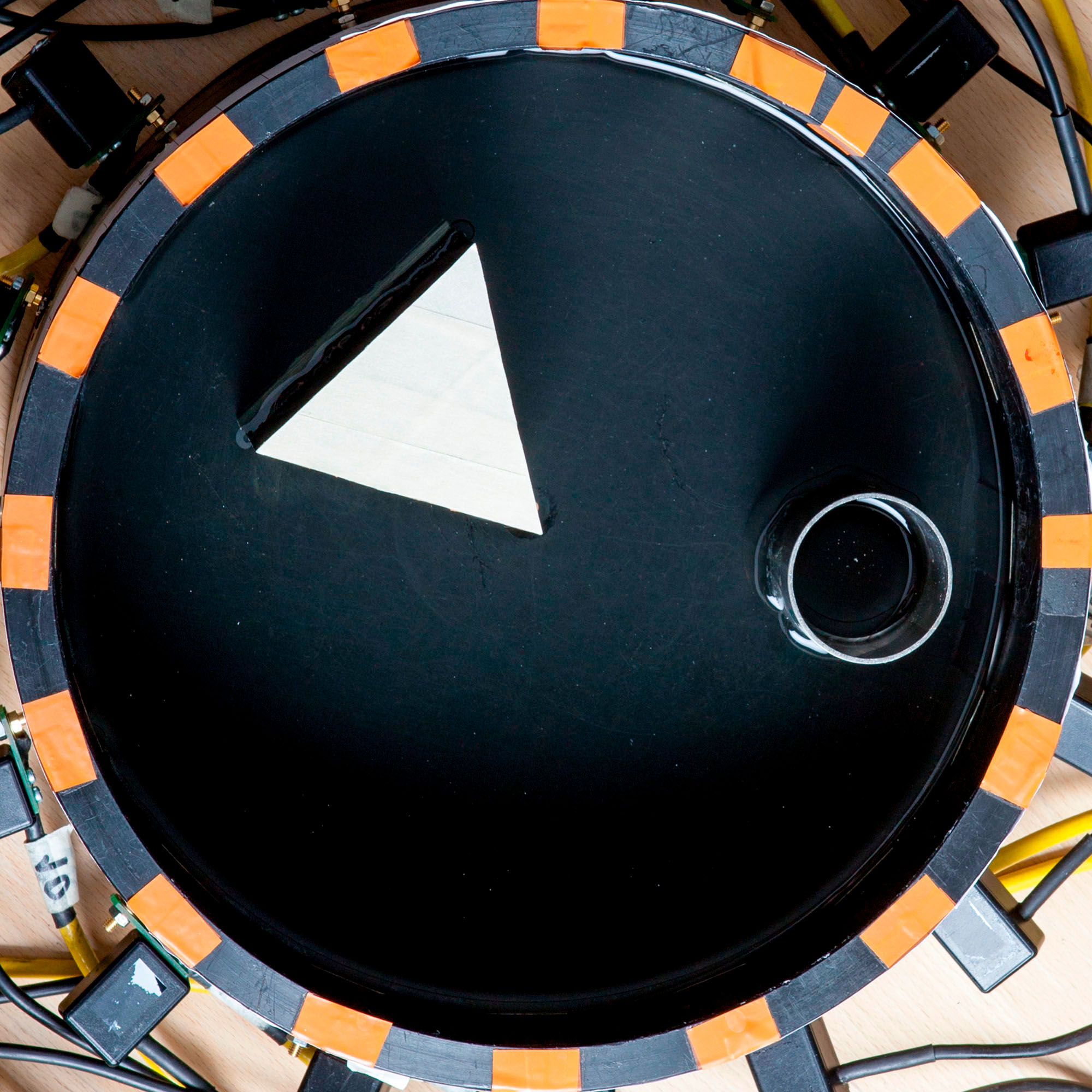} \newline
        \end{minipage}
        &\begin{minipage}{0.18\textwidth}
        \centering \includegraphics[height=2.3cm]{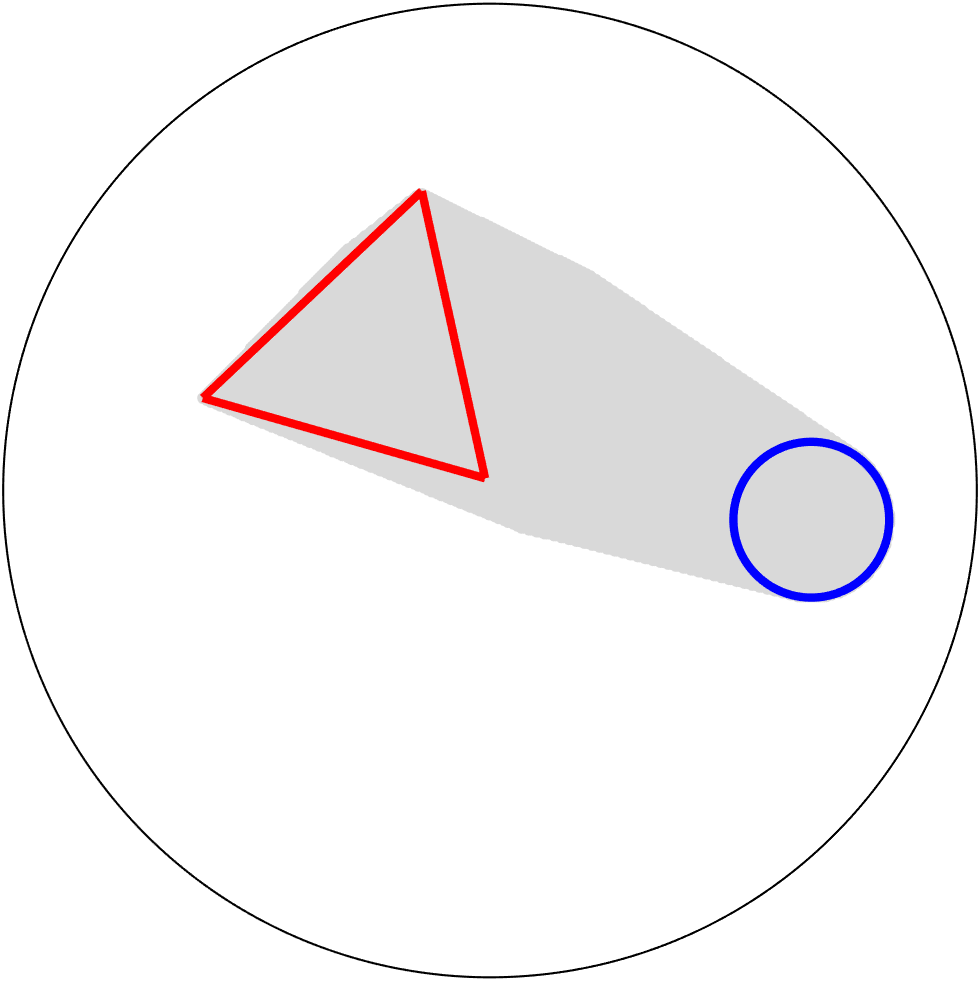} \newline
        \end{minipage}
        & \begin{minipage}{0.18\textwidth}
        \centering \includegraphics[height=2.3cm]{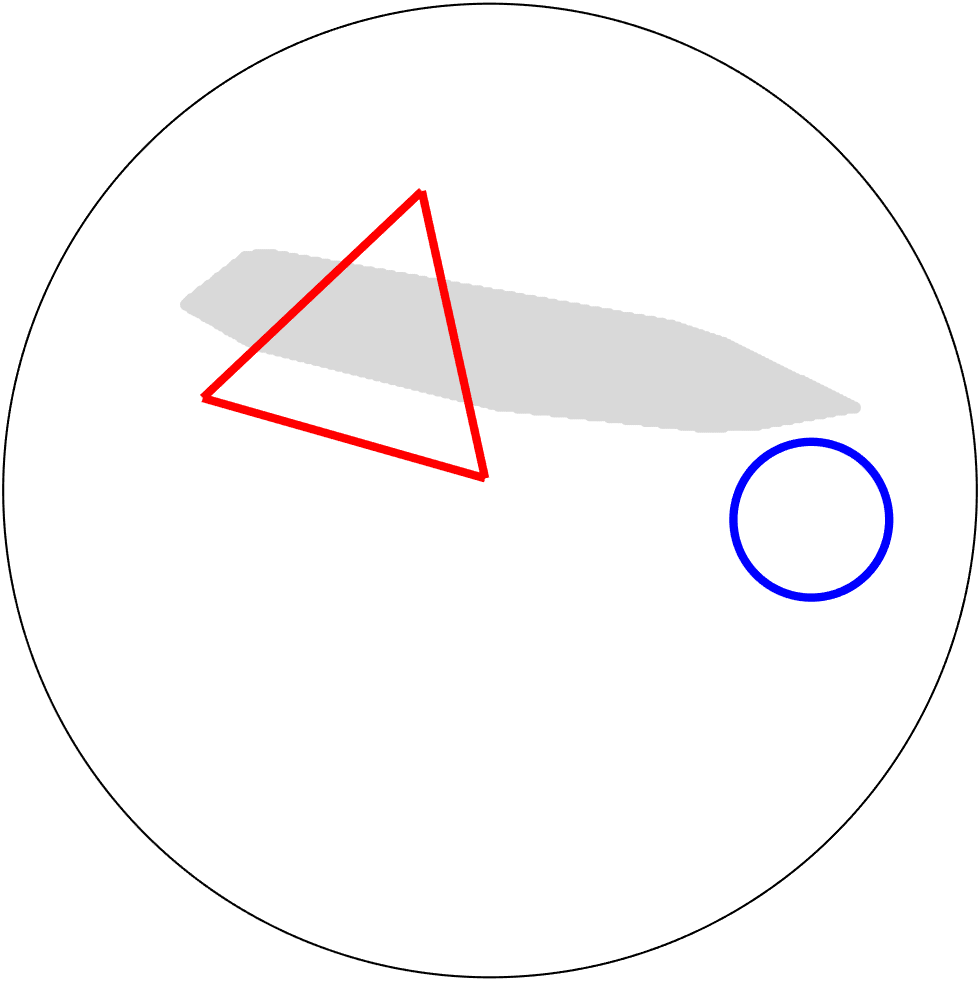} \newline 15.7\%
        \renewcommand{\thefigure}{4.1}
         \captionlistentry{} 
         \label{fig:experimental_results_4-1}
        \end{minipage}
        & \begin{minipage}{0.18\textwidth}
        \centering \includegraphics[height=2.3cm]{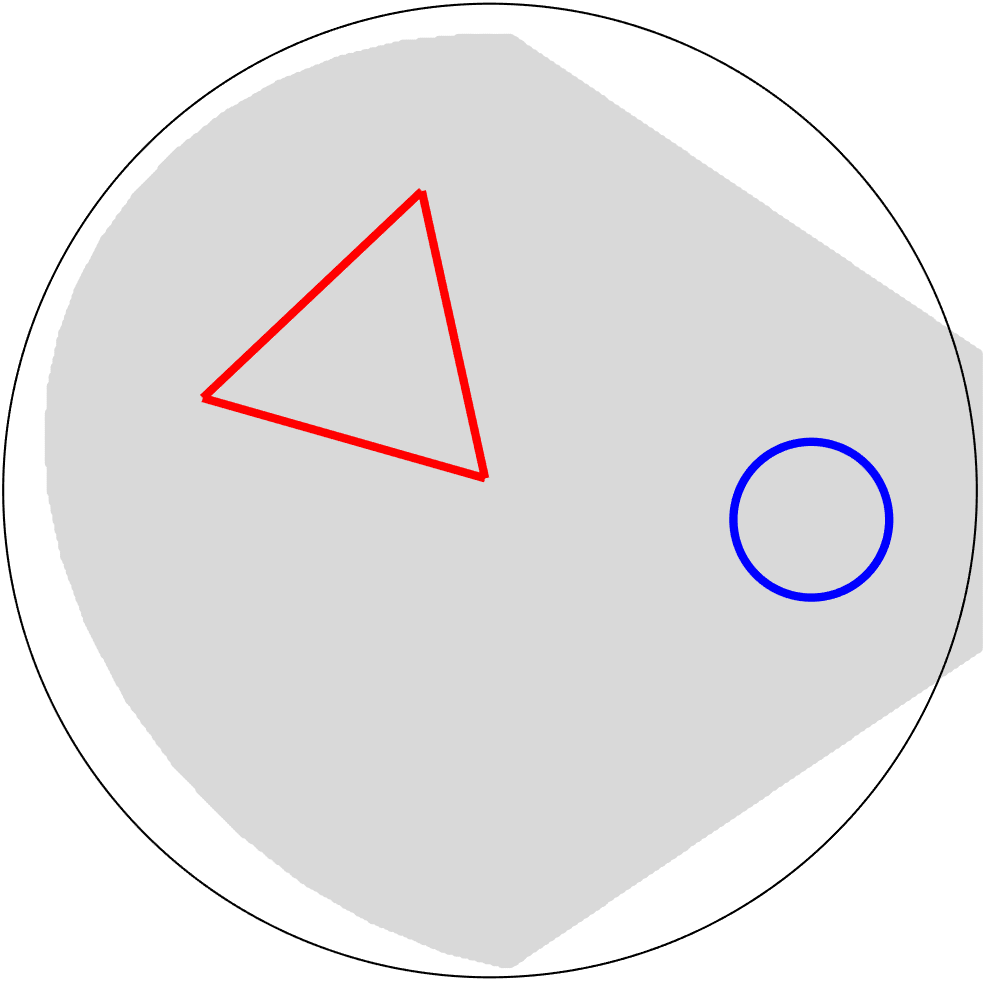} \newline 69.6\%
        \end{minipage}\\[1.3cm]
        4.2 
        &
        \begin{minipage}{0.25\textwidth}
        \centering \includegraphics[height=2.8cm]{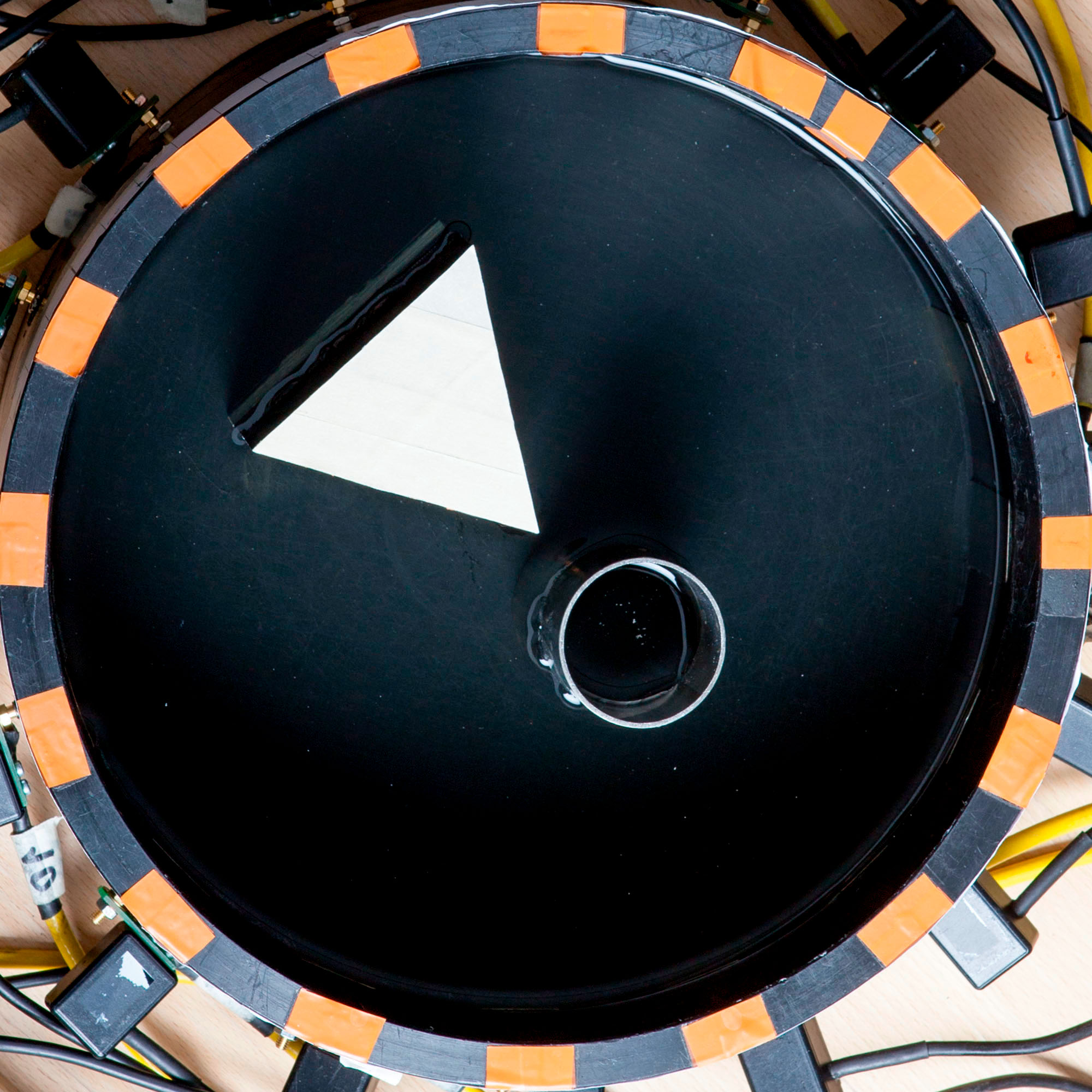} \newline
        \end{minipage}
        &\begin{minipage}{0.18\textwidth}
        \centering \includegraphics[height=2.3cm]{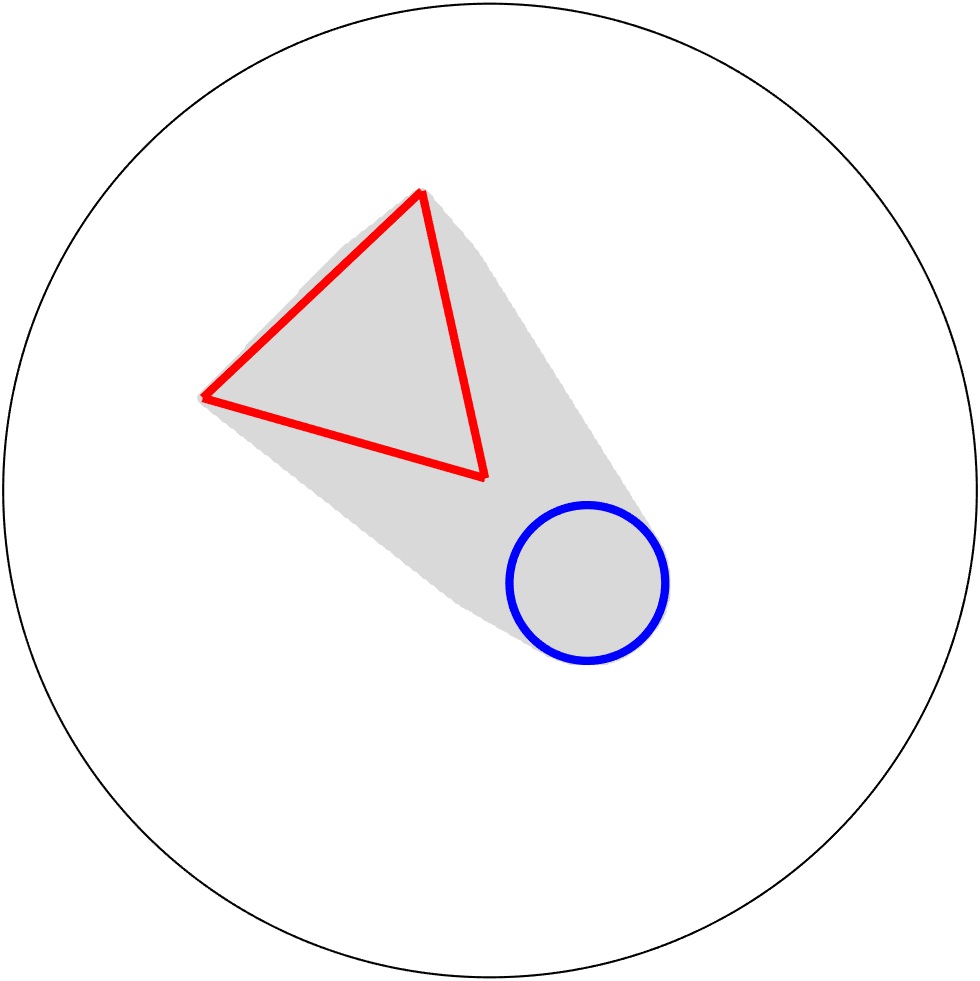} \newline
        \end{minipage}
        & \begin{minipage}{0.18\textwidth}
        \centering \includegraphics[height=2.3cm]{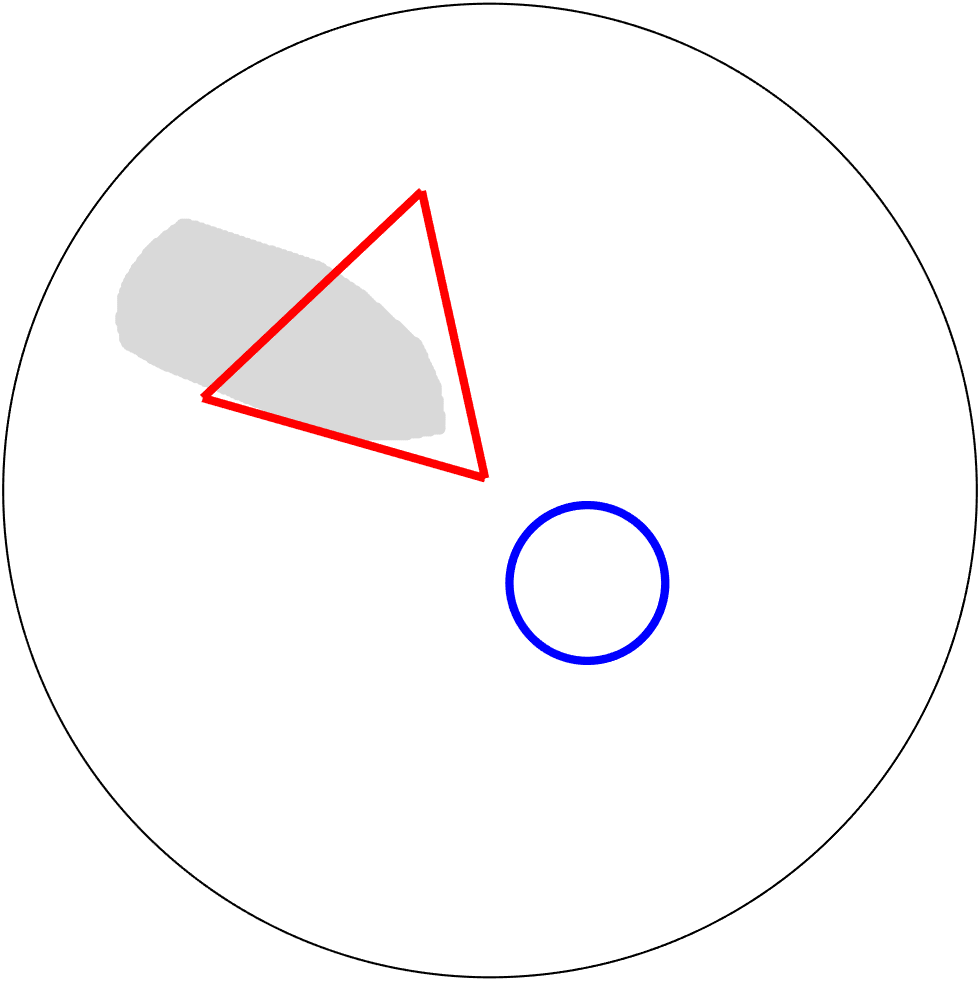} \newline 15.5\%
        \renewcommand{\thefigure}{4.2}
         \captionlistentry{} 
         \label{fig:experimental_results_4-2}
        \end{minipage}
        & \begin{minipage}{0.18\textwidth}
        \centering \includegraphics[height=2.3cm]{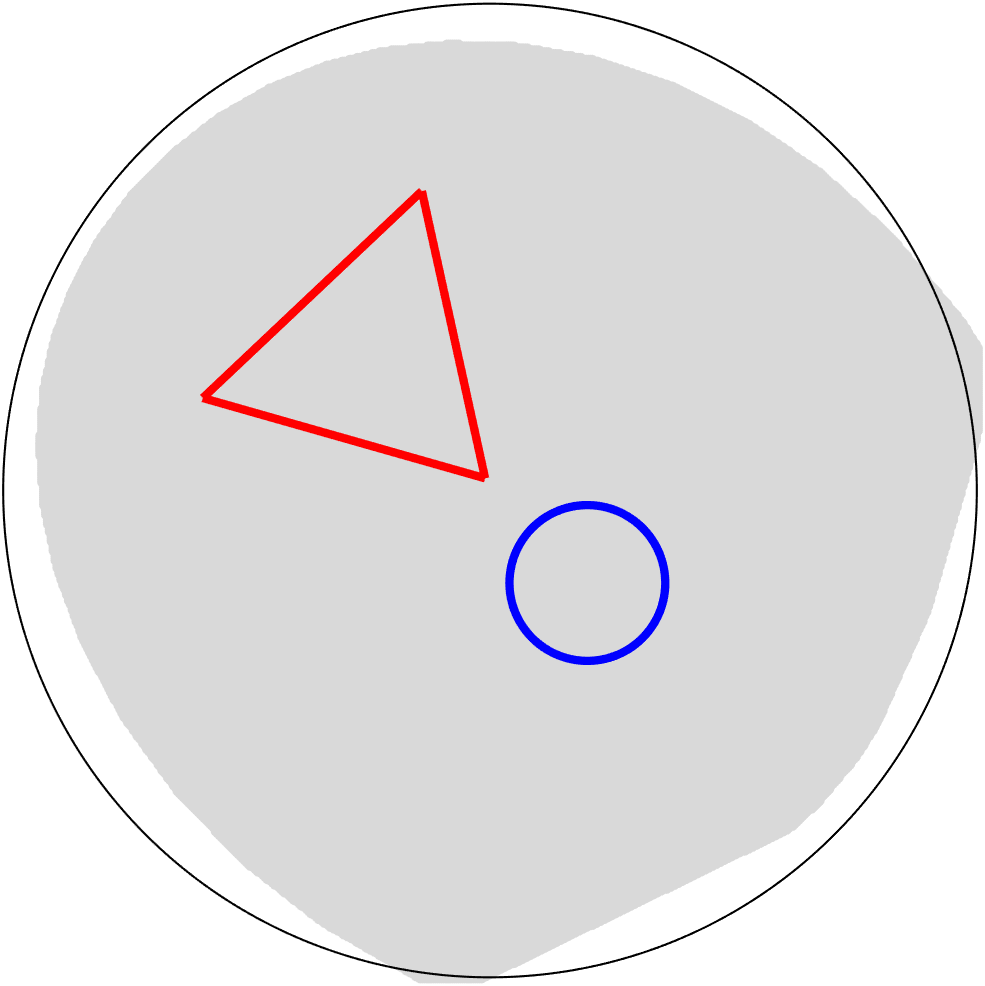} \newline 80.5\%
        \end{minipage}\\[1.3cm]
        4.3 
        &
        \begin{minipage}{0.25\textwidth}
        \centering \includegraphics[height=2.8cm]{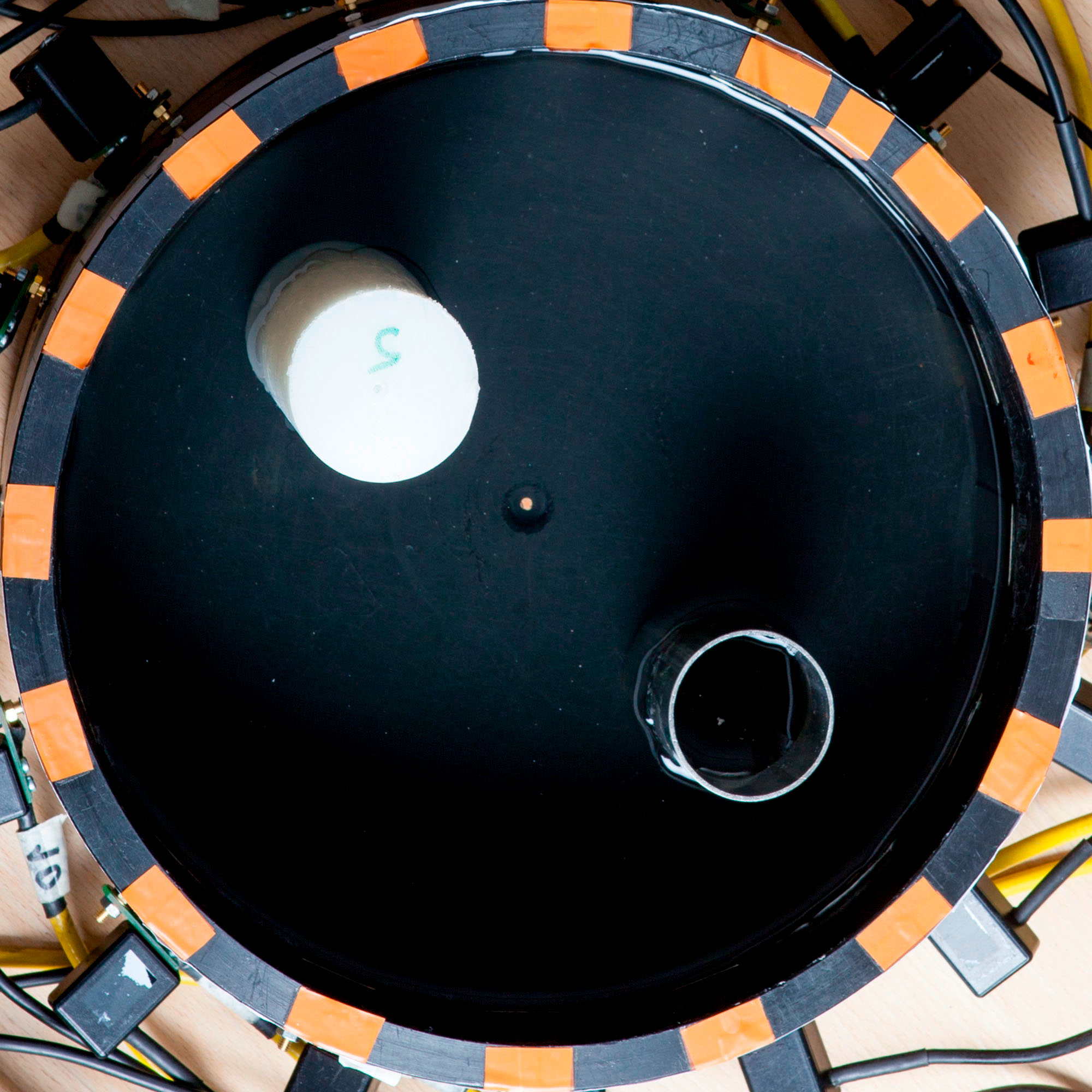} \newline
        \end{minipage}
        &\begin{minipage}{0.18\textwidth}
        \centering \includegraphics[height=2.3cm]{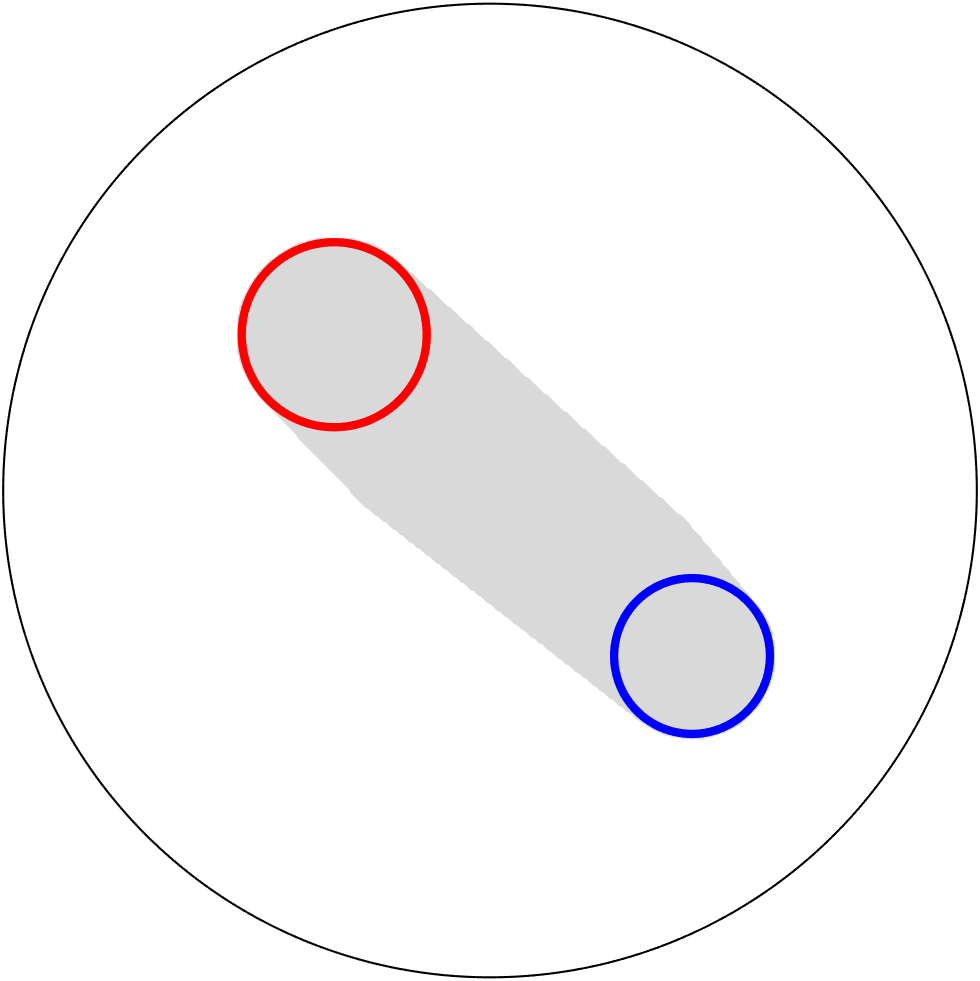} \newline
        \end{minipage}
        & \begin{minipage}{0.18\textwidth}
        \centering \includegraphics[height=2.3cm]{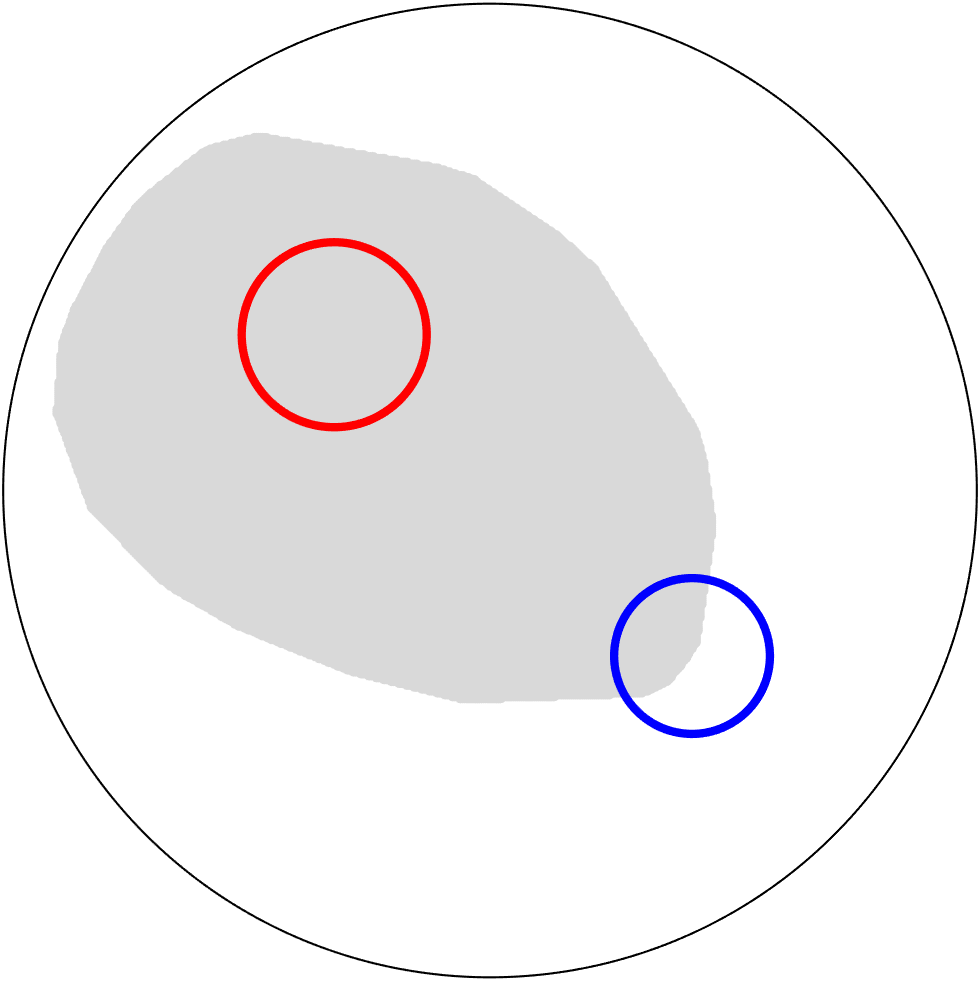} \newline 27.9\%
        \renewcommand{\thefigure}{4.3}
         \captionlistentry{} 
         \label{fig:experimental_results_4-3}
        \end{minipage}
        & \begin{minipage}{0.18\textwidth}
        \centering \includegraphics[height=2.3cm]{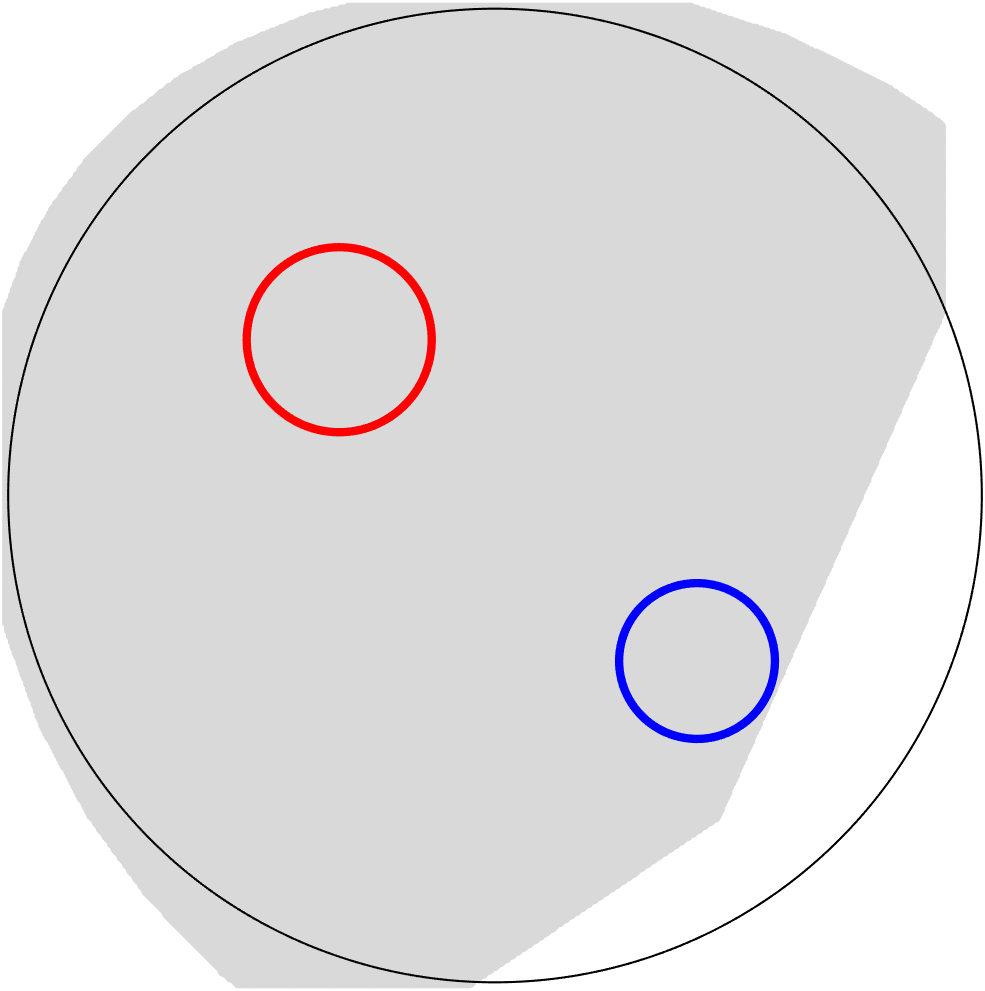} \newline 90.0\%
        \end{minipage}\\[1.3cm]
        4.4 
        &
        \begin{minipage}{0.25\textwidth}
        \centering \includegraphics[height=2.8cm]{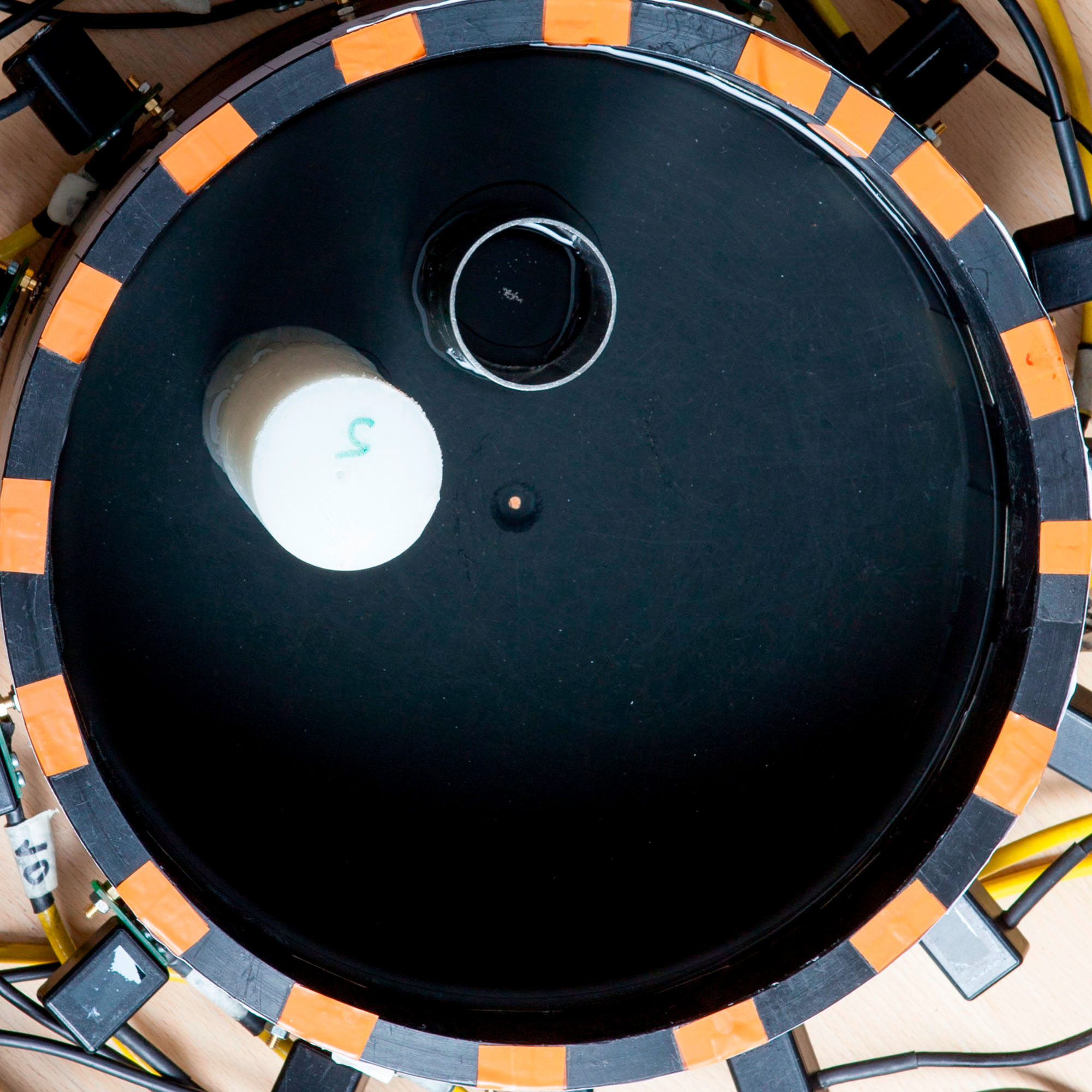} \newline
        \end{minipage}
        &\begin{minipage}{0.18\textwidth}
        \centering \includegraphics[height=2.3cm]{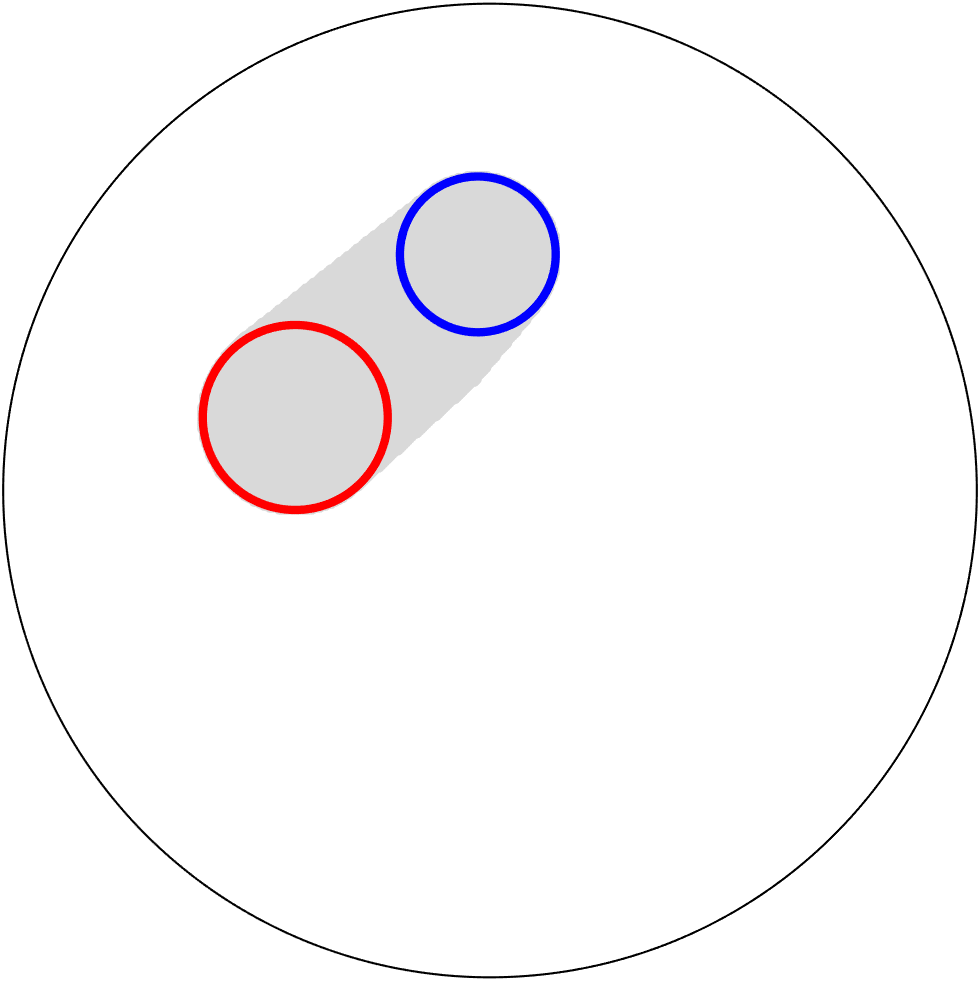} \newline
        \end{minipage}
        & \begin{minipage}{0.18\textwidth}
        \centering \includegraphics[height=2.3cm]{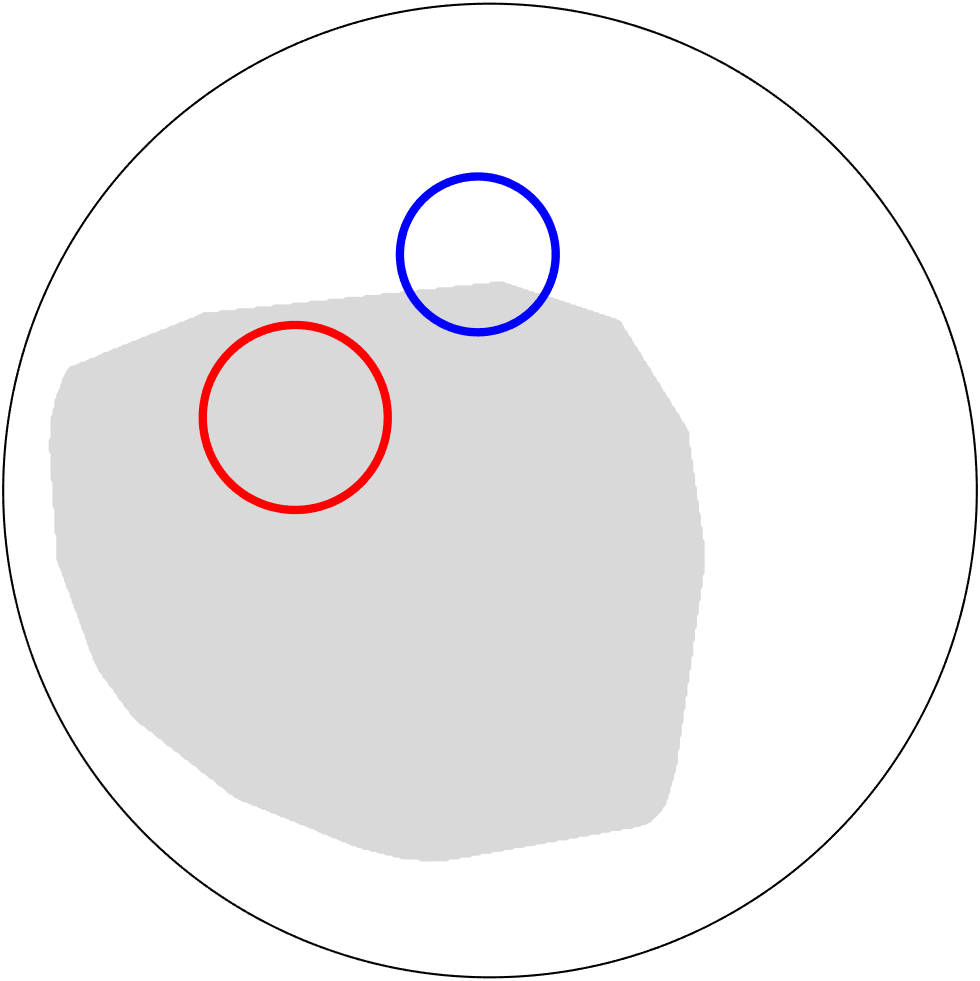} \newline 40.5\%
        \renewcommand{\thefigure}{4.4}
         \captionlistentry{} 
         \label{fig:experimental_results_4-4}
        \end{minipage}
        & \begin{minipage}{0.18\textwidth}
        \centering \includegraphics[height=2.3cm]{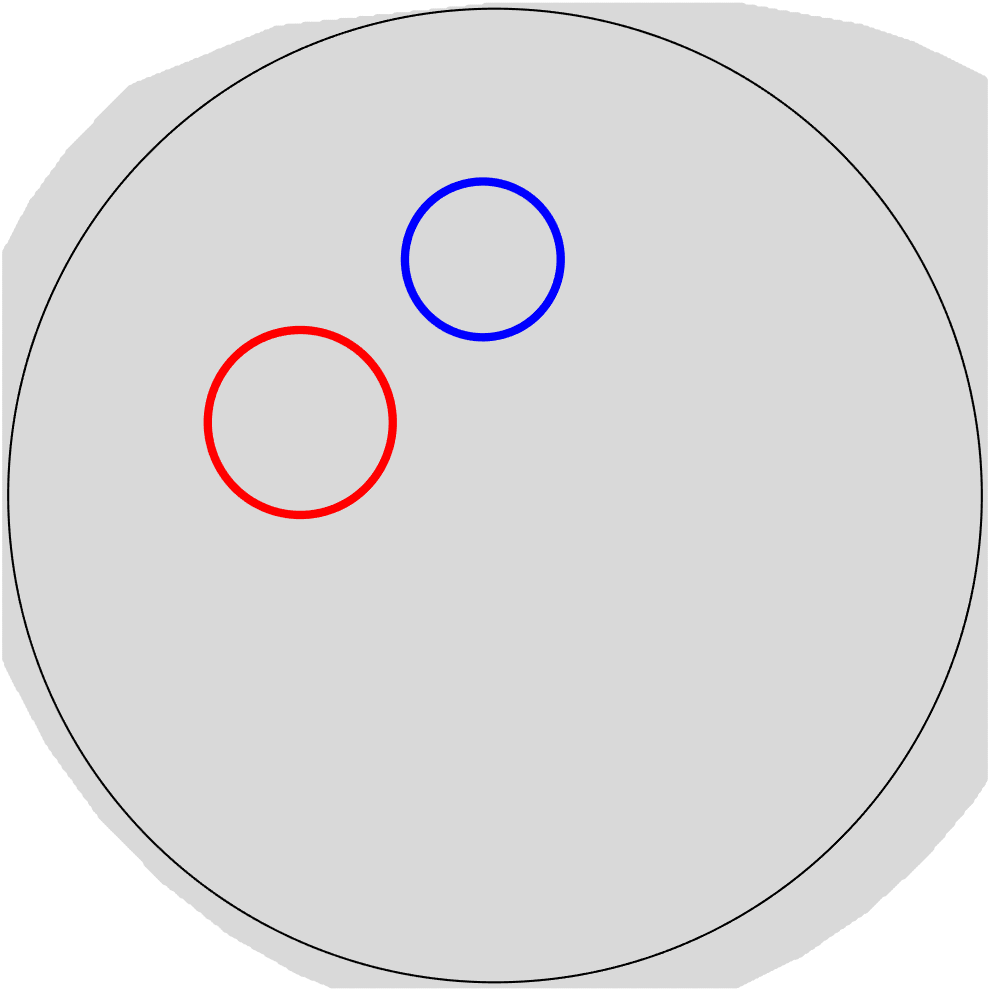} \newline 117.4\%
        \end{minipage}\\[1.3cm]
        5.1 
        &
        \begin{minipage}{0.25\textwidth}
        \centering \includegraphics[height=2.8cm]{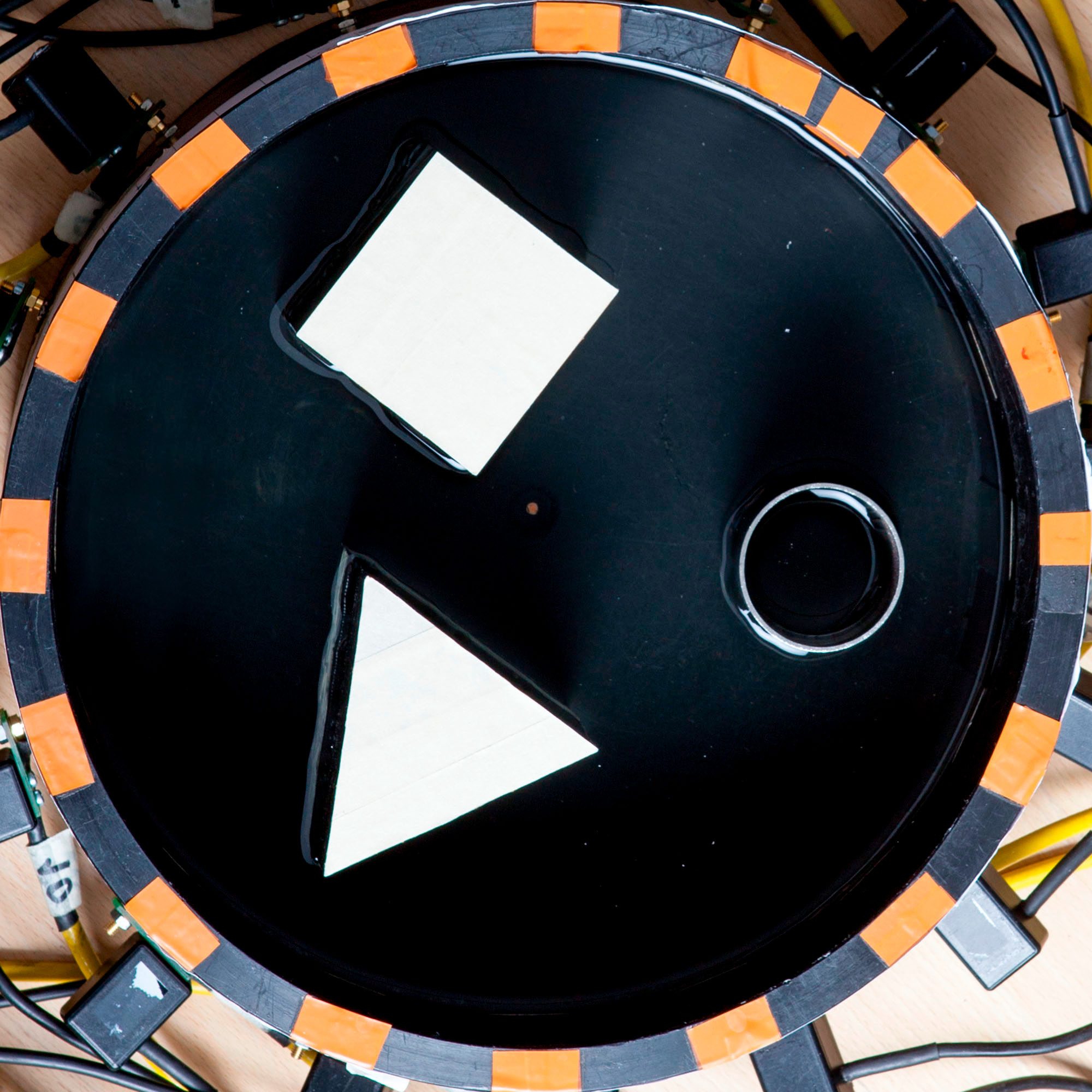} \newline
        \end{minipage}
        &\begin{minipage}{0.18\textwidth}
        \centering \includegraphics[height=2.3cm]{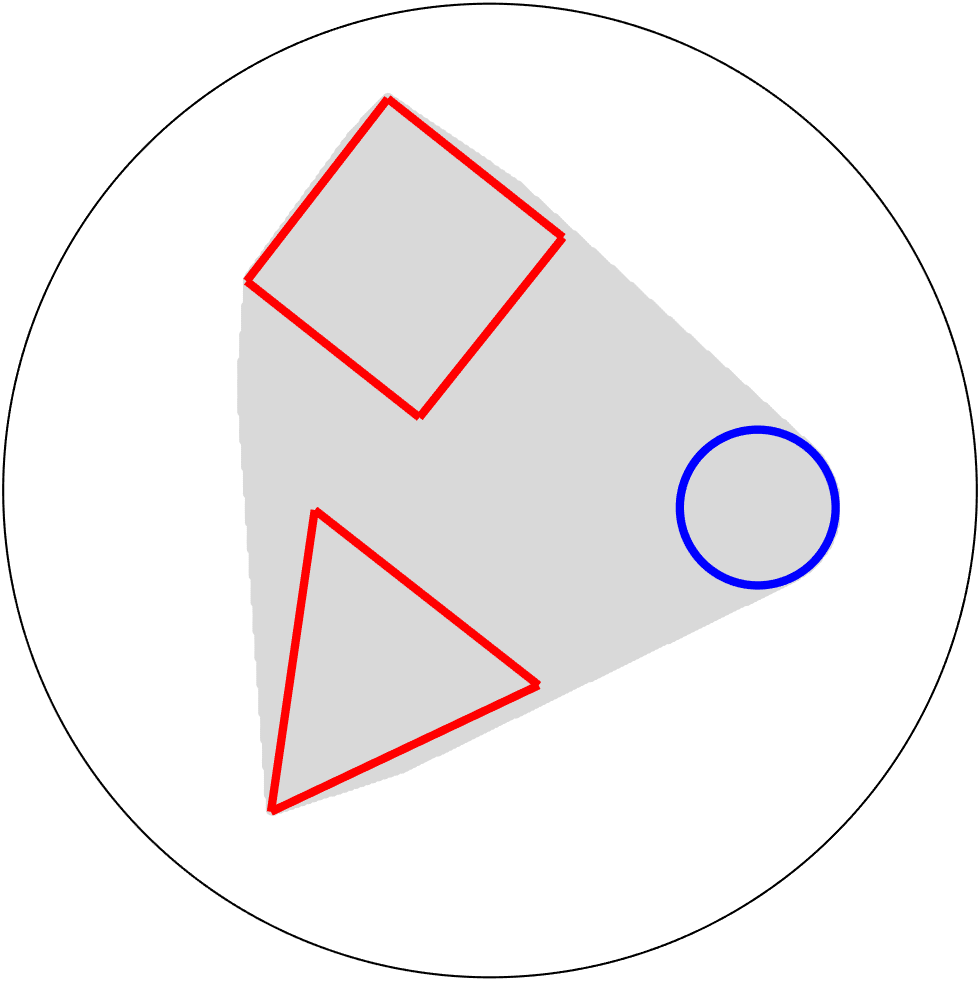} \newline
        \end{minipage}
        & \begin{minipage}{0.18\textwidth}
        \centering \includegraphics[height=2.3cm]{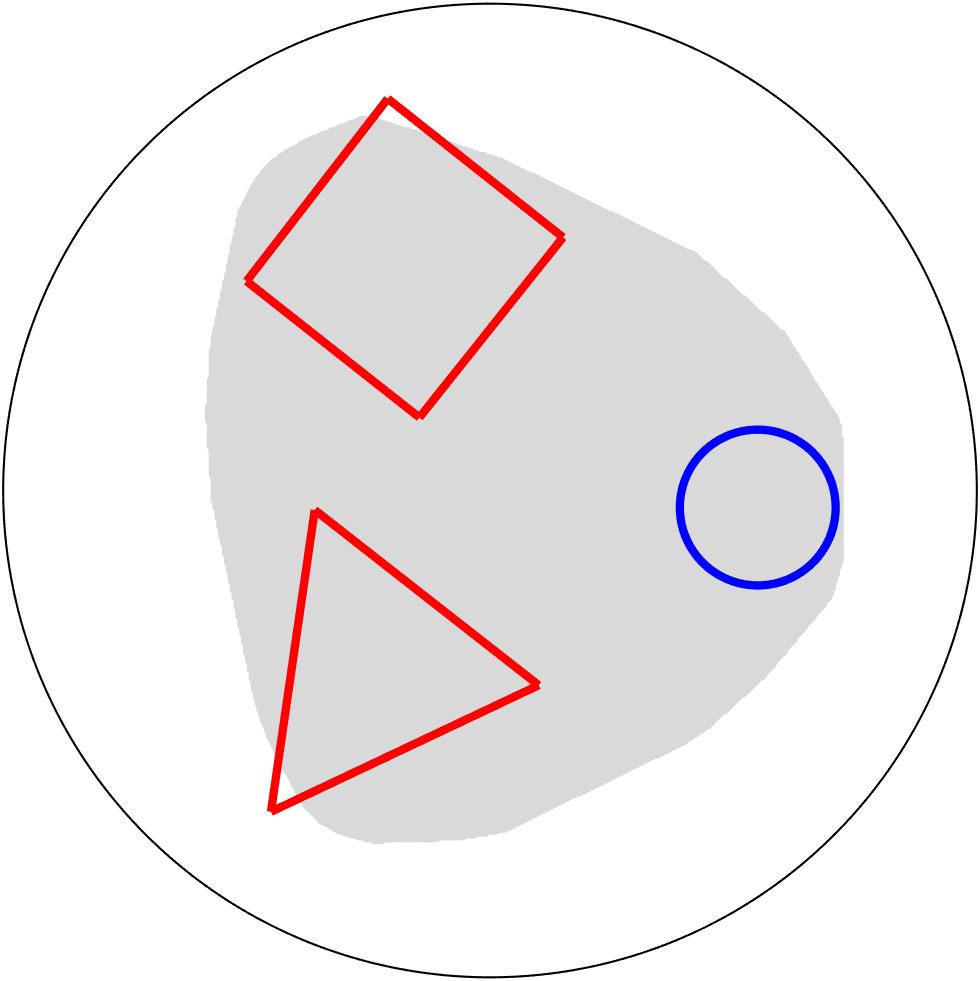} \newline 12.9\%
        \renewcommand{\thefigure}{5.1}
         \captionlistentry{} 
         \label{fig:experimental_results_5-1}
        \end{minipage}
        & \begin{minipage}{0.18\textwidth}
        \centering \includegraphics[height=2.3cm]{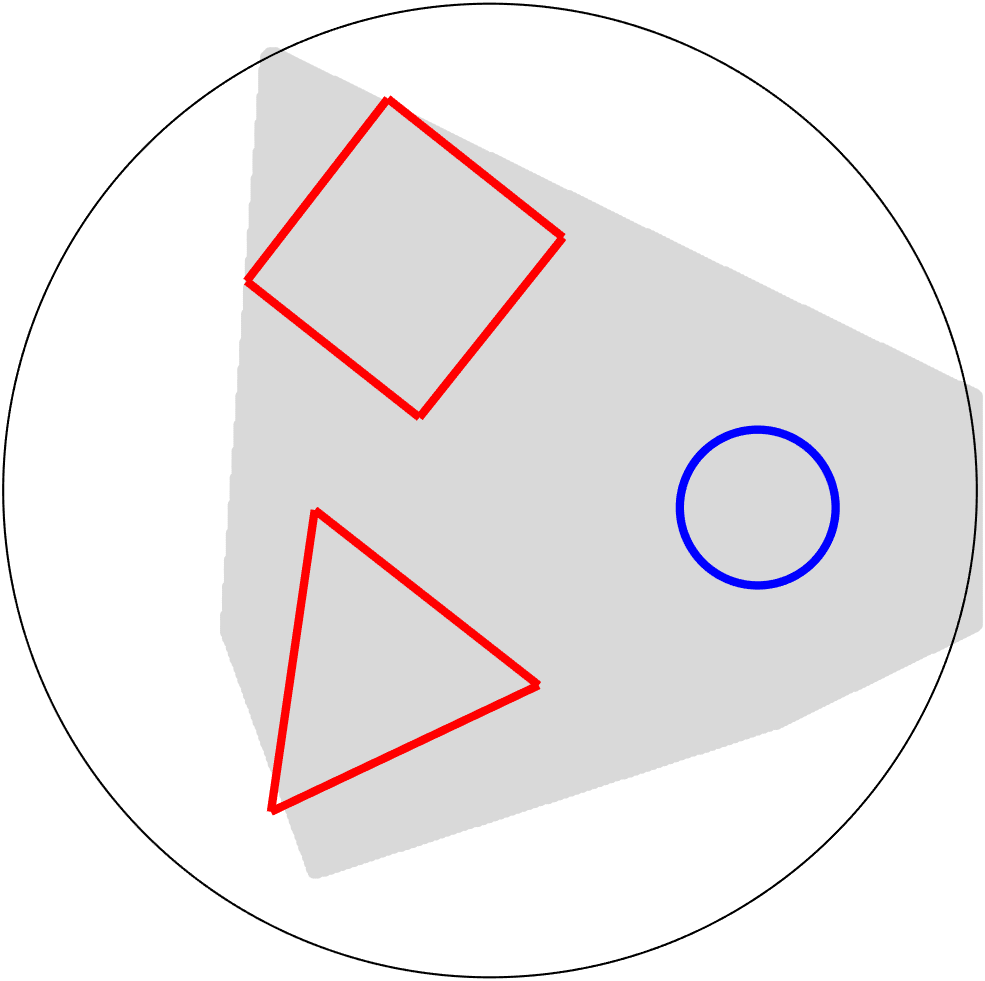} \newline 21.7\%
        \end{minipage}\\[1.3cm]
        5.2 
        &
        \begin{minipage}{0.25\textwidth}
        \centering \includegraphics[height=2.8cm]{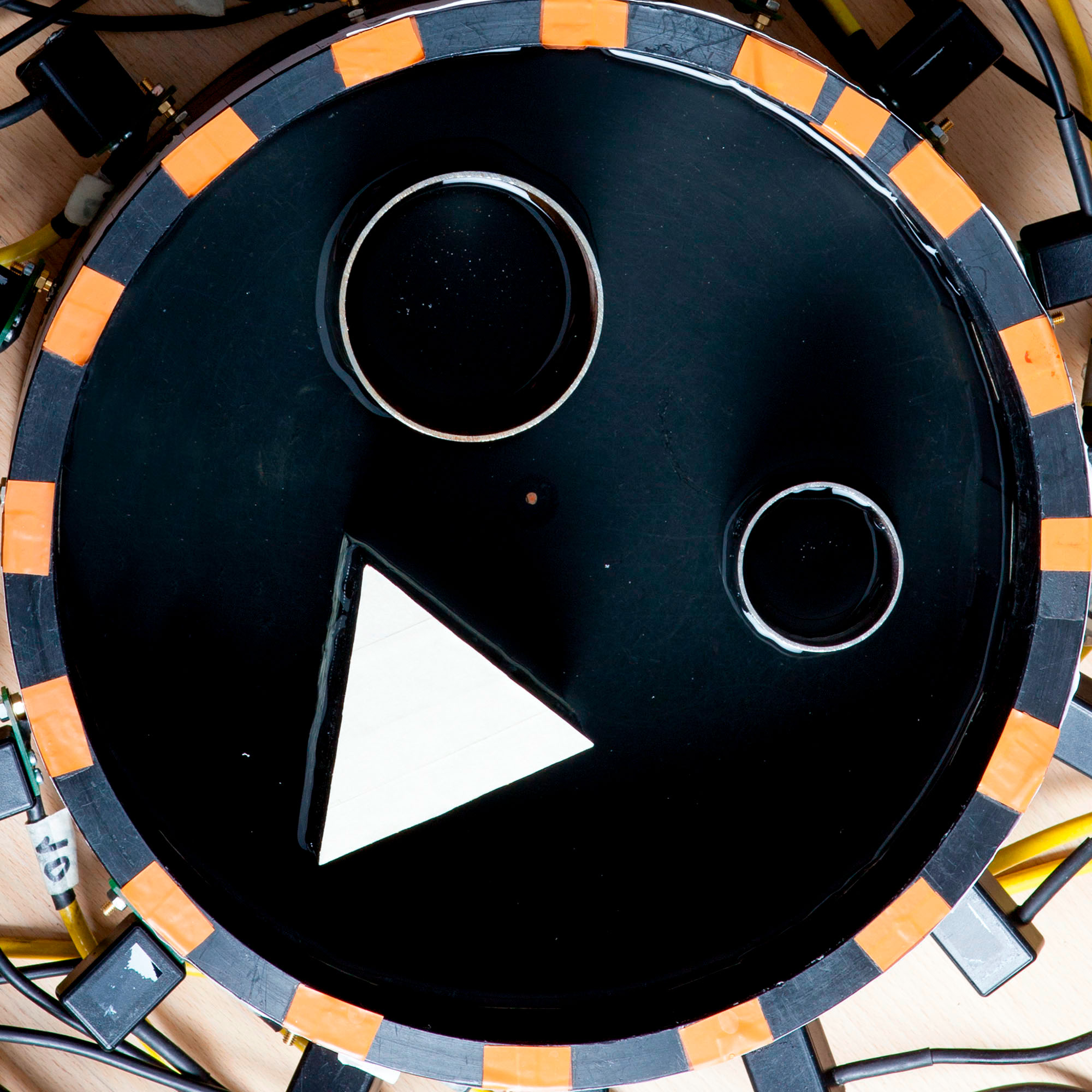} \newline
        \end{minipage}
        &\begin{minipage}{0.18\textwidth}
        \centering \includegraphics[height=2.3cm]{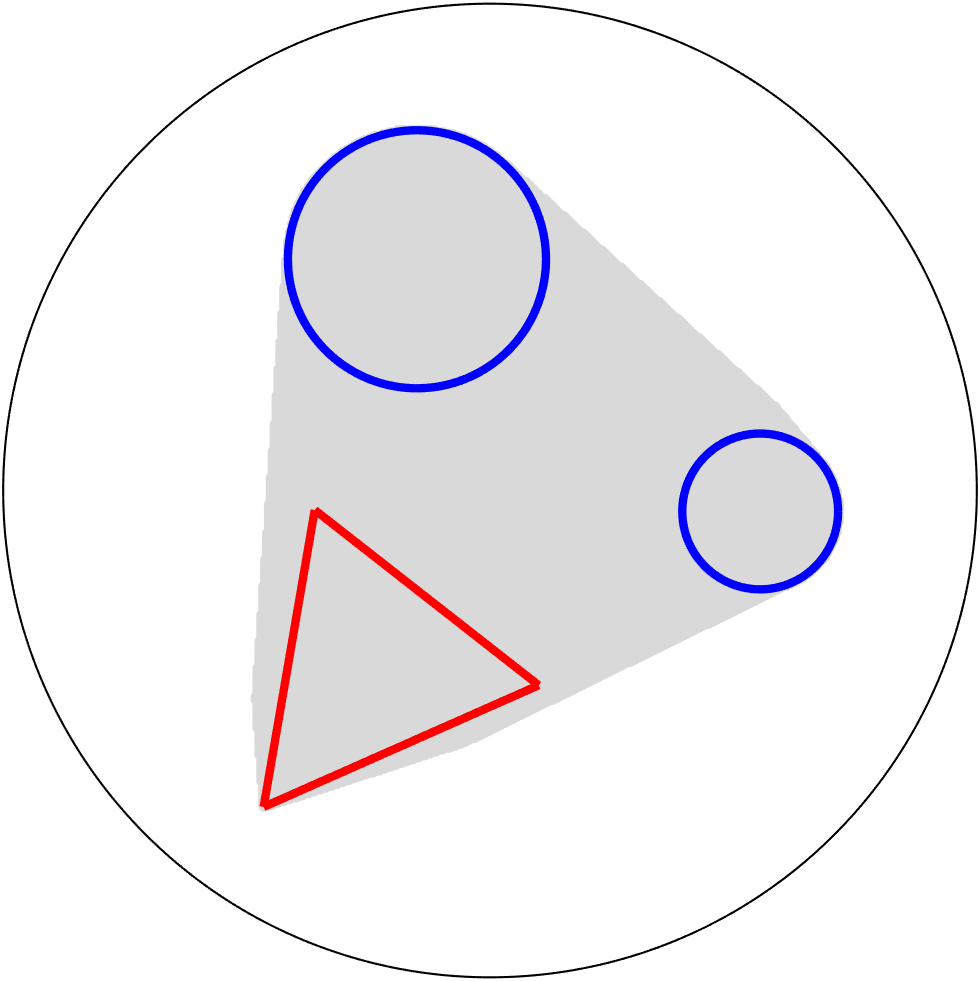} \newline
        \end{minipage}
        & \begin{minipage}{0.18\textwidth}
        \centering \includegraphics[height=2.3cm]{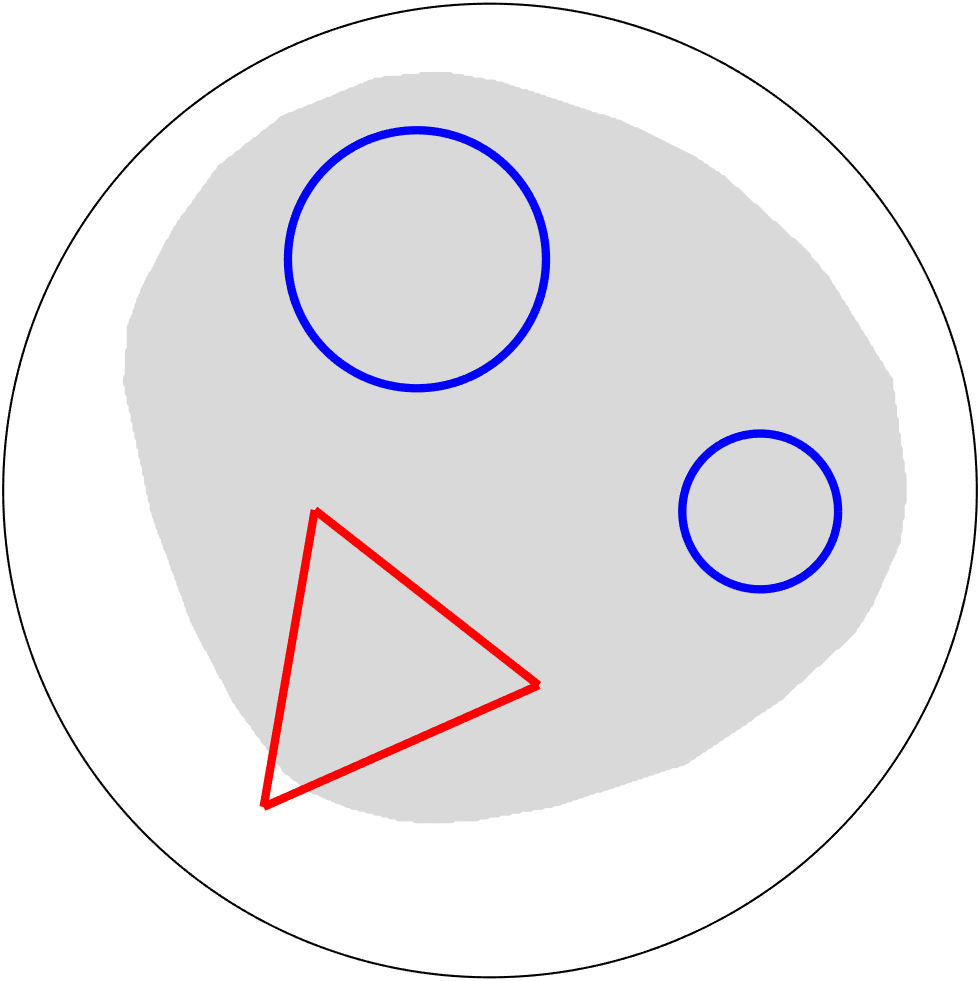} \newline 27.5\%
        \renewcommand{\thefigure}{5.2}
         \captionlistentry{} 
         \label{fig:experimental_results_5-2}
        \end{minipage}
        & \begin{minipage}{0.18\textwidth}
        \centering \includegraphics[height=2.3cm]{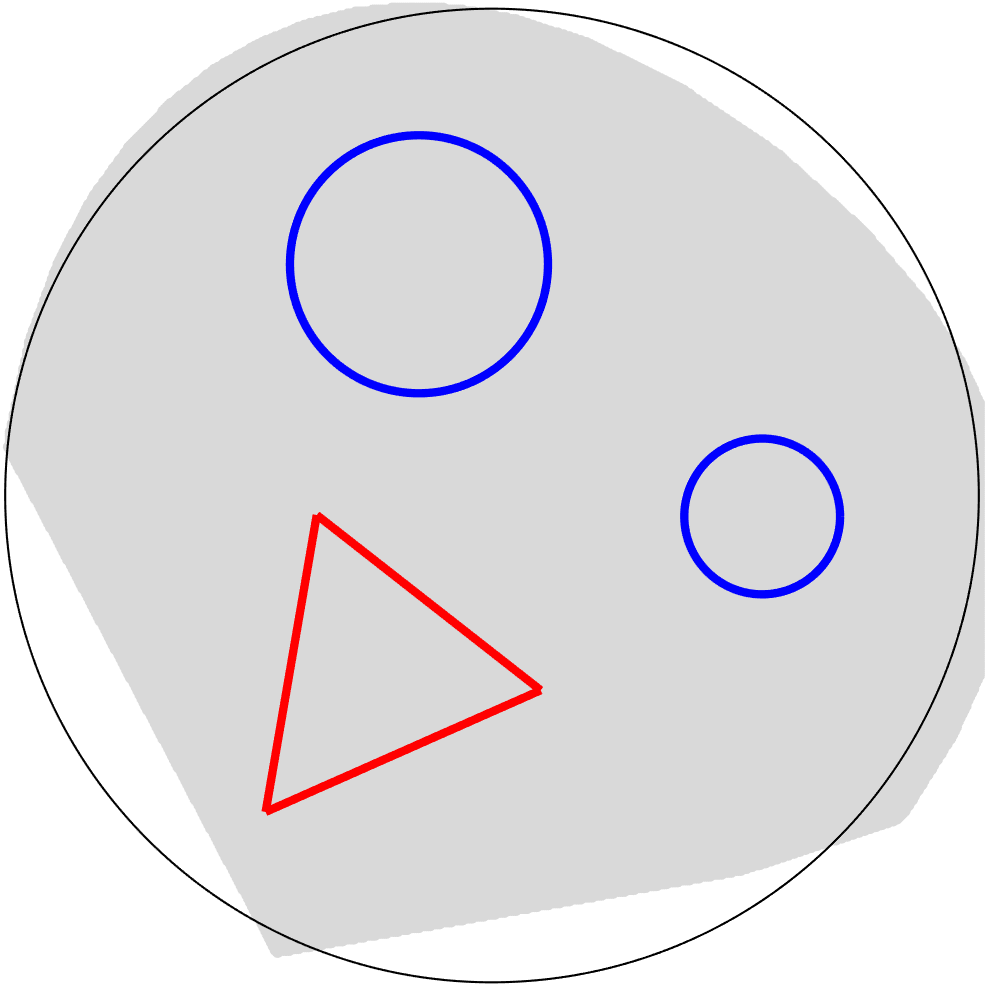} \newline 64.6\%
        \end{minipage}
    \end{tabular} 
  \addtocounter{figure}{-6}
    \caption{Comparison of the ground truth, learned and least squares hulls of the experimental phantoms. The error relative to the ground truth is shown below each phantom.}
    \label{fig:experimental_results_3}
\end{figure}

\clearpage
\begin{figure}[ht]
    \begin{tabular}{lccccc}
        Case & Phantom & Ground truth & Learned & Least squares \\ [.2cm]
        1.1 
        &
        \begin{minipage}{0.25\textwidth}
        \centering \includegraphics[height=2.8cm]{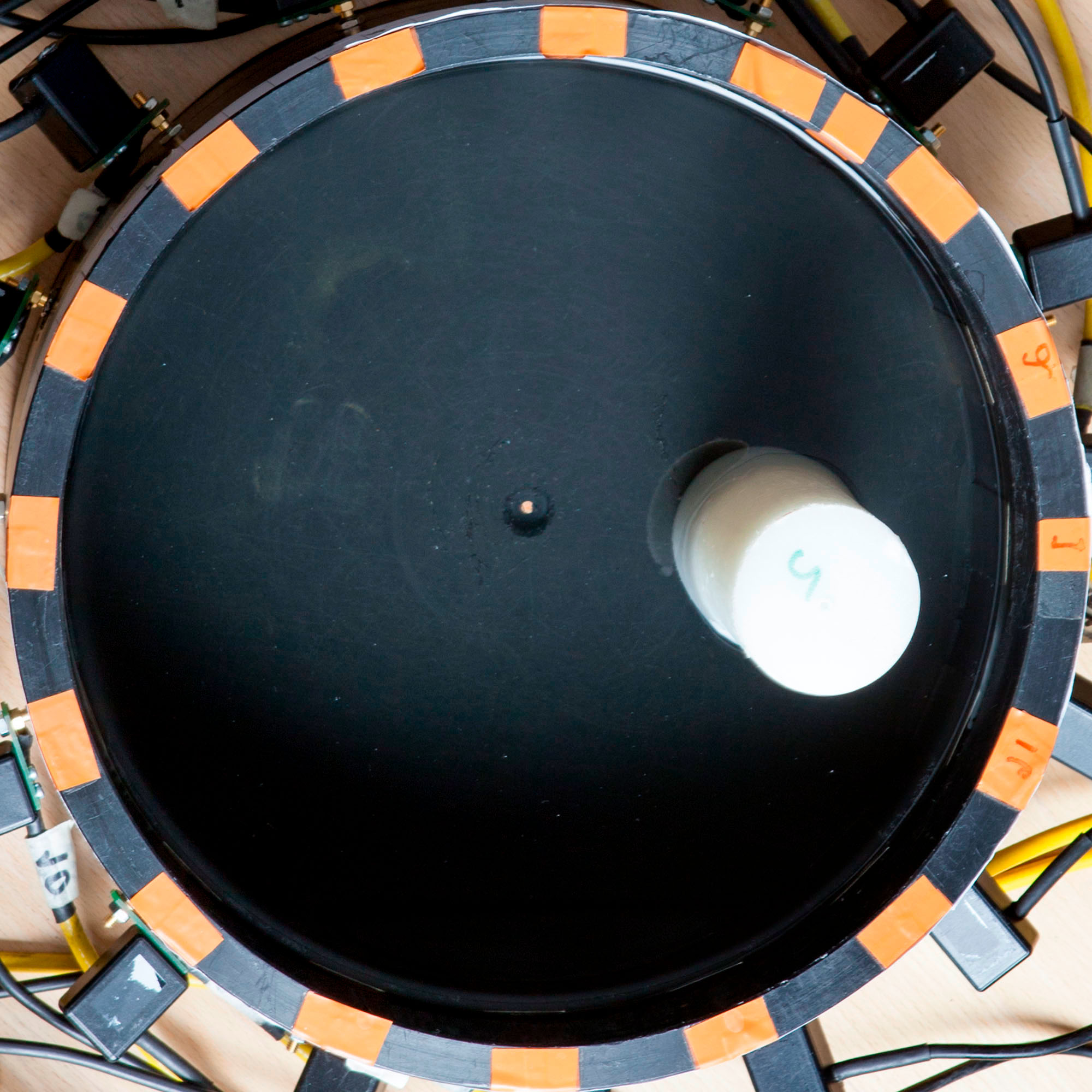} \newline
        \end{minipage}
        &\begin{minipage}{0.18\textwidth}
        \centering \includegraphics[height=2.3cm]{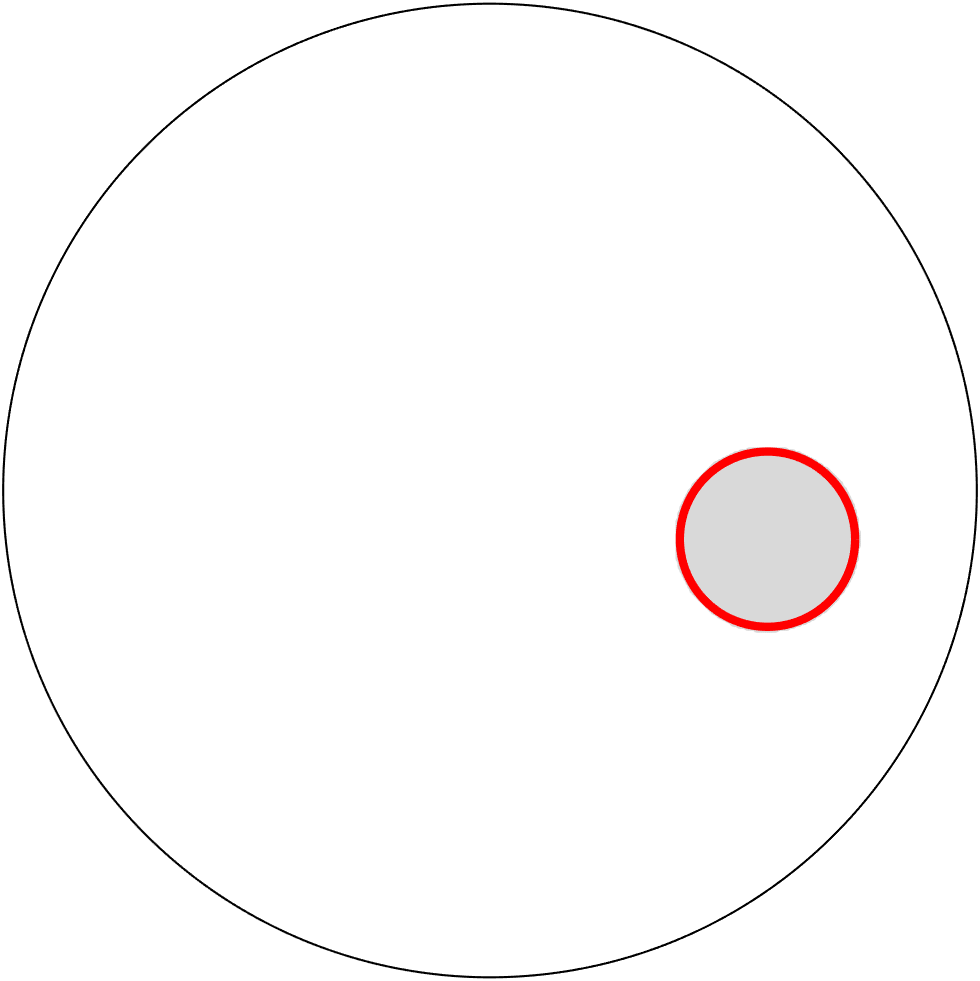} \newline
        \end{minipage}
        & \begin{minipage}{0.18\textwidth}
        \centering \includegraphics[height=2.3cm]{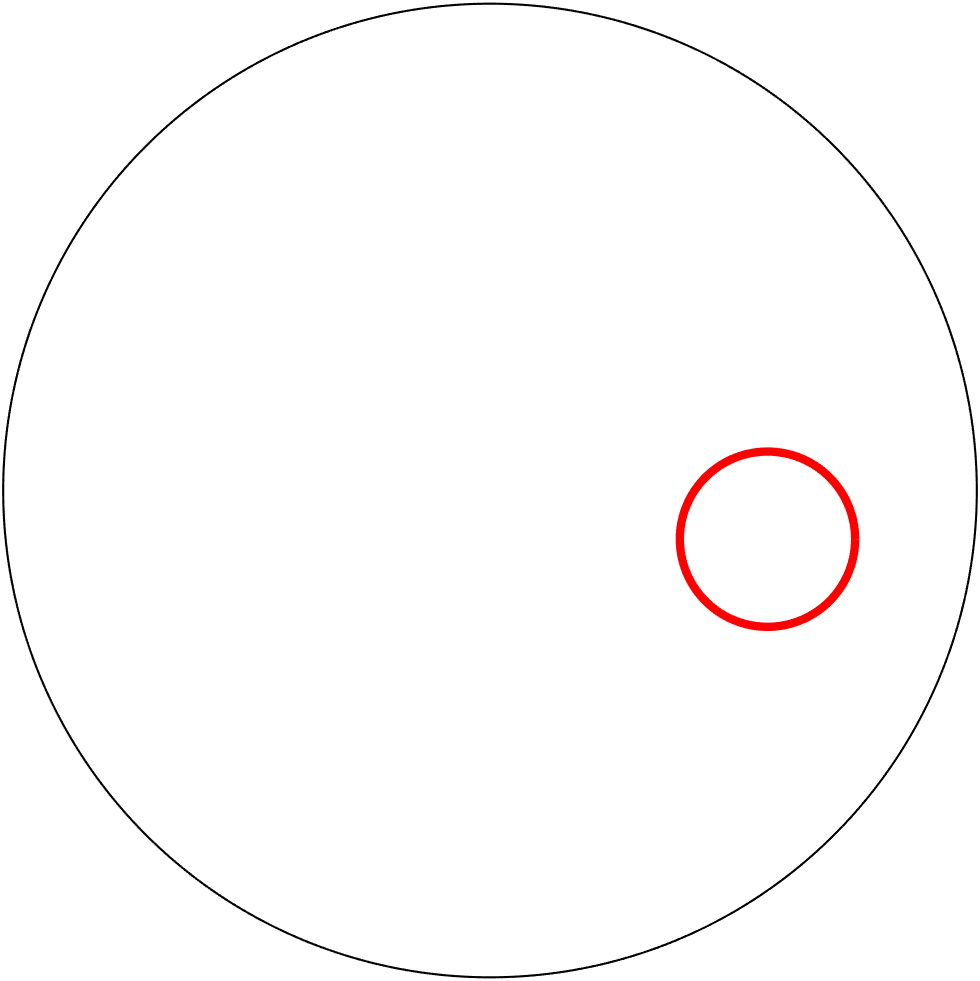} \newline 3.6\%
        \renewcommand{\thefigure}{1.1}
         \captionlistentry{} 
         \label{fig:experimental_results_1-1}
        \end{minipage}
        & \begin{minipage}{0.18\textwidth}
        \centering \includegraphics[height=2.3cm]{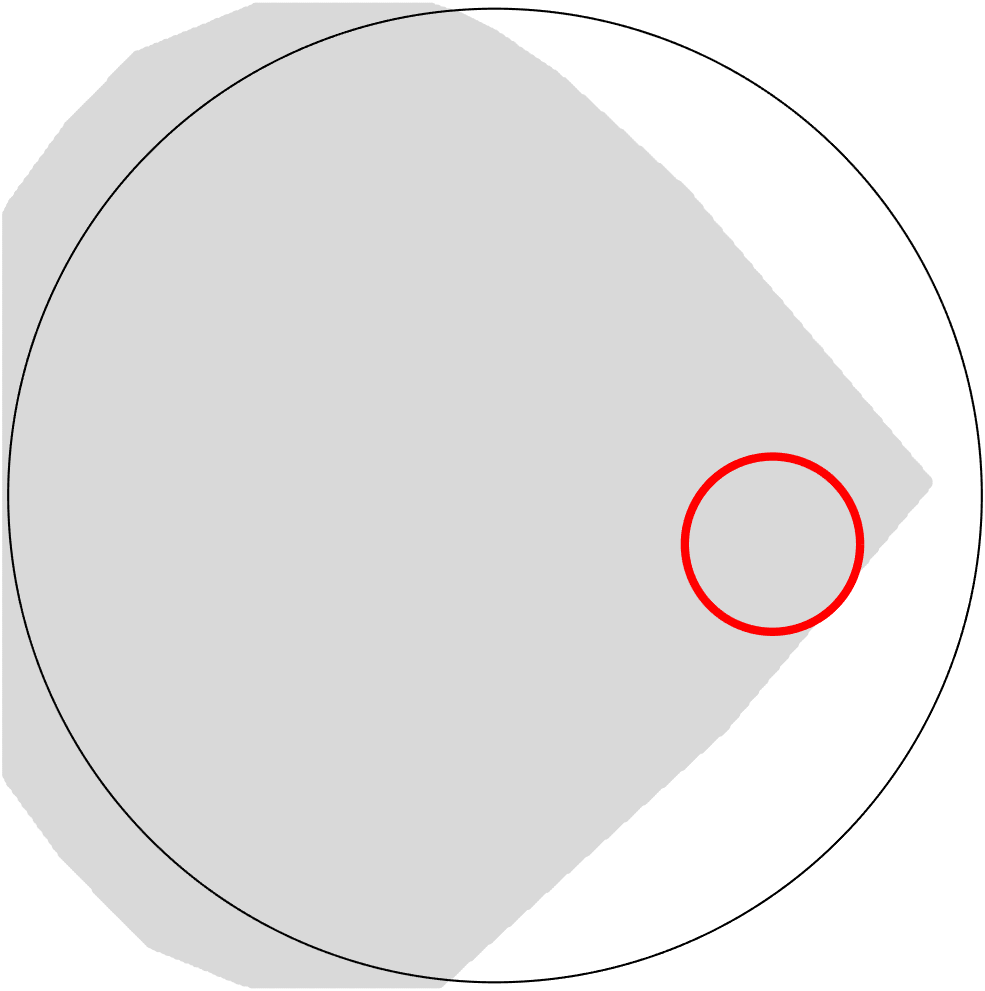} \newline 92.6\%
        \end{minipage}\\[1.3cm]
        1.2 
        &
        \begin{minipage}{0.25\textwidth}
        \centering \includegraphics[height=2.8cm]{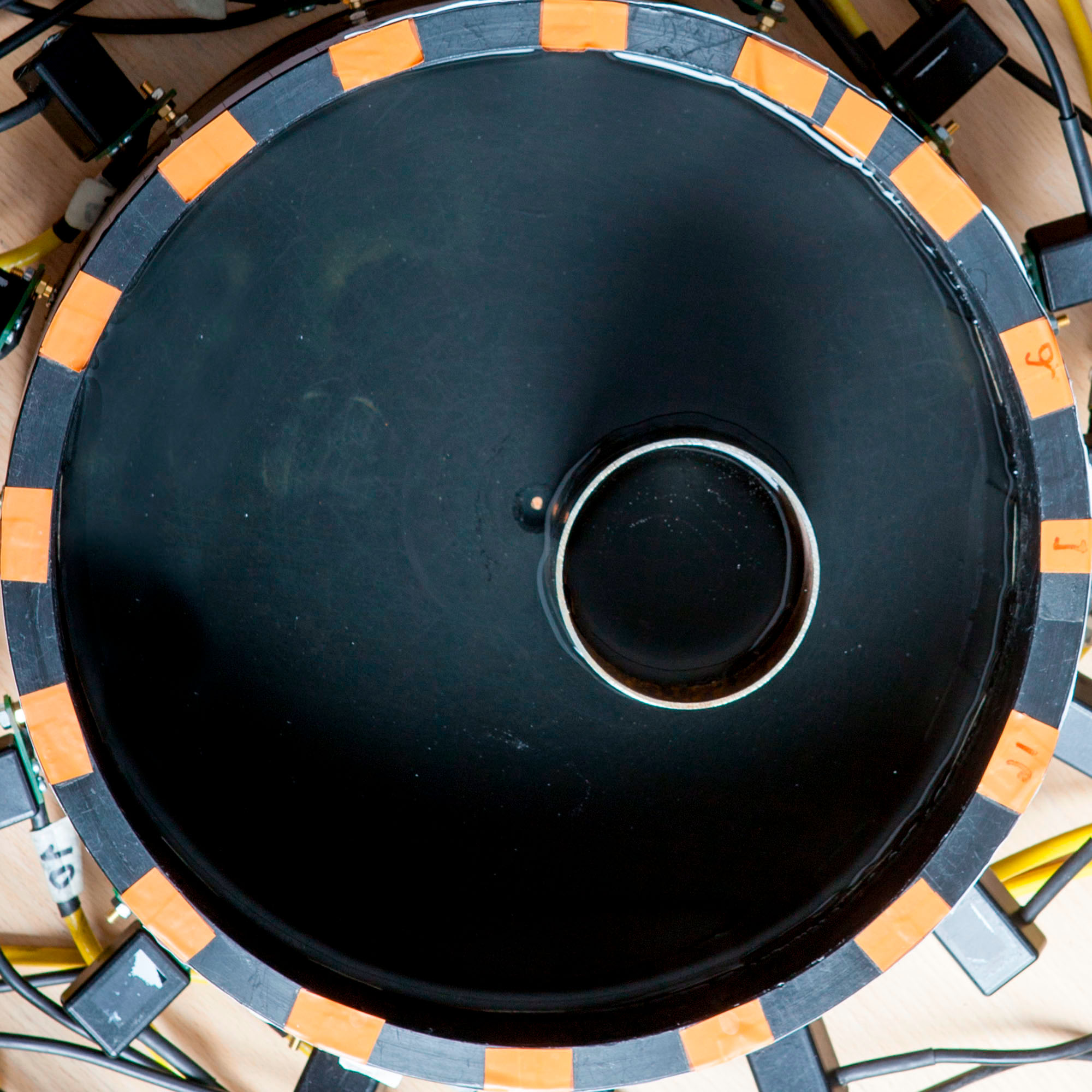} \newline
        \end{minipage}
        &\begin{minipage}{0.18\textwidth}
        \centering \includegraphics[height=2.3cm]{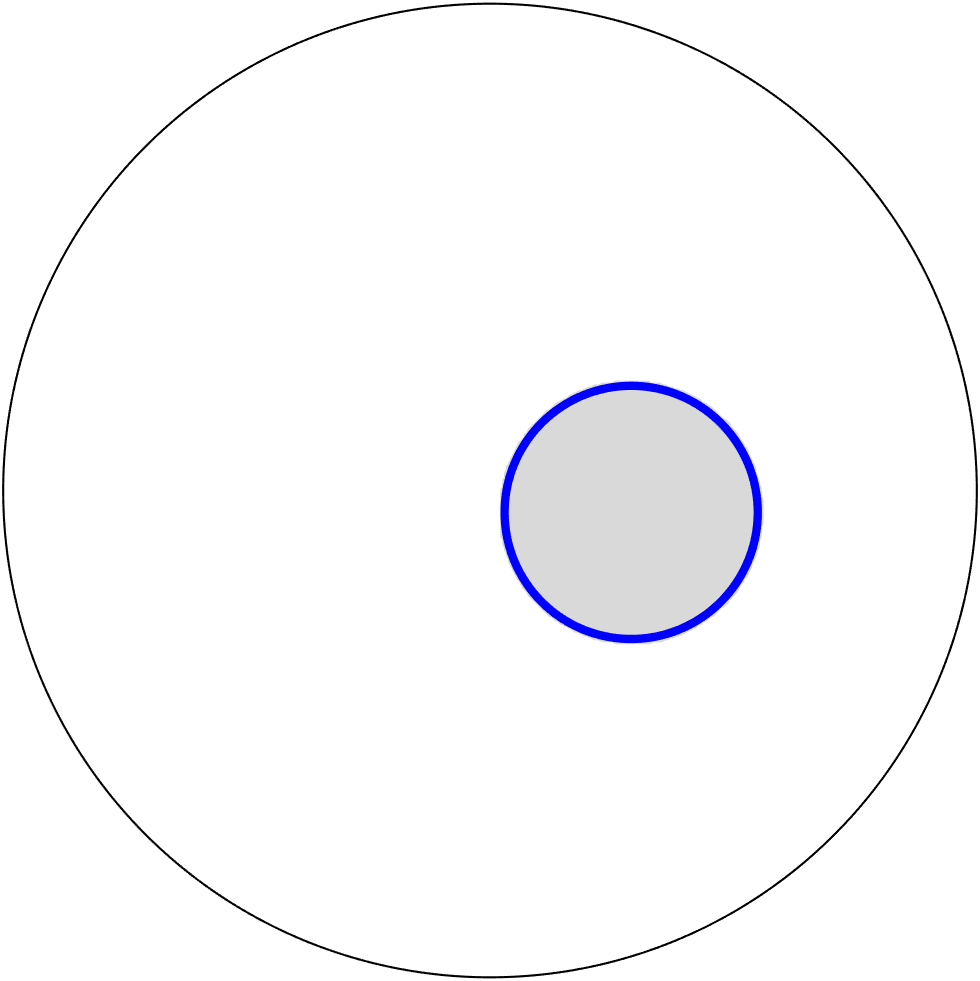} \newline
        \end{minipage}
        & \begin{minipage}{0.18\textwidth}
        \centering \includegraphics[height=2.3cm]{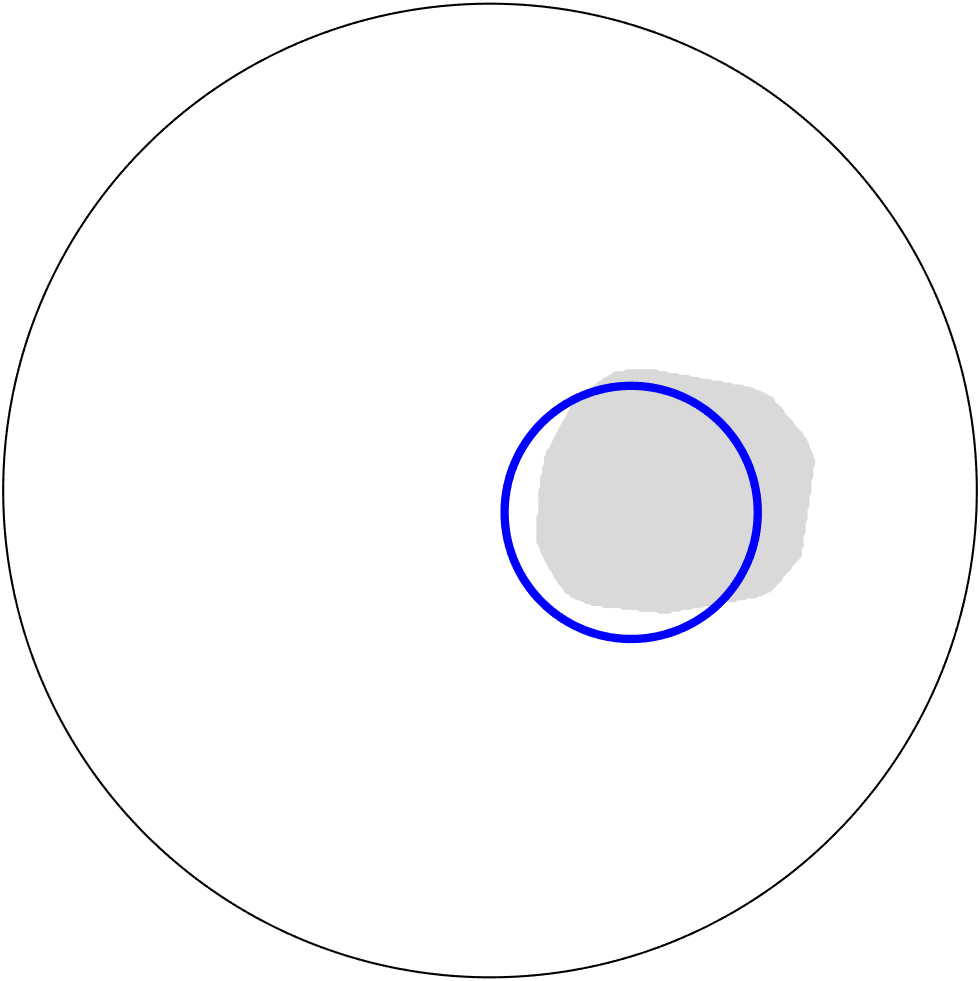} \newline 3.5\%
        \renewcommand{\thefigure}{1.2}
         \captionlistentry{} 
         \label{fig:experimental_results_1-2}
        \end{minipage}
        & \begin{minipage}{0.18\textwidth}
        \centering \includegraphics[height=2.3cm]{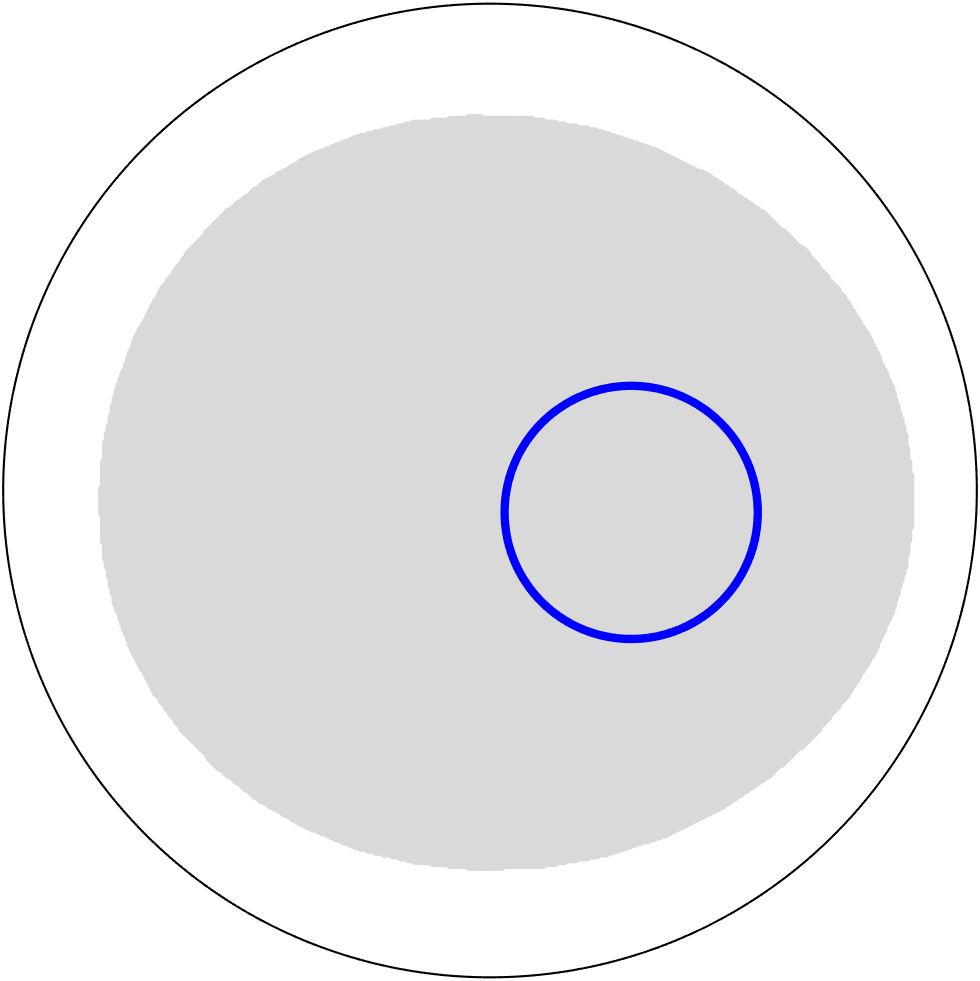} \newline 62.6\%
        \end{minipage}\\[1.3cm]
        1.3 
        &
        \begin{minipage}{0.25\textwidth}
        \centering \includegraphics[height=2.8cm]{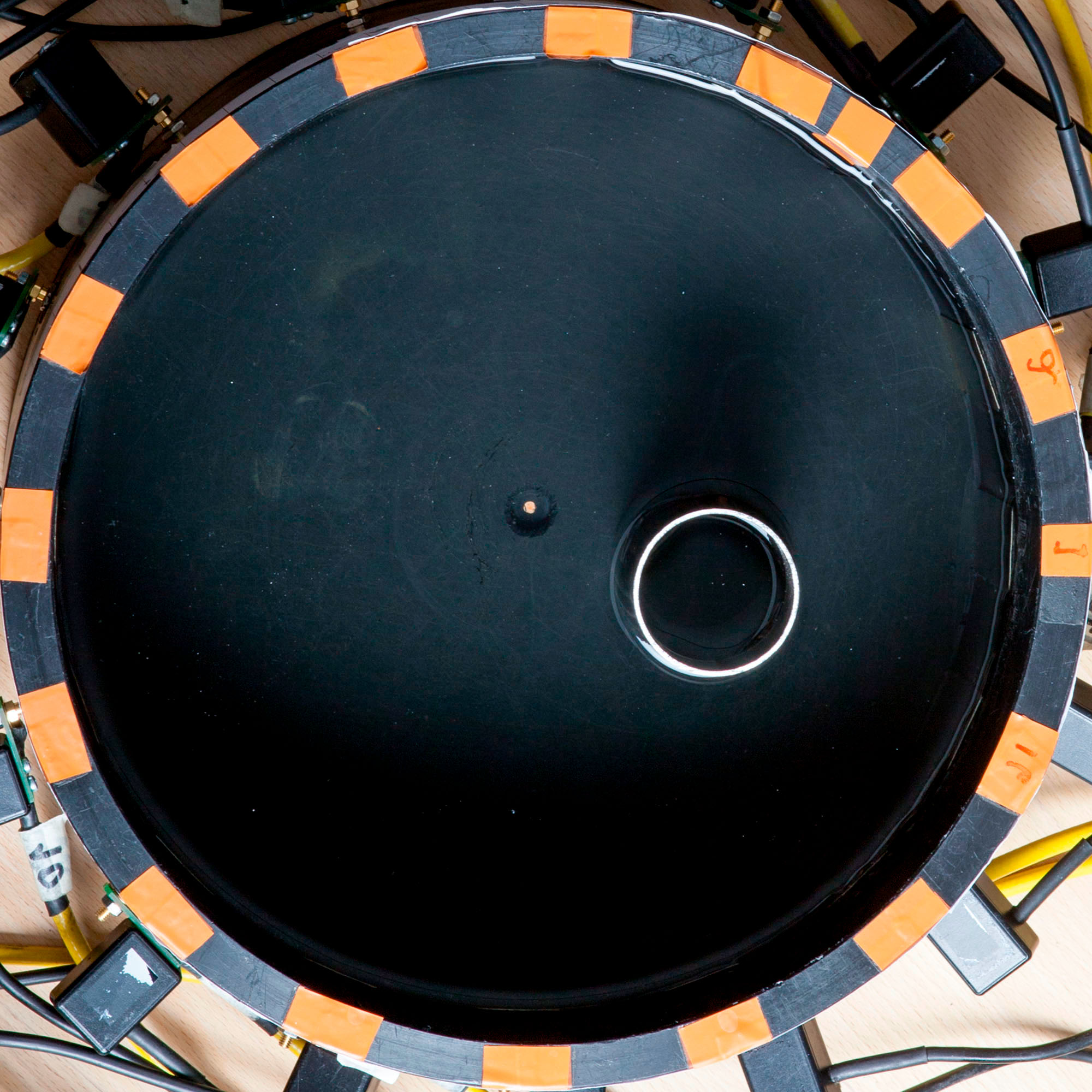} \newline
        \end{minipage}
        &\begin{minipage}{0.18\textwidth}
        \centering \includegraphics[height=2.3cm]{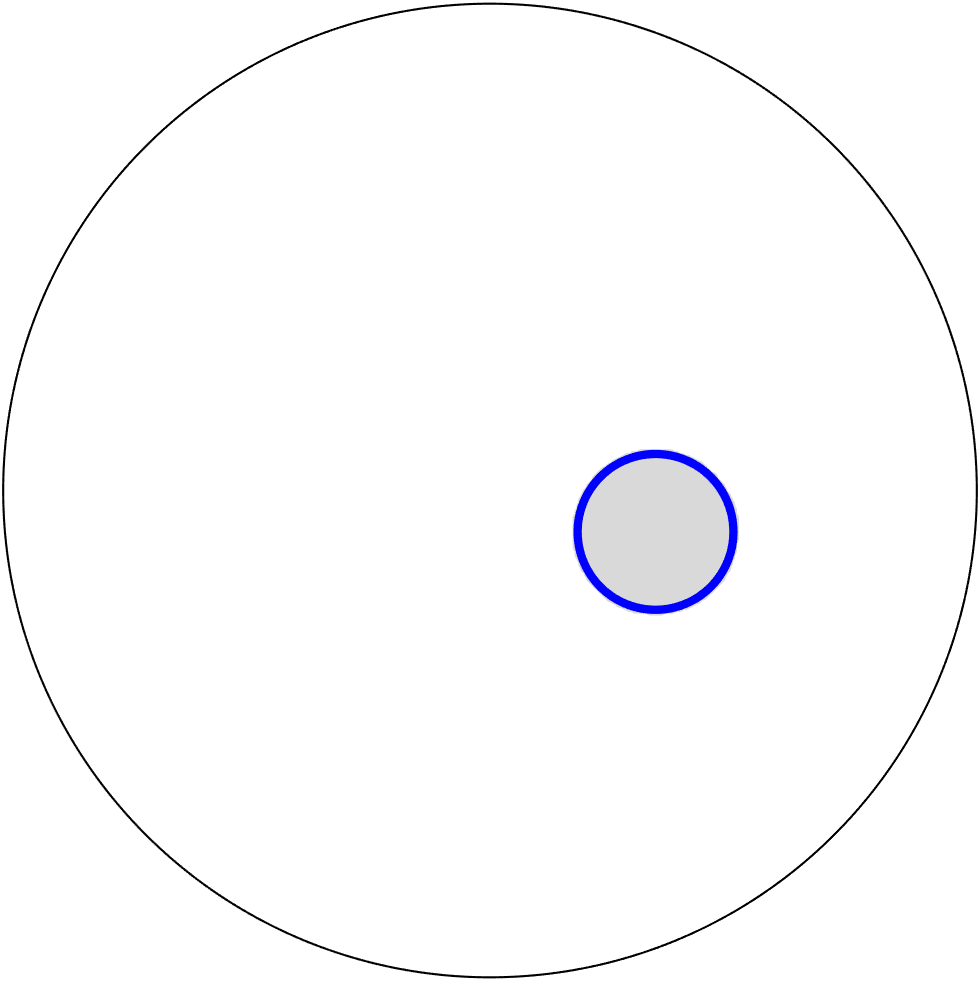} \newline
        \end{minipage}
        & \begin{minipage}{0.18\textwidth}
        \centering \includegraphics[height=2.3cm]{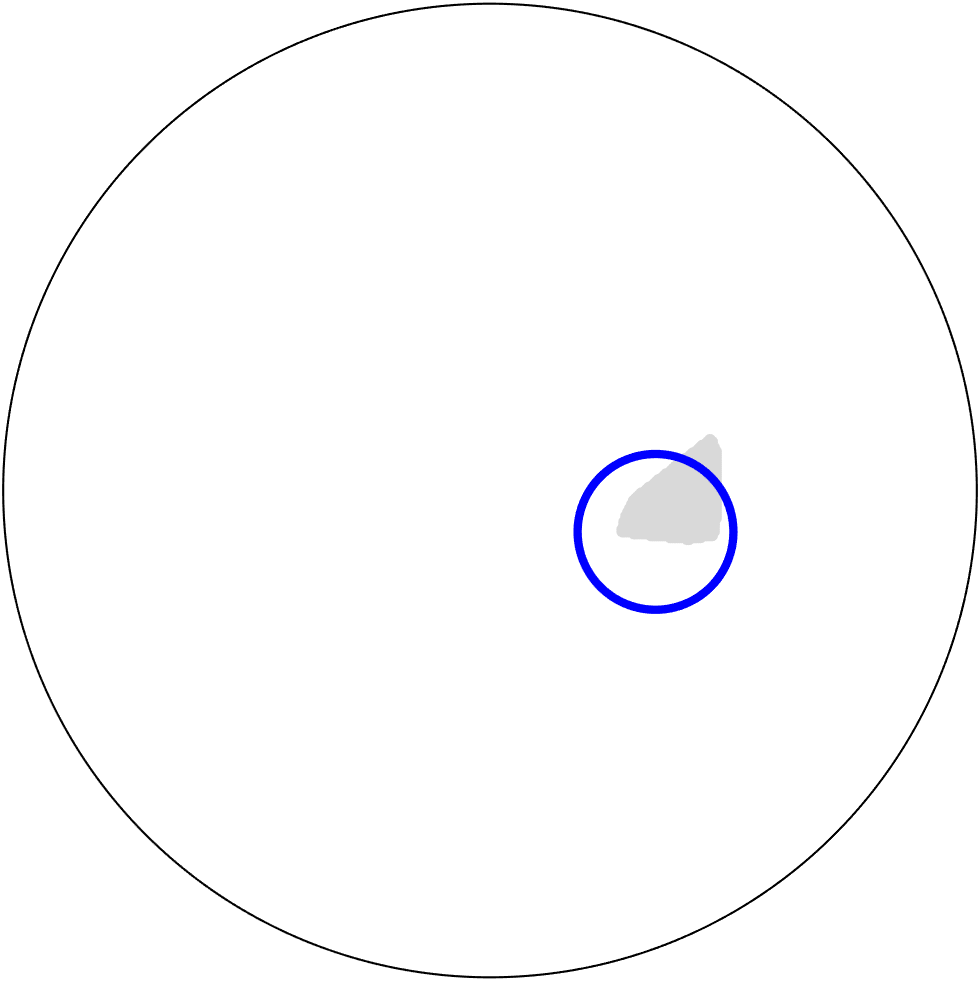} \newline 2.2\%
        \renewcommand{\thefigure}{1.3}
         \captionlistentry{} 
         \label{fig:experimental_results_1-3}
        \end{minipage}
        & \begin{minipage}{0.18\textwidth}
        \centering \includegraphics[height=2.3cm]{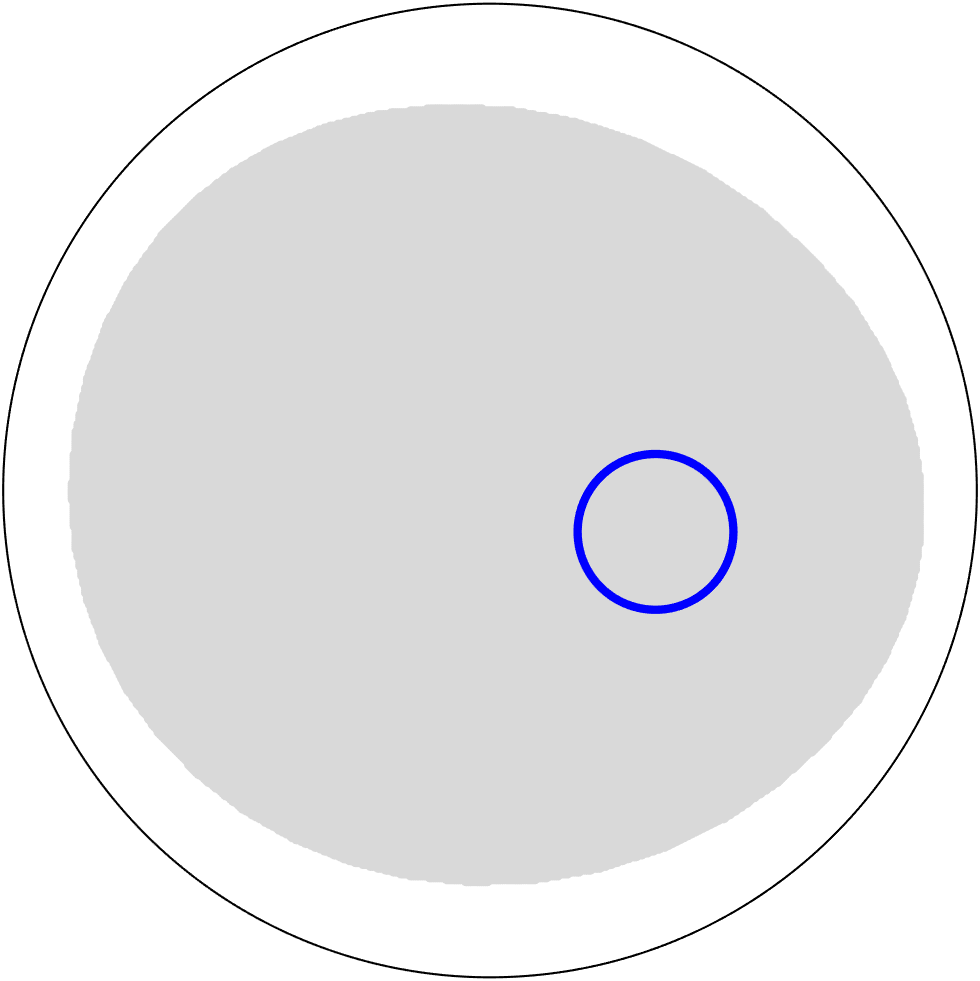} \newline 73.2\%
        \end{minipage}\\[1.3cm]
        1.4 
        &
        \begin{minipage}{0.25\textwidth}
        \centering \includegraphics[height=2.8cm]{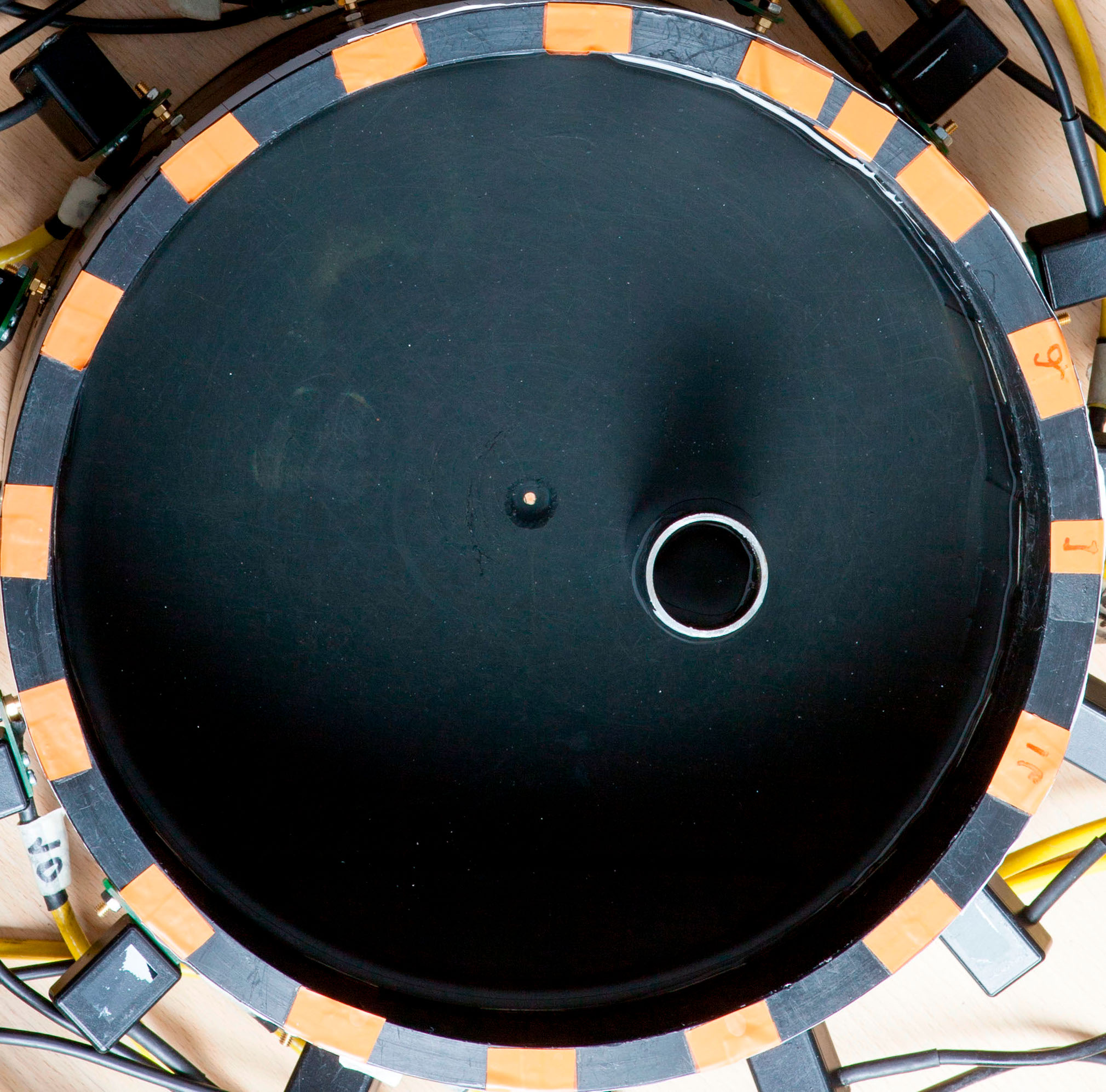} \newline
        \end{minipage}
        &\begin{minipage}{0.18\textwidth}
        \centering \includegraphics[height=2.3cm]{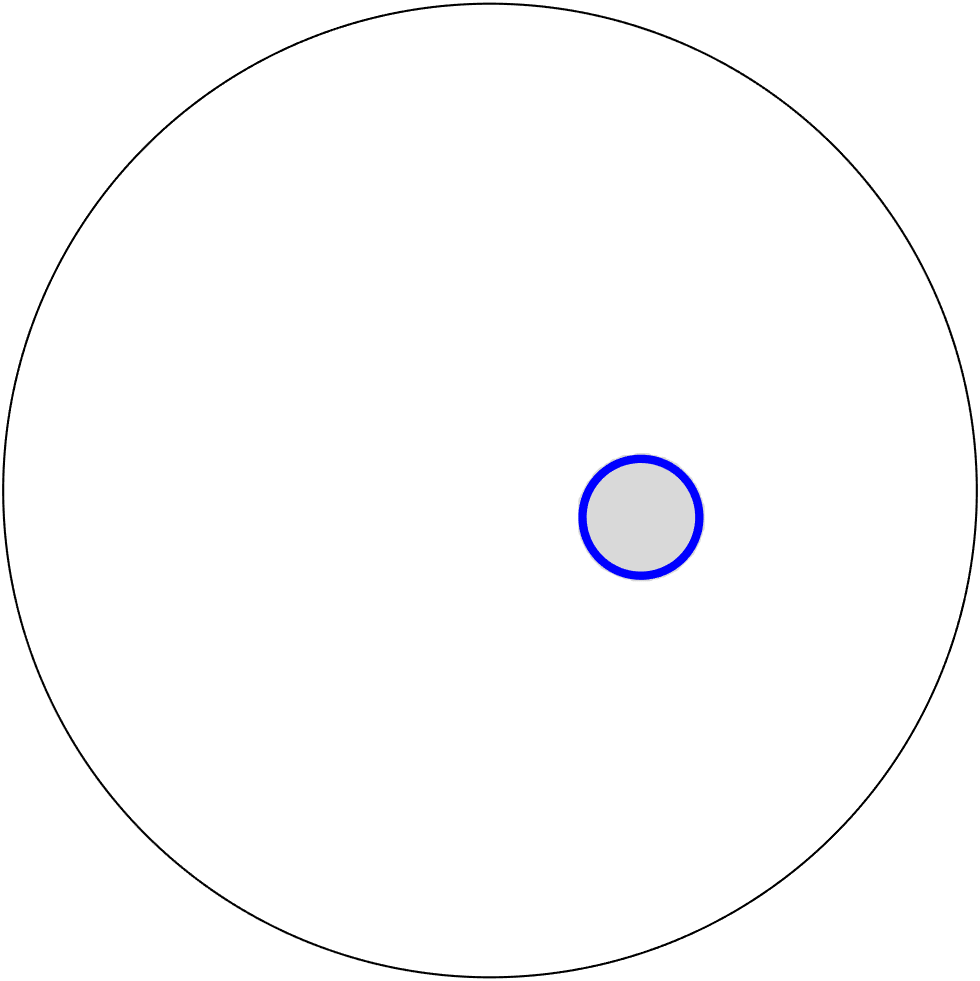} \newline
        \end{minipage}
        & \begin{minipage}{0.18\textwidth}
        \centering \includegraphics[height=2.3cm]{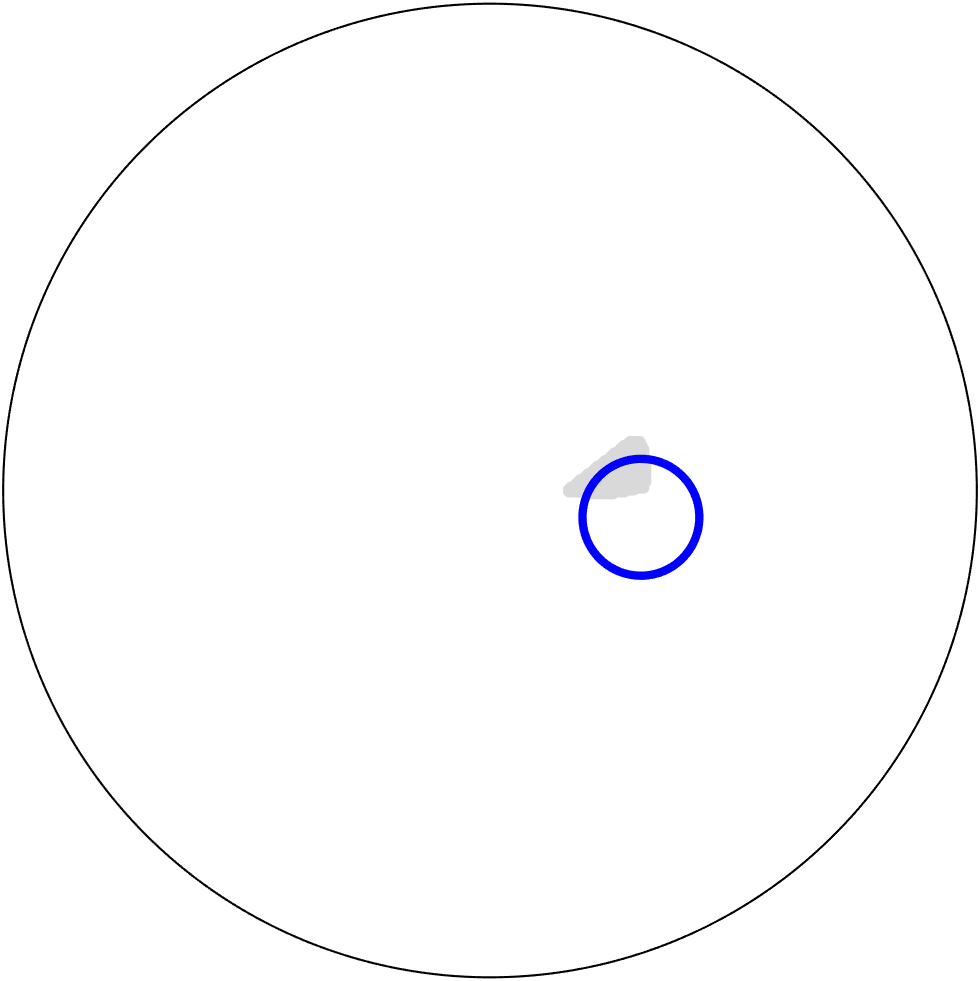} \newline 1.5\%
        \renewcommand{\thefigure}{1.4}
         \captionlistentry{} 
         \label{fig:experimental_results_1-4}
        \end{minipage}
        & \begin{minipage}{0.18\textwidth}
        \centering \includegraphics[height=2.3cm]{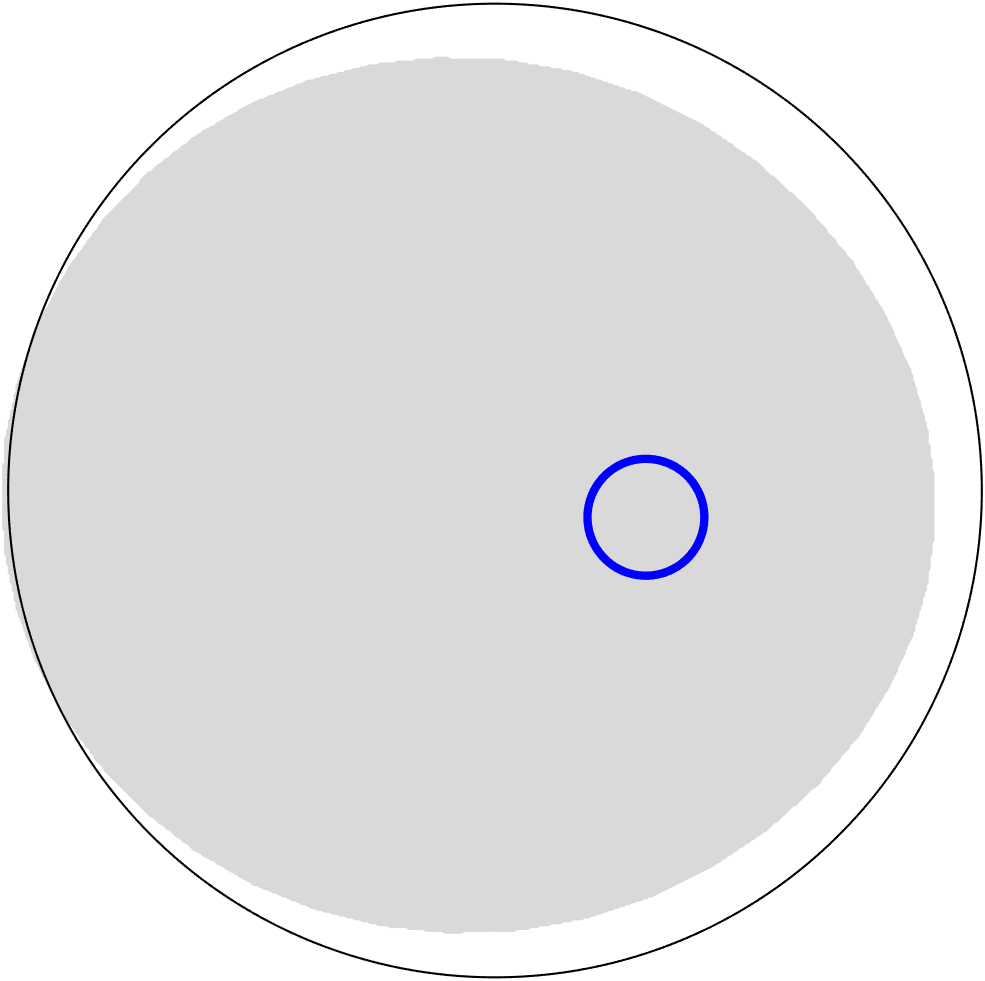} \newline 91.1\%
        \end{minipage}
    \end{tabular} 
  \addtocounter{figure}{-4}
    \caption{Comparison of the ground truth, learned and least squares hulls of the experimental phantoms. The error relative to the ground truth is shown below each phantom.}
    \label{fig:experimental_results_4}
\end{figure}

\section{Discussion}

\subsection{Performance on simulated data} For simulated data, the learned hull surpasses the least squares hull in accuracy for all 1000 conductivity phantoms in the test set. Based on the error analysis in the histograms, using the learned hull significantly decreases the relative errors within the simulated testing set (Figure \ref{fig:relerr_histograms}), even though the testing set contains cases that violate the jump condition and are therefore not covered by theory.

As can be seen in the histogram of false positives (Figure \ref{fig:relerr_histograms_b}), the least squares approach is highly prone to overestimation of the hull. This can already be seen in Figure $\ref{fig:LS_example}$ with a noise-free example, and adding noise to the DN maps further exacerbates the issue. 
On the other hand, the learned hull captures the inclusions quite well, but sometimes underestimates the size of the hull, leaving some parts of the inclusions outside, as can be seen in Figure \ref{fig:simulated_examples} in the examples \ref{fig:simulatedresults_III} and  \ref{fig:simulatedresults_IV} where the jump condition fails, and interestingly also in example \ref{fig:simulatedresults_V}. However, based on the histogram of false negatives (Figure \ref{fig:relerr_histograms_c}), this problem is also present and more serious with the least squares approach for some cases in the test set. 

Angles in the inclusions seem to be difficult for the learned hull to capture, most likely because the model was trained with only elliptical inclusions. 

\subsection{Performance on experimental data}
Even though the network was trained with simulated continuum model data, it generalizes well for experimental data, as can be seen in the results in Figures \ref{fig:experimental_results_1}, \ref{fig:experimental_results_2}, \ref{fig:experimental_results_3}, and \ref{fig:experimental_results_4}. The figures show that the learned hull is a more accurate reconstruction of the ground truth compared to the least squares hull for all phantoms except for the case \ref{fig:experimental_results_3-1}, where both reconstructions underestimate the hull. It could also be argued that the least squares approach performs better in the case \ref{fig:experimental_results_1-1}, because the learned hull predicts that there is no inclusion at all. Since the least squares hull is also having trouble with the case \ref{fig:experimental_results_1-1}, the cause of the inaccuracy is most likely already present in the indicator function, and thus in the input given to the neural network. Resistive inclusions are generally more difficult to detect, which can be a factor in the poor results for both approaches.

Due to distinguishability issues \cite{isaacson2007distinguishability}, inclusions further away from the boundary are exponentially more difficult to detect from the DN map, which might contribute to the lackluster results in cases \ref{fig:experimental_results_1-3} and \ref{fig:experimental_results_1-4}. The inclusions are also very small, which adds to the difficulty.

When there are multiple inclusions that satisfy the jump condition, it appears that both the learned and the least squares approach tend to underestimate the hull of resistive inclusions and overestimate the hull of conductive inclusions. This is best visualized in Figure \ref{fig:experimental_results_2}. For phantoms with both plastic and metal inclusions where the jump condition fails, the learned hull is able to find the location of the inclusions quite well, except in case \ref{fig:experimental_results_4-4} where the conductive and resistive inclusions are very close to each other. These cases are not covered by the theory and the least squares hull also produces poor results, especially for the aforementioned case. Thus, it is quite remarkable to have even coarse estimates for these cases with experimental data.

\begin{figure} [!ht]
    \centering \hfill
    \includegraphics[width=0.4\linewidth]{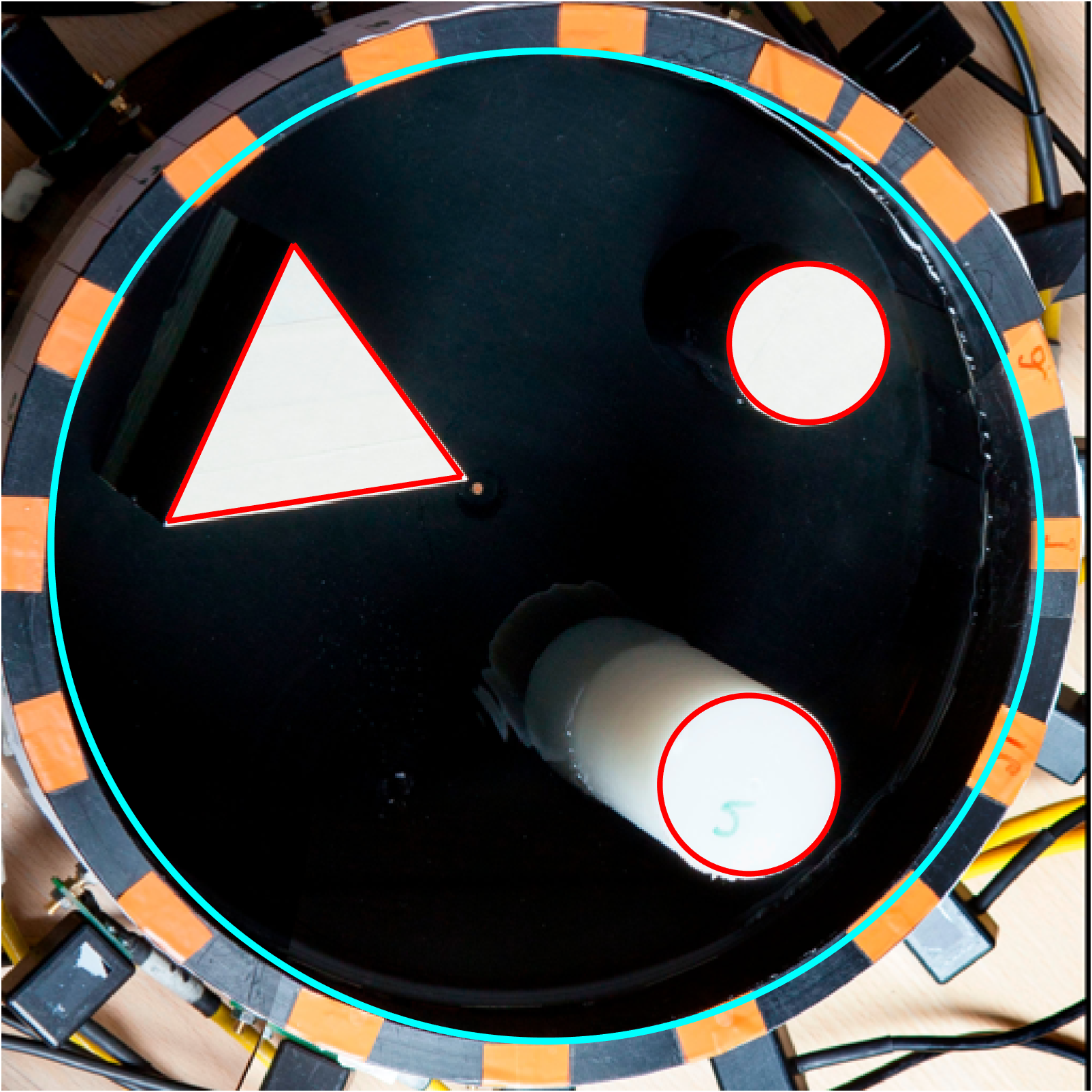}
    \hfill
    \includegraphics[width=0.4\linewidth]{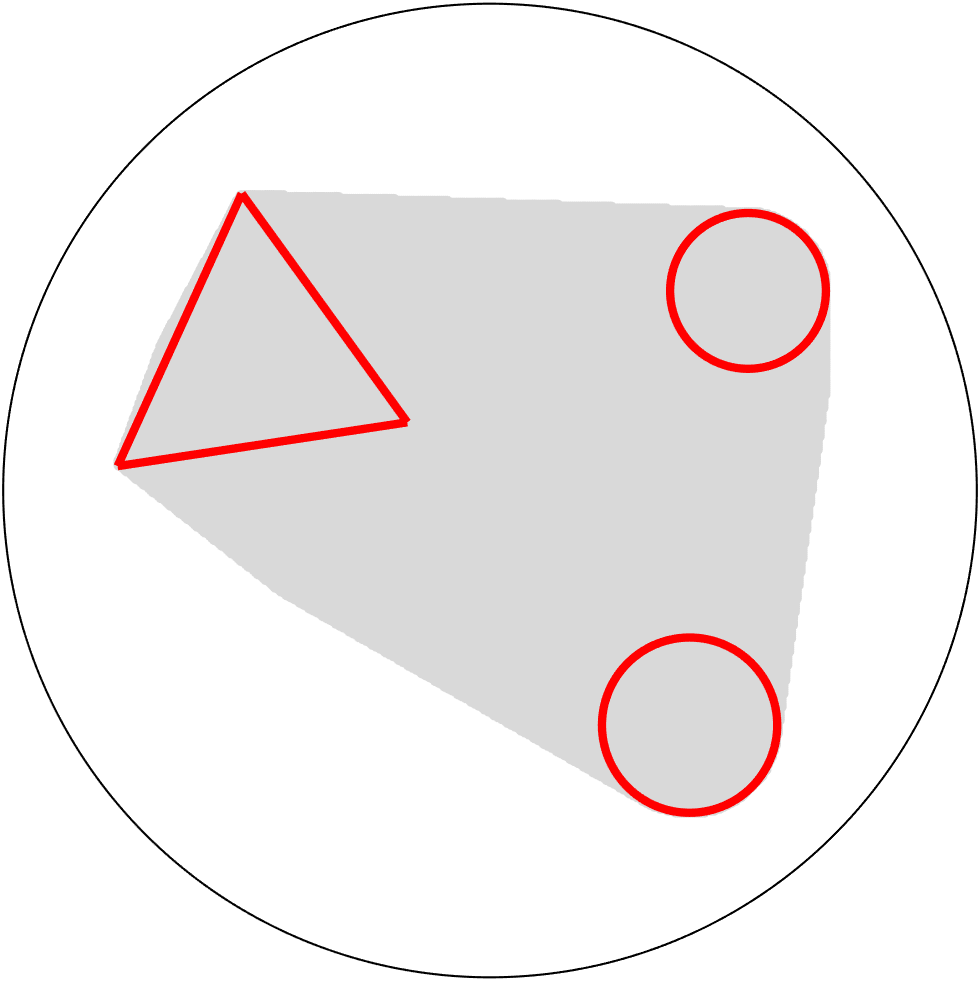} \hfill
  \caption{The image on the left shows an example of the segmentation and the image on the right shows the resulting convex hull that is used as ground truth.}
    \label{fig:segmentation_example} 
\end{figure}

Segmenting the objects according to their highest point makes the segmentation consistent across images, but it introduces errors into the ground truth since the objects have different heights. This is especially true for the cases \ref{fig:experimental_results_3-3}, \ref{fig:experimental_results_4-3}, \ref{fig:experimental_results_4-4}, and \ref{fig:experimental_results_1-1} that feature a plastic cylinder that is significantly taller than the other segmented objects. As an example, a closer look at the segmentation of the case 3.3. is shown in Figure \ref{fig:segmentation_example}. However, due to the way the relative error is calculated (\ref{relerr}), a small error in the location of one inclusion is likely to have a minimal effect on the relative error over the whole tank.

\subsection{Influence of network size and training data}
When training the neural network, we tested different architectures and training set sizes. Having more convolutional layers improved the results on simulated data, but made the network overfit. When testing with experimental data, the larger networks made the learned hulls too small or even completely disappear. Increasing the amount of training data did not improve the results, likely because the number of trainable parameters is quite small. This is expected, since we don't need high expressivity in the network as the learned task is fairly low-dimensional.

\section{Conclusion}

The enclosure method using a learned hull shows promise in simple experimental data and thus opens up several directions for future studies and ways to extend the method. To improve the method, simulated training data could be generated with CEM, instead of the continuum model, and it could also include inclusions with non-smooth shapes. 

Interesting directions for extending the method include probing with triangle corners or disks \cite{ide2007probing} instead of the half-planes, or extending the method to 3D \cite{ide2010local} for experimental data measurements.

\section*{Acknowledgments}
A. Hauptmann acknowledges support from Research Council of Finland (projs. 353093, 338408, 359186). T. Ide acknowledges support from Josai University Presidential Grants. S.Rautio and S. Siltanen acknowledge support from the Research Council of Finland (CoE 353097 and FAME Flagship 359182). S. Sippola acknowledges support from the Finnish Ministry of Education and Culture’s Doctoral Education Pilot for Mathematics of Sensing, Imaging and Modelling.

\bibliographystyle{plain}
\bibliography{main}

\end{document}

%% file: NN_arch.tex
\centering
\begin{tikzpicture}[
    arrow/.style={->, thick},
    squarelayer/.style={
        draw=black, fill=gray!20,
        font=\fontsize{7}{8}\selectfont,
        align=center,
        minimum width=2.2cm, minimum height=1.5cm
    }
]

\node[squarelayer, minimum width=2.5cm, minimum height=1.5cm] (input) {Input\\$30\times 50\times 1$};

\foreach \i in {0,...,10} {
    \node[squarelayer, right=1cm+\i*0.06cm of input, yshift=-\i*0.06cm] (conv\i) {};
}
\node[font=\fontsize{7}{8}\selectfont] at (conv10) {$31\times 47\times 60$};

\foreach \i in {0,...,10} {
    \node[squarelayer, right=1cm+\i*0.06cm of conv0, yshift=-\i*0.06cm] (convB\i) {};
}
\node[font=\fontsize{7}{8}\selectfont] at (convB10) {$32\times 44\times 60$};

\node[squarelayer, right=1cm of convB10, minimum width=0.5cm, minimum height=2.25cm] (fc) {$45$\\$\times$\\$1$\\$\times$\\$1$};

\draw[arrow,red] (input) -- (conv0);
\draw[arrow,red] (conv10) -- (convB10);
\draw[arrow,blue] (convB10) -- (fc);

\node[below= 15mm of conv10] (xx) {};
\node[above= 5mm of xx] (L1) {};
\node[right= 6mm of L1] (L2) {};
\node[right= 0mm of L2, font=\fontsize{7}{8}\selectfont] (L3) {6 x 4 conv, leaky ReLU};

\node[below= 2mm of L1] (L4) {};
\node[right= 6mm of L4] (L5) {};
\node[right= 0mm of L5, font=\fontsize{7}{8}\selectfont] (L6) {fully connected, tanh};

\node[fit=(L1)(L3) (L6), draw, inner sep=1mm] (fit1) {};

\draw[->,red] (L1)--(L2);
\draw[->,blue] (L4)--(L5);

\end{tikzpicture}
\caption{Neural network architecture, with two convolutional layers and a fully connected output layer.}

%% file: main.bbl
\begin{thebibliography}{10}

\bibitem{alessandrini1988stable}
Giovanni Alessandrini.
\newblock Stable determination of conductivity by boundary measurements.
\newblock {\em Applicable Analysis}, 27(1-3):153--172, 1988.

\bibitem{alsaker2024ct}
Melody Alsaker, Siiri Rautio, Fernando Moura, Juan~Pablo Agnelli, Rashmi
  Murthy, Matti Lassas, Jennifer~L Mueller, and Samuli Siltanen.
\newblock Ct scans without x-rays: parallel-beam imaging from nonlinear current
  flows.
\newblock {\em arXiv preprint arXiv:2408.12992}, 2024.

\bibitem{bruhl2001explicit}
Martin Br{\"u}hl.
\newblock Explicit characterization of inclusions in electrical impedance
  tomography.
\newblock {\em SIAM Journal on Mathematical Analysis}, 32(6):1327--1341, 2001.

\bibitem{bruhl2000numerical}
Martin Br{\"u}hl and Martin Hanke.
\newblock Numerical implementation of two noniterative methods forlocating
  inclusions by impedance tomography.
\newblock {\em Inverse Problems}, 16(4):1029, 2000.

\bibitem{Calderon1980}
A.-P. {C}alder{\'o}n.
\newblock On an inverse boundary value problem.
\newblock In {\em Seminar on {N}umerical {A}nalysis and its {A}pplications to
  {C}ontinuum {P}hysics ({R}io de {J}aneiro, 1980)}, pages 65--73. Soc. Brasil.
  Mat., Rio de Janeiro, 1980.

\bibitem{Cheney1999}
M.~Cheney, D.~Isaacson, and J.~C. Newell.
\newblock Electrical impedance tomography.
\newblock {\em SIAM Review}, 41(1):85--101, 1999.

\bibitem{garde2021mimicking}
Henrik Garde and Nuutti Hyv{\"o}nen.
\newblock Mimicking relative continuum measurements by electrode data in
  two-dimensional electrical impedance tomography.
\newblock {\em Numerische Mathematik}, 147:579–609, 2021.

\bibitem{guo2021construct}
Ruchi Guo and Jiahua Jiang.
\newblock Construct deep neural networks based on direct sampling methods for
  solving electrical impedance tomography.
\newblock {\em SIAM Journal on Scientific Computing}, 43(3):B678--B711, 2021.

\bibitem{Hallaji_2014}
Milad Hallaji, Aku Seppänen, and Mohammad Pour-Ghaz.
\newblock Electrical impedance tomography-based sensing skin for quantitative
  imaging of damage in concrete.
\newblock {\em Smart Materials and Structures}, 23(8):085001, jun 2014.

\bibitem{hamilton2018deep}
Sarah~Jane Hamilton and Andreas Hauptmann.
\newblock Deep d-bar: Real-time electrical impedance tomography imaging with
  deep neural networks.
\newblock {\em IEEE transactions on medical imaging}, 37(10):2367--2377, 2018.

\bibitem{hauptmann2018revealing}
Andreas Hauptmann, Masaru Ikehata, Hiromichi Itou, and Samuli Siltanen.
\newblock Revealing cracks inside conductive bodies by electric surface
  measurements.
\newblock {\em Inverse Problems}, 35(2):025004, 2018.

\bibitem{hauptmann2017open}
Andreas Hauptmann, Ville Kolehmainen, Nguyet~Minh Mach, Tuomo Savolainen, Aku
  Sepp{\"a}nen, and Samuli Siltanen.
\newblock Open 2d electrical impedance tomography data archive.
\newblock {\em arXiv preprint arXiv:1704.01178}, 2017.

\bibitem{ide2010local}
T~Ide, H~Isozaki, S~Nakata, and S~Siltanen.
\newblock Local detection of three-dimensional inclusions in electrical
  impedance tomography.
\newblock {\em Inverse problems}, 26(3):035001, 2010.

\bibitem{ide2007probing}
Takanori Ide, Hiroshi Isozaki, Susumu Nakata, Samuli Siltanen, and Gunther
  Uhlmann.
\newblock Probing for electrical inclusions with complex spherical waves.
\newblock {\em Communications on Pure and Applied Mathematics: A Journal Issued
  by the Courant Institute of Mathematical Sciences}, 60(10):1415--1442, 2007.

\bibitem{Ikehata1999a}
Masaru Ikehata.
\newblock How to draw a picture of an unknown inclusion from boundary
  measurements. {T}wo mathematical inversion algorithms.
\newblock {\em Journal of Inverse and Ill-Posed Problems}, 7(3):255--271, 1999.

\bibitem{Ikehata2000c}
Masaru Ikehata.
\newblock Reconstruction of the support function for inclusion from boundary
  measurements.
\newblock {\em Journal of Inverse and Ill-Posed Problems}, 8:367--378, 2000.

\bibitem{ikehata2000numerical}
Masaru Ikehata and Samuli Siltanen.
\newblock Numerical method for finding the convex hull of an inclusion in
  conductivity from boundary measurements.
\newblock {\em Inverse Problems}, 16(4):1043, 2000.

\bibitem{ikehata2004electrical}
Masaru Ikehata and Samuli Siltanen.
\newblock Electrical impedance tomography and mittag-leffler's function.
\newblock {\em Inverse Problems}, 20(4):1325, 2004.

\bibitem{isaacson2007distinguishability}
David Isaacson.
\newblock Distinguishability of conductivities by electric current computed
  tomography.
\newblock {\em IEEE transactions on medical imaging}, 5(2):91--95, 2007.

\bibitem{isaacson2004reconstructions}
David Isaacson, Jennifer~L Mueller, Jonathan~C Newell, and Samuli Siltanen.
\newblock Reconstructions of chest phantoms by the d-bar method for electrical
  impedance tomography.
\newblock {\em IEEE Transactions on medical imaging}, 23(7):821--828, 2004.

\bibitem{kourunen2008suitability}
J~Kourunen, T~Savolainen, A~Lehikoinen, M~Vauhkonen, and LM~Heikkinen.
\newblock Suitability of a pxi platform for an electrical impedance tomography
  system.
\newblock {\em Measurement Science and Technology}, 20(1):015503, 2008.

\bibitem{mueller2012linear}
Jennifer~L Mueller and Samuli Siltanen.
\newblock {\em Linear and nonlinear inverse problems with practical
  applications}.
\newblock SIAM, 2012.

\bibitem{pathiraja2020clinical}
Angela~A Pathiraja, Ruwan~A Weerakkody, Alexander~C von Roon, Paul Ziprin, and
  Richard Bayford.
\newblock The clinical application of electrical impedance technology in the
  detection of malignant neoplasms: a systematic review.
\newblock {\em Journal of Translational Medicine}, 18:1--11, 2020.

\bibitem{seppanen2001state}
A~Seppänen, M~Vauhkonen, PJ~Vauhkonen, E~Somersalo, and JP~Kaipio.
\newblock State estimation with fluid dynamical evolution models in process
  tomography-anapplication to impedance tomography.
\newblock {\em Inverse Problems}, 17(3):467, 2001.

\bibitem{siltanen2020electrical}
Samuli Siltanen and Takanori Ide.
\newblock Electrical impedance tomography, enclosure method and machine
  learning.
\newblock In {\em 2020 IEEE 30th International Workshop on Machine Learning for
  Signal Processing (MLSP)}, pages 1--6. IEEE, 2020.

\bibitem{smyl2018detection}
Danny Smyl, Mohammad Pour-Ghaz, and Aku Sepp{\"a}nen.
\newblock Detection and reconstruction of complex structural cracking patterns
  with electrical imaging.
\newblock {\em NDT \& E International}, 99:123--133, 2018.

\bibitem{somersalo1992existence}
Erkki Somersalo, Margaret Cheney, and David Isaacson.
\newblock Existence and uniqueness for electrode models for electric current
  computed tomography.
\newblock {\em SIAM Journal on Applied Mathematics}, 52(4):1023--1040, 1992.

\bibitem{tallman2020structural}
Tyler~N Tallman and Danny~J Smyl.
\newblock Structural health and condition monitoring via electrical impedance
  tomography in self-sensing materials: a review.
\newblock {\em Smart Materials and Structures}, 29(12):123001, 2020.

\bibitem{tanyu2023}
Derick~Nganyu Tanyu, Jianfeng Ning, Andreas Hauptmann, Bangti Jin, and Peter
  Maass.
\newblock Electrical impedance tomography: a fair comparative study on deep
  learning and analytic-based approaches.
\newblock {\em arXiv preprint}, (arXiv:2310.18636), 2023.

\bibitem{wei2019dominant}
Zhun Wei, Dong Liu, and Xudong Chen.
\newblock Dominant-current deep learning scheme for electrical impedance
  tomography.
\newblock {\em IEEE Transactions on Biomedical Engineering}, 66(9):2546--2555,
  2019.

\bibitem{zablah2025feasibility}
Jenny~E Zablah, Catalina Vargas-Acevedo, Nilton da~BarbosaRosa~Jr, Omid~Rajabi
  Shishvan, Gary Saulnier, David Isaacson, Gareth~J Morgan, and Jennifer~L
  Mueller.
\newblock Feasibility of electric impedance tomography in the assessment of
  lung perfusion and ventilation in congenital pulmonary vein stenosis.
\newblock {\em Pediatric Cardiology}, pages 1--8, 2025.

\end{thebibliography}
